\documentclass[11pt]{book}
\usepackage{rotating, float, caption}
\usepackage{graphicx}
\usepackage{amsmath}
\usepackage{amsfonts}
\usepackage{amssymb}
\usepackage{dsfont,mathrsfs}
\usepackage{geometry}
\usepackage{setspace}
\usepackage{cancel}
\usepackage{bm}
\usepackage{color}
\usepackage{booktabs}
\usepackage{threeparttable, tablefootnote}
\usepackage{subfig}
\usepackage{makecell}
\usepackage{longtable}
\usepackage[titletoc]{appendix}
\usepackage{fancyhdr, blindtext}
\usepackage{multirow}
\usepackage{array}
\usepackage{siunitx}
\usepackage{cite}
\usepackage[version=4]{mhchem}
\usepackage[justification=centering]{caption}
\usepackage[colorlinks=true, linkcolor=blue, citecolor=blue, urlcolor=blue, breaklinks=true]{hyperref} 

\usepackage[utf8]{inputenc}

\usepackage{mathtools} 
\usepackage{booktabs}
\usepackage{epsfig}
\usepackage{placeins}
\usepackage{subfig}
\usepackage{graphicx}
\usepackage{color}
\usepackage[dvipsnames]{xcolor}
\usepackage{enumerate}   
\usepackage{listings}
\usepackage{tabularx}
\usepackage{booktabs}
\usepackage{multirow}
\usepackage{colortbl}

\usepackage{pdfpages}
\usepackage{multibib}
\usepackage{titlesec}
\usepackage[sectionbib]{chapterbib}
\usepackage{bibunits}
\usepackage[bottom]{footmisc} 
\usepackage{titletoc}%
\titlecontents{chapter}
[0pt]
{\bfseries}
{\chaptername\ \thecontentslabel:\quad}
{}
{\hfill\contentspage}

\titleformat{\paragraph}
{\normalfont\normalsize\bfseries}{\theparagraph}{1em}{}
\titlespacing*{\paragraph}
{0pt}{3.25ex plus 1ex minus .2ex}{1.5ex plus .2ex}

\renewcommand*{\thesubsubsection}{\thesubsection.\alph{subsubsection}}

\definecolor{dkgreen}{rgb}{0,0.6,0}
\definecolor{gray}{rgb}{0.5,0.5,0.5}
\definecolor{mauve}{rgb}{0.58,0,0.84}
\definecolor{GreenYellow}{HTML}{DFE674}
\definecolor{Yellow}{HTML}{FFF200}
\definecolor{YellowOrange}{HTML}{FAA21A}
\definecolor{antiquewhite}{rgb}{0.98, 0.92, 0.84}
\definecolor{apricot}{rgb}{0.98, 0.81, 0.69}
\definecolor{bubbles}{rgb}{0.91, 1.0, 1.0}
\definecolor{buff}{rgb}{0.94, 0.86, 0.51}
\usepackage{amssymb}
\newcommand{\ket}[1]{\left|#1\right\rangle}

\def \({\left(}
\def \){\right)}
\def \[{\left[}
\def \]{\right]}
\usepackage{adjustbox}
\usepackage{array}
\definecolor{nicered}{rgb}{.647,.129,.149}
\usepackage[Lenny]{fncychap}
\ChNameVar{\color{black}\fontsize{14}{16}\usefont{OT1}{phv}{m}{n}\selectfont}
\ChNumVar{\color{black}\fontsize{60}{62}\usefont{OT1}{ptm}{m}{n}\selectfont}
\ChTitleVar{\color{black}\Huge\bfseries}
 \ChRuleWidth{2pt}

\titleformat{\section}
{\color{black}\normalfont\Large\bfseries}{\thesection}{1em}{}
\titleformat{\subsection}
{\color{black}\normalfont\large\bfseries}{\thesubsection}{1em}{}
\titleformat{\subsubsection}
{\color{black}\normalfont\normalsize\bfseries}{\thesubsubsection}{1em}{}

\begin{document}

\begin{spacing}{1.4}

\frontmatter

\thispagestyle{empty}

\begin{center}

\bf{
\vspace*{1cm}
{\huge Development of a RPC system with detailed performance simulation} \\
\vspace{1.5cm}
{\large \emph{By}} \\
\vspace{0.2cm}
{\LARGE TANAY DEY} \\ 
\vspace{0.4cm}
{\large Enrollment No. PHYS01201604020} \\
\vspace{0.4cm}
{\Large Bhabha Atomic Research Centre, Mumbai}
\vspace{1.0cm}

{\large \emph{A thesis submitted to \\
The Board of Studies in Physical Sciences }\\
\vspace{0.3cm}
\emph {In partial fulfillment of requirements}\\
\vspace{0.3cm}
\emph{for the Degree of}\\}
\vspace{0.3cm}
{\Large DOCTOR OF PHILOSOPHY} \\
\vspace{0.3cm}
{\large \emph{of}}\\
\vspace{0.3cm}
{\Large HOMI BHABHA NATIONAL INSTITUTE}
\vspace{1.2cm}

\begin{center}
\includegraphics[scale=0.3]{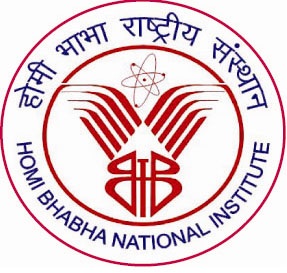} \\
\end{center}

\vspace{.5cm}
{\large December, 2023}

}	
\end{center}

\newpage
~~ 


\newpage
~~ 

\newpage

\begin{center}
\vspace*{2cm}
{\Huge \textbf{STATEMENT BY AUTHOR}}
\end{center}

\vspace{2cm}

This dissertation has been submitted in partial fulfillment of requirements for an advanced degree at 
Homi Bhabha National Institute (HBNI) and is deposited in the Library to be made available to borrowers 
under rules of the HBNI.

\vspace{1.5cm}

Brief quotations from this dissertation are allowable without special permission, provided that accurate 
acknowledgement of source is made. Requests for permission for extended quotation from or reproduction of 
this manuscript in whole or in part may be granted by the Competent Authority of HBNI when in his or her 
judgment the proposed use of the material is in the interests of scholarship. In all other instances, however, 
permission must be obtained from the author.

\vspace{4cm}

\begin{figure}[H]
	\hspace{10cm}\includegraphics[scale=0.5]{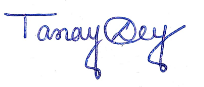}
\end{figure}
\hspace{10cm}  Tanay Dey

\newpage
~~ 

\newpage

\begin{center}
\vspace*{2cm}
{\Huge \textbf{DECLARATION}}
\end{center}

\vspace{2cm}

I, hereby declare that the investigation presented in the thesis has been carried out by me. 
The work is original and has not been submitted earlier as a whole or in part for a degree / diploma at 
this or any other Institution / University. 

\vspace{4cm}

\begin{figure}[H]
\hspace{10cm}\includegraphics[scale=0.5]{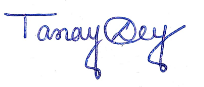}
\end{figure}
\hspace{10cm}  Tanay Dey

\newpage
~~ 

\newpage
\begin{center}
\vspace*{0cm}
{\LARGE \textbf{List of Publications arising from the thesis}}
\end{center}

\vspace{1cm}

\textbf{ {\LARGE  \underline{Journal}}}

\begin{enumerate}
	
	\item
	{\bf Numerical study of space
		charge electric field
		inside resistive plate
		chamber}
	\\{}\underline{Tanay Dey et al 2020 JINST 15 C11005}
	\\{}\textcolor{blue}{https://doi.org/10.1088/1748-0221/15/11/C11005.}

	\item
	{\bf Numerical study of
		effects of electrode
		parameters
		and image charge on the
		electric field
		configuration of RPCs}
	\\{}\underline{Tanay Dey et al 2022 JINST 17 P04015}
	\\{}\textcolor{blue}{https://doi.org/10.1088/1748-0221/17/04/P04015.}

	\item
	{\bf Parallelization of Garfield++ and neBEM to simulate space-charge effects in RPCs}
	\\{}\underline{ Tanay Dey, et al., Computer Physics Communications, Volume 294, 2024,}
	\\{} \textcolor{blue}{https://doi.org/10.1016/j.cpc.2023.108944}
	
	

\end{enumerate}

\vspace{0.2cm}

\textbf{ {\LARGE \underline{Symposium and Conference Proceedings}}}

\begin{enumerate}

	\item
	{\bf Numerical simulation of the response of single gap timing RPCs with the space charge effects and Garfield++}
	\\{}\underline{Tanay Dey}, et.al., 
	 \\{}\textcolor{blue}{ https://doi.org/10.48550/arXiv.2212.05345}

\end{enumerate}
\vspace{0.2cm}
\textbf{ {\LARGE \underline{Conference Presentations}}}
\begin{enumerate}
	\item Study of an unknown scintillator-like material, National Symposium on Particles, Detectors and instrumentation (NSPDI), October 4-7, 2017, Mumbai, TIFR.
	\item Numerical study of space
	charge electric field
	inside resistive plate
	chamber, XV Workshop
	on Resistive Plate Chambers and Related Detectors, \textbf{RPC 2020}, February 10-14, 2020, University of Rome, Tor Vergata, Italy.
	\item Parallelization of Garfield++ and neBEM to simulate space charge effects in RPCs, XVI Workshop on Resistive Plate Chambers and Related Detectors, \textbf{RPC 2022},
	September 26-30, 2022, 
	CERN.
\end{enumerate}

\vspace{2cm}
\textbf{ {\LARGE \underline{Others}}}
\begin{enumerate}
	\item { \bf A compact cosmic muon
	veto detector and
	possible use with the
	Iron Calorimeter
	detector for neutrinos}
	\\{} N. Panchal, S. Mohanraj, A. Kumar, \underline{T. Dey}, G. Majumder, R. Shinde, P. Verma, B. Satyanarayana and V.M. Datar, 2017 JINST 12 T11002,
\\{}\textcolor{blue}{DOI 10.1088/1748-0221/12/11/T11002.}
\\{}
\item {\bf Performance of a
prototype bakelite RPC
at GIF++ using self-
triggered electronics for
the CBM Experiment at
FAIR}
\\{} Mitali Mondal and \underline{Tanay Dey} and Subhasis Chattopadhyay and Jogender Saini and Zubayer Ahammed,
\\{} Nuclear Instruments and
Methods in Physics Research Section A:
Accelerators, Spectrometers, Detectors and
Associated Equipment,
Volume 1025,2022,166042, ISSN 0168-9002,
\\{}\textcolor{blue}{https://doi.org/10.1088/1748-0221/12/11/T11002.}
\end{enumerate}
\begin{figure}[H]
	\hspace{10cm}\includegraphics[scale=0.5]{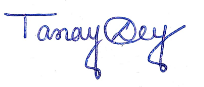}
\end{figure}
\hspace{10.5cm} Tanay Dey


\vspace*{10cm}
\begin{center}
\vspace*{4cm}
{\Huge \textbf{DEDICATION}}

\vspace{2cm}

  \bf{

\emph{
	Dedicated to my family,} \\
\emph{My Grand Mother - Mrs. Sadhana Rani Dey}\\
\emph{My Father - Mr. Dilip Dey} \\
\emph{My Mother - Mrs. Tara Dey} \\
\emph{My Wife - Dr. Tusita Sau (Dey)} \\
\emph{My Brother - Mr. Pranoy Dey}\\
\emph{\&}\\
\emph{My Physics Teacher - Shri Utpal Chattopadhyay}

}

\end{center}

\newpage
~~ 

\begin{center}
\newpage
{\Huge \textbf{ACKNOWLEDGEMENTS}}
\end{center}

\vspace{1cm}
I would like to take this opportunity to express my deepest gratitude and appreciation to all those who have contributed to the completion of this thesis.

First and foremost, I wish to convey my profound gratitude to my Ph.D. supervisor, Prof. Subhasis Chattopadhyay, for his invaluable guidance, support, and encouragement throughout my Ph.D. journey. His unwavering faith in my abilities has been a great motivation and inspiration to me.

I am equally grateful to Prof. Supratik Mukhopadhyay for his significant contribution to this research. His guidance, feedback, and expertise have been instrumental in shaping this thesis.

I would also like to express my heartfelt thanks to Dr. Rajesh Ganai for his mentorship and guidance as my senior in the lab. He has been a true pioneer and inspiration to me, showing me how to publish work and teaching me every aspect of research. Thank you, Rajesh da, for all of your invaluable help and support.

Furthermore, I am grateful to the members of my thesis committee, Dr. Chandana Bhattacharaya, Prof. Gobinda Majumdar, Dr. Zubayer Ahammed, and Dr. Tilak Ghosh, for their insightful comments, constructive criticism, and suggestions, which have helped me to refine my research and improve the quality of this thesis.

I am also grateful to my esteemed teachers in the PhD courses, namely Prof. G. Majumdar, Dr. B. Satyanarayana, Prof. S. Uma Shankar, Prof. V. Nanal, Prof. V. M. Datar, Prof. S. Banerjee, Dr. P.C Rout, and Prof. Sridhar K, for providing valuable insights into various fields. I would like to thank Mr. R. R. Shinde for his help and support during the course work. I am also indebted to Dr. Deepak Samuel and Dr. Sumanta Pal for helping me in the data analysis. Additionally, I am thankful to the INO Graduate Training Program for providing me with the opportunity to be a part of it. I would also like to express my gratitude to Tata Institute of Fundamental Research (TIFR), Variable Energy Cyclotron Center (VECC), and Saha Institute of Nuclear Physics (SINP) institutes for providing me with the research infrastructure.

I would like to express my deepest appreciation to Mr. Partha Bhaskar, Dr. Shuaib Ahmmed, Sh. Yuvraj, and Mr. Somnath Dalal for their generous help and support during my difficulties with electronics. I extend my thanks to Prof. Nayana Majumdar, Dr. Jhilam Sadukan, Dr. Anand Kumar Dubey, and Mr. Jogender Saini, Dr. Vikas Singhal for their valuable contributions in providing me with fruitful discussions whenever I encountered challenges. I would like to express my appreciation to all my lab colleagues, including Dr. Mitali Mondal, Ganesh da, Tushar da, Sukumar Da, Jayanta Ji, Khokon Da, Sanat Da, Tirthankar Da, Nabarun da, Bhaskar da, Santu da, Amit da, and Kamal da, for their valuable contributions during detector fabrication, electronics assembly, and testing. I am also grateful to my colleagues at VECC grid, particularly Abhishek, Prasun da, and Ashique, for their invaluable assistance whenever I faced computer or software-related issues. I am grateful to have friends and lunch partners like Rajesh da, Mitali, Shuvo, Abhishek, Sinjini, Gitesh, Shreyasi, Sridhar, Ekata di, and Souvik, arranged in the order of our first meetings at VECC. Their companionship and support have been sources of great strength and inspiration for me.

I would like to express my gratitude to my colleagues and friends at INO and TIFR, including Saikat, Anil, and Mohan, as well as my juniors Roni, Hariom, Mamta, Jim, and Honey, for their unwavering support. I would also like to extend my thanks to the esteemed seniors at INO, including Dr. Jaydeep, Dr. Neha Panchal, Dr. Abhijit Garai, Dr. Apoorva Bhatt, Dr. Md. Nizam, Dr. S. Pethuraj, and Dr. Suryanarayan Mondal, whose guidance and expertise have been invaluable to me.

I would also like to take this opportunity to acknowledge and express my gratitude to my childhood friends, Sujoy and Tirthankar, who have been essential parts of my life and growth. I consider myself fortunate to have had Shri Utpal Chattopadhyay as my first physics teacher in school. He not only taught me the fundamentals of physics, but also brought me insights and flavors of the subject, igniting my passion for it. I would like to thank my friends and former roommates, Dr. Tushar Kanti Bhowmik and Mr. Siddhartha Garain, as well as my permanent roommate, Dr. Tusita Sau, for holding me straight when a tornado came into my life during my PhD.

\par Lastly, I want to extend my thanks to my entire family, with a special thanks to my first teacher and mother, Mrs. Tara Dey, my Santa Claus and father, Mr. Dilip Dey, my backbone and puchku brother, Mr. Pranoy Dey, and finally, my sweetheart and friend who plays countless positive roles in my life, my wife, Mrs. Tusita Sau, for their unconditional love and unwavering support in every step of my life.
\par I am forever grateful to all of you who have played a significant role in my academic journey.

\begin{figure}[H]
	\hspace{10.2cm}\includegraphics[scale=0.4]{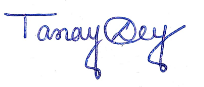}
\end{figure}
\hspace{10cm}  Tanay Dey

\newpage
~~ 

\newpage

\begin{center}
	\vspace*{2cm}
	{\Huge \textbf{Summary}}
\end{center}

\vspace{2cm}

Neutrinos, elementary particles of great interest to scientists, are challenging to detect due to their weak interactions with other matter., are challenging to detect due to their weak interactions with other matter, despite their widespread presence in the universe. Ongoing experiments like Super-Kamiokande, Daya Bay, and Ice Cube aim to unravel the mysteries surrounding neutrinos, focusing on properties such as mass and oscillations.

In India, the neutrino physics community is concentrating on studying neutrino oscillations in the atmospheric sector. To support this research, the country is developing the INO facility in the southern region, housing experiments like ICAL, NDBD, and DINO, each investigating distinct aspects of neutrinos. ICAL, a noteworthy instrument, is a large detector made up of active detectors using RPCs between magnetized iron plates. Its main goal is to decipher the genuine mass hierarchy of neutrinos by analyzing their interactions with Earth's matter. Additionally, ICAL contributes to precise measurements of $\theta_{23}$ and $\Delta m^2_{32}$, with potential implications for discovering new neutrino physics.

The active detector element of ICAL, known as the Resistive Plate Chamber (RPC), requires meticulous simulation techniques for its development. The RPC's functionality relies on signals generated during an Avalanche or streamer process, necessitating a comprehensive simulation model for avalanche or streamer development. While the study excludes the simulation of the streamer or avalanche-to-streamer transition, it emphasizes the working principles of RPCs and Gaseous detectors in a general overview.

The electric field inside an RPC, influencing electron-ion generation during an avalanche, undergoes continuous changes due to dynamic electric fields of electrons and ions, referred to as the space charge effect. A straight-line model, discussed without specific chapter references, is introduced for calculating the space charge field, highlighting its benefits such as fast estimation without numerical integrations. The polarization field due to dielectric electrodes is incorporated.

Analysis of the electric field inside an RPC, exploring various electrode parameters like permittivity, thickness, and gas gap, reveals observations related to saturation points and the impact of electrode thickness on field variation. The study introduces the class pAvalancheMC in the Garfield++ software, integrating straight-line and image charge models to consider the space charge effect during avalanche generation. Parallel computing using OpenMP enhances efficiency, resulting in significant speed-ups with and without space charge effects.

In the application realm, the development of an eight-RPC stack at VECC showcases the practical use of RPC detectors for determining the scattering angle of cosmic muons in high-density materials. A testbench is established for necessary evaluations, and a reconstruction algorithm is devised and tested using data from the IICHEP stack in Madurai, with plans for application in the VECC stack.

\newpage
~~ 

\end{spacing}

\newcommand{\changefont}{%
	\fontsize{9}{11}\selectfont
}
\fancyhead[LE]{\changefont \slshape \rightmark} 
\fancyhead[RE]{\changefont}
\fancyhead[RO]{\changefont \slshape \leftmark} 
\fancyhead[LO]{\changefont}
\fancyfoot[C]{\changefont \thepage} 
\pagestyle{fancy}

\begin{spacing}{2.0}

\tableofcontents

\listoffigures
\addcontentsline{toc}{chapter}{List of Figures}
\listoftables
\addcontentsline{toc}{chapter}{List of Tables}
\mainmatter

\chapter{Introduction To The Neutrino Physics}\label{ch1.Introduction}
\section{Introduction}

The India-based Neutrino Observatory (INO) is a major scientific research project will be located in Theni district of Tamil Nadu, India. It is designed to study the properties of neutrinos, which are fundamental particles that belong to the lepton family, which are extremely elusive subatomic particles that can travel through matter with little or no interaction. There are three types of neutrinos, with one associated with electrons and the other two associated with their heavier relatives, muons and taus. According to the standard model of particle physics, these charge-neutral particles have no mass. However, recent experiments suggest that they do have a finite, yet small mass that is yet to be determined. As they travel, neutrinos oscillate between their different types or flavors. One of the most critical unsolved issues in physics is determining the masses and mixing parameters of these elusive particles. 

\par Cosmic rays are energetic particles that come from the depths of outer space. They consist of small, electrically charged particles that collide with the Earth's atmosphere, breaking into even smaller fragments. These particles are influenced by magnetic fields present in our galaxy, causing them to move in directions that do not necessarily point back to their source. Hence, scientists rely on indirect methods to determine the source and propagation path of cosmic rays. The chemical makeup of cosmic rays provides valuable data, revealing insights into the chemical evolution of the universe. While studying the chemical composition of cosmic rays, scientists discovered elements that are incredibly rare in other stars, making these rays a direct sample of matter from beyond our solar system. Comparatively, rocks found on Earth have undergone significant changes over time, making it difficult to gain accurate data from them. Understanding cosmic rays is significant because they offer a glimpse into the chemical evolution of the universe, enabling researchers to learn more about the elements that make up our world \cite{cosmicrays}.

\par Based on energy of neutrinos, the India Based Neutrino Observatory (INO) focuses on atmospheric neutrinos, which are produced by the decay of secondary cosmic rays. In this chapter, we will first discuss the characteristics of primary and secondary cosmic rays. Then, we will explore the fundamentals of neutrino physics. After that, we will examine the experimental design of the Iron Calorimeter (ICAL) and how it contributes to achieving the objectives of the INO experiment. Finally, we will discuss the scope of the thesis.

\section{History of cosmic ray discovery}
The history of cosmic rays includes the study of unknown and seemingly inconsequential effects from the sky \cite{dorman1978history,dorman1978history2,dorman1978nature,dorman2013history} . The search for cosmic rays began (arround 1900) when physicists became interested in finding the source of constant and weak ionization of ambient air. They realized that the cause of the ionization was not the spontaneous disintegration of air molecules due to thermal excitations, but rather it may be induced by unknown radiations. In 1900, Julius Elster and Hans Geitel attempted to find and measure the sources of air ionization and concluded that the basic sources were particles from radioactive emissions and radioactive substances in the Earth's crust \cite{elster1900weitere,geitel1900ueber}. Charles Thomson Rees Wilson also independently discovered the unknown source of ions in air at the Cavendish laboratory in 1900\cite{wilson1900leakage,malcolm2014c,wilson1901ionisation}, using an electroscope to measure the conductivity of dust-free air in a closed vessel. He found that the air inside the vessel was always ionized and that about 20 ions were formed per second in 1 cubic meter of air. He also observed that the rate of leakage of positive and negative charges was the same and proportional to the pressure in the vessel, independent of whether the measurement was done in light or darkness.
\par In 1910, D. Pacini \cite{pacini1909,pacini1910} conducted simultaneous measurements of air ionizations at two different locations, one on the ground and one a few kilometers away from the coast. The results were not distinguishable based on the hypothesis that the ``origin of the penetrating radiations is in the soil." In June 1911, he made arrangements to measure the rate of radiation under 3 meters below the water of the sea and 300 meters from the shore of the Naval Academy of Livorno. He found that the ionization rate under the water was 20\% lower than that at the surface of the sea. His findings were published in the scientific journal Nuovo Cimento in 1912 \cite{Pacini1912}. 

\par Alfred Gockel was a Swiss scientist who went on three balloon flights between 1909 and 1911. He went up to a height of 4500 meters in these flights. He found that there was less reduction of ionization in the air than what was expected if the cause of ionization was radiation from the Earth's surface. However, it is important to note that the pressure in the device that he used fell during the flights. This could be the reason why there was less ionization. The reduction in pressure led to a decrease in the number of atoms in the air, rather than a decrease in ionizing radiation. He also suggested that the presence of cosmic radiation could be responsible for the observed ionization. However, he could not draw any definitive conclusions based on his observations.

\par
\begin{figure}
\centering\includegraphics[scale=0.4]{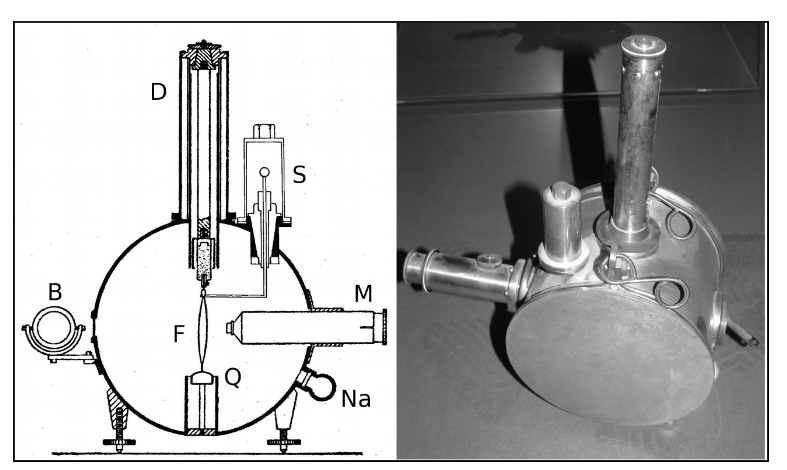}
\caption[]{Wulf electrometer \cite{wolf-spectrometer} \label{fig:electrometer}}
\end{figure}

 In 1911, Victor F. Hess conducted an experiment to measure the absorption factor of $\gamma$-radiation in air\cite{hess1911absorption}. He used a modified electroscope designed by Fr. Theodor Wulf in 1909 \cite{wulf1909atmosphare,wulf1910beobachtungen}, which was an improved version of the well-known Elster-Geitel design. The unique feature of this device was that the air pressure inside the vessel remained constant, regardless of the atmospheric pressure. It consisted of two thin metal wires held taut by a thin quartz thread, as shown in Figure. \ref{fig:electrometer}.
  As the ionization of the gas in the chamber increased, the wires discharged and the distance between the wires changed, which was measured by a traveling microscope. Thus, the speed of the discharge of the electrometer was measured. As a result of his experiment, he observed that initially the ionization level decreases with height, but after a certain height, it increases again. The variation of ionization rate with height from Hess' experiment is tabulated in Table \ref{tab:res_hes_experiment}.
\begin{table}
	\centering\begin{tabular}{|c|c|}
		\hline 
		$\begin{array}{c} \textrm{Average }\\ \textrm{height}\\ \textrm{from  the}\\\textrm{ground,m} \end{array}$ & $\begin{array}{c} \textrm{Observable} \\ \textrm{ionisation rate,}\\\textrm{ion.cm}^-3\,\textrm{s}^-1 \end{array}$
		\tabularnewline
		\hline 
		\hline 
		0 & $\begin{array}{c}
		16.3 (18)^*
		\end{array}$
		\tabularnewline
		\hline 
		$<$200 & $\begin{array}{c}
		15.4 (13)
		\end{array}$ 
		\tabularnewline
		 		\hline 
	200-500 & $\begin{array}{c}
	15.5 (6)
	\end{array}$
	\tabularnewline
			\hline 
	500-1000 & $\begin{array}{c}
	15.6(3)
	\end{array}$
	\tabularnewline
			\hline 
	1000-2000 & $\begin{array}{c}
	15.9 (7)
	\end{array}$
	\tabularnewline
			\hline 
	2000-3000 & $\begin{array}{c}
	17.3(1)
	\end{array}$
	\tabularnewline
			\hline 
	3000-4000 & $\begin{array}{c}
	19.8 (1)
	\end{array}$
	\tabularnewline	
		\hline 
	4000-5200 & $\begin{array}{c}
	34.4 (2)
\end{array}$
\tabularnewline	 
		\hline 
	\end{tabular}
	
		\caption[]{Results from Hess experiment \cite{dorman2014cosmic}. \\
			      *The numbers in brackets mean the
			      number of observations
		         \label{tab:res_hes_experiment}}
	
\end{table} 
From the results, Hess concluded that the influence of $\gamma$ radiation is significantly reduced above a height of 500 m and increases again above the height of 1000 m. Therefore, above this height, radiation is coming from outside of the Earth's atmosphere. To eliminate the possibility that the radiation is coming from UV rays from the sun, he conducted the same experiment during a partial solar eclipse on April 7, 1912 \cite{hess1912beobachtungen}, and found the same results as before. In 1925, the American experimentalist Millikan named this radiation "cosmic rays"\cite{milikan,millikan1930history}. In 1936, Hess received the Nobel Prize for the discovery of cosmic rays.

\par Initially, it was believed that the cosmic rays were only streams of high-energy photons. However, Bothe and Kolhörster first showed that the cosmic tracks in a cloud chamber were curved by a magnetic field. The Dutchman Jacob Clay observed the ``latitude effect" \cite{Clay1932,Clay1933}, which is that the cosmic ray intensity depends on the (geomagnetic) latitude. This proves that cosmic rays are not photons but charged particles deflected by the Earth's magnetic field.

\section{Nature of Primary and Secondary cosmic rays}
\subsection{Primary cosmic rays}
Primary cosmic rays are stable charged particles with a broad range of energies whose source can be somewhere in the universe. The present studies predict that in proportion to cosmic rays, 95\% are protons or hydrogen nuclei, 4\% are helium nuclei, and 1\% are stellar-synthesized elements up to iron \cite{dorman2014cosmic,gaisser_engel_resconi_2016,GRIEDER2001}.

\par Besides the charged particles, there are several X-ray and gamma-ray point sources overlaid on the spectrum of a modest diffuse flow of gamma rays, which displays a general power law behavior with some structure. This field of study is called gamma ray astronomy, where the energy range of study is from less than 0.5 MeV to multi-PeV.

\par
\begin{figure}
\centering\includegraphics[scale=0.15]{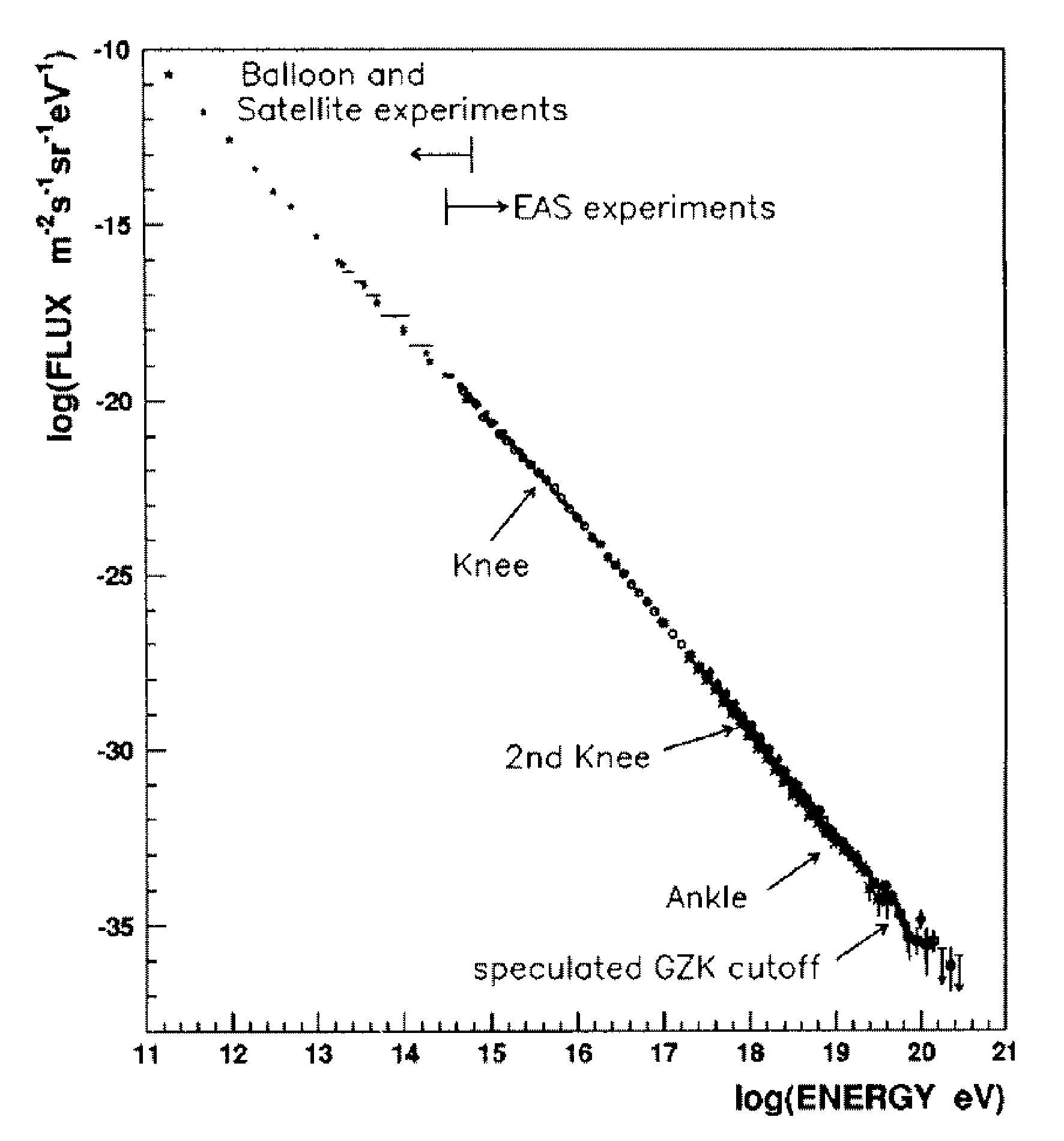}
\caption[]{Energy spectrum of primary cosmic rays\cite{HORANDEL2003193}\label{fig:energy_prime_cosmic}}
\end{figure}

 In 1963, a small flux of energetic electrons was also identified. The primary hadronic component's collisions with the interstellar medium and its interactions with the background radiation field may be the cause of this electron flux (decay chain of pions, e.g $\pi\rightarrow \mu+e$). Additionally, there must be a massive flux of extraterrestrial origin neutrinos and antineutrinos of all varieties. A power law can be used to characterize the differential energy spectra of all cosmic radiation components in space with kinetic energy greater than a few GeV, which is given below\cite{GRIEDER2001}:
\begin{equation}
I(E)\propto E^{-\gamma},
\end{equation}
where I is the intesity, E is the kinetic energy per nucleon and $\gamma$ is the spectral index. The value of the $\gamma$ is arround 2.7 for the hadronic component till the energy range $3\times10^6$ GeV (the knee point see Figure \ref{fig:energy_prime_cosmic}), discovered by Kulikov and
Khristiansen (1958). At the energy of $\approxeq 10^9$ GeV the value of $\gamma$ reaches 3.14 and at ultra high energies (beyond the ankel  $\approxeq 10^{10}$ GeV) the curve flatten and again $\gamma$ reaches the value 2.7.
\subsection{Secondary cosmic rays}
\begin{figure}
	\centering\includegraphics[scale=0.55]{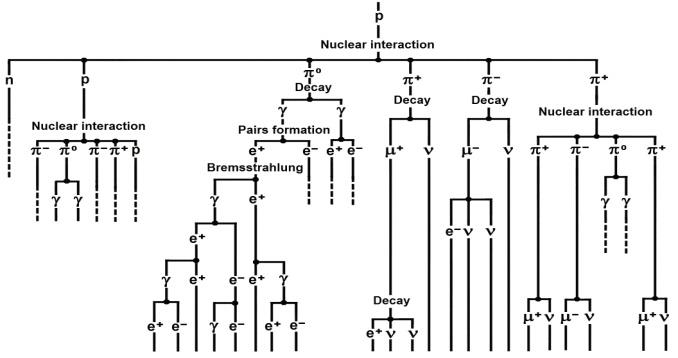}
	\caption[]{Example of cosmic ray shower \cite{dorman2014cosmic} \label{fig:cosmic-shower}}
\end{figure}

Primary cosmic rays with high energy were striking the upper atmosphere and ejecting high energy particles from nuclei, which in turn ejected other particles from other nuclei. This entire process results in a shower of particles, which is generally called a cosmic ray shower. A typical picture of a cosmic ray shower is shown in Figure \ref{fig:cosmic-shower}.
The maximum of the primary heavy nuclei is fragmented at a higher altitude due to the much larger interaction cross section $\sigma_n$ [cm$^2$].. The nucleon interaction length $\lambda_n$ [g/cm$^2$] in air can be written as \cite{GRIEDER2001}:
\begin{equation}
\lambda_n=\frac{A\,m_p}{\sigma^{air}_n},
\end{equation} 
where A is the mass number of air (target nuclei) ,$m_p$ is the mass of proton, $\sigma_n$ is interaction crossection.
\par High energy strong interactions and electromagnetic processes are involved in the production of secondary particles. Charged pions, mesons, hyperons, and nucleon-antinucleon pairs created due to strong interactions of energetic primaries with atmospheric target nuclei continue to propagate and contribute to the flux of hadrons in the atmosphere. Among all the secondaries, pions ($\pi^+,\pi^0,\pi^-$) are the most abundant. The secondary hadrons again start new hadronic interactions, producing new secondaries. This process continues and produces a hadronic shower. The unstable particles like pions and kaons can also decay before they interact with air nuclei. The mean life, energy of the particles, and the density of the propagation medium determine the type of processes, such as interaction with air nuclei or self-decay. Most of the secondary particles generated from hadronic interactions can be unstable and can decay during their flight. Therefore, decay probabilities must be known and accounted for in the calculation of the final particle fluxes and energy spectra. The decay probability $W$ for particle incidence at an angle $\theta$ from the vertical can be written as follows\cite{GRIEDER2001}:
\begin{equation}
W\simeq \frac{m_0\,X\,sec(\theta)}{\rho\,\tau_0\,p}
\end{equation} 
where,\\
$m_0$ = Rest mass of the particle (GeV/c$^{2}$).\\
X = Thickness traversed in the medium (g/cm$^2$).\\
$\tau_0$ = Mean life of particle at rest [s].\\
$p$ = Momentum of the particle [GeV/c]. \\
$\rho$= Density of the medium [g/cm$^3$]. \\
$\theta$ = Zenith angle.\\
\subsection{Decay modes of secondaries}\label{sec:decayModes}
  \subsubsection{Muons}
 Muons ($\mu^\pm$) are produced from the decay of pions ($\pi^{\pm}$) \cite{perkins_2000,Griffiths2008}. The mean lifetime of muons in its rest frame is 2.2 microseconds. The majority of muons are believed to be produced at a height of around 15 km. Muons can reach the Earth's surface through the air medium by losing its energy at a rate of 2 MeV per $g/cm^2$. The average surface energy of muons at the sea level is 4 GeV. With an energy of 4 GeV, the time dilation factor ($\Gamma$) and velocity are approximately 38.8 and 0.99c, respectively. Due to time dilation, the mean lifetime of muons in the Earth's frame becomes approximately 85 microseconds. Therefore, the range of a 4 GeV muon is 25 km, which is why muons are able to reach the Earth's surface from the point of generation at a height of 15 km. However, most of the muons can decay to electrons and neutrinos during their flight. The main decay modes of muons ($\mu^\pm$) are \cite{Griffiths2008}:
 \begin{align} \label{eqn:muondecay}
 \mu^{\pm}&\rightarrow e^{\pm}+\nu_{e}(\bar{\nu}_{e})+\bar{\nu}_{\mu}(\nu_{\mu}).
 \end{align}
 \subsubsection{Mesons}
The mean life time of Kaon ($K^{\pm}$) is $\approx1.24\times10^{-8}\,s$. Main decay modes of Kaon ($K^{\pm}$) are \cite{perkins_2000}:
\begin{align}
K^{\pm}&\rightarrow\mu^{\pm}+\nu_{\mu}(\bar{\nu}_{\mu}) ,\\
K^{\pm}&\rightarrow\pi^{\pm}+\pi^0 ,\\
K^{\pm}&\rightarrow\pi^{\pm}+\pi^{\pm}+\pi^{\mp} ,\\
K^{\pm}&\rightarrow\pi^{\pm}+\pi^{0}+\pi^{0} ,\\
K^{\pm}&\rightarrow\pi^{0}+e^{\pm}+\nu_e (\bar{\nu}_e) ,\\
K^{\pm}&\rightarrow\pi^{0}+\mu^{\pm}+\nu_\mu (\bar{\nu}_\mu).
\end{align}
The Decay of neutral Kaon ($K^0$) is complicated than others. The neutral kaon can be represent as short lived Kaon ($K_S^0$) and long lived Kaon ($K_L^0$). The mean life of $K_S^0$ and $K_L^0$ are $9\times10^{-11}\,s$ and  $5\times10^{-8}\,s$ respectively. The main decay modes are as follows\cite{perkins_2000,Griffiths2008}: 

\begin{align}
K^{0}_S&\rightarrow\pi^{+}+\pi^{-}, \\
K^0_S&\rightarrow\pi^{0}+\pi^{0}, \\
K^0_L&\rightarrow\pi^{+}+\pi^{-}+\pi^{0}, \\
K^0_L&\rightarrow\pi^{0}+\pi^{0}+\pi^{0}. 
\end{align} 
The mean life time of charged an neutral pions at rest are $\approx2.6\times 10^{-8}$ seconds and $8.4\times10^{-17}$ seconds respectively. The main decay modes of charged pions ($\pi^{\pm}$) and neutral pions $\pi^0$ are as follows:
\begin{align} \label{eqn:piondecay}
\pi^{\pm}&\rightarrow\mu^{\pm}+\nu_{\mu}(\bar{\nu}_{\mu}),\\
\pi^0&\rightarrow\gamma+\gamma 
\end{align} 
   \section{Story of neutrinos}
James Chadwick discovered the continuous $\beta$-spectrum in 1914, and it was confirmed by Charles Drummond Ellis and William Wooster in 1927. To explain this, Wolfgang Pauli proposed a new neutral particle called the neutron\cite{Pauli2000}. In 1932, after its discovery by Chadwick, Enrico Fermi named it the neutrino \cite{Fermi1933,Perrin1933}. Initially, the mass of the neutrino was considered to be zero. In 1934, Hans Bethe and Rudolf Peierls predicted the strength of neutrino interactions in their paper "On the Stopping of Fast Particles and on the Creation of Positive Electrons". However, they also mentioned in their paper that the predicted interactions might be too weak to be observed\cite{Bethe1934}. However, in the early 1950s, physicist Bruno Pontecorvo, Frederick Reines and Clyde L. Cowan independently proposed the use of inverse beta decay to detect antineutrinos \cite{Cowan1957,Reines1996}. Inverse beta decay is a process in which an antineutrino interacts with a proton, producing a neutron and a positron. Reines and Cowan successfully used this method to detect antineutrinos in 1956, through the detection of the positrons produced in the inverse beta decay process. This was the first experimental evidence of neutrino interactions. In 1962, the discovery of the muon neutrino ($\nu_\mu$) was announced by a team of physicists led by Leon M. Lederman, Melvin Schwartz, and Jack Steinberger at Brookhaven National Laboratory in Long Island, New York \cite{Danby1962}. They used a beam of high-energy protons to produce pions, which decayed into muons and muon neutrinos. They detected the muon neutrinos by studying the interactions of the neutrinos with a target made of iron and scintillating material. This was an important discovery as it confirmed the existence of more than one type of neutrino, known as flavors, and it also established the field of neutrino physics. The tau neutrino ($\nu_\tau$) was first detected by the DONUT experiment at Fermilab in July 2000. The discovery of the tau neutrino, the third flavor of neutrino, was a great achievement, as it confirmed the existence of the three flavors of neutrinos: electron neutrino, muon neutrino, and tau neutrino, which is also known as the three-neutrino paradigm.
 \section{Sources of neutrinos}
\begin{figure}
  	\centering\includegraphics[scale=0.35]{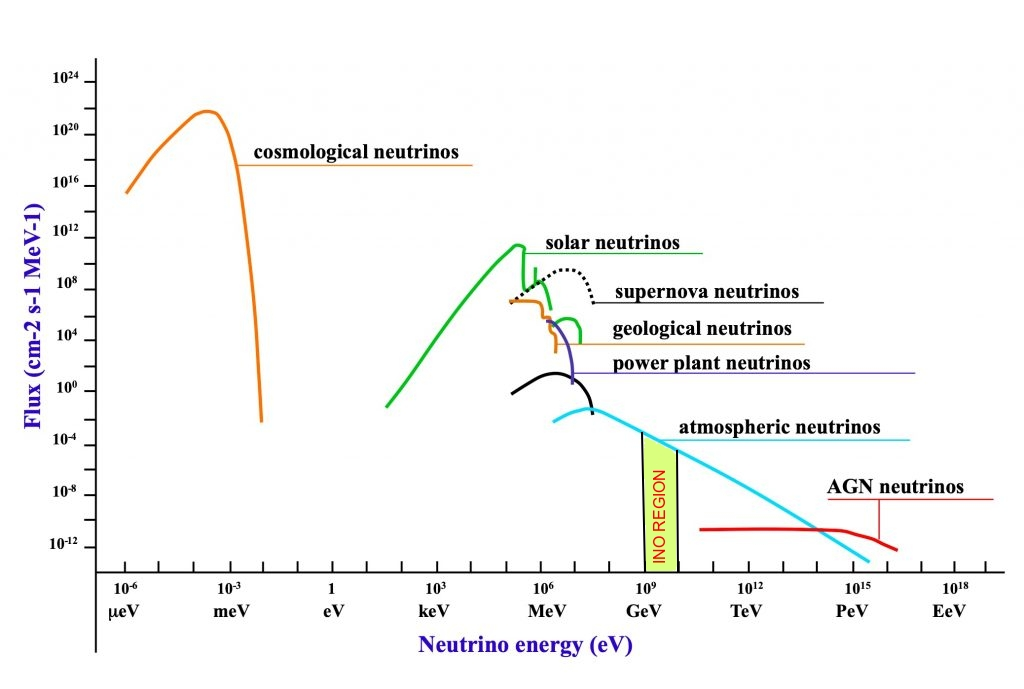}
  	\caption[]{Flux of neutrinos from different sources\cite{Spiering2012}\label{fig:neuEnAll}}
  \end{figure}

 Neutrinos can be generated from several processes, which can be broadly classified as natural and man-made. The energy, flux, and process of generation of neutrinos depend on their corresponding sources. Due to their low interaction cross-section, neutrinos are able to convey information about their sources and are thus referred to as messenger particles. This section discusses important sources of neutrinos. The energy spectrum of neutrinos from different sources is shown in Figure \ref{fig:neuEnAll}.
  \subsection{Solar neutrinos}
\begin{figure}
	\centering\includegraphics[scale=0.4]{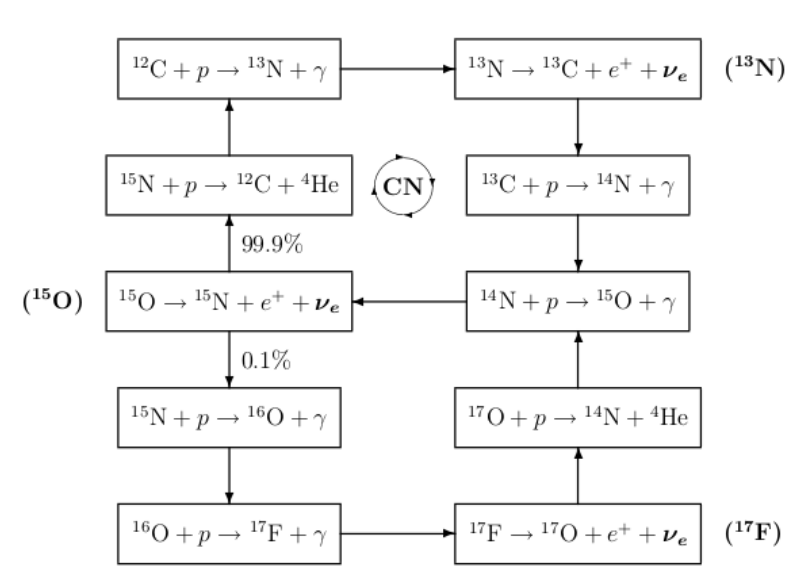}
	\caption[]{CNO-cycle reactions \cite{GuintiKim}\label{fig:cno}}
\end{figure}

\begin{figure}
	\centering\includegraphics[scale=0.4]{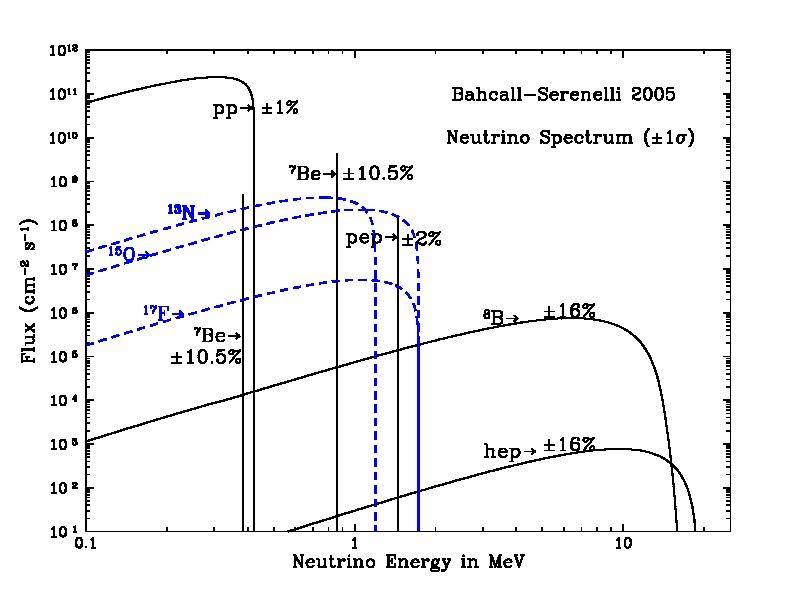}
	\caption[]{Flux of solar neutrinos\cite{Bahcall2005}\label{fig:solarnueflux}}
\end{figure}

 The structure and evolution of the Sun can be described by the Standard Solar Model (SSM) \cite{Bahcall2001}, which predicts the reactions inside the Sun, energy, and high-intensity electron neutrino flux due to nuclear fusion inside the Sun. Neutrinos coming from the Sun are produced from the proton-proton chain reactions and the
 \begin{figure}
 	\centering\includegraphics[scale=0.4]{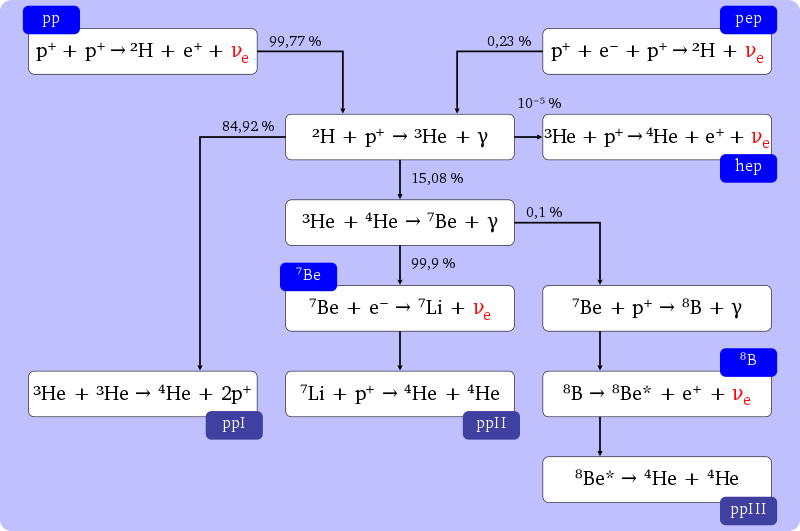}
 	\caption[]{pp-chain reactions \cite{ppchainwiki}\label{fig:pp}}
 \end{figure}
  Carbon-Nitrogen-Oxygen (CNO) cycle inside the Sun. The majority of neutrinos are generated from the pp chain reaction. The reactions corresponding to the pp-cycle and CNO cycle can be found in Figures \ref{fig:pp} and \ref{fig:cno} respectively. Figure \ref{fig:solarnueflux} shows the solar neutrino energy spectrum along with the uncertainties.
\par The Homestake experiment \cite{Lande1992,Cleveland1998}, which was the first to detect solar neutrinos in 1970, observed that the rate of neutrino interactions in the detector was less than half of what was expected based on the solar model. This discrepancy persisted despite the fact that experiments using different detection techniques and at different locations confirmed the original result.

Later experiments such as the Kamiokande, Super-Kamiokande, SAGE, GALLEX, and GNO experiments \cite{Abdurashitov1999,Hampel1999,Hosaka2006,Cravens2008}, all observed similar deficits in the number of solar neutrinos. 

\subsection{Supernovae neutrinos}
Supernova neutrinos are emitted from the collapse of massive stars. The first observation of neutrinos from a supernova occurred in 1987. During the collapse, protons and electrons that make up the core are packed together so tightly that they combine to form neutrons. This process, known as electron capture, produces large numbers of neutrinos. As neutrinos are weakly interacting particles, they can escape the star almost immediately, allowing scientists to study the earliest moments of the supernova explosion. This makes neutrinos a powerful tool for studying supernovae. The intense heat of the proto-neutron star core, which reaches around 40 MeV, results in the creation of neutrinos of all flavors via the process of electron-positron pair annihilation. The neutrino burst from SN1987A is the only supernova detection made through neutrinos, leading to notable advancements in the field of neutrino astronomy \cite{Hirata1987,Bionta1987,Alekseev1987}.
\subsection{Reactor neutrinos}
Reactor neutrinos, primarily in the form of electron anti-neutrinos, are generated in large quantities by nuclear reactors through the beta-decay of fission products. The number of these neutrinos generated is directly proportional to the power output of the reactor. Detection of reactor electron anti-neutrinos is achieved by the inverse beta decay process, which was the first method that allowed detection of this type of neutrino \cite{Cowan1957,Reines1996}. The energy range of neutrinos from reactors is from 0.1 MeV to around 10 MeV.
\subsection{Geo-neutrinos}
Geo-neutrinos are neutrinos that are produced by the natural radioactive decay of certain elements in the Earth's crust and mantle. These neutrinos are created when the nuclei of elements such as uranium and thorium, as well as potassium, undergo beta decay. The resulting neutrinos have very low energies, on the order of a few tens of kilo-electron volts \cite{Bellini2010,Bellini2013,Araki2005}. These emissions are the result of the following decay sequences:
\begin{align}
\prescript{238}{}{U}&\rightarrow \prescript{206}{}{Pb}+8\alpha+8e^-+6\bar{\nu_e}+51.7MeV\\
\prescript{232}{}{Th}&\rightarrow \prescript{208}{}{Pb}+6\alpha+4e^-+4\bar{\nu_e}+42.7MeV\\
\prescript{40}{}{K}&\rightarrow\prescript{40}{}{Ca}+e^-+\bar{\nu_e}+1.31MeV
\end{align}
The flux of  produced neutrinos is about $5\times10^{10}$/m$^2$/s.

\subsection{Atmospheric neutrinos}
It is discussed in section \ref{sec:decayModes} that neutrinos can be generated in the atmosphere due to the reaction of hadrons and leptons with the air nuclei. The maximum contribution of the atmospheric neutrino flux comes from the decay of pions into muons. At sufficiently high energies, kaons also contribute to the muon flux. Then, muons based on their charges can decay into an electron (positron), an electron neutrino (electron anti-neutrino) and a muon anti-neutrino (muon neutrino) before reaching the Earth's surface, as shown in equation \ref{eqn:muondecay}. From equation \ref{eqn:muondecay} and \ref{eqn:piondecay} the ratio (R) of total muon neutrino flux and electron neutrino flux can be calculated as follows \cite{ICALpotential}:
\begin{equation}
R=\frac{\phi(\nu_\mu)+\phi(\bar{\nu}_\mu)}{\phi(\nu_e)+\phi(\bar{\nu}_e)}\approx 2,
\end{equation} 
where the neutrino flux is represent as $\phi$. . The ratio is not entirely accurate as high energy muons can reach the surface of the Earth without undergoing decay into electrons..
\par
\begin{figure}[H]
	\centering\includegraphics[scale=0.5]{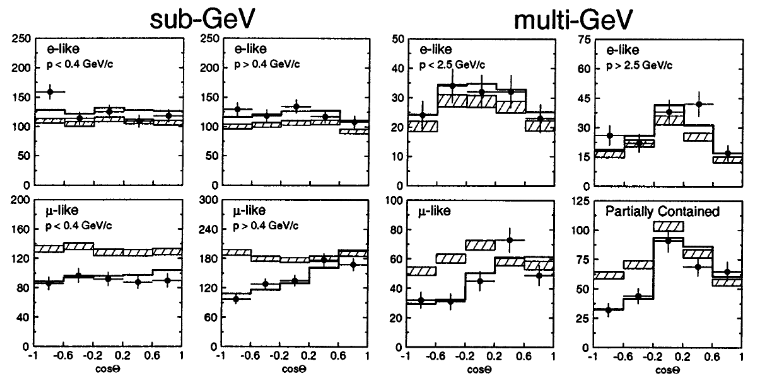}
	\caption[]{Distribution of Zenith angle for $\mu$-like and e-like events for different energy ranges (sub-GeV and multi-GeV). The condition for upward and downward going particles are cos$(\Theta)<0$ and  cos$(\Theta)>0$. \cite{atmosneutrino}.
		 \label{fig:anomaly}}
\end{figure}

 A zenith angle deficit has been observed in the data of the neutrino experiment Super-Kamiokande (see Figure \ref{fig:anomaly}) \cite{atmosneutrino}. The flux of up-going muon neutrinos showed a deficit, while the downward going muon neutrino flux did not. This is because upgoing neutrinos travel through longer path lengths through the earth (from a few hundred kilometers to around 13,000 kilometers) from the point of production to the detector. These results suggest that the deficit is dependent on the path length, a phenomenon known as the atmospheric neutrino anomaly. This was the first experimental evidence of neutrino oscillation.
\section{Neutrino Oscillation}
Neutrino oscillation is the phenomenon where neutrinos of a specific flavor (such as electron neutrinos) can transform into another flavor (such as muon neutrinos) as they propagate through space. In the late 1950s, physicist Bruno Pontecorvo proposed the concept of neutrino mixing, which is now considered a key aspect of neutrino oscillation \cite{Pontecorvo1957,Pontecorvo1958}. The theory of neutrino oscillation is based on the idea that neutrinos have mass, which is a property not predicted by the Standard Model\cite{standardmodel} of particle physics. In the Standard Model, neutrinos are assumed to be massless, and therefore only interact via the weak force. However, experimental evidence for neutrino oscillation suggests that neutrinos do have mass \cite{Ahmad2002}. The theory of neutrino oscillation explains this discrepancy by proposing that there are actually three different types of neutrinos, each with its own mass. These three types are known as ``neutrino mass eigenstates," and they correspond to the three flavors of neutrinos: electron, muon, and tau. The theory also predict that neutrinos are not produced or detected in their mass eigenstates, but rather in their ``flavor eigenstates". This means that when a neutrino is produced, it is associated with a specific flavor (such as electron), but its true state is actually a superposition of the three mass eigenstates. As the neutrino propagates through space, the mass eigenstates ``oscillate" to form different flavor eigenstates. The probability of a neutrino in a specific flavor eigenstate oscillating into another flavor eigenstate is determined by the neutrino's mass and the distance it has traveled. The mass difference between the three mass eigenstates and the distance traveled in space are among the parameters used to determine the oscillation probability. The oscillation probability is also affected by the neutrino's mixing angles. These angles determine the ``mixing" of the mass eigenstates with the flavor eigenstates, and they are determined by the relative strengths of the weak force interactions that govern the production and detection of neutrinos.
\subsection{Neutrino oscillation in vacuum}

 A specific flavor state $\nu_\alpha$ is represented as a combination of mass eigenstates $\nu_i$ , with masses $m_i$, that are orthonormal and linear in nature. Therefore the realation between flavour and mass eigen state can be written as follows\cite{GuintiKim,neutrinophys-rnp}:
\begin{equation}\label{eqn:state_eqn}
|\nu_\alpha\rangle = \sum_{i=1}^{3} U_{\alpha i}^* |\nu_i\rangle \,\,\,\,\,\, (\alpha = e, \mu,\tau ) \,\,\,
and \,\,\,
(i = 1, 2, 3)
\end{equation}
where, the unitary matrix, $U$, commonly referred to as the Pontecorvo-Maki-Nakagawa-Sakata (PMNS) matrix or the lepton mixing matrix. The PMNS matrix can be written as \cite{GuintiKim,AtharSingh2021,Zuber2021,AtharSinghnote},
\begin{align}\label{eqn:pmns1}
U_{PMNS}&= U_{23}(\theta_{23},0)U_{13}(\theta_{13},\delta)U_{12}(\theta_{12},0) \\\nonumber
&= \begin{pmatrix}
1 & 0 & 0 \\
0 & c_{23} & s_{23} \\
0 & -s_{23} & c_{23}
\end{pmatrix}\times \begin{pmatrix}
c_{13} & 0 & s_{13}e^{-i\delta} \\
0 & 1 & 0 \\
-s_{13}e^{i\delta} & 0 & c_{13}
\end{pmatrix} \times \begin{pmatrix}
c_{12} & s_{12} & 0 \\
-s_{12} & c_{12} & 0 \\
0 & 0 & 1
\end{pmatrix} 
\end{align}
Where c$_{ij}$ =cos $\theta_{ij}$ and s$_{ij}$ =sin $\theta_{ij}$ are the mixing angles, $\delta$ is the CP-violating phase. The three matrices that make up the PMNS matrix, $U_{12}$, $U_{13}$ and $U_{23}$, represent rotations in the 12, 13, and 23 flavor spaces, respectively. The composit form of equation \ref{eqn:pmns1} can be written as:
\begin{align}
U_{PMNS}=\begin{pmatrix}
c_{12}c_{13} & s_{12}c_{13} & s_{13}e^{-i\delta} \\
-s_{12}c_{23}-c_{12}s_{23}s_{13}e^{i\delta} & c_{12}c_{23}-s_{12}s_{23}s_{13}e^{i\delta} & s_{23}c_{13} \\
s_{12}s_{23}-c_{12}c_{23}s_{13}e^{i\delta} & -c_{12}s_{23}-s_{12}c_{23}s_{13}e^{i\delta} & c_{23}c_{13}
\end{pmatrix}
\end{align}

 The sector associated with the mass-squared difference labeled "atmospheric" ($\Delta m^2_{atm}$) corresponds to the mixing of the second and third generation of neutrinos, which is represented by the $U_{23}$ matrix. The sector associated with the mass-squared difference labeled ``solar" ($\Delta m^2_{\odot}$) corresponds to the mixing of the first and second generation of neutrinos, which is represented by the $U_{12}$ matrix. The $U_{13}$ matrix, which describes the mixing of the first and third generation of neutrinos, is responsible for electron neutrino flavor transitions at the atmospheric scale, which are yet to be observed. The CP-violating phase, represented by $\delta$, allows for the possibility of violation of the symmetry between matter and antimatter in the appearance mode of neutrino oscillation. 

\par The probability of a neutrino flavor oscillating from flavor $\alpha$ to flavor $\beta$ in three-flavor mixing can be written as:
\begin{equation}\label{eqn:osc1}
P(\nu_{\alpha} \rightarrow \nu_{\beta}) = \sum_{i=1}^{3} \sum_{j=1}^{3} U_{\alpha i}^* U_{\beta i} U_{\alpha j} U_{\beta j}^* exp\(-2\pi i\frac{L}{L_{ji}^{osc}}\)
\end{equation}
Where U is the Pontecorvo-Maki-Nakagawa-Sakata (PMNS) mixing matrix, L is the distance between the neutrino source and detector, and $L_{ji}^{osc}$ is the oscillation length or the distance at which a complete oscillation occurs. $L_{ji}^{osc}$ can be written in terms of the mass-squared difference between mass eigenstates i and j (i.e. $\Delta m_{ji}^2$), and the neutrino energy E, as follows \cite{GuintiKim}:
\begin{equation}
L_{ji}^{osc}=\frac{4\pi E}{\Delta m_{ij}^2}	 
\end{equation}
 Alternatively the equation \ref{eqn:osc1} can be written as:
\begin{align}\label{eqn:osc2}
P(\nu_{\alpha} \rightarrow \nu_{\beta}) = \delta_{\alpha \beta} - 4 \sum_{i>j}^{3} \text{Re}(U_{\alpha i}^* U_{\beta i} U_{\alpha j} U_{\beta j}^*) \sin^2 \left( \frac{\Delta m_{ij}^2 L}{4E} \right) +\\\nonumber
 2 \sum_{i>j}^{3} \text{Im}(U_{\alpha i}^* U_{\beta i} U_{\alpha j} U_{\beta j}^*) \sin \left( \frac{\Delta m_{ij}^2 L}{2E} \right)
\end{align}
For $\alpha=\beta$ the probability $P(\nu_{\alpha} \rightarrow \nu_{\alpha})$ is called survival probability and for  $\alpha \neq \beta$ $P(\nu_{\alpha} \rightarrow \nu_{\beta})$ is called transition probability.  
\subsection{Neutrino interactions with matter}\label{sec:numatter}
\begin{figure}
	\centering\includegraphics[scale=0.4]{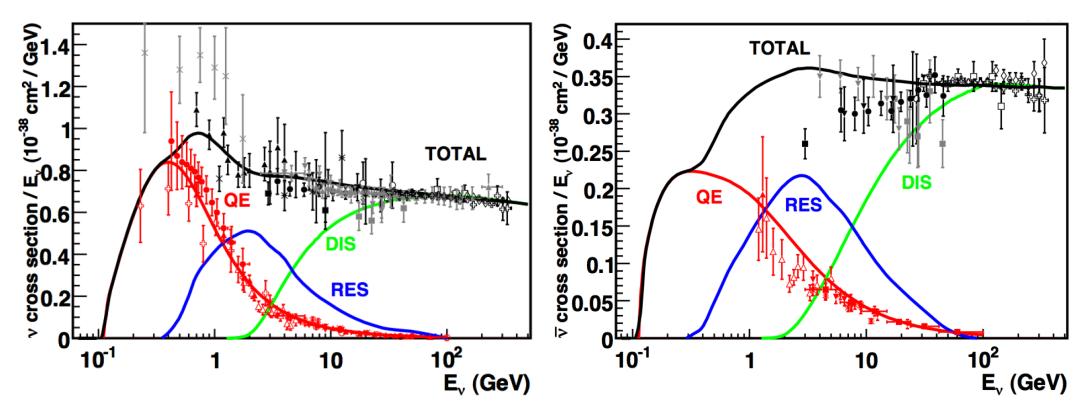}
	\caption[]{Interaction Crossection of Muon neutrino (left) and muon antineutrino (right) as a function of energy \cite{ppchainwiki}\label{fig:neu_cross}}
\end{figure}

The physics processes of neutrino interaction with matter are governed by the weak nuclear force, which is one of the four fundamental forces of nature. During the propagation through the matter neutrinos can interact with electrons as well as with neucleons. The basic neutrino interactions with matter are Neutral Current(NC) and Charged Current (CC). More details on neutrino ineractions can be found in 
\cite{Zuber2021,AtharSingh2021,AtharSinghnote,BargerMarfatiaWhisnant2012,Scully2013}. Some few process of neutrino interactions are as follows:
 \begin{enumerate}
 	\item \textbf{Neutrino-Electron Scattering:}
 	  All flavour of neutrinos can interact with electron with neutral current (NC) and only electron neutrino has an extra charged current (CC) mode. The examples are given below:  
 		\begin{itemize}
 			\item $\nu_l+e^- \xrightarrow[\text{NC}]{Z_0} \nu_l+e^- \,\,\,(l=e,\mu,\tau) $
 			\item $\nu_e+e^-\xrightarrow[\text{CC}]{W^-}\nu_e+e^-$ 
 		\end{itemize}
 	\item \textbf{Neutrino-Neucleon Scattering:}
 	 		\begin{itemize}
 	\item \textbf{Elastic Scattering of Neutrinos on Nucleons:} Neutrinos can scatter off of the nuclei of atoms in matter via the weak nuclear force. The process is similar to elastic scattering of electrons in a diffraction pattern. In this process, the neutrino transfers a small amount of energy and momentum to the nucleus, and the scattered neutrino continues on its way with a slightly different direction and energy. An example of elastic scattering is given below.
 	\begin{align}
 	 \nu_l+N \xrightarrow[\text{NC}]{Z_0}\nu_l+ N \,\,\,(l=e,\mu,\tau,\,\text{N=Nucleus})
       \end{align}
 	\item \textbf{Quasielastic Scattering (QES) of Neutrino and Nucleons:} In this process with slightly more energetic neutrino (anti neutrino) can interact with neucleons (neutron n, proton p) and the final state will contain a charged lepton and neucleon different from initial state. An example of this ineraction for neutrino is given below:
 	 \begin{align}
 	 \nu_l+n \xrightarrow[\text{CC}]{W^-}l^-+ p \,\,\,(l=e,\mu,\tau)
 	 \end{align}
 	\item \textbf{Resonance Elastic Scattering (RES):}
     In this process neutrino has a sufficient energy to interact with a neucleon. Due to the energy transfer, the target neucleon is ``knoked" into a baryonic resonance and then decay into messons and baryons. For example a $\nu_l$ interact with a neutron (n)  produces $l^-$ and $\Delta^+$ where $\Delta^+$ can be decayed into either $p+\pi^0$ or $n+\pi^{+}$ :
        \begin{align}
       \nu_l+n \xrightarrow[\text{CC}]{W^+}l^-+ \Delta^{+}\,^{\nearrow}_{\searrow}\, ^{^{p+\pi^0}}_{_{n+\pi^+}}\,\,\,(l=e,\mu,\tau)
       \end{align}
 	\item \textbf{Deep Inelastic Scattering (DES):}
 	As the energy levels increase, neutrinos possess the capability to impart a significant amount of momentum which enables the discernment of the substructural composition of nucleons. This has led to the discovery of a phenomenon known as deep inelastic scattering (DES), in which neutrinos can engage in direct interactions with the subatomic quarks that constitute the nucleons. An example of DIS is given below:
 	  \begin{align}
 	\nu_l+N \xrightarrow[\text{}]{CC}l^-+ X\,\,\,(l=e,\mu,\tau) \\ \nonumber
 	\nu_l+N \xrightarrow[\text{}]{NC}\nu_l+ X\,\,\,(l=e,\mu,\tau), \\ \nonumber
 	\end{align}
 	where X is the Jet of particles.
 	 		\end{itemize}	
 	 \end{enumerate}   
The interaction crossection of muon (anti) neutrino interactions as a function of its energy $E_\nu$ for different scattering process has been shown in Figure \ref{fig:neu_cross}.
\subsection{Matter Effect}
In 1978, Wolfenstein discovered that when neutrinos travel through matter, they interact with the electrons and nucleons in the medium. This interaction causes them to change how they mix, and can make the mixing more pronounced. This is particularly true when there are two types of neutrinos and the matter density is high. In 1985, Mikheev and Smirnov discovered that when neutrinos pass through a medium with changing density, they can switch identities in a specific way. This happens when the density of the medium changes in a certain way along the path of the neutrinos, and the mixing angle reaches its maximum value of $\pi/4$. This effect is known as resonant flavor transition. This discovery, known as the MSW effect, has important implications for our understanding of neutrinos and their behavior. Neutrinos traversing a medium are subject to both coherent and incoherent interactions with the particles within it. While the former, coherent forward elastic scattering, plays a significant role in their behavior, the latter, incoherent scatterings, are typically of minimal magnitude and can be disregarded in most scenarios. The effective potential for NC forward coherent scattering can be writen as \cite{GuintiKim}:
\begin{align}
V_{NC}^{\nu}= -\frac{1}{2}G_F\sqrt{2}  N_{n}
\end{align}
Where $G_F$ is the Fermi constant, and $N_{n}$ is the number density of neutral-current scattering centers (such as neutrons). For antineutrinos the NC potential can be written as $V_{NC}^{\bar{\nu}}=-V_{NC}^{\nu}$.
The effective potential for forward coherent charged-current (CC) scattering for neutrinos is given by \cite{GuintiKim}:
\begin{align}
V_{CC}^{\nu}= \sqrt{2} G_F N_e
\end{align}
where, $N_e$ is the electron number density inside matter.  For antineutrinos the CC potential can be written as $V_{CC}^{\bar{\nu}}=-V_{CC}^{\nu}$.
\subsubsection{Flavour oscillation under the effect of matter}
The relation between the flavour state and the mass state is shown in equation \ref{eqn:state_eqn}. Let us consider that mass state $\ket{\nu_k}$ is the eigen state of the vacuum Hamiltonian $H_0$. Therefore, we can write the following relation for momentum $\vec{p}$ and mass $m_k$:
\begin{align}
H_0\ket{\nu_k}=E_k\ket{\nu_k},
\end{align}
where $E_k=\sqrt{ \left|\vec{p}\right|^2+m_k^2}$.
\par The Hamiltonian in matter is different from the vacuum. Hence the total Hamiltonian in the flavour basis can be written as \cite{GuintiKim}:
\begin{align}
H_F^{eff} =
\frac{1}{2E}U\begin{bmatrix}
0 & 0 & 0 \\
0 & \Delta m^2_{21} & 0 \\
0 & 0 & \Delta m^2_{31}
\end{bmatrix}U^\dagger + \begin{bmatrix}
V_{CC} & 0 & 0 \\
0 & 0 & 0 \\
0 & 0 & 0
\end{bmatrix}.
\end{align}
For two flavour oscillation case $\nu_e\rightarrow \nu_\mu$ the effective Hamiltonian $H_F^{eff}$ can be written as follows:
\begin{align}
H_F^{eff} &= \frac{1}{2E} \begin{pmatrix}
\cos\theta & \sin\theta \\
-\sin\theta & \cos\theta \\
\end{pmatrix}\begin{pmatrix}
0 & 0 \\
0 & \Delta m^2 \\
\end{pmatrix} \begin{pmatrix}
\cos\theta & -\sin\theta \\
\sin\theta & \cos\theta \\
\end{pmatrix}+ \begin{pmatrix}
V_{CC} & 0 \\
0 & 0 \\
\end{pmatrix} \\\nonumber
&=\frac{1}{2E}\begin{pmatrix}
\Delta m^2 \sin^2\theta + 2EV_{CC}  & \Delta m^2 \sin\theta \cos\theta\\
\Delta m^2 \sin\theta \cos\theta & \Delta m^2 \cos^2\theta \\
\end{pmatrix}
\end{align}
The neutrino mixing matrix in matter can be calculated by finding the diagonal form of the Hamiltonian ($H_F^{eff}$) in the flavor state using an orthogonal rotation matrix $U_M$. Therefore we can write:
\begin{align}
U_M=\begin{pmatrix}
\cos\theta_M & \sin\theta_M \\
-\sin\theta_M & \cos\theta_M \\
\end{pmatrix}
\end{align} 
So that,
\begin{align}
U_M^TH_F^{eff}U_M=\begin{pmatrix}
m^2_{i,M} & 0 \\
0 & m^2_{j,M} \\
\end{pmatrix},
\end{align} 
where,
\begin{align} \label{eqn:matter_mix_angle}
\tan 2\theta_M=\frac{\sin 2\theta}{\cos 2\theta-\frac{2\text{E} V_{CC}}{\Delta m^2}}
\end{align}
and,
\begin{align}
\Delta m^2_M&=m^2_{i,M}-m^2_{j,M}\\\nonumber
&=\sqrt{(\Delta m^2 \sin 2\theta)^2+(\Delta m^2 \cos 2\theta -2 E V_{CC})^2}
\end{align}
The parameters $\theta_M$ and $\Delta m^2_M$ are the mixing angle and the mass-squared difference in matter respectively. For $\Delta m^2 \cos 2\theta=2 E V_{CC}$, $\theta_M=45^{\circ}$ , which is maximal mixing. This phenomenon is called the
MSW resonance effect. If 2EV$_{CC}<<\Delta m^2$ in equation \ref{eqn:matter_mix_angle} then Matter effect is negligible. If 2EV$_{CC}>>\Delta m^2$ then $\theta_M\rightarrow\frac{\pi}{2}$, therefore oscillations are suppressed.

\subsection{Measurement of Oscillation parameters}
Neutrinos are everywhere and have always amazed people. The existance of non zero oscillation parameters are proof that neutrino has mass. The magnitude and sign of the mass-squared difference $\Delta m^2_{21}$ has been determined, but for the mass-squared difference $\Delta m^2_{31}$ (or $\Delta m^2_{32}$), only the magnitude has been established. This results in two possible configurations of the mass states; in the first configuration, the mass state $m_3$ is heavier than $m_2$ which is heavier than $m_1$, which is known as the Normal Hierarchy (NH). In the second configuration, the mass state $m_2$ is heavier than $m_1$ which is heavier than $m_3$, which is known as the Inverted Hierarchy (IH). It's worth mentioning that the mass ordering is one of the main unknown parameters of the neutrino oscillation and many ongoing and future experiments are designed to determine it.
\begin{table}[h]
	\centering\includegraphics[scale=0.5]{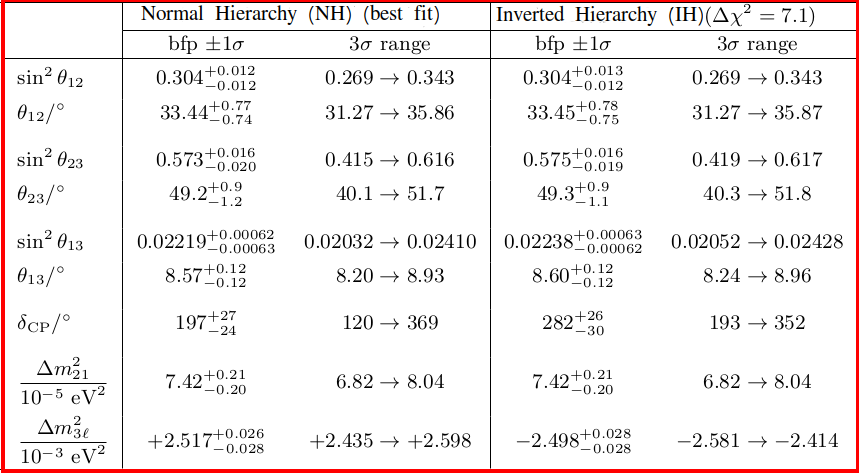}
	\caption[]{Three-flavor oscillation parameters from fit of global data \cite{Esteban2020}. \label{tab:globalfit}}
\end{table} 
\par The determination of oscillation parameters has done by several experiments. Scientists have determined the parameters of the oscillation of solar neutrinos ($\Delta m^2_{21}$ , $\theta_{12}$) by analyzing data from a variety of experiments, including KamLAND, Chlorine, Gallex/GNO, SAGE, SNO, Borexino, and all four phases of Super Kamiokande. Similarly, they have determined the parameters of atmospheric neutrino oscillation ($\Delta m^ 2_{31}$ , $\sin^2\theta_{23}$) by analyzing data from experiments such as Super-Kamiokande (SK), T2K, MINOS, and other experiments focused on atmospheric neutrinos \cite{NEUTRINO_OSCILLATION_REVIEW}. The combination of data from these experiments has allowed scientists to make precise measurements of the properties of neutrinos and our understanding of them greatly advanced. The best-fit value from the global analysis is given in Table \ref{tab:globalfit}.

\section{INO-ICAL Detector Design and Experimental Goal}
The India Based Neutrino Observatory (INO) is a proposed research facility that is set to be built in the Bodi West Hills region of the Theni district in Tamil Nadu, India. The main instrument that will be used at INO is the magnetized Iron CALorimeter (ICAL) detector, which has a massive weight of approximately 51,000 metric tons. The purpose of the ICAL is to investigate the properties of neutrinos by studying atmospheric neutrinos that have varying energies and traversal distances. One of the main objectives of this research is to determine the hierarchy of neutrino masses. To achieve this, the ICAL will observe the matter effects that occur when neutrinos travel through the Earth. The ICAL's capability to differentiate between neutrinos and antineutrinos will be instrumental in achieving this goal. The ICAL detector will be furnished with active detector elements known as Resistive Plate Chambers (RPCs). The ICAL has been engineered to be highly responsive to atmospheric muon neutrinos in the energy range of 1-15 GeV. The ICAL's architecture, with its horizontal layers of iron interspersed with RPCs, grants it near-total coverage of the incoming direction of neutrinos, save for those that produce muons traveling at an almost horizontal trajectory. This blueprint enables a thorough detection of various path lengths traveled by neutrinos as they pass through the Earth, and the atmospheric neutrino flux presents a vast spectrum of energy levels. Furthermore, The ICAL's ability to detect muon neutrinos is particularly valuable as the atmospheric muon neutrino flux is significantly higher than that of other flavors, thus providing a large number of events to study neutrino properties.
\subsection{ICAL detector geometry and design}
The implementation of the laboratory beneath the Bodi Hills entails the creation of a horizontal tunnel, measuring approximately 1900 meters in length, to gain access to the laboratory that is situated beneath a mountain peak. The laboratory will consist of one primary and three secondary caverns, all encased in a rock layer of at least 1000 meters or more in depth, with a vertical rock coverage of approximately 1300 meters. These measures are necessary to accommodate the experimental activities that will take place in the laboratory.
\par
\begin{figure}[h]
	\centering\includegraphics[scale=0.33]{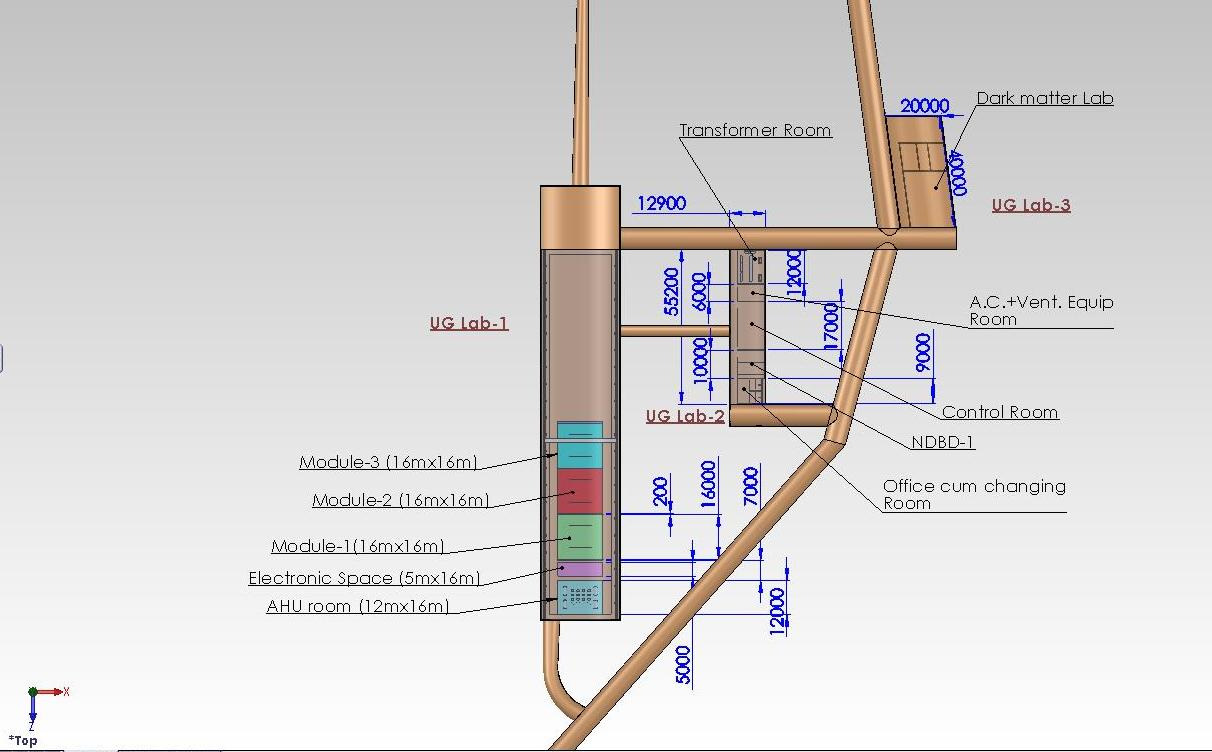}
	\caption[]{A graphical representation of the subterranean caverns' layout, depicting the spatial extent of the proposed experiments and other pertinent elements \cite{ICALpotential}. \label{fig:icalprint}}
\end{figure}

\begin{figure}[h]
	\centering\includegraphics[scale=0.6]{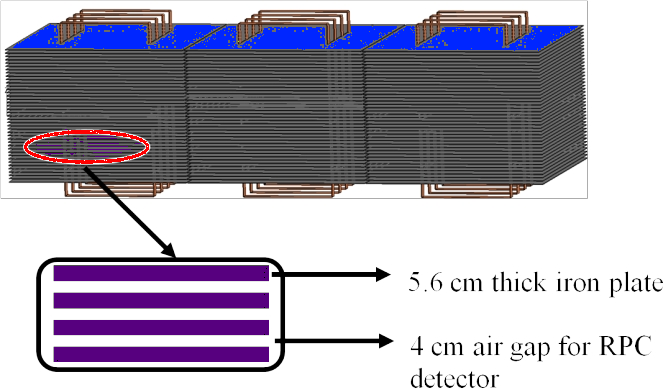}
	\caption[]{Schematic view of the 50 kt ICAL detctor \cite{ICALpotential}. \label{fig:ical-det}}
\end{figure}

 The current blueprint of the laboratory caverns is depicted in Figure \ref{fig:icalprint}. The paramount cavern, which will serve as the primary site for the installation of the main iron calorimeter detector (ICAL), exhibits dimensions of 132 meters in length, 26 meters in width, and 32.5 meters in height. This cavern, known as "UG-Lab 1," is designed to hold a 50 kiloton ICAL (planned) and a potential second ICAL-II neutrino detector of similar size. Each ICAL comprises of three modules, each measuring 16 meters in length, 16 meters in width, and 14.5 meters in height, resulting in a total area of 96 meters in length and 16 meters in width for both detectors. Each module features two vertical slots that have been precisely cut to accommodate the placement of current-carrying copper coils. These coils, when energized, generate magnetic fields that permeate the iron plates, as illustrated in Figure \ref{fig:ical-det}. Simulation studies have revealed that it is possible to magnetize the iron plates to a field strength of around 1.5 Tesla. Fields greater than 1 Tesla present in a substantial majority of the detector's total volume, specifically over 85\%.
\par The ICAL detector, shown in Figure \ref{fig:ical-det}, will have a composite structure made up of 151 layers of magnetized iron plates each of 5.6 cm thick. These layers will be arranged horizontally, with 4 cm spaces in between, creating a total height of 14.5 meters. These spaces will serve as the housing for the active detector layers. Iron spacers, which will provide structural support, will be placed at regular 2-meter intervals along both the X and Y axes. The active detection components, specifically the resistive plate chambers (RPCs), will comprise a pair of glass plates that are 3 mm thick, measuring 2 meters by 2 meters, and separated by 2 mm spacers. These will be placed within the gaps between the iron layers. The operating voltage of the RPCs will be 10 kV, so they will operate in avalanche mode. 
\par When an energetic particle traverses the resistive plate chambers, it generates signals that are captured by X and Y pickup strips on either side of the RPCs. These strips are about 3 cm wide. Extensive research and development has revealed that the RPCs exhibit a high level of efficiency, with an estimated range of 90-95\%. Additionally, the RPCs possess an exceptional time resolution, measuring in at approximately one nanosecond. This high efficiency and time resolution make the RPCs an ideal choice for high energy particle detection. The ability to determine the X and Y coordinates from the hit strips in conjunction with the Z coordinate from the layer number, allows for the estimation of the trajectory of charged particles passing through the RPCs. The RPC time resolution of around 1 nano second as observed, allows to distinguish between particles that are going up and particles that are going down. By analyzing the pattern of hits in the resistive plate chambers, the energy and trajectories of the charged particles generated by neutrino interactions can be reconstructed with high precision. Specifications of ICAL detector is summarised in Table \ref{tab:ical-spec}.
\begin{table}
\centering
	\begin{tabular}{|c|c|} 
		\hline
		\rowcolor{apricot}\textbf{Parameter} & \textbf{Value} \\ [0.5ex] 
		\hline
		\rowcolor{cyan}\textbf{ICAL} &  \\
		\rowcolor{buff}No. of modules & 3 \\ 
		\rowcolor{bubbles}Module dimension & 16 m × 16 m × 14.5 m \\ 
		\rowcolor{buff}Detector dimension & 48 m × 16 m × 14.5 m \\ 
		\rowcolor{bubbles}No. of layers & 151 \\ 
		\rowcolor{buff}Layer material & Iron \\
		\rowcolor{bubbles}Iron plate thickness & 5.6 cm \\ 
		\rowcolor{buff}Total weight & 50,000 tonnes \\
		\rowcolor{bubbles}Gap for RPC trays & 4.0 cm \\ 
		\rowcolor{buff}Magnetic field & 1.5 Tesla \\ 
		\rowcolor{cyan}\textbf{RPC} &  \\
		\rowcolor{bubbles}RPC material & Glass \\
		\rowcolor{buff}RPC Gas mixture    & Freon (95.2\%) : Isobutane (4.5\%) : SF6 (0.3\%) \\
		\rowcolor{bubbles}RPC Operating Voltage & 10 kV \\
		\rowcolor{buff}RPC unit dimension & 2 m x 2 m \\ 
		\rowcolor{bubbles}Readout strip width & 3 cm \\ 
		\rowcolor{buff}No. of RPC per Layer and per Module & 64 \\ 
		\rowcolor{bubbles}Total no. of RPC & 30000 \\ 
		\rowcolor{buff}No. of electronic readout channels & 3.9 x 10$^6$ \\

		\end{tabular}
	\caption{Specifications of ICAL detector \cite{ICALpotential}.\label{tab:ical-spec}}
\end{table}
\subsection{Physics analysis at INO}
The ICAL detector is capable of detecting a wide range of neutrino energies, between 1 and 25 GeV. The interactions of neutrinos with matter are discussed in section \ref{sec:numatter}. The ICAL detector is able to detect neutrinos as they interact with electrons and the nucleus of iron, through both neutral current (NC) and charged current (CC) interactions. However, it is not possible to directly observe NC and CC interactions with electrons. Instead, the final state particles, such as leptons and hadrons, resulting from QES, RES, and DES processes can be observed. Muons produced from CC interactions play a crucial role in the analysis, as they possess a minimum ionizing property, allowing them to travel a longer path through the ICAL detector, resulting in more precise reconstruction of their path and energy. These muon events with long tracks are referred to as ``track-like" events, while hadrons that give rise to shower-like features are called ``shower-like" events. By analyzing the direction of bending of the muon tracks, it is possible to determine the charge of muons, and thus distinguish between neutrinos and anti-neutrinos. 
\par The ICAL detector, which is yet to be constructed, has had its potential properties simulated using the CERN GEANT4 \cite{Agostinelli2003} software package. Atmospheric neutrino events have been generated by the NUANCE \cite{Casper_2002} neutrino generator, utilizing the Honda 3D \cite{Honda_2011} fluxes for the Kamioka site in Japan. A simple track-finding algorithm is employed to identify the longest track or ``track-like event", in an event. It has been found that around 90\% of total muons with true momenta of 2 GeV/c are reconstructed as ``muon-like tracks," and for muons with true momenta of 5 GeV/c, this number increases to around 95\%. These ``muon-like tracks" are then fitted using a Kalman-filter based algorithm, and the best-fitted track is used to estimate the vertex position, momentum, and charge of the muon.

\par Main physics goal of INO is as follows:
\begin{itemize}
	\item measure precisely magnitude of $\Delta m^2_{23}$
	\item Identification of mass hierarchy or sign of $\Delta m^2_{23}$ .
	\item value of $\theta_{23}$ and correct octant of it.
\end{itemize}
In addition to its ability to measure neutrino oscillations, the ICAL detector also possesses a unique capability to explore new areas of physics. Among these, it can study the potential violation of CPT symmetry in the leptonic sector, which could help to explain the matter-antimatter discrepancy. Additionally, the detector can be used to search for the possible existence of sterile neutrinos, which are hypothetical neutrino states that may not interact via the weak force. It can also be used to probe non-standard interactions (NSI) in neutrino oscillations, which are deviations from the standard model predictions. Furthermore, it could be used to search for dark matter, and study solar, supernova and geo-neutrinos.
\begin{figure}[h]
	\centering\includegraphics[scale=0.4]{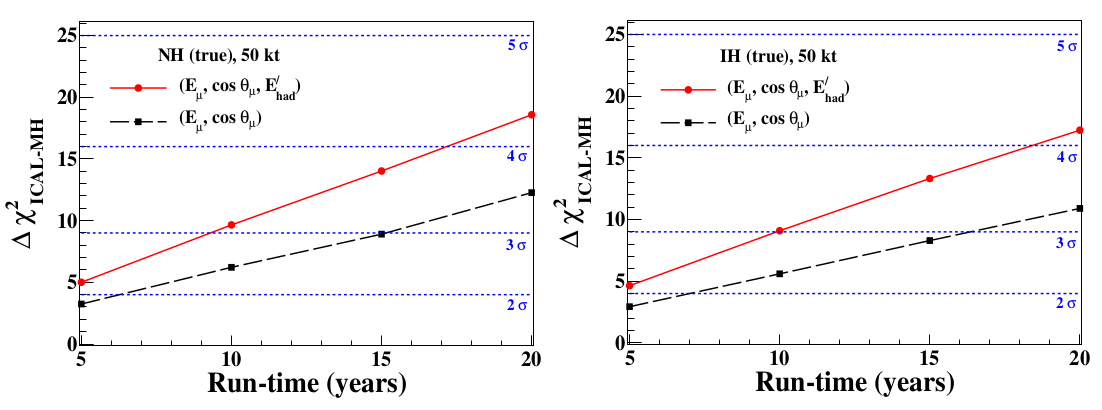}
	\caption{The left panel shows the dependence of $\Delta \chi^2_{ICAL-MH}$ on the run time assuming Normal Heirarchy (NH) as the true hierarchy, while the right panel displays the same dependence but assuming Inverted Heirarchy (IH) as the true hierarchy. \cite{ICALpotential}. \label{fig:ical-pre1}}
\end{figure}
\begin{figure}[h]
	\centering\includegraphics[scale=0.3]{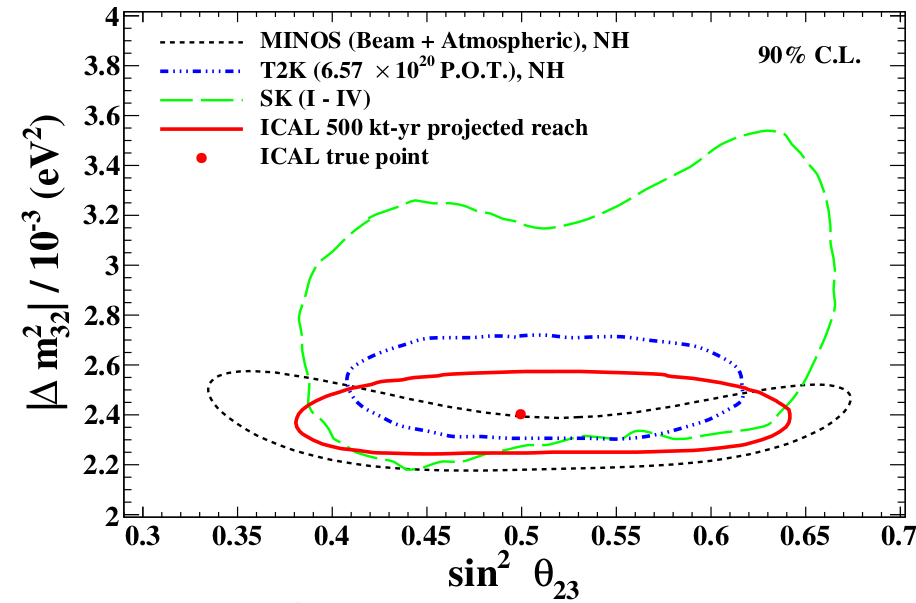}
	\caption{90\% C.L. contours in the $\text{sin}^2\theta_{23}-\,|\Delta m^2_{32}|$  plane (2 dof): Presently, the restrictions derived from the findings of SuperKamiokande \cite{Super-Kamiokande}, MINOS \cite{MINOS}, and T2K \cite{T2K} have been shown alongside with ICAL's capacity to analyze neutrinos given an exposure of 500 kiloton-years under the "true NH" premise. ICAL's selected parameters, assumed to be accurate, have been indicated on the graph with a singular mark. \cite{ICALpotential}. \label{fig:ical-pre}}
\end{figure}
\subsubsection{ICAL Precision measurement of oscillation parameters}
\begin{table}[h]
	\centering\begin{tabular}{|c|c|c|c|c|c|}
		\hline
		$\Delta m^2_{21}$ (eV$^2$) & $\Delta m^2_{eff}$ (eV$^2$) & $\sin^2\theta_{12}$ & $\sin^2\theta_{23}$ & $\sin^2 2\theta_{13}$ & $\delta_{CP}$ \\
		\hline
		7.5 × 10$^{-5}$ & 2.4 × 10$^{-3}$ & 0.3 & 0.5 & 0.1 & 0$^\circ$ \\
		\hline
	\end{tabular}
	\caption{True values of the input oscillation parameters used in the analyses \cite{ICALpotential}\label{tab:icaltrue}}
\end{table}
To conduct the physics analysis, a significant number of unoscillated events are created using NUANCE, then adjusted to a specific exposure, and oscillations are incorporated through a reweighting algorithm. Realistic resolutions and efficiencies obtained from GEANT-4 based simulation studies of ICAL are integrated in order to ensure the accuracy of the analysis. 
To simulate the behavior of the measured quantities with greater precision, the energies and directions of the relevant particles are smeared based on the resolutions that have been previously ascertained. This method provides an approximation of the expected average behavior of the particles. At present, the simulations assume an idealized scenario, wherein the muon track and hadron shower can be identified with utmost efficiency, and any spurious signals stemming from stochastic hits in the vicinity of the signal events can be discounted. Although these assumptions are deemed sound, empirical evidence is still necessary to validate them. The acquisition of such evidence could be accomplished through the use of a prototype detector. In the simulation of ICAL data the true values shows in the Table \ref{tab:icaltrue} have been used, where $\Delta m_{eff}^2= \Delta m_{32}^2-(\text{cos}^2 \theta_{12}-\text{cos}\,\delta_{CP}\,\,\text{sin}\,2\theta_{12}\,\,\text{tan}\,\theta_{23})\Delta m^2_{21} $.
\par To quantify the effectiveness of the ICAL experiment in determining the neutrino mass hierarchy, a specific hierarchy (either normal or inverted) is selected as the true (input) hierarchy. CC muon neutrino events are categorized according to predetermined criteria, and a $\chi^2$ analysis is carried out while accounting for systematic uncertainties. A range of 3$\sigma$ is used to marginalize the parameters $|\Delta m_{eff}^2|$, $\text{sin}^2 \theta_{23}$, and $\text{sin}^2,2\theta_{13}$. The degree of significance of the result is evaluated based on the value of $\Delta \chi^2_{ICAL-MH}$,(MH stands for Mass Hierarchy) which reflects the level of confidence in rejecting the incorrect hierarchy, where $\Delta \chi^2_{ICAL-MH}$ can be written as follows:.
 \begin{align}
 \Delta \chi^2_{ICAL-MH}=\chi^2_{ICAL}(\text{false MH})-\chi^2_{ICAL}(\text{true MH})
 \end{align}
The value of $\chi^2_{ICAL}(\text{false MH})$ and $\chi^2_{ICAL}(\text{true MH})$ can be found from the fit of the observed data. Figure \ref{fig:ical-pre1} displays the dependence of $\Delta \chi^2_{ICAL-MH}$ on run time for Normal Hierarchy as the true hierarchy in the left panel and for Inverted Hierarchy in the right panel. The results obtained with hadron energy information are marked in red, while those without are marked in black. The figure clearly indicates that with hadron energy information, ICAL can reject the incorrect hierarchy with 3$\sigma$ significance using 10 years of data.

\par The comparison of the 10-year reach of 51 kt ICAL in the $\sin^2 2\theta_{23}–\Delta m^2_{32}$ plane with the current limits from other experiments is presented in Figure \ref{fig:ical-pre}. It is anticipated that the precision of ICAL's $\Delta m^2_{32}$ would be more accurate than that of atmospheric neutrino experiments that employ water Cherenkov detectors, due to its superior ability to measure energy.
\section{Scope of this thesis}
The active detector element of ICAL detector is RPC. Hence, for the long run it is expected that all RPCs runs effectively to capture every events. This thesis is focused on the detailed simulation and experimental of RPC physics.
\par In Chapter \ref{ch2} a short review on the detector physics of RPC and other gaseous detectors is given.
\par It is known that during the avalanche generation process inside an RPC the applied electric field can be distorted due to the electric field of electrons and ions (space-charges), which is called space charge field. The model to calculate space charge field is discussed in the Chapter \ref{space_charge_field_calculation}.

\par The effect of electrode parameters and image charge on the electric field configurations in presence of space-charge is discussed in Chapter \ref{ch4}.
\par Simulation of avalanche inside an RPC with the help of Garfield++ software is time consuming. Therefore, parallel computation is used to solve this issue, which is discussed in Chapter \ref{chp:chap_fast_aval}.
\par In Chapter \ref{ch:tomography} an applicaion of RPC detector a stack of RPCs is made to determine scattering angle of cosmic muon in presence of high density materials.
\par Finally, the summary and outlook of this thesis is discussed in Chapter \ref{ch:summary}.  
\chapter{Detector Physics of RPCs and Other Gaseous Detectors}\label{ch2}
\section{Introduction}
The 1890s marked a significant milestone in the field of particle physics with the discovery of X-rays and radioactivity. These groundbreaking discoveries motivated scientists to develop particle detectors. The earliest known particle detector was likely the photosensitive paper used by French physicist Henri Becquerel \cite{satyaThesis}. He discovered that uranium salt emitted a form of radiation that blackened the paper, indicating the presence of particles.
\par Ernest Rutherford and Ernest Marsden were two scientists who made significant contributions to the field of nuclear physics. In 1909, they performed an experiment to study the structure of atoms using alpha particles \cite{rutherford1,rutherford2}. The alpha particles were emitted by a radioactive source and directed at a thin sheet of gold. They used a scintillating screen to detect the scattered alpha particles. Rutherford and Marsden found that the majority of alpha particles passed through the gold foil undeflected, but a small fraction were scattered at large angles. This led Rutherford to propose the nuclear model of the atom, in which the positive charge and most of the mass of an atom are concentrated in a small, dense nucleus. This experiment was a key piece of evidence for the existence of the atomic nucleus and provided insight into the structure of atoms. In 1908, Geiger and his colleague Ernest Rutherford developed the first version of the Geiger counter to detect ionizing radiation, such as alpha or beta particles. In 1928, Geiger and Mueller improved the design of the detector by using a tube filled with a more sensitive gas, such as helium or argon, and adding a metal electrode, called the ``Geiger-Mueller tube" \cite{geiger1928elektronenzahlrohr}. This allowed for more efficient detection of ionizing radiation and made the Geiger counter more sensitive and reliable. This was a time when the development of the field of experimental particle physics began. The spark chamber (1930s, Bruno Touschek), cloud chamber (1920s, Charles Thomson Rees Wilson), and bubble chamber (1950s, Donald A. Glaser) were also used to reconstruct the tracks of charged particles. The path of the incident particle created inside the chambers was photographed. These photographs provided detailed information about the particle's trajectory and energy. However, it was very difficult to gather a large amount of track data using photographs, and analyzing those tracks was also time-consuming. Therefore, scientists developed detectors in which the information of the track can be stored electronically, and data can be analyzed with the help of computer programming. Such modern gaseous particle detectors can be classified into two types: wired-based and parallel-plate-based detectors. This thesis focuses on the Resistive Plate Chamber (RPC) detector, so a short review of other gaseous detectors, including the Proportional Counter, Multiwire Proportional Counter, Drift Chamber, Time Projection Chamber, Micromegas, and Gas Electron Multiplier, is provided in Section \ref{sec:review}, and the physics of the RPC detector is also discussed briefly in Section \ref{sec:RPC}.
\section{Short review of gaseous particle detectors}\label{sec:review}
\subsection{Wired based detectors}

\subsubsection{Cylindrical proportional counter:}
\begin{figure}[H]
	\centering\includegraphics[scale=0.4]{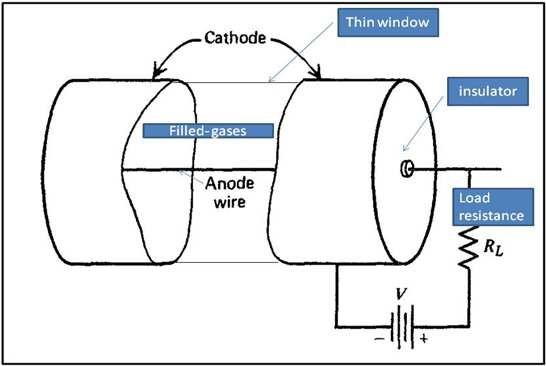}
	\caption[]{Schematic diagram of a proportional counter\label{fig:prop}\cite{Hemn2014}}
\end{figure}

The basic geometry of a proportional counter consists of a cylindrical chamber filled with a gas (such as argon or neon) and a central wire electrode that is maintained at a high voltage with respect to the chamber walls, as shown in Figure \ref{fig:prop}. The electric field inside the chamber varies inversely with the radial distance from the central wire, following a $1/r$ relationship \cite{leo1994techniques}. Ionizing radiation enters the chamber and causes the ionization of gas atoms, creating ion-electron pairs. The electrons are accelerated towards the central wire electrode by the electric field created by the high voltage, creating a secondary current that is proportional to the number of ion-electron pairs formed. The electrons rapidly pass through the wire ($\approx1,ns$), and due to their greater mass, positive ions slowly move towards the cathode. This ion drift causes a flow of current, which generates the signal observed on the electrodes.
\subsubsection{Multi wire proportional chamber (MWPC):} 
The field of experimental particle physics requires the ability to accurately track and measure the trajectories of particles. Prior to 1970, the tools used for this purpose were mainly optical, such as photographic emulsions and various chambers that recorded particle tracks on film. These optical methods have limitations, including the need for manual analysis of each frame and the inability to handle a large number of events.
\par
\begin{figure}[H]
	\centering\includegraphics[scale=0.4]{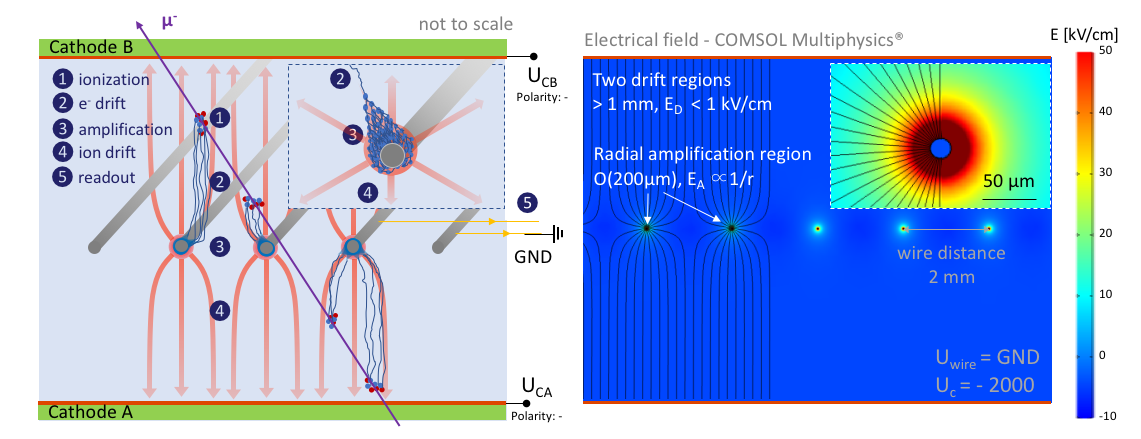}
	\caption[]{The diagram on the left illustrates the design and signal generation within a Multi-Wire Proportional Chamber (MWPC). When a muon (depicted in purple) travels through the chamber, it causes primary ionization, freeing electrons (shown in blue). These electrons then drift towards the anode wires, where they are amplified. The movement of ions (depicted by red arrows) resulting from this amplification is what is measured. The diagram on the right illustrates the electric field in the area near an MWPC anode wire grid, specifically highlighting the transition from a uniform drift field to regions of radial field enhancement. \cite{micromegas}\label{fig:mwpc}}
\end{figure}

 Scientists sought an electronic device that would overcome these limitations and allow for more efficient and accurate tracking. One option was to use arrays of proportional counters, but this approach was not practical due to mechanical constraints. In 1968, Georges Charpak revolutionized the field with the invention of the multiwire proportional chamber (MWPC) \cite{CHARPAK1979405}. The MWPC uses an array of closely spaced anode wires within the same chamber, each acting as an independent proportional counter. Additionally, with the use of transistorized electronics, each wire has its own amplifier integrated onto the chamber frame, making it a practical option for position sensing. The gas mixture used in MWPCs, consisting of 75\% Argon, 24.5\% Isobutane, and 0.5\% Freon, is known as the ,``magic gas mixture". A typical field map of an MWPC, which shows the variation in electric field strength across the chamber, is shown in Figure \ref{fig:mwpc}. This map is important for understanding the behavior of the device.
\par The pulse generated on an anode wire is caused by positive ions from the avalanche moving towards the cathode. As a charged particle passes through the chamber, it leaves ionization along its path, causing electrons to continue arriving over a prolonged period. Similar to the proportional counter, the electric field strength is stronger near the wire grid, which results in maximum multiplication of the electron signal. This occurs because the electrons are accelerated more strongly in this region, as shown in Figure \ref{fig:mwpc}. As a result, the leading edge of the pulse increases gradually. However, the electrons that arrive first are those closest to an anode wire, which cannot be more than half a wire spacing away, resulting in a timing uncertainty of no more than the time required for electrons to travel that distance, approximately 25-30 ns for a typical 2 mm wire spacing. A significant portion, around 8\%, of this signal occurs within the first 10 nanoseconds of the pulse, resulting in a relatively fast leading edge. In addition to rate capability, the main limitation of the MWPC is the variation in time between the occurrence of the event and the initiation of the avalanche at the anode.

\subsubsection{Drift Chamber:}
\begin{figure}[H]
	\centering\includegraphics[scale=0.5]{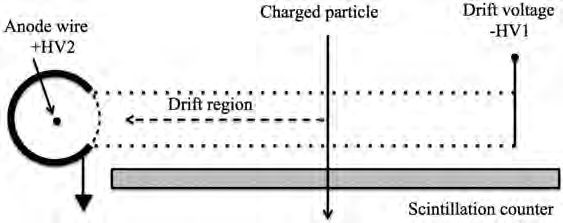}
	\caption[]{Diagram of single cell drift chamber \cite{micromegas}\label{fig:drift1}}
\end{figure}

The diagram in Figure \ref{fig:drift1} illustrates the basic operation of a drift chamber. The drift cell is defined by a high voltage electrode on one end and the anode of a simple proportional counter on the other. To create a constant electric field, a series of cathode field wires held at specific voltages are placed along the drift region. To receive a trigger signal upon the arrival of a particle, a scintillation counter or any detector with better efficiency and time resolution can be placed before or after the chamber to cover the entire sensitive area. When a particle passes through the chamber, it liberates electrons in the gas which then begin drifting towards the anode. At the same time, the fast signal from the scintillator starts a timer. The signal generated at the anode when the drifting electrons arrive is then used to stop the timer, providing the drift time. With a trigger signaling the arrival of a particle and knowledge of the drift velocity u(E), the distance, x, at uniform field can be calculated using the equation 
x = u(E) (t1-t0), 
where t0 represents the arrival time of the particle and t1 represents the time at which the pulse is detected at the anode. 
\par
\begin{figure}[H]
	\centering\includegraphics[scale=0.5]{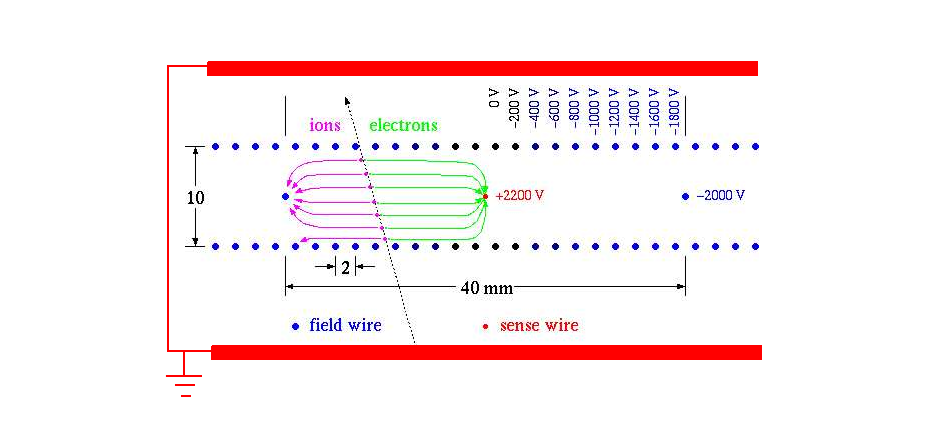}
	\caption[]{Diagram of multi cell drift chamber \cite{micromegas}\label{fig:drift2}}
\end{figure}

 A typical multi-wire proportional chamber (MWPC) can be used as a drift chamber, however, for large wire spacing, the non-uniformity of the electric field can cause issues with linearity and slow electron drift. To address this problem, modifications were made to the design by H. Walenta and colleagues in 1971. They added thicker field wires in between the anode wires at appropriate potentials to increase the strength of the electric field in crucial regions. However, this design still had limitations, as the chamber needed to be thick to maintain a uniform field, affecting the packaging density. To overcome these limitations, a different design was proposed by Charpak et al. in 1973. This design consisted of two sets of parallel cathode wire planes connected to increasingly high negative potentials, symmetrically from the center of the cell as shown in Figure \ref{fig:drift2}
. The anode wire was kept at a positive potential, and additional field wires were placed at the potential of the adjacent cathode wires to sharpen the transition between cells. Two grounded screening electrodes were added to protect the internal field structure from external influences. For a constant uniform field The spatial resolution mostly depend on the Diffusion constant of gas and drift path. A spatial resolution of 100 $\mu \text{m}$ can be achieved for a 5 cm drift path.  
\subsubsection{Time Projection Chamber (TPC):}
\begin{figure}[H]
	\centering\includegraphics[scale=0.25]{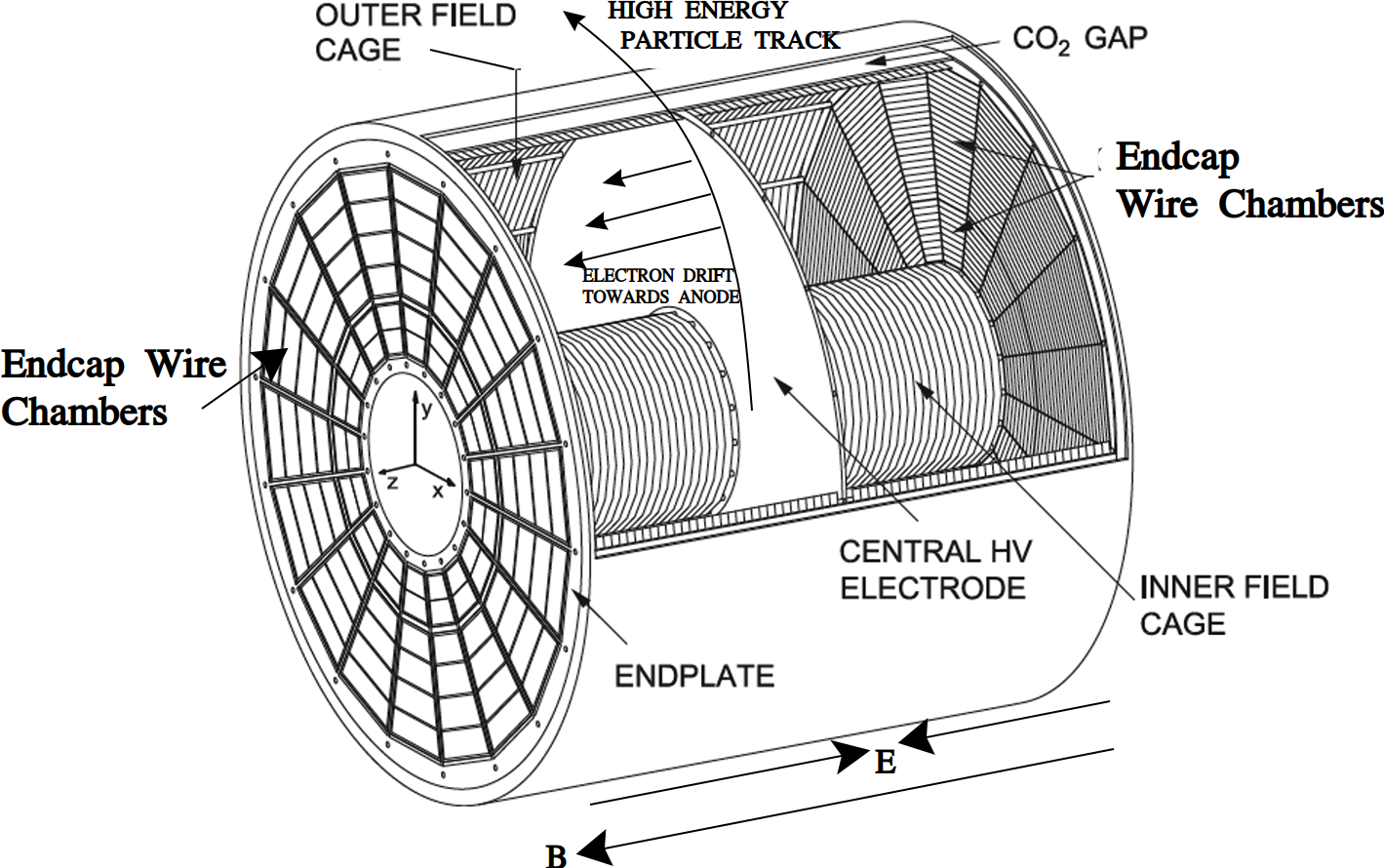}
	\caption[]{Diagram of Time Projection chamber (ALICE TPC)\cite{MUNZER2020162058}\label{fig:tpc1}}
\end{figure}
\begin{figure}[H]
	\centering\includegraphics[scale=0.35]{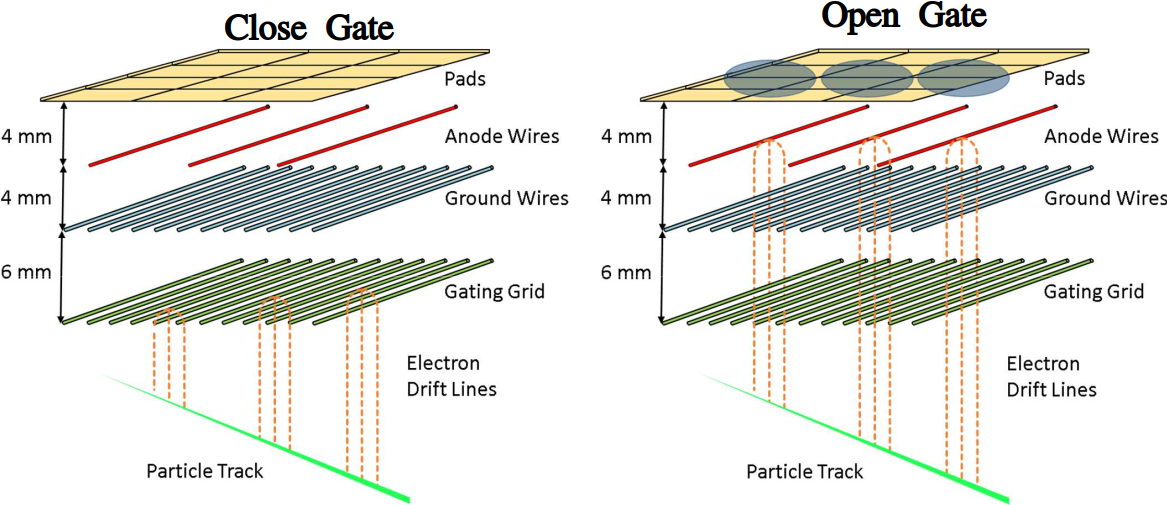}
	\caption[]{Scematic diagram of readout of TPC, LEFT: Close gate condition (Neither ions nor electrons go to the avalanche region), RIGHT: Open Gate condition (Drifting electrons can pass into the avalanche region).\cite{gatinggrid}\label{fig:tpc2}}
\end{figure}

The Time Projection Chamber (TPC) \cite{MUNZER2020162058} is a large cylindrical device that combines features from both the Multi-Wire Proportional Chamber (MWPC) and the drift chamber, as shown in Figure \ref{fig:tpc1}. The TPC is typically placed at the interaction point in a collider machine, and its cylindrical shape provides full coverage of the solid angle close to 4$\pi$. The gas mixer, similar to the MWPC, can be utilized in the TPC. The gas is contained in the hollow area outside the beam pipe. To produce a uniform electric field for drifting inside the TPC, a ring-shaped cathode is placed at the center. To ensure an electric field of consistent intensity, the walls of the chamber are fitted with field strips that form a field cage. These strips, made of conductive substances such as copper, are of equal width and evenly spaced. Their voltage gradually decreases from the cathode to the anode. To produce a more uniform field close to the wall, additional strips, referred to as mirror strips, may be utilized. These strips are placed in the spaces between two field strips and located on the opposite side of an insulating layer. They should be set to a potential that is between the cathode and anode. The ends of the TPC are covered by arrays of anode wires arranged in sectors. Parallel to each anode wire is a cathode strip that is divided into rectangular segments, or cathode pads, as shown in Figure \ref{fig:tpc2}. To maintain a uniform electric field throughout the drift volume, a secondary grid composed of grounded wires is employed to protect this area from the field in the amplification region, as shown in Figure \ref{fig:tpc2}.
\begin{figure}[H]
	\centering\includegraphics[scale=0.3]{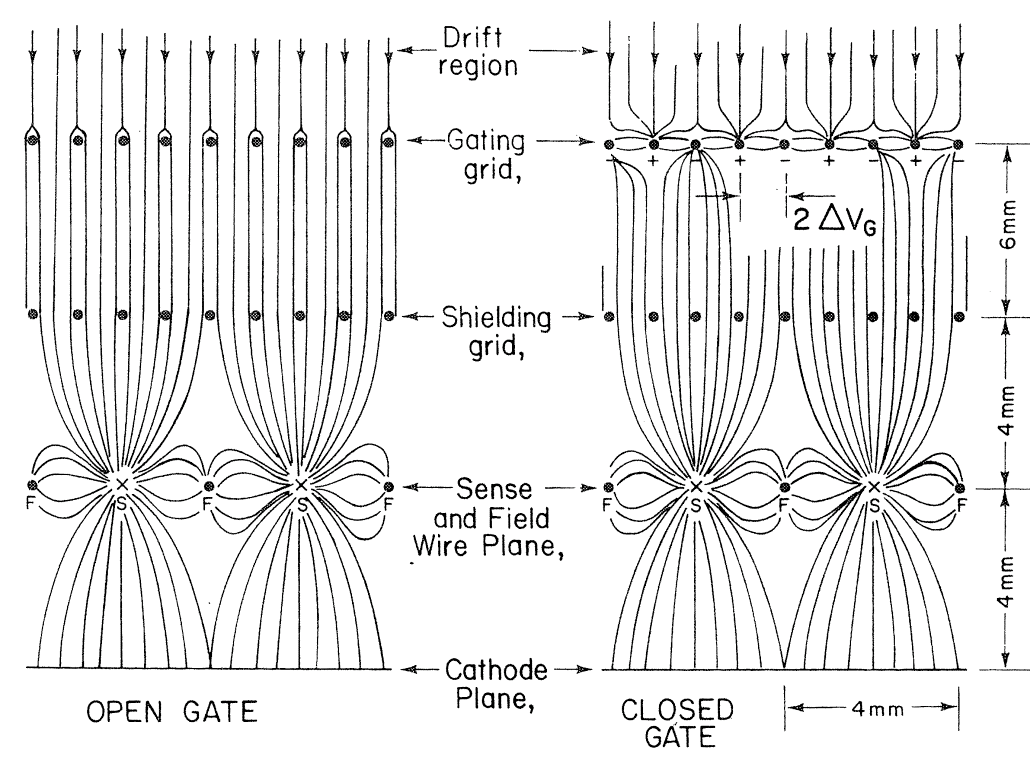}
	\caption[]{Field Map of TPC.\cite{Janssen2008}\label{fig:tpc3}}
\end{figure}
When particles pass through the TPC, they produce free electrons that drift towards the endcaps, where they are detected by the anode wires. This allows the location of a space point to be projected onto the endcap plane, with one coordinate given by the position of the firing anode wire and the second by the signals induced on the cathode pads. The third coordinate, along the axis of the TPC, is determined by the drift time of the ionization electrons.
\par To prevent the buildup of ions during amplification from disrupting the sensitive volume and distorting the electric field, a gating grid has been installed. When the gating grid is open (see Figure \ref{fig:tpc3}), the voltage on each wire is set to a common value ($V_a$) to align with the electric field in the chamber and allow for optimal transmission of drifting electrons. When the gating grid is closed in the bipolar mode (see Figure \ref{fig:tpc3}), the voltage of alternate wires is adjusted to two different values: $V_h = V_a + \Delta V$ and $V_l = V_a - \Delta V$. This grid acts as a barrier, trapping the ions in its closed position. With the proper voltage selection, the drift lines of electrons will end at the positively charged wires (connected to $V_h$), and positive ions will end at the negatively charged wires (connected to $V_l$). This effectively seals the gating grid, preventing both unwanted electrons (which do not correspond to a trigger) and ions from passing through.
\par The charge collected from the readout is proportional to the energy loss of the incident particle. Therefore, with proper calibration, it is possible to extract information about the energy loss rate, dE/dX. More details on the particle identification method can be found in \cite{deDx}.
\begin{figure}[H]
	\centering\subfloat[\cite{Lin2014SignalCO}\label{fig:megas1-sd}]{\includegraphics[scale=0.4]{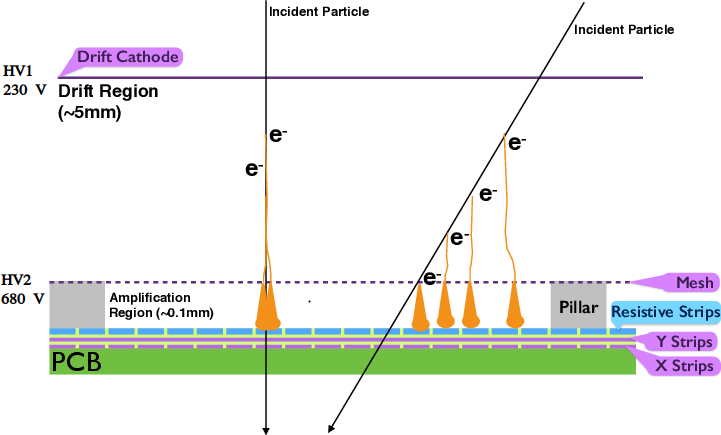}}
	
	\subfloat[\cite{FabjanSchopper_2023}\label{fig:megas1-fmap}]{\includegraphics[scale=0.4]{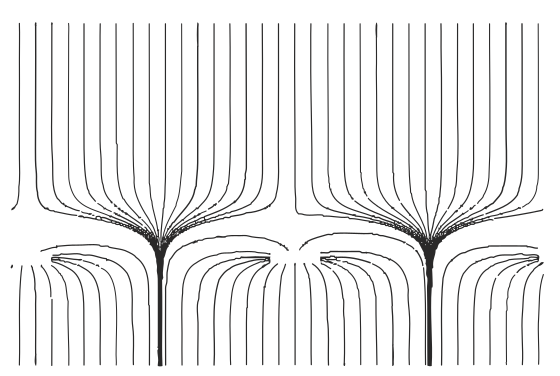}}
	\caption[]{(a) Schematic diagram  and (b)field map of micromegas.\label{fig:megas1}}
\end{figure}
  
 \begin{figure}[H]
	\centering\subfloat[\cite{GIOMATARIS199629}\label{fig:megas3-sd}]{\includegraphics[scale=0.45]{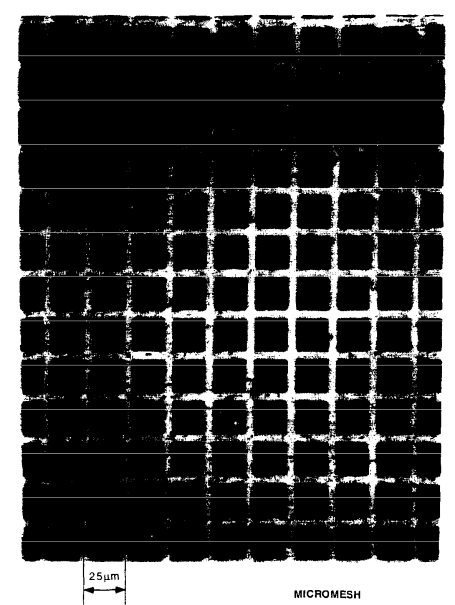}}
	
	\subfloat[\cite{GIOMATARIS199629}\label{fig:megas4-fmap}]{\includegraphics[scale=0.45]{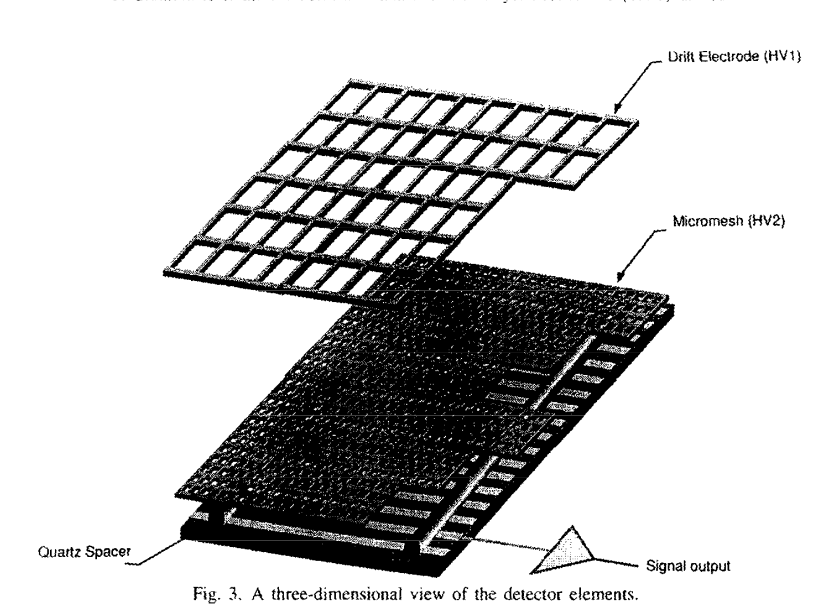}}
	\caption[]{(a) Microscopic view of micromesh  and (b) 3D view of micromegas.\label{fig:megas4}}
\end{figure}
\vspace{3cm}
 \subsection{ Parallel Plate Based Gaseous Detectors}
\subsubsection{Micromegas:}
The main limitation of MWPCs is the accumulation of a large amount of positive ions and backflow of ions to the drift regions. The drift velocity of the ions is in the order of microseconds, leading to a strong space charge effect that limits the gain of the signal and decreases the rate capability. Moreover, the spatial resolution of MWPCs depends on the separation of wires, which is approximately 1 mm.
\par To overcome these limitations, a new detector called MICROMEGAS (MICRO-MEsh-GAseous Structure) was developed and invented in 1995 by I. Giomataris and collaborators. It has a thin micromesh with a hole diameter of 25 $\mu$m (see Figure \ref{fig:megas4}) that splits the gas volume into two regions: (a) drift region and (b) amplification region, as shown in Figure \ref{fig:megas1-sd}. A gas mixture of Ar (93\%), CO$_2$ (5\%), and $\text{iC}4\text{H}{10}$ (2\%) can be used inside the gas volume \cite{Vogel}. Spacers made of insulating material maintain a distance between the anode and mesh. By applying potentials to the drift cathode, mesh, and anode, electric fields are generated in the two regions, creating a two-stage parallel plate setup.
\par The region above the mesh (ionization region) produces primary electrons via X-ray or high-energy charged particle interaction with gas. The field in this region ranges from 100 V/cm to 10 kV/cm. A high electric field of the order 68 kV/cm at the amplification region can be achieved due to the narrow gap of about 100 $\mu$m between the mesh and anodes. The field configuration creates a funnel-shaped path for field lines (see Figure \ref{fig:megas1-fmap}) through holes, attracting electrons to follow and enter the amplification region. As electrons are drawn into the amplification region, a robust field propels an electron avalanche until it reaches the anode. Remaining ions then drift towards a mesh, where the majority undergo neutralization, and a select few are discharged into the drift volume. The electrical current resulting from these movements can be measured using readout electronics located at the anode.
\par Micromegas chambers play a significant role in several physics experiments, including as a kaon beam spectrometer in the KABES experiment \cite{peyaud2004}, X-ray detectors in the CAST axion search experiment \cite{ferrerribas2007}, and as a near neutrino detector in the T2K experiment \cite{kudenko}.
    \begin{figure}[H]
 	\centering{\includegraphics[scale=0.45]{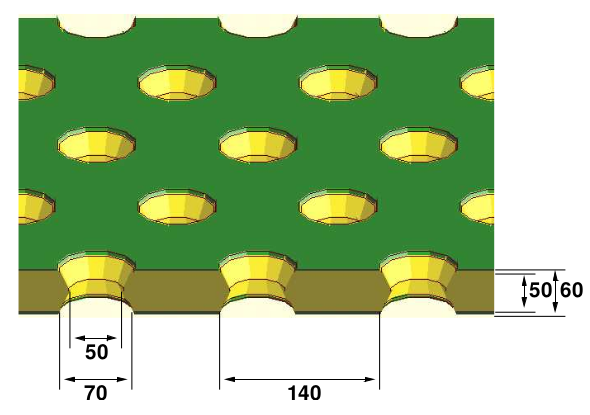}}
 	\caption[]{Dimensions of GEM foil.\label{fig:gemfoil}\cite{Hasterok2011}}
 \end{figure}
 
 \subsubsection{Gas Electron Multipliers (GEM):} 
\begin{figure}[H]
	\centering\subfloat[\label{fig:SchematicofGEM}]{\includegraphics[scale=0.4]{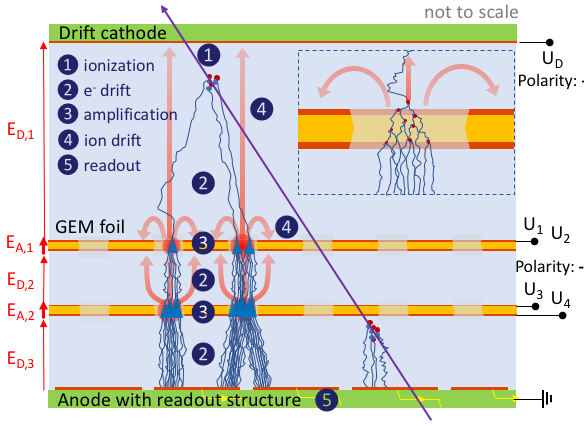}}\subfloat[\label{fig:fieldmapGEM}]{\includegraphics[scale=0.4]{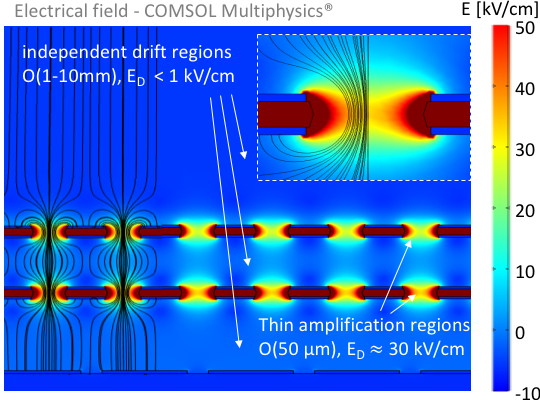}}
	\caption[]{(a) Schematic of double stage GEM  and (b) Field Map of double stage of GEM.\label{fig:fieldGEM}\cite{micromegas}}
\end{figure}

\begin{figure}[H]
	\centering\subfloat[\label{fig:readGEM}]{\includegraphics[scale=0.2]{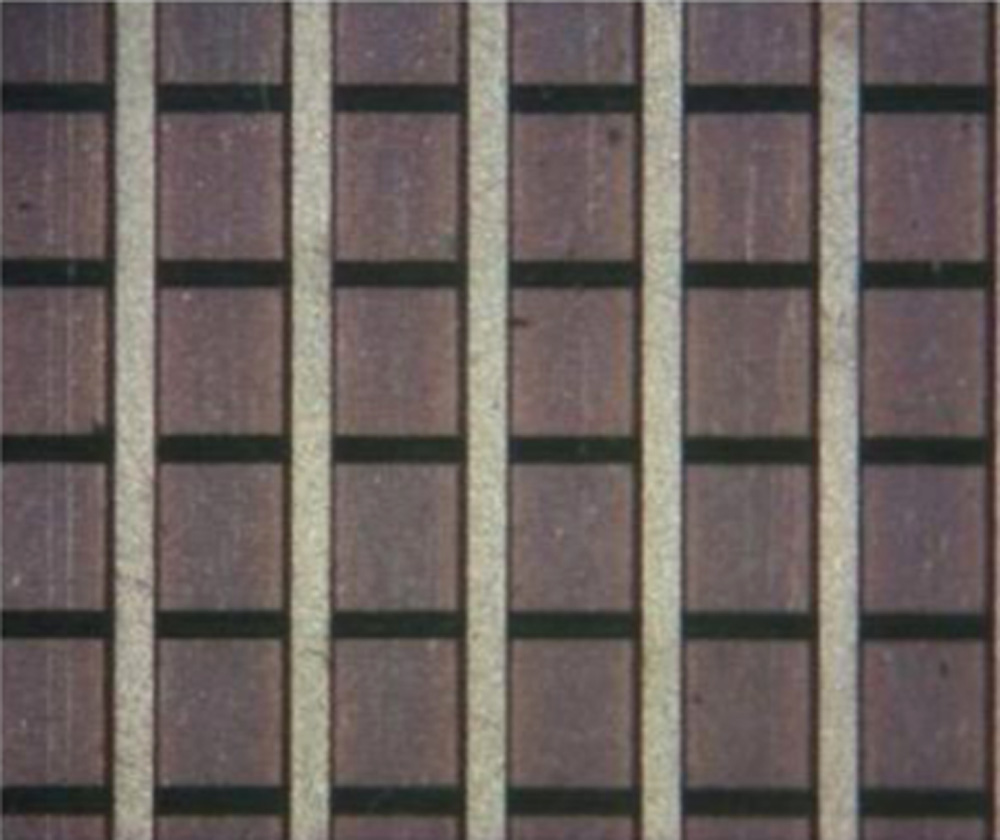}}~~\subfloat[\label{fig:sigGEM}]{\includegraphics[scale=0.25]{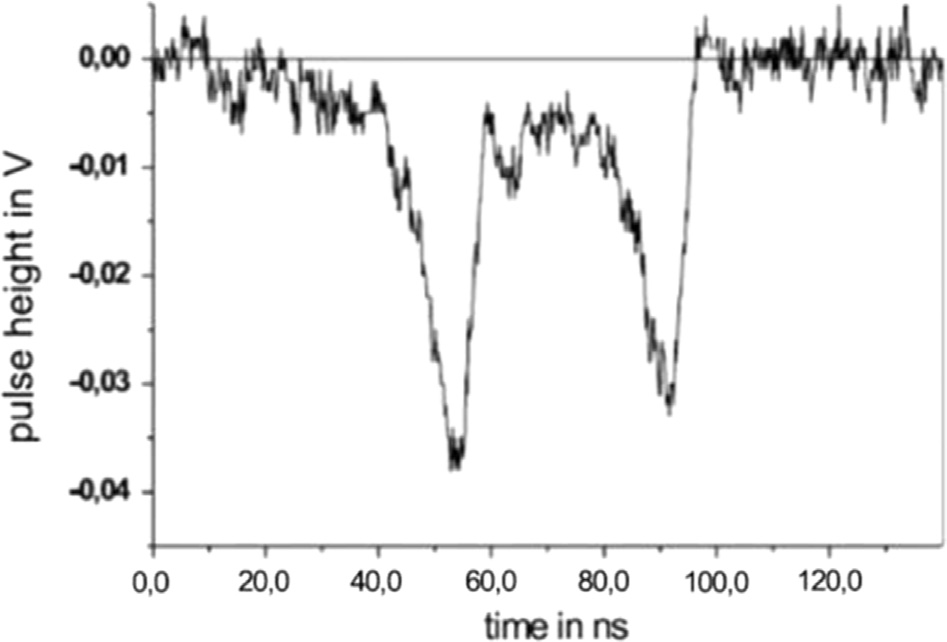}}
	\caption[]{(a)Readout strips of GEM.(b) Example of signal induced in GEM.\label{fig:gemsig}\cite{SAULI20162}}
\end{figure}

The Gas Electron Multiplier (GEM), invented by Fabio Sauli in 1997 \cite{SAULI20162}, is a particle detector used in high energy physics experiments to identify and track particles, such as in experiments like HERA-B \cite{zeuner2000}, COMPASS \cite{ketzer2004}, and LHCb \cite{raspino2007}. It consists of a thin polymer foil coated with metal and containing a high density of microscopic holes (typically 100 holes per square millimeter). The GEM foil is made of 50-micron Kapton$^{\tiny{\textcircled{R}}}$ material with copper on both sides, and has holes (50-70 microns in diameter and 100-150 microns apart) etched into it using photolithography, as shown in Figure \ref{fig:gemfoil}. The foil is placed between a drift and a collection electrode, and a voltage is applied to create an electric field. By applying specific potentials, the GEM electrode creates strong electric fields near the holes, as shown in Figure \ref{fig:fieldmapGEM}, resulting in equipotential lines and high electric fields in the holes. 
A typical gas mixture of Ar (70\%) and CO$_2$ (30\%) can be used inside the gas volume \cite{Guida}. When a charged particle enters the GEM, it can leave primary ionizations in the drift region (electric field E $<1$ kV/cm). Then, the primaries drift towards the GEM foil and are guided to the holes or amplification region by the field lines. Since the electric field inside the hole is very high (E $\approx$ 30 kV/cm), the primaries begin to multiply. The multiplication stops when the electrons leave the amplification region and reach the next section of the detector, where the electric field is relatively low (E $\approx$ 6kV). After this process, the secondary electrons are either collected at the anode for single GEM or move to the next amplification region for double GEM, as shown in Figure \ref{fig:fieldGEM}. The typical gain per single-stage GEM is on the order of $10^3$. The signal (see Figure \ref{fig:sigGEM}) is collected with the help of two-dimensional readout, as shown in Figure \ref{fig:readGEM}.
\section{Resistive Plate Chamber (RPC)}\label{sec:RPC}
\begin{figure}[H]
	\centering\includegraphics[scale=0.4]{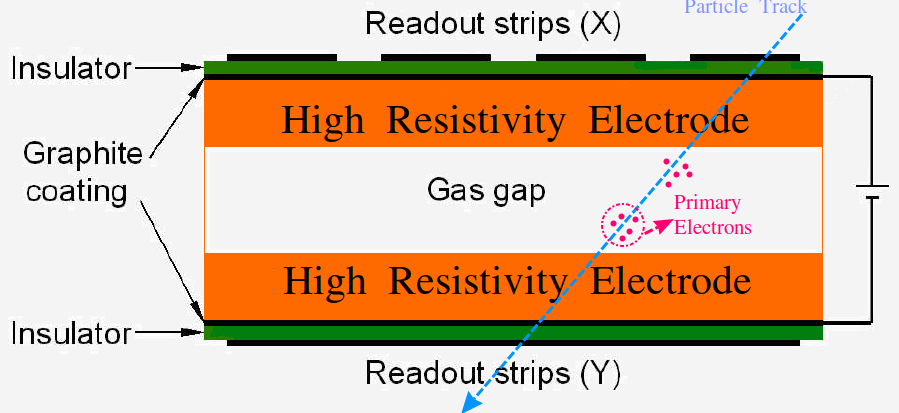}
	\caption[]{Schematic diagram of an Resistive Plate Chamber (RPC)\label{fig:schRPC}}
\end{figure}

The parallel-plate geometry detectors, with their minimal timing fluctuations, have renewed interests. Spark counters were the first parallel-plate detectors. In the 1980s, scientists Rinaldo Santonico and Roberto Cardarelli in Rome invented an effective design or prototypes of the modern Resistive Plate Chamber (RPC) \cite{cardeli-1, cardeli-2}. The RPC consists of two parallel resistive bakelite or glass plates, as shown in Figure \ref{fig:schRPC}. The space between the electrodes contains a particular gas mixture with a high absorption coefficient for ultraviolet light, and a high voltage (say $\pm5$kV) is applied to each plate. When a high-energy charged particle passes through the RPC, it can knock out primary electrons from gas molecules along its path, as shown in Figure \ref{fig:schRPC}. Due to the high electric field inside the gas gap of the RPC, those primaries are accelerated toward the electrodes and produce secondaries by the ionization process. This process continues and develops an avalanche of numerous electrons and ions, which are also called space charges. A signal pulse is induced due to the movement of electrons in  pick-up strips placed on one or both sides of the detector \cite{Sauli:2014cyf}.
During avalanche generation, the electric field around the point of the avalanche decreases. The presence of high resistive electrodes limits the spread of the affected region. As a result, the discharge does not propagate throughout the entire volume of gas, and only the avalanche region is affected. This is the fundamental difference between a spark chamber and a Resistive Plate Chamber (RPC).
\section{Working Principle of RPC}
\subsection{Interaction of particle with matter}
It is known that when a high-energy charged or neutral particle passes through matter, it can interact with the electrons and nuclei of the matter through various interaction processes. The main interactions for gaseous detectors are Coulombic and other different electromagnetic interactions. The energy loss of the incoming particle takes the following forms:
\begin{itemize}
	\item Electromagnetic interactions refer to the 
	processes by which charged particles, such as electrons, lose energy through various means. These processes include:
	\begin{itemize}
		\item Excitation: Electrons collide with atomic electrons, transferring energy and exciting the target atom.
		\item Ionization: Electrons collide with atomic electrons with enough energy to remove one or more electrons from the target atom.
		\item Bremsstrahlung: Charged particles emit radiation as they scatter off the nuclei of atoms.
	\end{itemize}
\item In addition to electromagnetic interactions, strongly interacting particles can also lose energy through hadronic interactions. These interactions involve the exchange of strong force-carrying particles, such as gluons, between the strongly interacting particles and atomic nuclei. 
\item Photons, or particles of light, can lose energy or disappear through various processes of interaction with matter. These include:
\begin{itemize}
\item Compton scattering: Photons collide with atomic electrons, transferring some of their energy and altering their wavelength in the process.

\item Photoelectric effect: Photons collide with atomic electrons, transferring all of their energy and causing the electrons to be emitted from the atom.

\item Pair production: Photons collide with atomic nuclei, resulting in the creation of an electron-positron pair and the disappearance of the photon.
\end{itemize}
\end{itemize} 
The energy loss of charged particles through ionization and excitation is a key aspect of particle detection, including the Resistive Plate Chamber (RPC) which is the subject of this thesis.
\par The average differential energy loss per unit length, due to Coulomb interactions, can be calculated using the formula established by Bethe and Bloch in the context of relativistic quantum mechanics \cite{leo1994techniques}. 
\begin{equation} \label{eqn:bethe}
-\frac{dE}{dx} = K \rho \frac{Z}{A} \frac{z^2}{\beta^2}  \left[\frac{1}{2} \ln \frac{2 m_e \gamma^2 v^2 W_{max}}{I^2} - 2\beta^2 - \delta-2\frac{C}{Z} \right]
\end{equation}

Where,

	 	K = 2$\pi N_a r_e^2 m_e c^2$ , $m_e$ = Electron mass, 	I = Mean exitation potential, 	$N_a$ =Avogadro's number, 	Z= Atomic number of target material, $z$= Charge of incident particle is in the unit of electronic charge e,  A= Atomic weight of target material, 	$r_e$=classical electron radius, $\rho$= Density of target material, 	$\beta$=v/c,  $\gamma$=$\frac{1}{\sqrt{1-\beta^2}}$, $\delta$= Density correction,  C= Shell correction, 	$W_{max}$=Maximum energy transfer in a single collision.

At head on or knok on collision the transfer of energy will be maximum, which is given below:
\begin{align}
W_{max}=\frac{2m_ec^2\eta^2}{1+2s\sqrt{1+\eta^2+s^2}}
\end{align}
Where, s=$\frac{m_e}{M}$, $\eta=\beta\gamma$ and M= Mass of incident particle.

\par
\begin{figure}[H]
	\centering\includegraphics[scale=0.45]{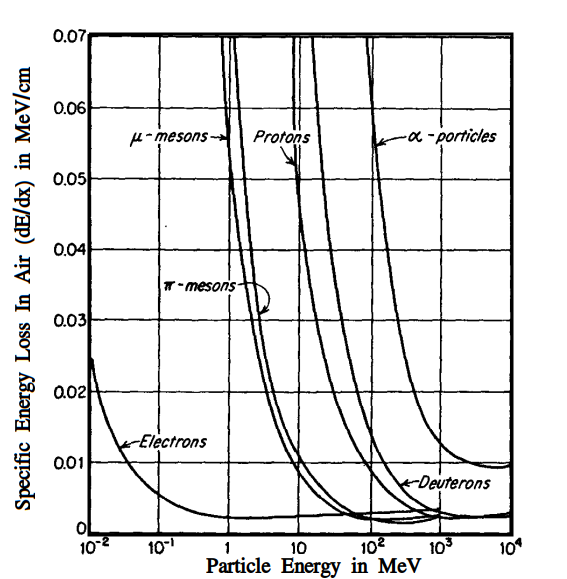}
	\caption[]{Variation of energy loss of different particle in air as a function of its energy\cite{knoll2010radiation}.\label{fig:energyloss}}
\end{figure}

 The variation of energy loss (dE/dx) as a function of incident particle energy is shown in Figure \ref{fig:energyloss}. From Figure \ref{fig:energyloss} and equation \ref{eqn:bethe}, it is clear that the value of dE/dx first decreases according to the law $\beta^{-2}$ and becomes almost constant around $\beta=0.97$, and slowly increases as the value of $\beta$ approaches 1. Therefore, at energies of a few hundred MeV, most heavy charged particles exhibit their minimum ionizing property, where the value of dE/dx for most materials is 2 MeV $g^{-1}$ $cm^2$. The lightest particle, the electron, also falls into this category at energies around 1 MeV.
\subsection{Choice of Gas mixture}
The working of RPC depends highly on the choice of the gas mixture. The criteria to decide a gas mixture for RPC is as follows:
\begin{enumerate}[(a)]
	\item High density of primary electron-ion clusters. It is important for better detection efficiency.
	\item The gas mixture in a detector can emit photons upon collision with a high-energy particle, leading to the creation of additional electron-ion pairs elsewhere in the detector. This can affect the accuracy of determining the position of the charged particle and spark in high space charge field regions. To prevent these events, the gas mixture must possess photon quenching properties.
	\item Electronegativity in the gas mixture improves the localization of discharges by reducing transversal spread.
	\item The production of chemicals such as hydrofluoric acid during electron multiplication should be limited to prevent damage to the detector and gas system components.
	\item The gas mixture used in electron multiplication should have low Global Warming Potential (GWP) to minimize its contribution to the greenhouse effect.
\end{enumerate}
In high energy physics, a commonly used RPC gas mixture consists of three gases in varying proportions to achieve the desired target rate capability and efficiency. For instance, the single-gap muon trigger system at LHC's CMS, ATLAS, and ALICE use different ratios of R134a, iso-butane, and SF6, such as 95.2\% R134a, 4.5\% iso-butane, 0.3\% $\text{SF}_6$; 94.7\% R134a, 5\% iso-butane, 0.3\% $\text{SF}_6$; and 89.7\% R134a, 10\% iso-butane, 0.3\% $\text{SF}_6$, respectively \cite{CHIODINI2006133,SColafranceschi_2012}. The gas mixture contains R134a as the primary source for secondary ionization, iso-butene as a photon quencher, and SF6 as an electron quencher. It is noted that the $\text{C}_2\text{H}_2\text{F}_4$ (R134a) and $\text{SF}_6$ gases, with a Global Warming Potential (GWP) of 1430 and 29800 respectively, must be replaced. HFO (Tetrafluoropropene) is being considered as a replacement due to its significantly lower GWP of 4. At present, it is not feasible to solely utilize HFO as a replacement in operational RPCs, as it falls short in terms of efficiency and quenching ability compared to R134a and SF6 \cite{Benussi_2015,Benussi_2015_2}.
\subsection{Primary Electron Generation}\label{sec:primary}
\begin{figure}[H]
	\centering\includegraphics[scale=0.45]{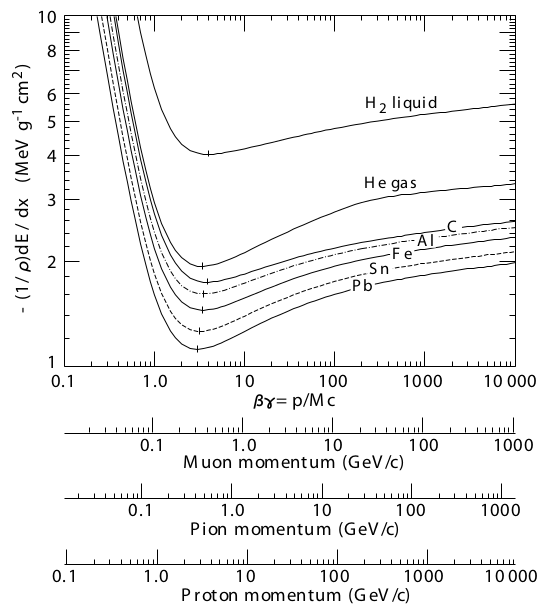}
	\caption{ Energy loss due to ionization and excitation has been studied in liquid hydrogen, gaseous helium, as well as various elements such as carbon, aluminum, iron, tin, and lead \cite{sauli_2014}. \label{fig:energyloss2}}
\end{figure}

In an RPC, as charged particles pass through the gas gap, they experience energy loss through ionization or excitation of gas atoms/molecules, which can be explaind with the help of the equation \ref{eqn:bethe}. When a charged particle collides with a gas atom in an RPC, it can cause ionization or excitation. If ionization occurs, free charge carriers are produced near the collision site. If the atom is only excited, it will quickly lose the excess energy through the emission of photons or Auger electrons. Photons with enough energy to cause Photo Electric Effect will be absorbed, while lower energy photons will escape undetected by the RPC. The amount of energy loss through ionization and excitation by a charged particle can vary based on the material, as shown in Figure \ref{fig:energyloss2}. This energy loss results in the formation of clusters of free charge carriers, or electron-ion pairs, along the particle's path. 
\par  Let us consider a single-gap Resistive Plate Chamber (RPC), where the journey of a high-energy particle through the gas gap results in the formation of primary electron-ion clusters. These clusters consist of electron-ion pairs created by the collision between the incoming particle and the gas molecules, and can potentially include additional pairs if the primary electron possesses enough energy to ionize the atoms or molecules in the gas mixture. The distribution of clusters over a volume of gas in the gap (g) is believed to follow Poisson statistics, where the interactions between the incoming particle and gas molecules are considered to be independent of each other.
If $\lambda$ is the average distance between two primary clusters  over a gas-gap g then the probability $P_{cl}$ of finding $k$ clusters can be written as follows \cite{LippmanThesis}:
\begin{align}\label{eqn:clus_pois}
P_{cl}(k)=\frac{(\frac{g}{\lambda})^k}{k!}e^{-\frac{g}{\lambda}}.
\end{align}
The intrinsic inefficiency of the detector can be found if one put k=0 in equation \ref{eqn:clus_pois} which is as follows  \cite{LippmanThesis}:
\begin{align}
P_{cl}(0)= e^{-\frac{g}{\lambda}}.
\end{align}
\par Let us consider that the distribution of primary cluster is exponential. Then the probability $P_{cl}^{1}$ of finding 1$^{st}$ cluster between $x$ and $x+dx$ is as follows:
\begin{align}
P_{cl}^{1}(x)=\frac{1}{\lambda}e^{-x/\lambda}.
\end{align}
Therefore, the probability $P_{cl}^{j}$ of finding j$^{th}$ cluster within $x$ and $x+dx$ can be written as follows \cite{RIEGLER2003144}:
\begin{align} \label{eqn:gammadist}
P_{cl}^{j}(x)=\frac{x^{j-1}}{(j-1)!\lambda^j}e^{-\frac{x}{\lambda}}
\end{align}
Therefore the spatial distribution of cluster follows Gamma ($\Gamma$) distribution with an average distance from the gas gap edge of $\bar{x}=j\lambda$.
\par
\begin{figure}[H]
	\centering\includegraphics[scale=0.45]{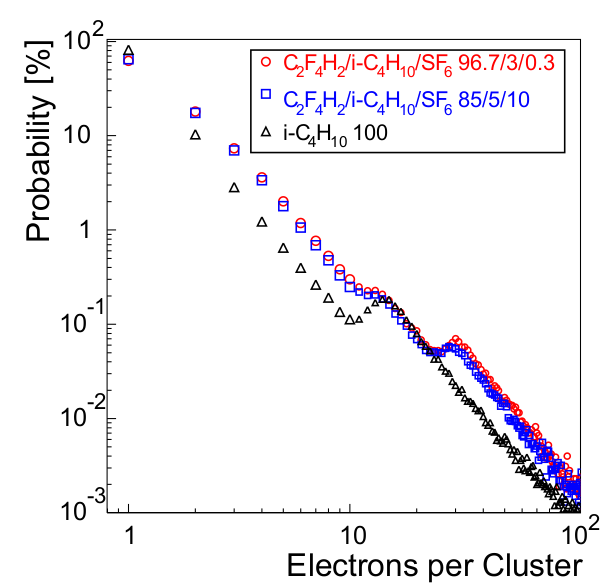}
	
	\caption[]{Primary Cluster size distribution in an RPC for three gas mixtures.\label{fig:clus_dist} \cite{RIEGLER2003144}}
\end{figure}

 The number of electrons released per cluster is determined by the level of energy exchanged during the collision, which can vary significantly. This variation is referred to as the ``cluster size" distribution. In this thesis, the primary clusters of electron-ion pairs were generated using the code HEED \cite{VEENHOF1998726,SMIRNOV2005474}. An average of 9.5 clusters/mm is found for an incident pion particle of energy 7 GeV within a single-gap RPC gas, resulting in an average inter-cluster distance of $\lambda=105 \,\mu \text{m}$. The cluster size distribution for the three gases is depicted in Figure \ref{fig:clus_dist}, calculated using ``HEED".

\subsection{Drift and Diffusion of electrons in gaseous medium}\label{sec:drift}
Electron transport in gaseous medium in an electric field is a complex process that is dependent on a variety of factors, including the strength of the electric field, the gas pressure and temperature, and the type of gas and its composition. It can be characterized by the electron mean free path ($\lambda$) and the electron drift velocity ($w^{-}$). The electron mean free path is the average distance an electron travels before colliding with another atom or molecule. The electron drift velocity is the average velocity of the electrons in the direction of the electric field. In a low electric field, the electron mean free path is relatively long, and the electrons move in a random walk pattern. However, in a high electric field, the mean free path is reduced, and the electrons move in a more direct path, resulting in a higher drift velocity. 
\par
\begin{figure}[H]
	\centering	\subfloat[\label{fig:drift-vel}]{\includegraphics[scale=0.3]{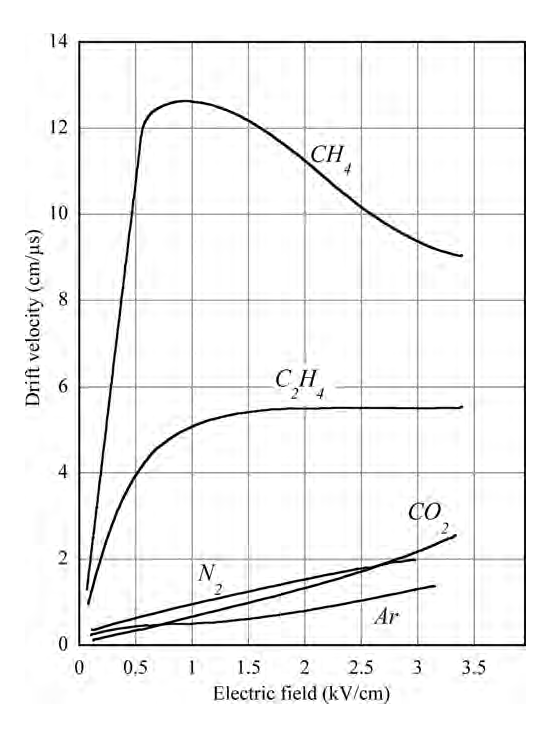}}	\subfloat[\label{fig:drift-vel2}]{\includegraphics[scale=0.35]{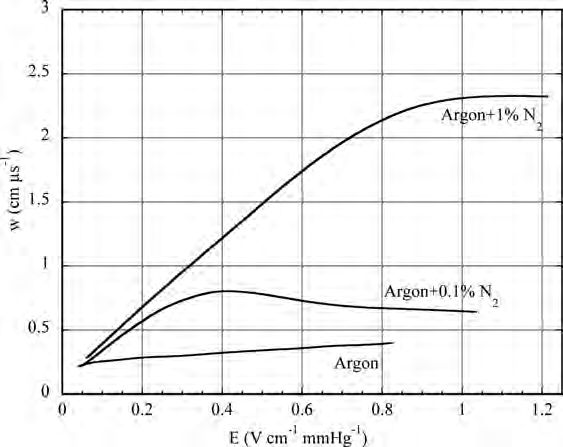}}
	\caption[]{(a)Variation of electron drift velocity with electric-ﬁeld in pure gases at NTP, (b) Effect on drift velocity of small nitrogen addition.\label{fig:drift-vels}\cite{sauli_2014}}
\end{figure}

 The formula for the drift velocity of electron of mass m and charge e in an electric field E can be written as follows \cite{sauli_2014}:
\begin{align}
w^{-}=k\frac{eE}{m}\tau,
\end{align}
where $\tau$ is the average time between collisions. The value of $k$ can range from 0.75 to 1. The graph in Figure \ref{fig:drift-vel} displays the variations in the drift velocity of various pure gases as the electric field strength changes. This graph highlights a wide range of values that the drift velocity can take on, as well as the distinct shapes of the curves that describe the relationship between drift velocity and electric field strength. When a small amount of another gas is added to another gas, it can alter the average energy and significantly impact the drift properties of the mixture. This phenomenon is demonstrated in Figure \ref{fig:drift-vel2}, which shows how the introduction of even a small amount of one gas can lead to significant changes in the drift properties of the gas mixture.
\par Electrons are known to randomly collide with gas molecules due to thermal agitation, resulting in diffusion. When an electric field is absent, diffusion is characterized by a Gaussian distribution and is isotropic in nature. If a cloud of electrons is initially concentrated at position $\vec{r}_0$ at time t=0, the distribution of its density at time t can be written as \cite{LippmanThesis}:
\begin{align}
\phi_{iso}=\frac{1}{(\sqrt{2\pi}\sigma(t)))^3} exp\left(-\frac{(r-r_0)^2}{2\sigma(t)^2}\right)
\end{align}
The sigma of this distribution is defined by
$\sigma^2 = 2Dt$. On a microscopic scale, electrons are affected by the electric field, and they interact with gas molecules through collisions. During the time between collisions, the electrons drift a distance of $\delta z$ and obtain kinetic energy denoted by $T = e_0 | E | \delta z$, where $e_0$ denotes the charge of the electron, and $|E|$ denotes the intensity of the electric field. Following each collision, the velocity of the electron reduces, leading to energy loss. However, the electron gains energy again and continues drifting until the next collision occurs. This process repeats itself as the electron slows down after each collision. 
\par The presence of an electric field has an impact on the diffusion process, resulting in the addition of a constant drift motion to the thermal diffusion and making it anisotropic. This anisotropic diffusion can be separated into two distinct terms: longitudinal and transverse. It can be expressed mathematically using cylindrical coordinates, assuming that the electron cloud exhibits rotational symmetry around the electric field axis. The equation is given by \cite{LippmanThesis}:
\begin{eqnarray}
\phi_L(z,l)&=&\frac{1}{\sqrt{2\pi l}D_L}exp\Big(-\frac{(z-z_0)^2)}{2D_L^2 l}\Big)\\
\phi_T(r,l)&=&\frac{1}{D^2_T l}exp \Big(-\frac{(r-r_0)^2}{2 D^2_T l}\Big)
\end{eqnarray}
Where,$\phi_L$ and  $\phi_T$ is longitudinal and transverse gaussian distributions, $D_L$ and $D_T$ is longitudinal and transvers diffusion constants which are calculated using MAGBOLTZ \cite{BIAGI1989716}, $z_0$ and $r_0$ is the position of center of mass of the distribution, $l$ is the drifted distance at time $t$, $r$ and $z$ are the position of electron in cylindrical co-ordinate system. The transport properties of ions are not discussed here because we have assumed that ions are stationary with respect to electrons since their drift velocity is much slower than that of electrons.
\subsection{Avalanche generation statistics in an RPC}\label{sec:avalanche}
After the primary electron-ion pairs are generated by an incident ionizing particle, if the detector has a sufficient bias electric field, the charges begin to move in the direction determined by the applied electric field and the sign of the charges. As a result, the charges acquire additional kinetic energy from the applied field and interact with other gas molecules, leading to further ionization and an avalanche.
\par It is considered that the occurrence of electron multiplication is not influenced by prior multiplication events. The progression of the avalanche is described by the Townsend coefficient ($\alpha$) and the attachment coefficient ($\eta$). The likelihood that an avalanche containing $n$ electrons at location $z$ will have $n+1$ electrons at $z+dz$ is represented as $n\alpha dz$. The probability of an electron being attached, forming a negative ion, in an avalanche of size n over a distance of $dz$ is expressed as $n\eta dz$. This results in the following relationships for the average number of electrons ($\bar{n}$) and positive ions ($\bar{p}$)\cite{raether1964electron,RIEGLER2003144}:
\begin{align}\label{eqn:avg_aval}
\frac{d\bar{n}}{dz}=(\alpha-\eta)\bar{n}, \,\,\, \frac{d\bar{p}}{dz}=\alpha \bar{n}.
\end{align}
With the boundary conditions $\bar{n}(0)=1$ and $\bar{p}(0)=0$ the solution of equation \ref{eqn:avg_aval} can be written as:
\begin{align}
\bar{n}(z)=e^{(\alpha-\eta)z},\,\, \bar{p}=\frac{\alpha}{\alpha-\eta}\left[e^{(\alpha-\eta)z}-1\right].
\end{align}
Thus the average number of negative ions is then $\bar{p} - \bar{n}$. The next task is to find different model to discuss the statistical fluctuation of the avalanche. There are some such model which is (a) The Riegler-Lippmann-Veenhof model, (b) Yule-Furry Model.
\subsubsection{The Riegler-Lippmann-Veenhof model}
\begin{figure}[H]
	\centering\includegraphics[scale=0.32]{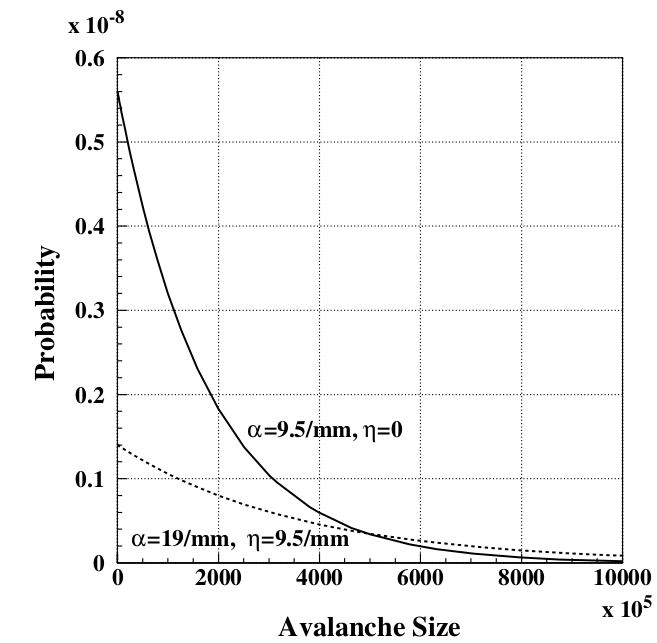}
	\caption[]{Distribution of avalanche charges with the same $\alpha_{eff}$ for both curves \cite{RIEGLER2003144}\label{fig:Aval_size}.}
\end{figure}

The determination of the probability $P(n,z)$, signifying the likelihood of an avalanche started by a single electron to consist of $n$ electrons after a distance $z$ has been covered, is presented as follows \cite{RIEGLER2003144}.
\begin{align}\label{eqn:aval_prob}
P(n,z+dz)=&\,P(n-1,z)(n-1)\alpha\,dz\,(1-(n-1)\eta dz) \nonumber\\
         +&\,P(n,z)(1-n\alpha dz)(1-n\eta\,dz)\\ \nonumber
         +&\,P(n,z)\, n\alpha\,dz\, n\eta\,dz \\\nonumber
         +&\,P(n+1,z)(1-(n+1)\alpha\,dz) \, (n+1)\eta\,dz.\nonumber
\end{align}
In equation \ref{eqn:aval_prob}, the probability of finding n electrons at position $z + dz$ is represented through four possible scenarios, as follows:
\begin{enumerate}
	\item The probability of $n-1$ electrons being located at $z$, with one of them duplicating and none being attached, is indicated in the first line.
	\item The probability of $n$ electrons being found at $z$, with no duplication or attachment, is depicted in the second line.
	\item The third line shows the probability of one electron multiplying and one being attached from a total of $n$ electrons.
	\item In the fourth line, the probability of one electron being attached and no duplication occurring from $n+1$ electrons is presented.
\end{enumerate}
The differential probability can be derived from equation \ref{eqn:aval_prob}, taking into account only the first order term of dz, as indicated below:
\begin{align}\label{eqn:diff_prob}
\frac{dP(n,z)}{dz}=&-P(n,z)\,n\,(\alpha+\eta)+P(n-1,z)\,(n-1)\alpha \nonumber \\
				   &+P(n+1,z)(n+1)\eta.	
\end{align}
A general solution of equation \ref{eqn:diff_prob} can be derived as follows:
\begin{align}\label{eqn:prob_solv}
P(n,z)=& k\frac{\bar{n}(z)-1}{\bar{n}(z)-k},\,\,\,&n=0 \\
	=&\bar{n}(z)(\frac{1-k}{\bar{n}(z)-k})^2(\frac{\bar{n}(z)-1}{\bar{n}(z)-k})^{n-1},\,\,\,&n>0,
\end{align}
where,
$k=\frac{\eta}{\alpha}$.
The variance $\sigma^2(z)$ of the distribution can be determined as follows:
\begin{align}
\sigma^2(z)=\left(\frac{1+k}{1-k}\right)\bar{n}(x)(\bar{n}-1).
\end{align}
For $\alpha=\eta$ and $\alpha=0$ the above equation is invalid. Therefore, for $\alpha=\eta$ we need to use following equation:
\begin{align}\label{eqn:prob_solv2}
P(n,z)=&\frac{\alpha z}{1+\alpha z}, &n=0 \nonumber\\
      =&\frac{1}{(1+\alpha z)^2}\left(\frac{\alpha z}{1+\alpha z}\right)^{n-1}, &n>0
\end{align}
and variance becomes,
\begin{align}
\sigma^2(z)=2\alpha z.
\end{align}
In case $\alpha=0$ we need to use following equations:
\begin{align}\label{eqn:prob_solv3}
P(n,z)=&1-e^{\eta z},&n=0 \nonumber\\
	  =&e^{-\eta z}, &n=1 \\\nonumber
	  =&0.            &n>1 \nonumber
\end{align}
The corresponding variance for this case is,
\begin{align}
\sigma^2(z)=e^{-2\eta z}(e^{\eta z}-1)
\end{align}
To simulate an avalanche we need to generate random number according to equation \ref{eqn:prob_solv} as described in \cite{RIEGLER2003144}. Therefore if one draw a uniform random number R between 0 and 1 then the value of n can be estimated form the following equation \cite{RIEGLER2003144}:
\begin{align}\label{eqn:rnd1}
n=&0, &R<k\frac{\bar{n}-1}{\bar{n}-k} \nonumber\\
 =&1+ \text{Trunc}[\frac{1}{\text{ln}(1-\frac{1-k}{\bar{n}-k})}\text{ln}(\frac{(\bar{n}-k)(1-R)}{\bar{n}(1-k)})], &R>k\frac{\bar{n}-1}{\bar{n}-k}
\end{align}
In the above equation \ref{eqn:rnd1} Trunc means the truncation of decimals.
Similarly, one can generate random number form equaiton \ref{eqn:prob_solv2} as follows \cite{RIEGLER2003144}:
\begin{align}
n=&0,	&R<\frac{\alpha z}{1+\alpha z} \nonumber\\
 =&1+\text{Trunc}\left[\frac{1}{\text{ln}(\frac{1}{\frac{\alpha z}{1+\alpha z}})}\text{ln}\left((1-R)(1+\alpha z)\right)\right], &R>\frac{\alpha z}{1+\alpha z}.
\end{align}
Again, with the equation \ref{eqn:prob_solv3} the generation of random number follows \cite{RIEGLER2003144}:
\begin{align}
n=&0, &R<e^{-\eta z}\nonumber \\
 =&1, &R>e^{-\eta z}
\end{align}
It is observed that the average number of electrons depends on the effective Townsend coefficient $\alpha_{eff} = \alpha - \eta$. However, the probability distribution depends on $k = \frac{\eta}{\alpha}$. In Figure \ref{fig:Aval_size}, the avalanche charge distribution for two different values of $\alpha$ and $\eta$ is shown. The values of $\alpha$ and $\eta$ have been chosen such that the value of $\alpha_{eff} = \alpha - \eta$ remains the same.
\subsubsection{Yule-Furry Model}\label{sec:ch2yuleFurry}
\begin{figure}[H]
	\centering\includegraphics[scale=0.4]{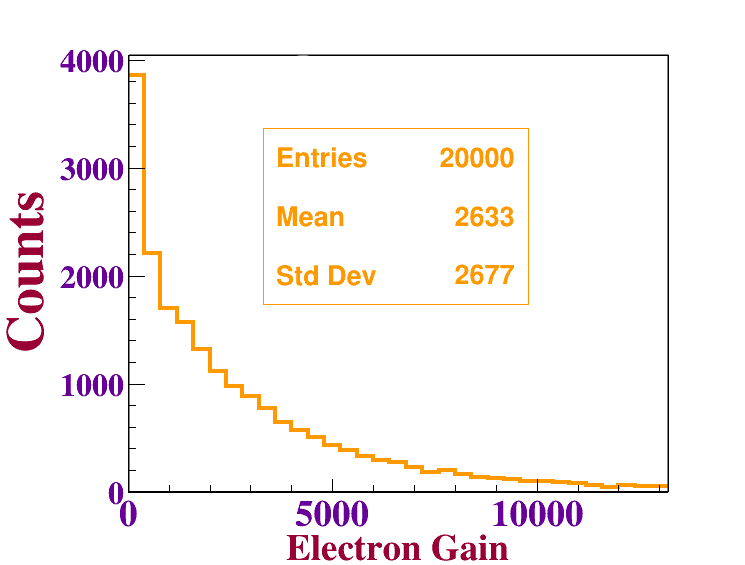}
	\caption[]{Distribution of avalanche charges calculated by using Garfield++ \label{fig:gain_garfield}.}
\end{figure}

\begin{figure}[H]
	\centering	\subfloat[\label{fig:aval}]{\includegraphics[scale=0.35]{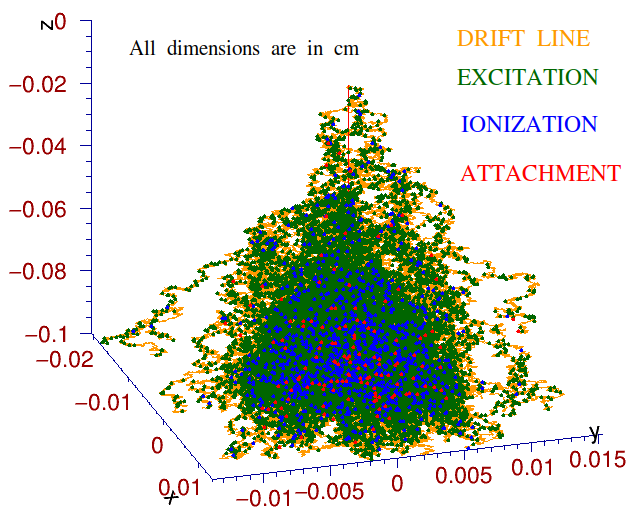}}	\subfloat[\label{fig:signal}]{\includegraphics[scale=0.3]{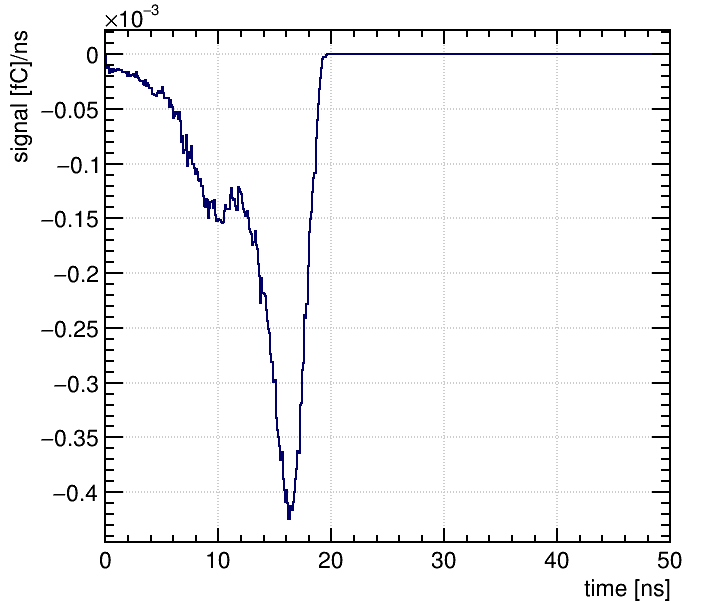}}
	\caption{(a) Drift line of avalanche electrons in an RPC, generated using Garfied++ (color signifies different phenomenon and quantities), (b) Corresponding signal of the same avalanche.\label{fig:avalSignal}}
\end{figure}

The Furry Model \cite{Furry} was first introduced in 1937  to examine the variation in the size of electromagnetic cascades from cosmic rays in lead. This model uses the same mathematical framework as the Townsend Avalanche. However, in this model, the contribution of the attachment coefficient has not been taken into account. In Garfield++, a modified version of this model has been used to generate avalanches, which considers the attachment phenomenon. The algorithm to calculate electron gain at each step of the avalanche is described below and is implemented in the AvalancheMC class of Garfield++ \cite{Garfield}:
\begin{enumerate}
	\item In the simulation a drift step (dx) can be further sub-divided such that each subdivision can compare to the inverse of the Townsend coefficient $\alpha^{-1}$. We need this subdivision to calculate gain within the interval of dx.
	\item The Townsend coefficient $\alpha$ and $\eta$ is function of the electric field at location dx. Therefore, the value of the same determined using MAGBOLTZ. 
	\item A length dp of subdivision is considered as 0.01. Therefore, the number of subdivision nDiv is:
	\begin{equation}
	 \text{nDiv}=\frac{\alpha +\eta}{dp}
	\end{equation}

	\item Probability of ionisation (p) and attachment (q) is calculated as follows:  
	 \begin{align}
      p=&\frac{\alpha}{\text{nDiv}}, \\
      q=&\frac{\eta}{\text{nDiv}}
	 \end{align}
	 \item A random number between 0 and 1 is selected (u $\in$ [0,1]). If this number is less than a certain value p i.e $u < p$, a new electron is created. Alternatively, if the random number is less than a different value q i.e $u<q$, an electron is attached.	 
\end{enumerate} 
The process being described is essentially an application of the Yule-Furry process using Monte Carlo methods. As a result of this process, the size distribution of the resulting avalanches is expected to follow an exponential distribution. We have used this method to generate avalanches throughout this thesis. An example of the distribution of the gain for a single-gap RPC with a 2 mm gas gap, gas mixture of $\ce{C_2H_2F_4}$ (85\%), $\ce{i-C_4H_{10}}$ (5\%), $\ce{SF_6}$ (10\%), and an applied field of 43 kV/cm calculated using the above algorithm has been shown in Figure \ref{fig:gain_garfield}. Garfield++ has the capability to remember the drift position, ionization point, excitation point, and attachment point of each avalanche electron and ion. An example of the profile (drift line, excitation, ionization, attachment) of an avalanche generated using Garfield++ in an RPC of the same configuration is shown in Figure \ref{fig:aval}, and the corresponding signal is shown in Figure \ref{fig:signal}.
\subsection{Signal generation in an RPC}\label{sec:signal}
\begin{figure}[H]
\centering	\subfloat[\label{fig:total_g}]{\includegraphics[scale=0.45]{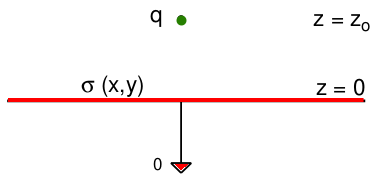}}	\subfloat[\label{fig:strip_g}]{\includegraphics[scale=0.45]{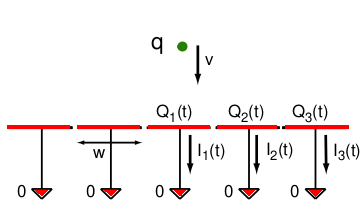}}
	\caption[]{A charge density $\sigma(x,y)$ on a grounded metal electrode is induced by a charge q located at $z=z_0$ with respect to the plane. (b) A moving charge q induces different currents on segmented metal electrodes.\label{fig:point_q_grounded_electrode} \cite{rieglerbook}}
\end{figure}

The electric field defines the direction of movement of the electron-ion pair generated during the avalanche inside the RPC. A signal current may be induced in the electrodes of the RPC due to the movement of charges inside it. To understand signal generation, let us first consider the simplest case. Let us consider a point charge located at M(0, 0, $z=z_0$) in front of a grounded electrode placed at z = 0, as shown in Figure \ref{fig:total_g}. By using the method of images, the electric field at any point P(x, y, 0) on the metal surface can be determined. The electric field is given by \cite{rieglerbook}:
\begin{align}
E_z(x,y) = -q\frac{2z_0}{2\pi\epsilon_0\left(x^2+y^2+z_0^2\right)^{\frac{3}{2}}},\,\,\, E_x=0,\,\,\,E_y=0,
\end{align}
where, $\epsilon_0$ is the free space permittivity.
The charge desity $\sigma (x,y)$ on the elctrode can be found using the formula $\sigma (x,y)=\epsilon_0 E_z(x,y)$. Therefore the total charge on the elecrode can be found as follows \cite{rieglerbook}:
\begin{align}
Q^{ind}=\int_{-\infty}^{-\infty} \int_{-\infty}^{-\infty} \sigma(x,y)\,dx\,dy=-q.
\end{align}
In the case when the elctrode is segmented into several strips of width $w$ as shown in Figure \ref{fig:strip_g}, the indudced charge on the central srip can be given as \cite{rieglerbook}:
 \begin{align}\label{eqn:ind_charge_strip}
 Q^{ind}_{strip}(z_0)=\int_{-\infty}^{-\infty} \int_{-w/2}^{-w/2} \sigma(x,y)\,dx\,dy=-\frac{2q}{\pi}\text{arctan}\left(\frac{w}{2z_0}\right).
 \end{align}
From equation \ref{eqn:ind_charge_strip}, it can be seen that the induced charge depends on the z position ($z_0$) of the charge. If the charge is moving with velocity $v$, then $z_0$ becomes a function of time $t$, expressed as $z_0 = z_0(t)$. As a result, the induced current flowing between the electrode and ground can be written as \cite{rieglerbook}:
\begin{align}
I^{ind}_{strip}(t)&=-\frac{d}{dt}Q^{ind}_{strip}(z_0(t))\\ \nonumber
&= -\frac{\partial Q^{ind}_{strip}(z_0(t))}{\partial z_0}\frac{d z_0(t)}{dt}\\ \nonumber
&=\frac{4\,q\,w}{\pi \left[4z_0(t)^2+w^2\right]}v.
\end{align}  
\subsubsection{Calculation of signal using Ramo's theorem} 
\begin{figure}[H]
	\centering	\subfloat[\label{fig:ramo-electrode1}]{\includegraphics[scale=0.45]{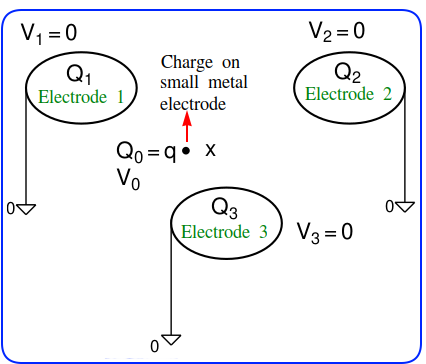}}	\subfloat[\label{fig:ramo-electrode2}]{\includegraphics[scale=0.45]{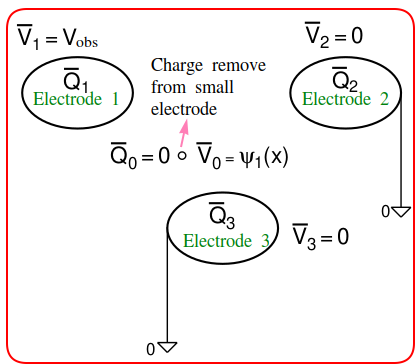}}
	\caption[]{(a) The point charge q induces charges $Q_n$ on the grounded electrodes. (b) A second electrostatic configuration with different voltages and charges is used to calculate the weighting potential. \cite{rieglerbook}.\label{fig:ramo-electrode}}
\end{figure}

We have a point charge $Q_0 = q$ at position x near three grounded electrodes, as shown in Figure \ref{fig:ramo-electrode1}. The point charge is on a infinitely small metal electrode, making a total of four metal electrodes. The voltages at the three electrodes are all zero ($V_1 = V_2 = V_3 = 0$). We want to find out the charges induced, $Q_1$, $Q_2$, and $Q_3$, due to the presence of $Q_0 = q$.
\par Another electrostatic configuration is selected, as depicted in Figure \ref{fig:ramo-electrode2}, where the charge q is removed from the infinitely small electrode, resulting in $\bar{Q}_0 = 0$. The potential $\bar{V}_0$ associated with the charge $\bar{Q}_0$ is called as the weighting potential. Electrode 1 is assigned a voltage of $V_{obs} (\bar{V}_1 = V_{obs})$, while the remaining electrodes are kept grounded ($\bar{V}_2 = \bar{V}_3 = 0$). With these conditions, from Green’s reciprocity theorem it can be written as \cite{rieglerbook}:
\begin{align}
q\bar{V}_0+Q_1\bar{V}_1+Q_2\bar{V}_2+Q_3\bar{V}_3&=\bar{Q}_0V_0 \\ \nonumber
q\bar{V}_0+Q_1V_{obs}&=0 \\ 
Q_1&=-q\frac{\bar{V}_0}{V_{obs}}. \label{eqn:q1charge}
\end{align}
When the charge is moving with a velocity $v_d$ then the weighting potential $\bar{V}_0$ will be function of the position. Therefore, it is considered that $\bar{V}_0=\psi(x(t))$ and equation \ref{eqn:q1charge} can be writen as \cite{rieglerbook}:
\begin{align}
Q_1(t)&=-q\frac{\psi(x(t))}{V_{obs}}.
\end{align}
The induced current on electrode 1 of Figure \ref{fig:ramo-electrode1}  can be found as follows \cite{rieglerbook}:
\begin{align}
I^{ind}_{1}(t)&=-\frac{d}{dt}Q_{1}(t)\\ \nonumber
&= \frac{q}{V_{obs}}\nabla \psi(x)\frac{d x(t)}{dt}\\ 
&=-\frac{q}{V_{obs}}E_1(x)v_d,
\end{align} 
where, $E_1(x)=-\nabla \psi(x)$, and $E_1$ is called the weighting field of electrode 1.  The induced current in the remaining electrodes can be determined using a similar approach.
\begin{figure}[H]
	\centering\includegraphics[scale=0.45]{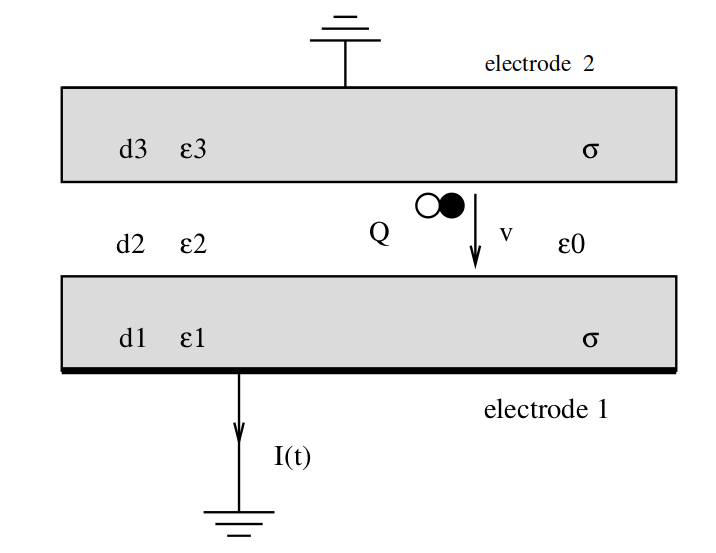}
	\caption[]{Current induced on the electrodes of an RPC due to movement of charge. \cite{RIEGLER2002258}.\label{fig:rpcSignal}}
\end{figure}
\subsubsection{Weighting field for a single gap RPC}
\begin{figure}[H]
	\centering\subfloat[\label{fig:fieldlines_w}]{\includegraphics[scale=0.3]{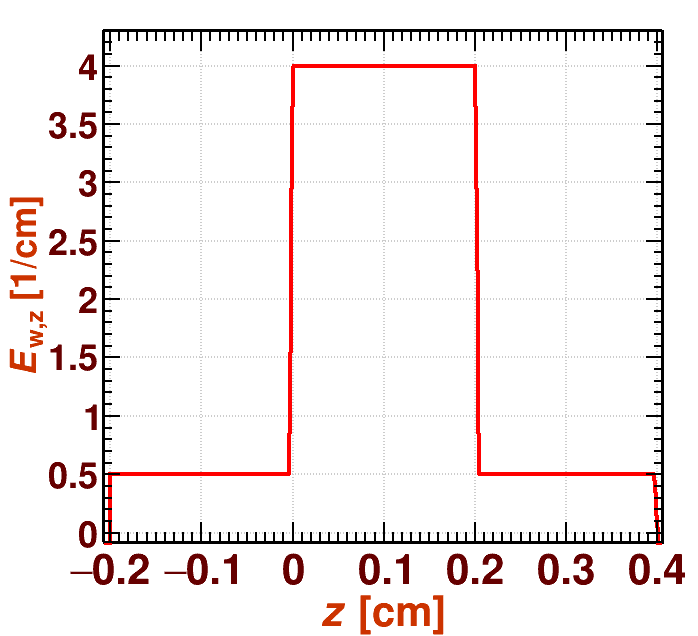}}
	
	\centering\subfloat[\cite{LippmanThesis}\label{fig:weightfield}]{\includegraphics[scale=0.25]{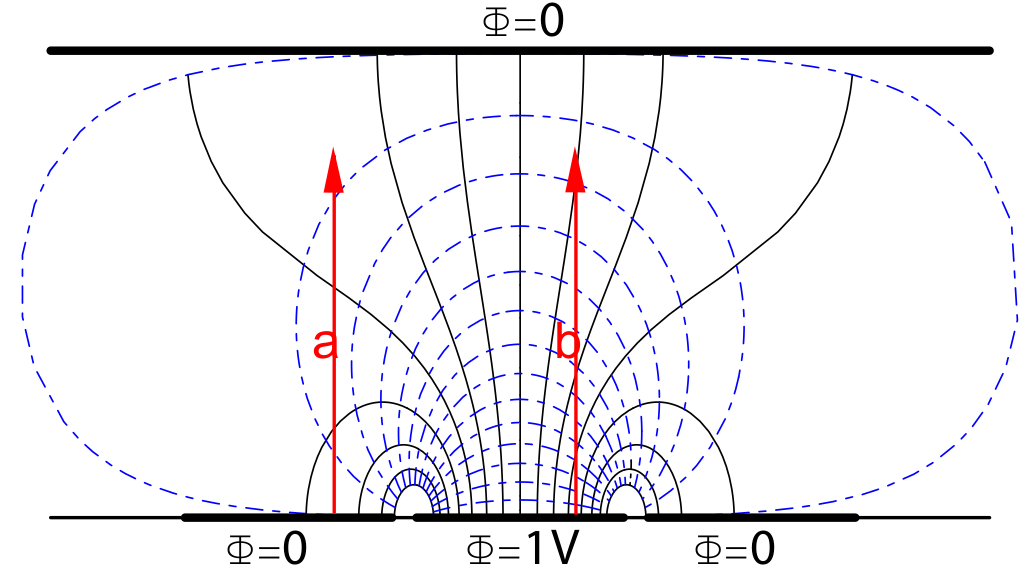}}
	\caption[]{(a) The weighting field for an RPC with a gas gap of 2 mm, calculated using neBEM. (b) Field lines and equipotential lines of the weighting field for segmented strip electrodes.\label{fig:weightrpc}}
\end{figure}

The calculation of z-component of weighting field $E^{weight}_z$ for an electrode 1 of RPC showed in Figure \ref{fig:rpcSignal} can be approximated as follows \cite{RIEGLER2002258}:
\begin{align}\label{eqn:weightingfieldrpc}
E^{weight}_z=\frac{v_0\,\epsilon_1\epsilon_3}{\epsilon_2\epsilon_3d_1+\epsilon_1\epsilon_3d_2+\epsilon_1\epsilon_2d_3},
\end{align}

where, $v_0$ represents the potential on electrode 1, $d_1$ and $d_3$ are the thicknesses of electrodes 1 and 2 respectively, $d_2$ is the thickness of the gas gap, $\epsilon_1$ and $\epsilon_3$ are the permittivities of electrodes 1 and 2, and $\epsilon_2$ is the permittivity of the gas within the gas gap.
Finding the analytical expression of the weighting field for a single gap Resistive Plate Chamber (RPC) with a three layer geometry, as given in equation \ref{eqn:weightingfieldrpc}, was straightforward. However, more complex designs with multiple layers require the use of numerical field solvers such as neBEM \cite{MAJUMDAR2008346,MAJUMDAR2009719}, COMSOL\cite{comsol} for the calculation of the weighting field. An example of  weighting field along z-axis of an RPC of 2 mm gas gap, calculated using neBEM, is shown in Figure \ref{fig:fieldlines_w}. In RPCs, pick-up or signal strips are typically segmented, as shown in Figure \ref{fig:weightfield}. The field lines and equipotential lines for a central strip are also displayed in this figure, with the central strip at a voltage of 1 V and the rest of the electrodes grounded. 
\subsubsection{Calculation of induced current due to an avalanche in an RPC}
It is understood that the current induced in the electrode is caused by the movement of electrons and ions generated during the avalanche process. However, the current induced by ions is weaker than the current induced by electrons due to the slower drift velocity of ions compared to electrons. If N(t) is the number of charges in a cluster moving with a velocity $v_d$, then the induced current $i(t)$ at time t can be determine using Ramo's theorem as follows \cite{LippmanThesis}:
\begin{align}
i(t)= E_z^{w}(x_j(t))\,v_d\,e_0\,N(t),
\end{align}
where $e_0$ is the unit charge and $E_z^{w}$ is the weighting field inside the gas-gap when the central strip is at $V_{obs}=1\,V$ (see Figure \ref{fig:weightfield}).
Again, if there present a number of such $n_{cl}$ clusters then the total current is the sum over all clusters, which is as follows \cite{LippmanThesis}:
\begin{align}
i(t)=\sum_{j=1}^{n_{cl}} E_z^{w}(x_j(t))\,v^j_d\,e_0\,N_j(t).
\end{align}

\subsection{Operation modes of RPC}\label{sec:modes}
RPCs, or Resistive Plate Chambers, can function in two different modes: (a) avalanche mode and (b) streamer mode.
\subsubsection{Avalanche Mode}
The avalanche mode, also called the limited proportional mode, refers to a process where the initial release of primary charge caused by ionizing radiation is followed by the multiplication and propagation of electrons, leading to what is known as a Townsend avalanche. The physics behind the formation of avalanches and their resulting signals have been explained in detail in sections from the primary charge release to the signal generation.

The behavior of the avalanche changes as the gas gain increases, and the electron gain in this mode is limited. Typically, the mean charge produced in this mode is around 1 pC, which may require a pre-amplifier to process the avalanche pulses.

Despite this, avalanche mode has several advantages. It provides good time resolution, experiences less distortion in the applied field due to the reduced space charge effect, and is suitable for high particle rate applications since it generates low charges.
\subsubsection{Streamer mode}
\begin{figure}[H]
	\centering\includegraphics[scale=0.4]{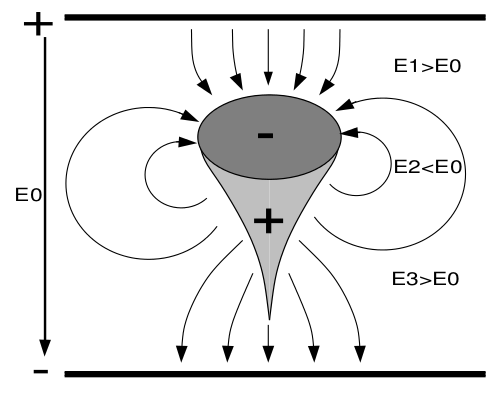}
	\caption{Distortion of applied electric field $E_0$ inside an RPC while an avalanche is developing \cite{Lippmann_1}.\label{fig:aval_elec}}
\end{figure}
The streamer mode Resistive Plate Chambers (RPCs) are capable of generating signals of relatively larger magnitude, ranging from 50 pC to a few nC. As a result, no amplification is required in detectors to detect the signals, and they can be directly discriminated against the detection threshold. This makes the readout process for streamer mode RPCs relatively simple. However, streamers also pose several issues, including a strong space charge effect, low time resolution, ageing of electrodes, and less rate capability. In general, a strong space charge effect is the key to starting streamers. As an avalanche develops, the field of avalanche charges distorts the detector field which is called space charge effect. As a result, a high-field region is developed at the tip of the avalanche and a low-field region is developed in the middle region of the avalanche as shown in Figure \ref{fig:aval_elec}. After a certain time, the avalanche becomes saturated, and the probability of gain and loss becomes almost equal. A detail discussions on space charge effect can be found in he next chapters.
\begin{figure}[H]
	\centering\subfloat[\label{fig:streamerpulse}]{\includegraphics[scale=0.32]{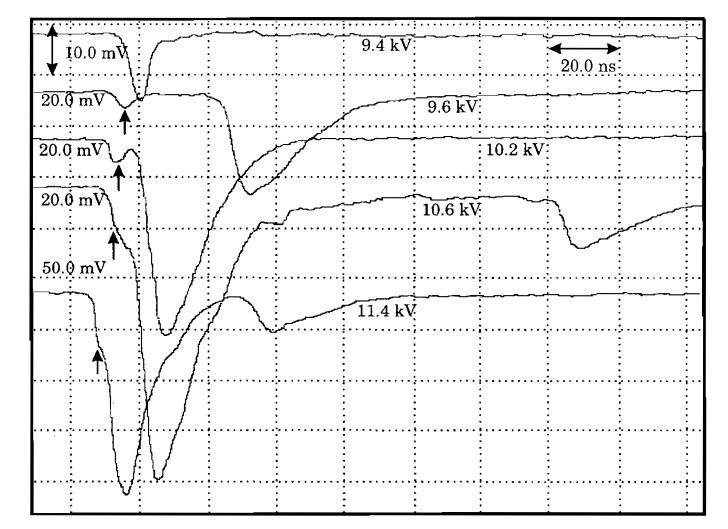}}\subfloat[\label{fig:streamerdelay}]{\includegraphics[scale=0.28]{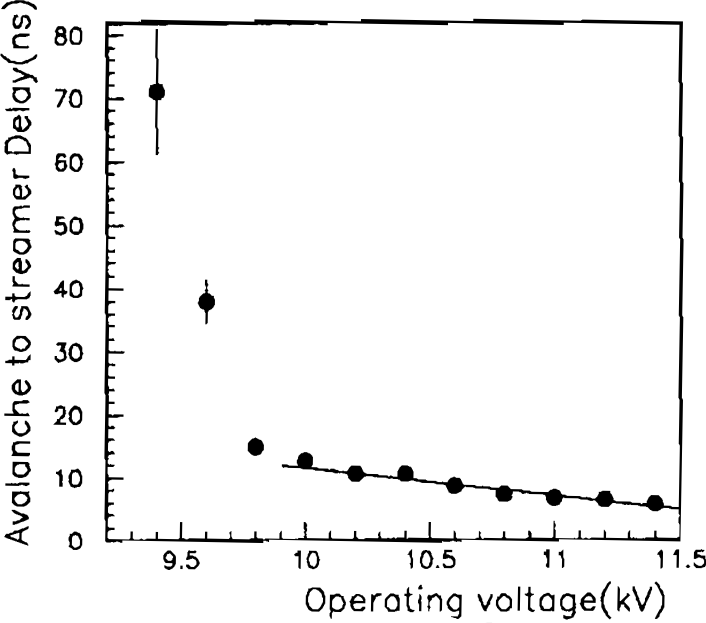}}
	\caption{(a) The experimentally measured RPC pulses for different voltages consist of two pulses. The first pulse corresponds to the precursor signal (indicated by a vertical up arrow), and the second pulse corresponds to the streamer. (b) Variation of the average delay between the precursor and streamer pulse. \cite{cardarelli2000avalanche}.\label{fig:streamer}}
\end{figure}
\par Streamer formations can arise from a sequence of avalanches occurring in a gap, resulting from a high influx of primary particles or secondary avalanches from photoelectric effect. There are two known possibilities to generate electrons near the high field regions and start a streamer. Firstly, when an antecedent avalanche emits UV-photons that aren't absorbed, it can knock out electrons from the cathode surface within a few millimeters of its radial direction. Secondly, UV-photons ionize the gas near the high field region, creating these electrons, which then generate successive avalanches that may eventually convert into a streamer. It has been experimentally observed that primary avalanches produce precursors before the formation of a streamer. The time delay between the precursor and the streamer depends on the applied voltage, as depicted in Figure \ref{fig:streamerpulse} and the average time delay at different voltages is shown in Figure \ref{fig:streamerdelay}.
\subsection{Various design of RPCs}
\begin{figure}[H]
	\centering\includegraphics[scale=0.25]{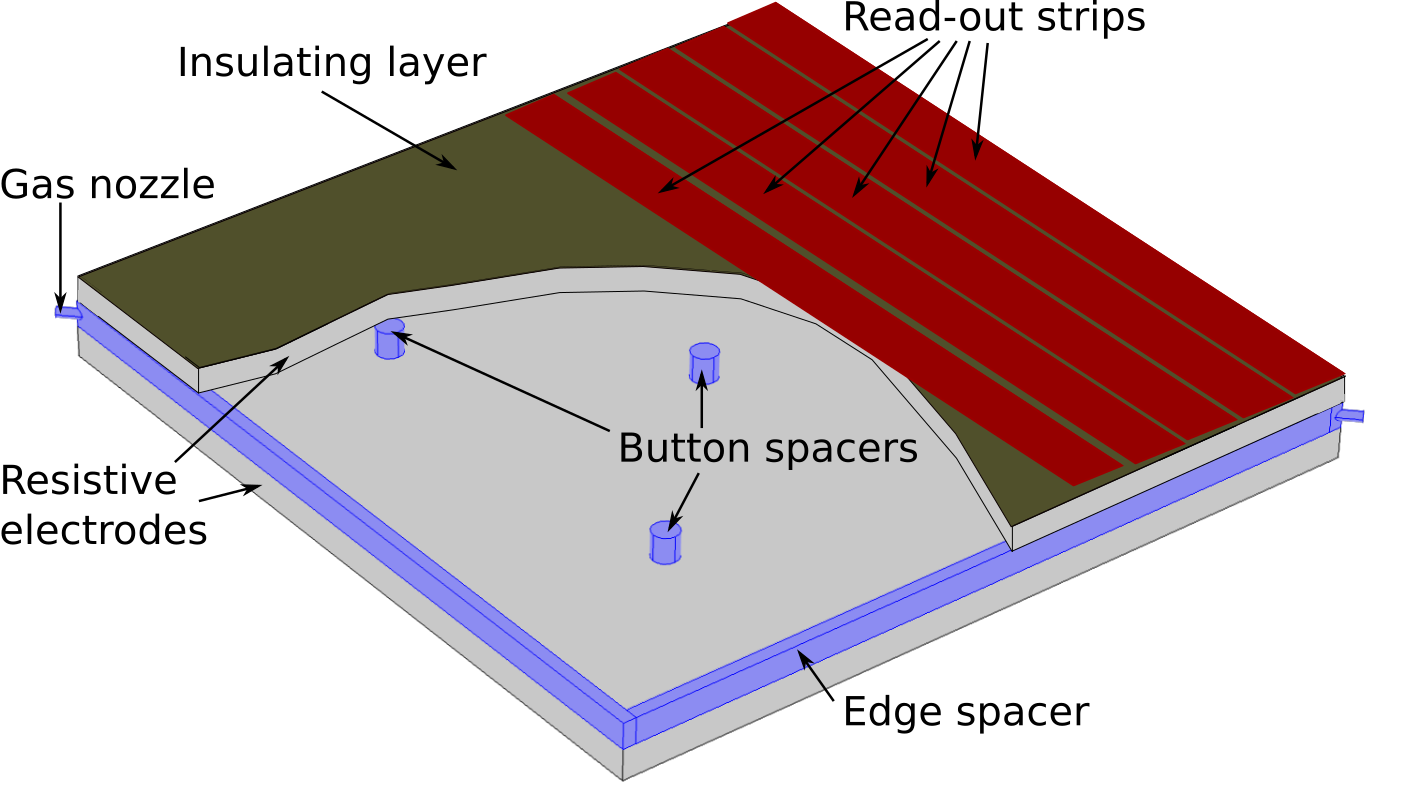}
	\caption{Full schematic of a single gap RPC \cite{jash2017studies}.\label{fig:fullrpc}}
\end{figure}

The full schematic diagram of the RPC is shown in Figure \ref{fig:fullrpc}. It consists of two electrodes made of glass or bakelite, which are separated by button and edge spacers. Two gas nozzles allow gas to flow between the electrodes, as shown in the previous figure. The outer surface of the electrodes is painted with resistive graphite paint (with a usual resistivity of approximately 10$^6$ $\Omega/\Box$), which allows the applied voltage to be spread over the surface. To collect signals due to avalanches inside the RPC, a panel of pickup strips is placed on both sides of the RPC. A mylar sheet is also introduced to separate the pickup panel and the graphite painted surface.
\par RPCs can be classified into two catagories (a) Trigger RPC, (b) Timing RPC. Timing RPCs are generally expected to offer more precise timing resolution compared to trigger RPCs. This is why they are typically employed as Time-Of-Flight (TOF) detectors in many experiments. On the other hand, trigger RPCs are typically used to detect the passage of charge particles like muons.
\subsubsection{Trigger RPCs}
The difference between a Trigger RPC and a timing RPC can be distinguished by using the gas gap. Usually, the gas gap in a single-gap Trigger RPC is 2 mm. Therefore, the primary can be generated anywhere within the 2 mm gap. The spatial spread of primary clusters determines the fluctuations in the signals. A thicker gap brings more fluctuations of time in the signal, which eventually leads to a decrease in time resolution. However, due to this thicker gap, the chances of generating a greater number of primaries are higher, and hence the efficiency will be higher. A typical single-gap Trigger RPC has an efficiency of greater than 98\%, depending on the applied voltage, environmental conditions, and the flux of the particle. The time resolution of the same configured RPCs can be 1 ns. Due to the ease of construction, one can make a very large area of RPC. In INO, the proposed size of the Trigger RPC is 2 m × 2 m \cite{ICALpotential}.
\subsubsection{Timing RPCs}
\begin{figure}[H]
	\centering\includegraphics[scale=0.3]{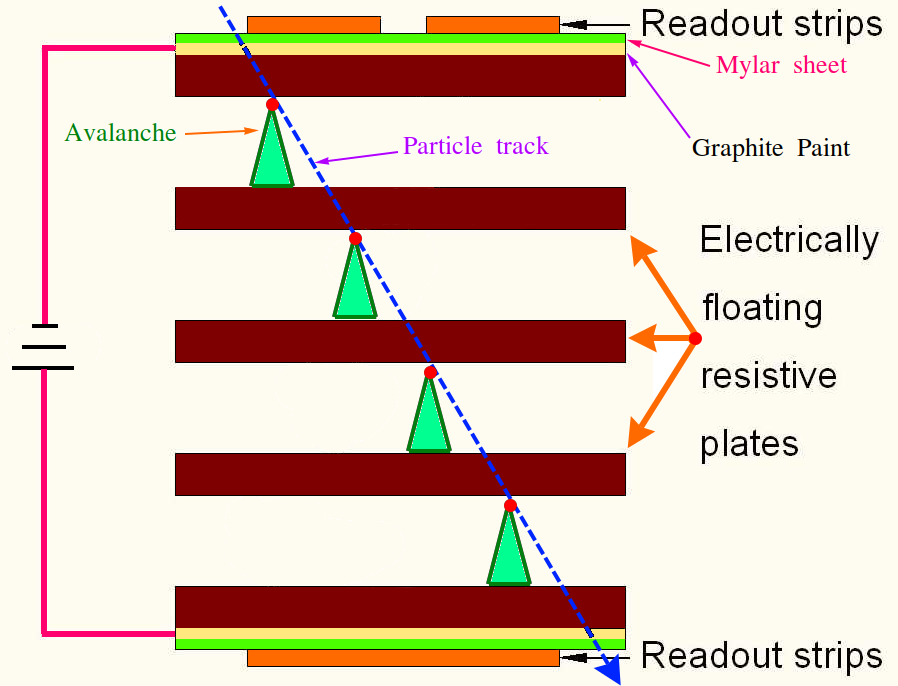}
	\caption{Schematic of a Multi Gap RPC (MRPC).\label{fig:multrpc}}
\end{figure}

A typical Timing RPC consist of the gas-gap of 0.2 mm to 0.3 mm. Due to such thin gap the generation of primaries are also less which in effect leads to a low efficiencies in single gap Timing RPCs. However, due to this thin gap the fluctuations in the signals are less hence a good time resolution 50 picosecond can be observed. To solve the issues in the efficiency timing RPCs are widely used in multi-gap configurations, which is called multi-gap-rpc or MRPC as shown in Figure \ref{fig:multrpc}.
\par In the year 1996 \cite{CERRONZEBALLOS1996132}, the introduction of the first MRPC marked a significant milestone in the field of RPC detectors. In MRPCs, a single gas gap RPC is divided into numerous gas sub-gaps using highly resistive electrodes. This division allows for the detection of particle interactions with greater accuracy and precision. The other beneficial properties of RPC detectors, such as efficiency, are still maintained in MRPCs. The application of high voltages (HV) is exclusively confined to the external surfaces of the top and bottom electrodes, while the intermediate electrodes remain electrically floating. To ensure adequate separation, very thin spacers measuring approximately 0.250 mm are positioned between each electrode pair. The readout panels or electrodes are positioned outside the stack and carefully insulated from the high voltage electrodes. When a charged particle traverses a Multi-gap Resistive Plate Chamber (MRPC), it can create an avalanche in any or all of the gas gaps. These avalanches exhibit similar temporal development across all gaps. Since the intermediate plates of the MRPC are transparent to the avalanche signals, the induced signals on the external anode and cathode represent the analog summation of all avalanches occurring in the different gaps. This process plays a vital role in the MRPC's ability to achieve high time resolution and efficiency in particle detection. If $\sigma_t$ and $\epsilon_t$ are the time resoluion and efficiency of a single gap RPC, then the time resolution of ``n" gap MRPC can be scaled as $\sigma_t/\sqrt{n}$ and the efficiency can be scaled as $1-(1-\epsilon_t)^n$ \cite{RIEGLER2003144}.  
\subsubsection{Resistive Cylindrical Chambers (RCC)}
\begin{figure}[H]
	\centering\subfloat[\label{fig:cylRpc}]{\includegraphics[scale=0.2]{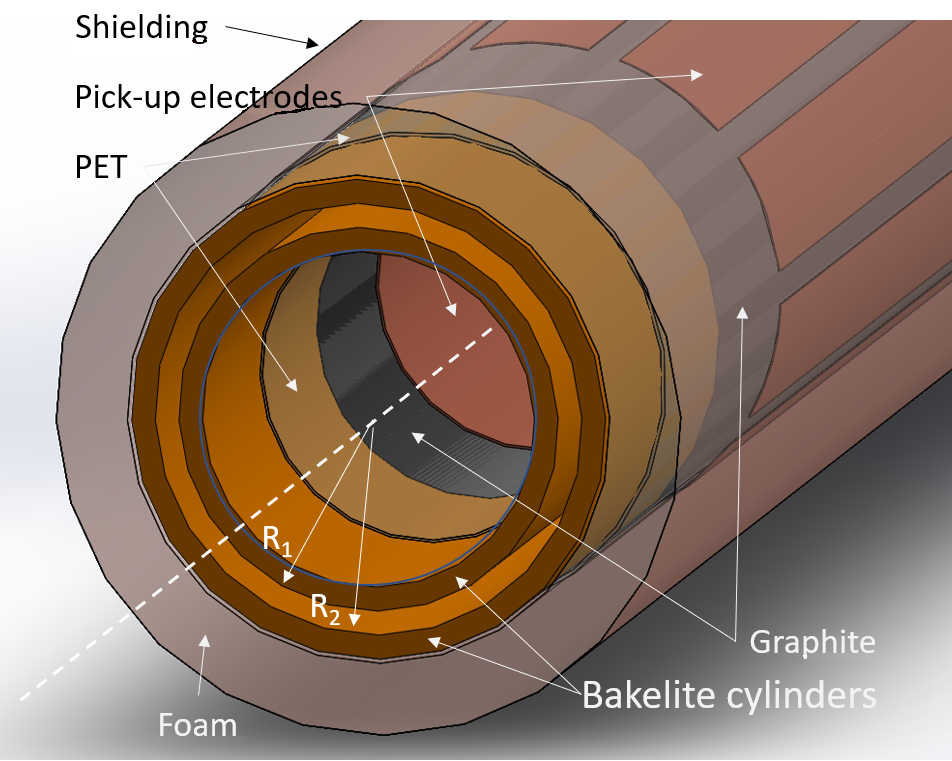}}\subfloat[\label{fig:RCCstrip}]{\includegraphics[scale=0.25]{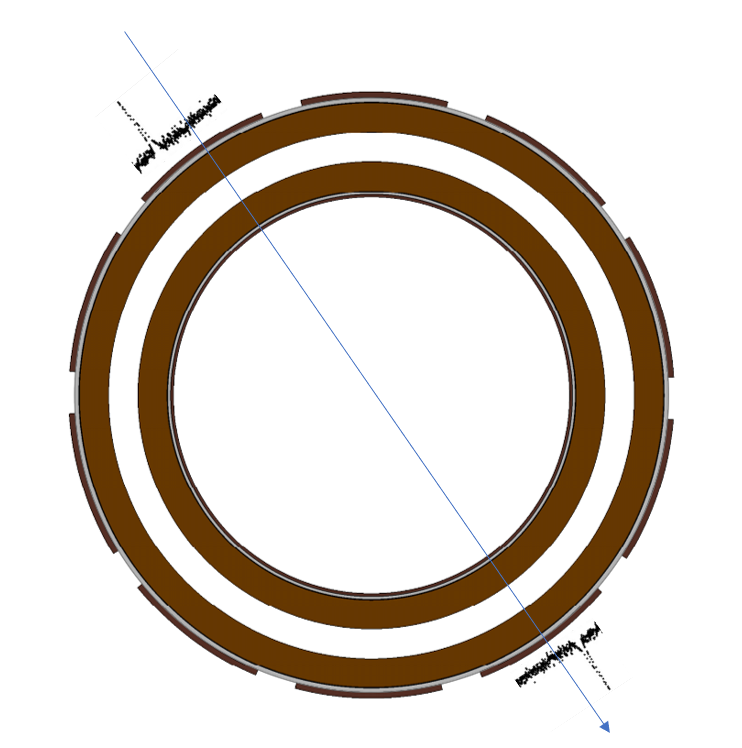}}

	\caption{(a) Schematic of a Resistive Cylindrical Chambers (RCC), (b) Signal readout scheme for RCC \cite{Cardarelli_2021}.}
\end{figure}

The figure \ref{fig:cylRpc} depicts a cylindrical structure comprising two pipes positioned concentrically with a gap $g$ filled with gas \cite{Cardarelli_2021}. The radius of the inner cylinder is $R_1$ and the radius of the outer cylinder is $R_2$. Therefore the gas-gap is $g=R_2-R_1$. The electric field inside the gas-gap g can be written as \cite{Cardarelli_2021}:
 \begin{align}
 E(r)=\frac{V}{r\,\log(k)},\,\,k=\frac{R_1}{R_2}
 \end{align}

\begin{figure}[H]
	\centering\includegraphics[scale=0.4]{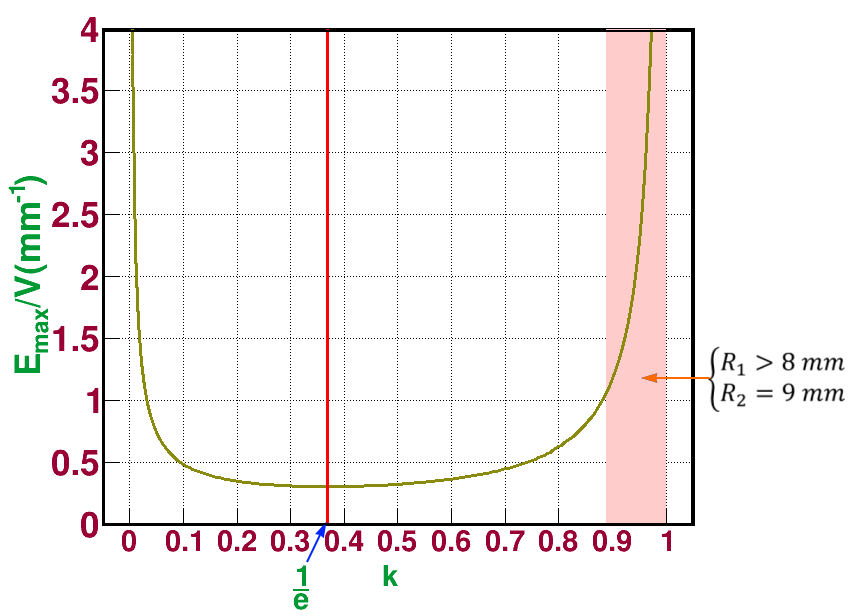}
	\caption{Variaion of maximum electric field at $r=R_1$ of an RCC with the $k=\frac{R_1}{R_2}$, where we $R_2$ is fixed at 9 mm \cite{Cardarelli_2021}.\label{fig:maxfieldRCC}}
\end{figure}

This geometry offers a unique way to study the behavior of gas discharge, which is heavily influenced by the ratio $k = R_1 /R_2$ of the radius of the facing surfaces. The maximum field within the gas-gap is achieved when the radial distance $r=R_1$. The behavior of this maximum field as a function of the parameter $k$, with $R_2$ being constant, is depicted in Figure \ref{fig:maxfieldRCC}. It is observed that the field's value varies with different values $k$, with the minimum value occurring at $k=\frac{1}{e}$.
Depending on the value of $k$, the gas discharge can either exhibit an approximately uniform electric field or a radial field. This provides the detector with the ability to operate under different conditions, ranging from a nearly constant electric field to a gradually changing radial field. For $R_2>R_1$ the nature of the field is radial and for $R_2\approxeq R_1$ or $k\approxeq1$ the field behaves as uniform field.
\par When the device geometry is such that $k<\frac{1}{e}$, the chances of spark discharge is less than when $k>\frac{1}{e}$ or in uniform field regions. However, for uniform field detectors, the temporal resolution is better than the radial field detectors. Therefore, one need to optimize the value of k in order to obtain good temporal and low number of spurious counts. The additional benifit of this shape is the cylindrical geometry has a unique property that makes it particularly robust to high gas pressure. This means unlike parallel plate detectors the cylindrical shape can withstand or resist pressure from gases without suffering damage or deformation. It is known that the construction of an MRPC is not as straightforward as that of a single gap RPC, and a single, narrow gap RPC is less efficient. It is argued that with greater pressure, the likelihood of primary ionizations and secondary multiplications increases. As a result, a thin gas gap in the Resistive Cylindrical Chambers (RCC) can offer good efficiency and temporal resolution, which is comparable to a thick-gap RPC and Multi gap RPC respectively. 
\par
\begin{figure}[H]
	\centering\includegraphics[scale=0.12]{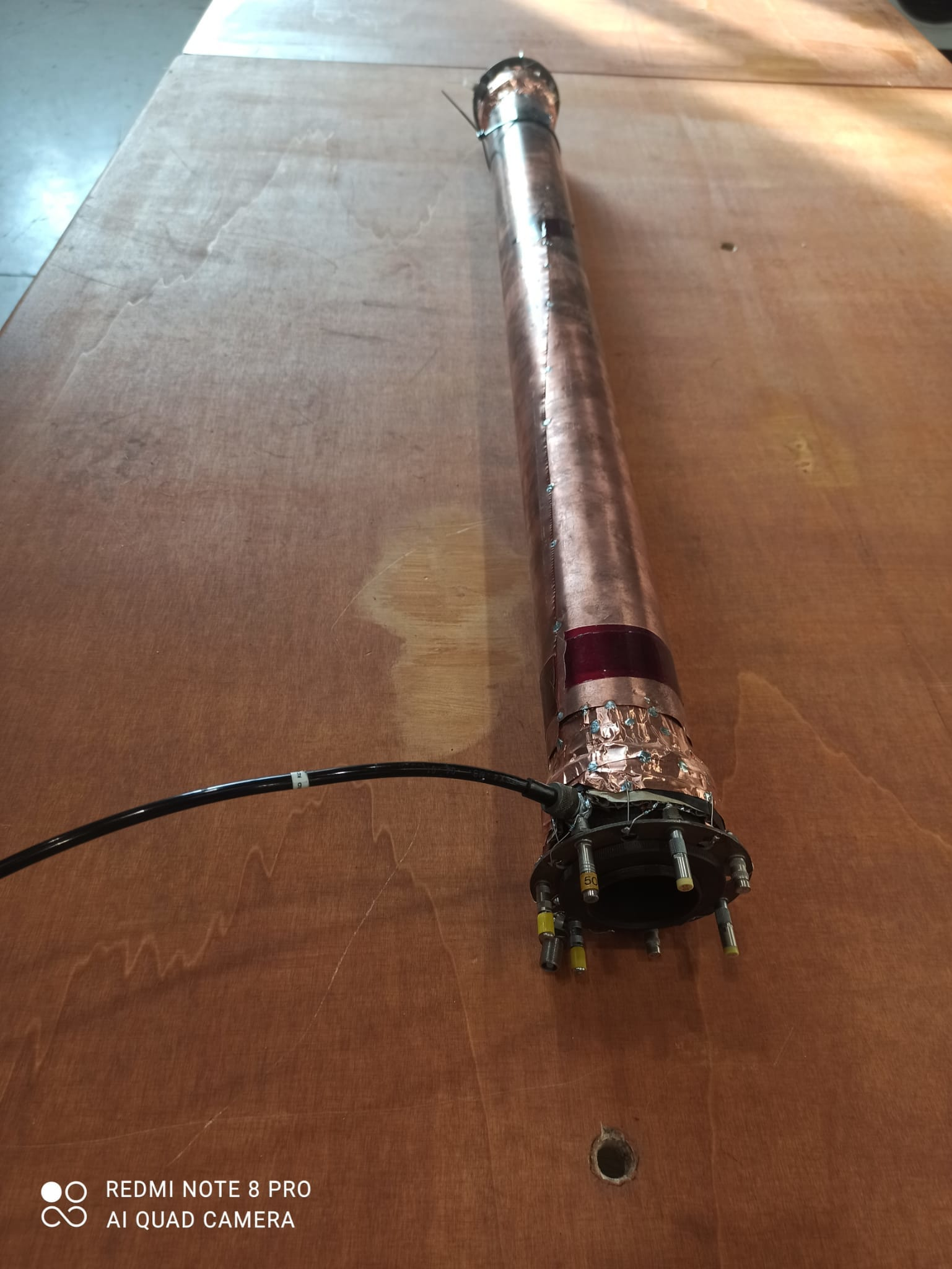}
	\caption{First working prototype of RCC \cite{Cardarelli_2021}.\label{fig:RCCreal}}
\end{figure}

 At the University of Rome Tor Vergata, the first RCC (shown in Figure \ref{fig:RCCreal}) with a uniform field has been tested, and the detector operates in a saturated avalanche regime. This prototype consist of two co-centric bakelite cylinders of radius $R_1=8$ mm and $R_2=9$ mm. To provide voltage, a graphite coat with a resistivity of 100 k$\Omega/\Box$ has been painted on the inner surface of the inner cylinder and the outer surface of the outer cylinder, as shown in Figure \ref{fig:cylRpc}. The thickness of the bakelite is taken as 1 mm. A gas mixture of R134a (95\%):Isobuten (4.5\%): SF6 (0.5\%) is used inside the gas-gap. Parallel pickup panels have been used to collect signals, as shown in Figure \ref{fig:cylRpc}. Therefore, when any charged particle passes through the RCC, signals can be generated in the opposite strips, as shown in Figure \ref{fig:RCCstrip}. However, this prototype is in developing stage and more study is needed before using it in a real application.
\subsection{Application of Resistive Plate Chambers}
Resistive Plate Chambers have become widely utilized gas detectors in high-energy physics experiments, owing to their remarkable success. The benefits of employing RPCs include their excellent performance and reliability. The few advantages of use of an RPC is geiven below:
\begin{enumerate}
	\item The cost of constructing an RPC is minimal, and it can be easily manufactured.
	\item In terms of position resolution, standard RPCs are capable of providing a relatively high accuracy of approximately 1 cm.
	\item In the case of a single gap RPC, they offer an excellent time resolution of approximately 1 nanosecond.
	\item RPCs can be produced with a large active area, typically in the range of several square meters.
	\item Depending upon the need of an experiment, the design and construction of RPCs can be
	tuned.
	\item The shape of RPCs can be customized to meet the specific needs of an experiment.	
\end{enumerate}
\begin{table}[H]
	\centering
	\begin{tabular}{|p{0.9in}|l|p{0.5in}|l|p{0.5in}|p{0.5in}|l|}
		\hline
		\textbf{Experiment} & \textbf{Application} & \textbf{Area in m$^2$} & \textbf{Electrodes} & \textbf{Gap in mm} & \textbf{No. of Gaps} & \textbf{Mode} \\ \hline
		BaBar & Trigger & 2000 & Bakelite & 2 & 1 & Streamer \\ \hline
		Belle & Trigger & 2200 & Glass & 2 & 2 & Streamer \\ \hline
		ALICE-Muon & Trigger & 140 & Bakelite & 2 & 1 & Streamer \\ \hline
		ATLAS & Trigger & 6550 & Bakelite & 2 & 1 & Avalanche \\ \hline
		CMS & Trigger & 2953 & Bakelite & 2 & 2 & Avalanche \\ \hline
		STAR & Timing & 50 & Glass & 0.22 & 6 & Avalanche \\ \hline
		ALICE-TOF & Timing & 150 & Glass & 0.25 & 10 & Avalanche \\ \hline
		OPERA & Trigger & 3200 & Bakelite & 2 & 1 & Streamer \\ \hline
		YBJ-ARGO & Trigger & 5630 & Bakelite & 2 & 1 & Streamer \\ \hline
		BESIII & Trigger & 1200 & Bakelite & 2 & 2 & Streamer \\ \hline
		HARP & Timing & 10 & Glass & 0.3 & 4 & Avalanche \\ \hline
		COVER-PLASTEX & Timing & 16 & Bakelite & 2 & 1 & Streamer \\ \hline
		EAS-TOP & Timing & 40 & Bakelite & 2 & 1 & Streamer \\ \hline
		L3 & Trigger & 300 & Bakelite & 2 & 2 & Streamer \\ \hline
		HADES & Timing & 8 & Glass & 0.3 & 4 & Avalanche \\ \hline
		FOPI & Timing & 6 & Glass & 0.3 & 4 & Avalanche \\ \hline
		PHENIX & Trigger & ? & Bakelite & 2 & 2 & Avalanche \\ \hline
		CBM TOF & Timing & 120 & Glass & 0.25 & 10 & Avalanche \\ \hline
		NeuLAND & Timing & 4 & Glass & 0.6 & 8 & Avalanche \\ \hline
	\end{tabular}
	\caption{A summary of Past, Present, and Future Experiments Utilizing RPCs \cite{fonte2000high,ganai2017probing}. \label{tab:rpcapp}}
\end{table}
Due to the above mentioned advantages, RPCs are preferred over other technologies for experiments involving accelerators, cosmic rays, neutrinos, and applications requiring timing or triggering. A summary of RPC applications in different experiments is shown in Table \ref{tab:rpcapp}.

\section{Summary}
The development of gaseous detectors has opened the door to new knowledge in the field of particle physics. In this chapter, we have discussed two families of gaseous detectors: (a) wire-based, and (b) plate-based.
\par A cylindrical proportional chamber is a wire-based gaseous detector whose electric field varies radially. The signal is induced on the central wire due to ionizing radiation and primary and secondary multiplication in the gas caused by the applied electric field.
\par A multi-wire proportional chamber (MWPC) is a collection of several proportional counters that were invented to track the position of incoming ionizing radiation. The main drawback of this detector is the rate capability, which is caused by the space charge effect due to the pile-up of ions, and uncertainty in the detection of the incoming particles.
\par A drift chamber is a combination of high voltage cathodes and a proportional counter. It consists of a drift region and an avalanche region. When a charged particle enters the drift region, it leaves primaries that drift towards the proportional counter. Near the anode wire, due to the high electric field, primaries are multiplied, generating a signal. An external triggering detector like scintillator is used to record the time of the incoming particle first. Therefore, the drift time is the difference between the avalanche time and the time at the trigger detector.
\par The Time Projection Chamber (TPC) is a combination of a drift tube and multi-wire proportional counter, which is used in accelerator experiments to track the particles coming out from the collision. The gridding mechanism of TPC is effective in suppressing the space charge effect due to pile-up of ions in the MWPC.
\par The alternate detector to MWPC is the Micromegas, which was invented to deal with the problems of space charge effect, ion backflow, etc. It consists of a thin micro-mesh with holes of 25 $\mu$m, which separates the drift region and amplification region. The avalanche occurs in the amplification region, and the signal is collected at the anode, with most of the ions being absorbed in the mesh.
\par The Gas Electron Multiplier (GEM) is made up of a 50-micron thin Kapton film with 5-micron copper foil on either side, which has periodic holes of varying diameters on both sides of the foil. This foil separates the drift and amplification regions, with the maximum multiplication happening in the hole area where the electric field is strong. Additionally, the gain of the detector can be increased by introducing more foil and creating a double or triple stage GEM.
\par An RPC (Resistive Plate Chamber) is made up of two electrodes with high resistance placed a certain distance apart to create a gas gap. Like other gaseous detectors, RPCs rely on gas ionization, which involves primary and secondary ionization followed by avalanche formation. The Yule-Furry algorithm can be used to simulate the avalanche in an RPC, while Ramo's theorem can be used to calculate the signal. RPCs can operate in either avalanche or streamer mode, with the gain being less limited in the latter. As a result, pre-amplifiers are required to detect the signal in avalanche mode. Single gap RPCs are generally used for tracking and triggering experiments, such as INO, whereas multigap RPCs are used for timing purposes. A list of experiments that use RPCs can be found in table \ref{tab:rpcapp}. Cylindrical resistive chambers are another type of detector in this family, which can be used for both timing and triggering purposes in experiments, but they are still in development. 
\chapter{Numerical Models to Calculate Space-Charge Field in an RPC}\label{space_charge_field_calculation}

\section{Introduction}
The Resistive Plate Chamber (RPC) is a widely used gaseous detector for detecting charged particles \cite{cardeli-1,cardeli-2}\footnote{The contents of this chapter are taken from the following publication by the author:
	
	\textbf{Numerical study of space charge electric field inside Resistive Plate Chamber, T. Dey et al, 2020, JINST 15 C11005, DOI 10.1088/1748-0221/15/11/C11005.}}. The detailed geometric configuration and working of an RPC has been discussed in Chapter 2, where two operational modes were described: (a) Avalanche and (b) Streamer. In avalanche mode, the average charge generated is approximately 1 pC, while in streamer mode, it can range from 50 pC to several nC. Additionally, there is another mode called saturated avalanche \cite{AIELLI20036,MOSHAII2012S168}, which occurs when the number of electrons created is almost equal to the number of electrons attached, in other words, when the gain of electrons is stopped due to a large number of electrons and ions screening the electric field. This results in a low electric field region where electrons do not have sufficient energy to multiply, and the probability of attachment is greater than that of ionization. However, the probability of ionization is still higher in the high field regions.
\par
\begin{figure}[H]
\centering	\subfloat[Primary generation\label{fig:forAval-a}]{\includegraphics[scale=0.25]{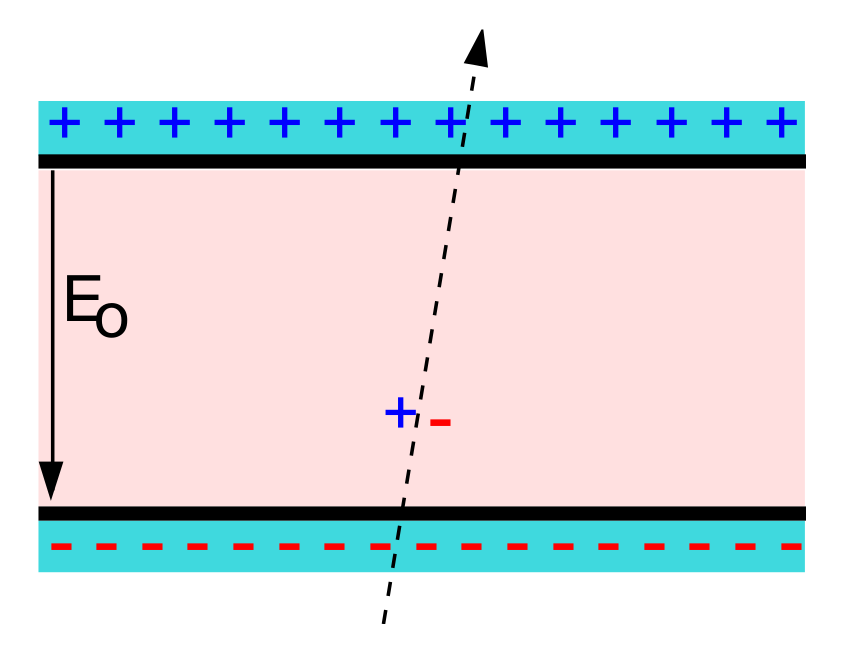}}\subfloat[Secondary multiplication and avalanche formation\label{fig:forAval-b}]{\includegraphics[scale=0.25]{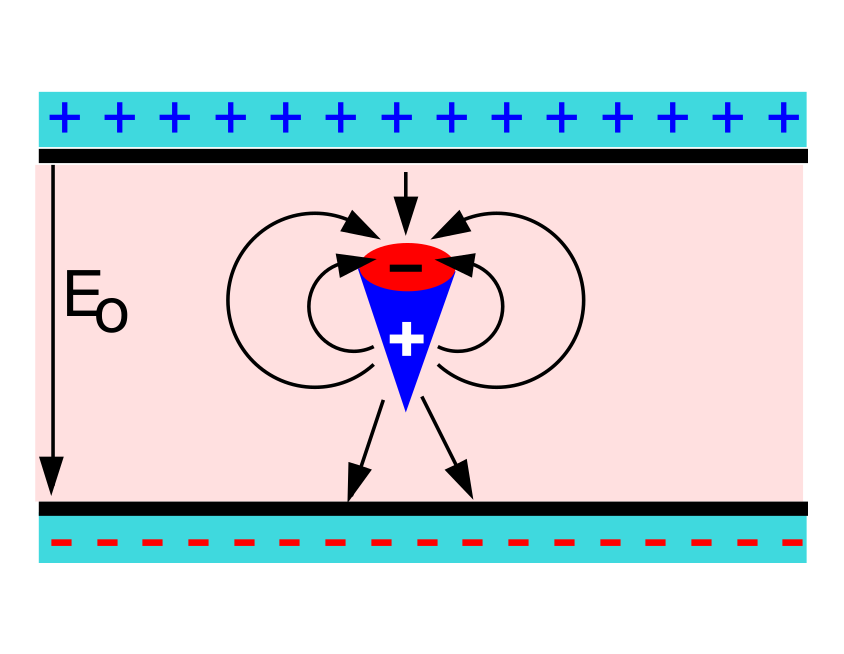}}
	
\centering	\subfloat[Ionisation due to photon generated in avalanche\label{fig:forAval-c}]{\includegraphics[scale=0.25]{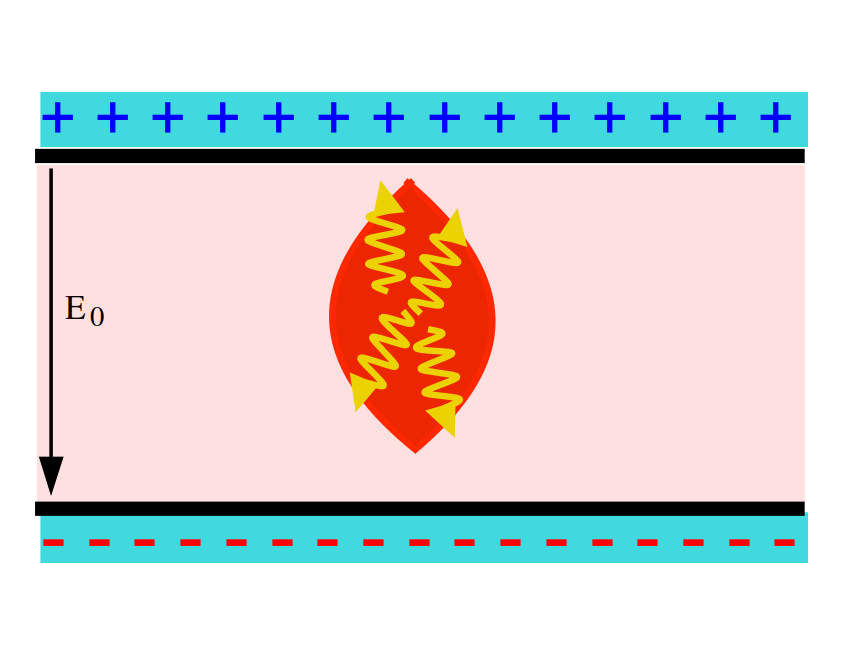}}\subfloat[Higher field at the top and bottom of the avalanche\label{fig:forAval-c1}]{\includegraphics[scale=0.25]{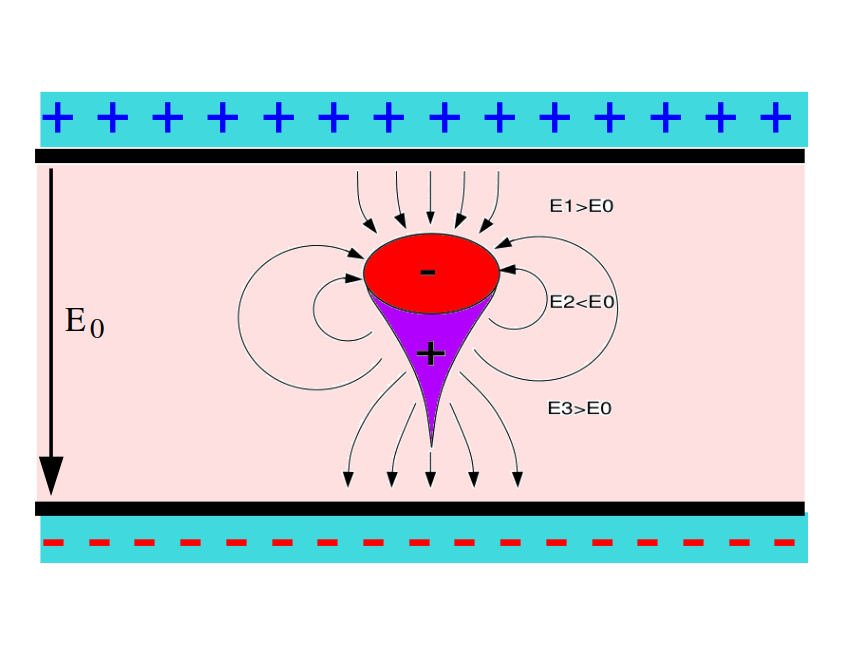}}
	
\centering	\subfloat[Streamer generation\label{fig:forAval-d}]{\includegraphics[scale=0.25]{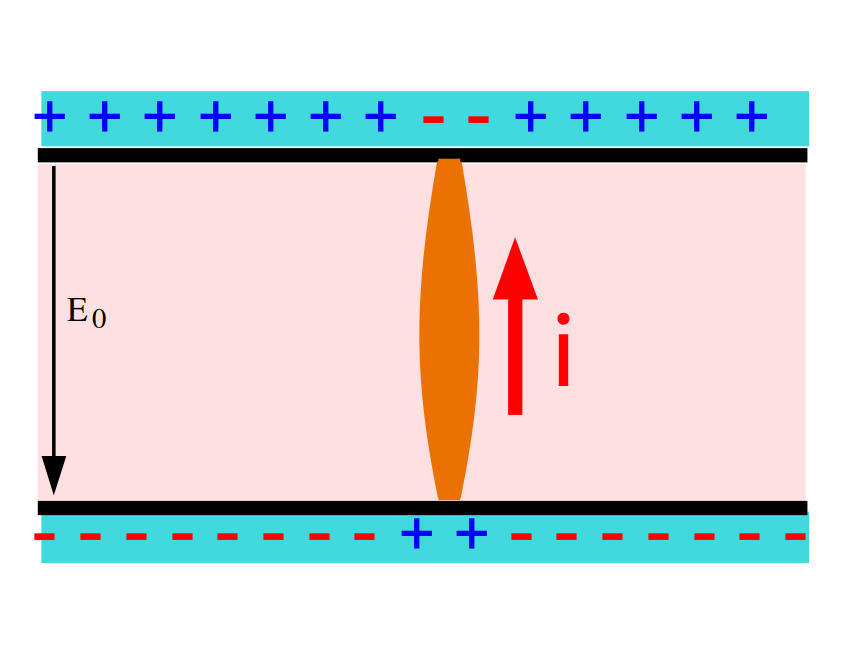}}\subfloat[Dead spot on electrodes due to pile up of ions and electrons on the electrodes\label{fig:forAval-e}]{\includegraphics[scale=0.25]{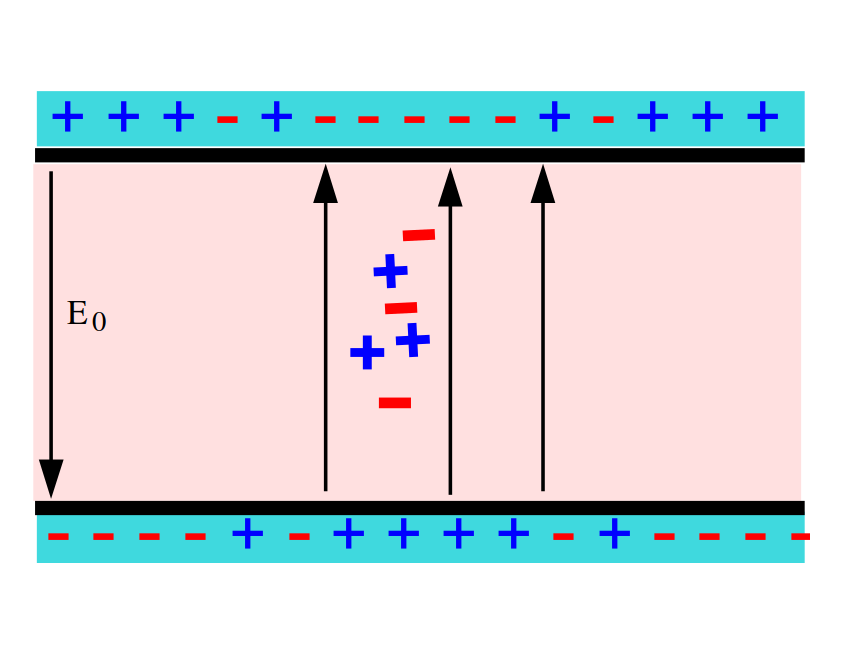}}
\caption{Steps of Avalanche formation and avalanche to streamer transition \label{fig:forAval}\cite{LippmanThesis}.}
	
\end{figure}

 In Chapter 2, it is discussed that when the size of the avalanche attains a certain value (our observed value is of the order of $10^5$), the space charge effect becomes significant. The step-by-step formation of an avalanche and its transition to a streamer is shown in Figure \ref{fig:forAval}. In Figure \ref{fig:forAval-a}, the primary particles are generated due to an incoming ionizing particle. In Figure \ref{fig:forAval-b}, the primaries gain energy from the applied electric field and multiply, eventually forming an avalanche. Depending on the applied voltage, either the electrons and ions will pass through the electrodes or there is a possibility that the photons generated in the avalanche can cause further ionization in the gas and generate secondaries near the avalanche region, as shown in Figure \ref{fig:forAval-c}. Now, at the tip where most of the ions are situated, the electric field below this or at the opposite end  where the electron density is high of the avalanche can be much higher than the applied field, which is shown in Figure \ref{fig:forAval-c1}. It is possible that the field at the tip can be reached a value twice of the applied field. Therefore, the electrons generated by avalanche photons can start a new avalanche process with high gain and eventually form a continuous path along the gas gap, which is called a streamer, as shown in Figure \ref{fig:forAval-d}. In this case, a dead spot can be created on the cathode near the streamer region due to the high density of ions, which prevents further incoming particles from being detected, as shown in Figure \ref{fig:forAval-e}.
Therefore, it is clear that to investigate the detector physics
of RPCs, we need to simulate that avalanche process very precisely.
We know that space charge plays a crucial role while an avalanche
is developing. So for simulating an avalanche, the electric field
due to space charge needs to be calculated dynamically. 

In this chapter, we will first give an overview of the Garfield++ software. Then, we will discuss three different methods (A, B, C) to calculate the electric field in the presence of space charge. Methods A and C (see sections \ref{subsubsec:ring-approximation} and \ref{subsec:Method-available-in}) involve modeling the space charge region as several concentric charged rings of gradually increasing radius, as described in reference \cite{Lippmann_1}. Additionally, a ring can be thought of as a collection of charged straight lines (see section \ref{subsec:Straight-line-approximation-with}) of equal length $S$, which is our method B. Now, two cases need to be discussed:

Case 1: The linear charge density ($\lambda$) of a ring is kept constant in methods A and C. The lines corresponding to a ring have been shared the equal amount of charge ($\lambda S$) with method B.

Case 2: The conditions for the rings remain identical to those in case 1. However, the charges of the lines have been calculated separately. We then compare the electric fields for these lines and rings in two cases. In this initial phase, we are ignoring the reflections of charges on the ground plates. This is a significant issue and will be discussed in Chapter 4.
\begin{figure}[H]
	\centering \subfloat[Solid Box centered at origin (0,0,0) of half length along x,y,z axis is $l_x,l_y,l_z$ respectively.]{\includegraphics[scale=0.35]{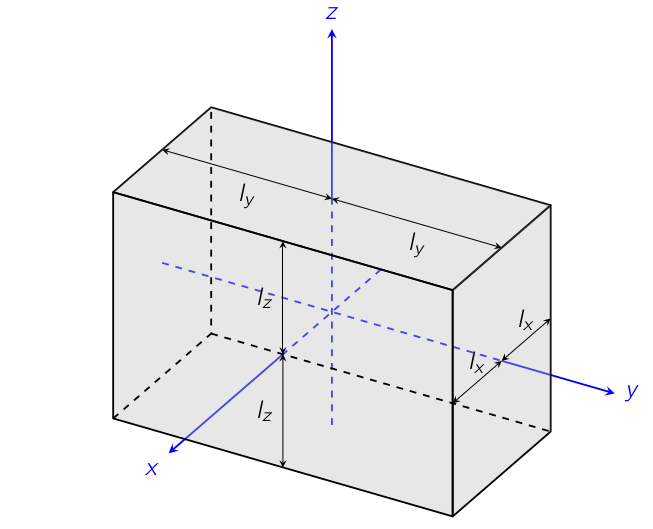}}\subfloat[Solid Sphere centred at (0, 0, 0)]{\includegraphics[scale=0.35]{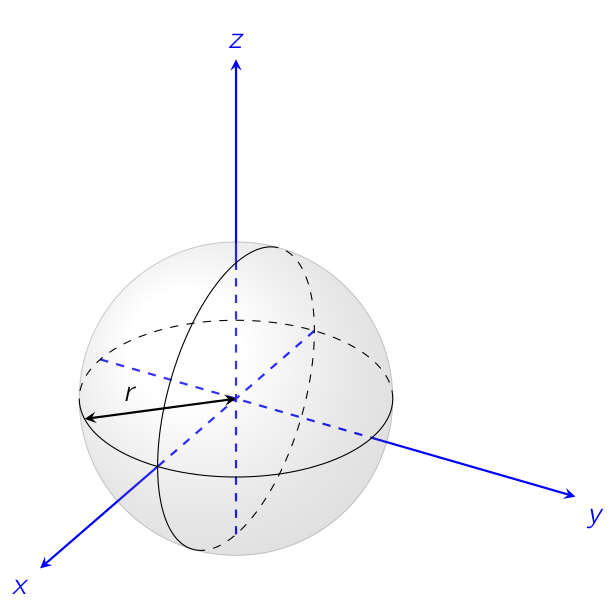}}
	
	\centering \subfloat[Solid tube centered at origin (0,0,0)]{\includegraphics[scale=0.35]{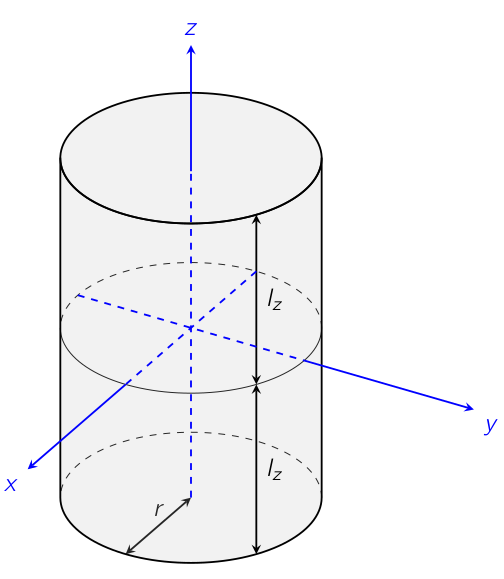}}
	\caption{Geometry defined in Garfield++ tools \cite{Garfield}.\label{fig:garfgeo}}
\end{figure}
\section{A short overview of Garfield++}
Garfield++ \cite{Garfield} is a C++-based numerical simulation tool that can be used to simulate the device physics of gaseous and semiconductor detectors. The workings of Garfield++ can be divided into several parts: (a) Detector Modeling, (b) Sensor Definition, (c) Generation of Incoming Particle Track and Primary Ionization, and (d) Charge Transport and Electron Multiplication. These points are elaborated below:
\subsection{Detector modeling}

	\subsubsection{Geometry and Materials } Garfield++ has built-in tools to define the geometry of detectors, such as the ``SolidBox", ``SolidTube", and ``SolidSphere" classes. Figure \ref{fig:garfgeo} shows a few examples of the numerical geometry created in Garfield++. All components of the geometry can be combined to form a single unit or solid using the ``AddSolid" method of the ``GeometrySimple" class. The 2D or 3D shape of the composite geometry can then be viewed using the ``ViewGeometry" class. However, complex geometries can be difficult to define using these tools, so they can be defined using external software such as COMSOL \cite{comsol}, GMSH \cite{geuzaine2009gmsh}, ANSYS\cite{ansys}, etc. and then fed into Garfield++.
	
	\par The properties of the materials used in the shapes can be defined using several classes, such as ``MediumMagboltz" to define gas mixtures, ``MediumConductor" to define conducting media, and for user-defined materials, the ``Medium" class can be used. To define a gas mixture, one can use the ``SetComposition" method of the ``MediumMagboltz" class. For example, this method can be invoked by the currect object of MediumMagboltz as ``SetComposition(``Ar", 93., ``CO2", 7.)", where a mixture of Argon and Carbon-dioxide is defined in a ratio of 93:7.
\subsubsection{Electric field maps } 
To define the electric field inside the detectors, the ``Component" class can be used. ``Component" is the base class, and its derived classes are ``ComponentConstant," ``ComponentNeBem3d", ``ComponentElmer", ``ComponentComsol", and so on. For simple geometries like RPC, one can use ``ComponentConstant" to provide a constant physical electric field and weighing electric field to the detector. For complex geometries and precise field maps, one needs to use field solvers. The neBEM field solver is already integrated with Garfield++. One can use neBEM with the help of the ``ComponentNeBem3d" class. However, one can borrow the field maps (both physical electric field and weighing electric field) from external field solvers like Elmer, COMSOL, etc., and feed them to Garfield++ using the ``ComponentElmer" and ``ComponentComsol" classes, respectively. It should be noted that the neBEM and Elmer field solvers are freely available, whereas COMSOL and ANSYS are not free.

\subsection{Sensor}
The main task of the Sensor class is to calculate signals and induced charge. Therefore, we need to define an object of the ``Sensor" class, which will contain all the information about the detectors, such as its geometry, material, physical and weighting field, and so on. In the first step, the total geometry is fed to the corresponding component class, and in the second step, the object of the component class is fed to the ``Sensor" using the ``SetComponent" method. However, for external solvers like COMSOL and ANSYS, the geometry is already defined in the field maps, so the first step can be skipped. The time window within which the user is interested in seeing the signals or event can be set using the ``SetTimeWindow" method of the ``Sensor" class. To collect signal one need to specify the one component of geometry as electrode using the method ``AddElectrode" of ``Sensor" class.
\subsection{Generation of incoming paricle track and primary ionisation}
Heed \cite{SMIRNOV2005474}  is a software program that utilizes the photo-absorption ionization (PAI) model. The PAI model helps in understanding the behavior of ionizing radiation in matter, by calculating the probabilities of interactions between particles and atoms. The software was developed by I. Smirnov and an interface to Heed is accessible through a class called TrackHeed.
\par After, defining the Sensor, the definition of incoming particle and its kinetic energy can be set using the methods ``SetParticle", ``SetKineticEnergy" of the class ``TrackHeed". Then for each event a Track of incoming particle can be simulated by using the method ``NewTrack". For an example if x0,y0,z0 is the initial position and t0 initial time of the particle and dx,dy,dz its direction, then a track can be simulate by invoking NewTrack(x0, y0, z0, t0, dx, dy, dz) using the current object of TrackHeed.
\par Primary clusters can be generated in the gas due to the ionizing particle. The ``GetClusters" method returns all the information, such as cluster position, transferred energy, and the number of electrons in each cluster. By looping over all clusters, one can obtain the information (position, energy, etc.) of each electron contained in each cluster using the ``GetElectron" method.
\subsection{Charge transport and electron multiplication}
Garfield++ offers several techniques to produce an avalanche, including Microscopic tracking, Monte Carlo integration, and Runge-Kutta-Fehlberg integration. These methods can be executed using specific classes within the program. The Microscopic tracking method, for instance, is performed through the ``AvalancheMicroscopic" class, while the Monte Carlo integration method can be implemented using the ``AvalancheMC" class. The Runge-Kutta-Fehlberg integration method, on the other hand, can be executed using the ``DriftLineRKF" class. Overall, Garfield++ provides diverse and effective approaches to simulate an avalanche, each offering its unique benefits and suitable for different purposes. The basic techniques of the Microscopic tracking and Monte Carlo integration is discussed below. 
\subsubsection{Microscopic tracking}
 The microscopic tracking approach \cite{Garfield}, which is currently only available for electrons, involves monitoring a particle's movement from collision to collision. To use this method, a table is required that specifies the collision rates $\tau^{i-1}(E_e)$ for each scattering process $i$ as a function of the electron's energy $E_e$. These data are typically provided by the class MediumMagboltz for gases. Between collisions, the electron is traced along a classical vacuum trajectory based on the local electric field. Optionally, a magnetic field may also be included. The duration $\Delta t$ of each free-flight step is determined by the total collision rate $\tau^{i-1}(E_e)=\sum_{i}\tau^{i-1}(E_e)$. To sample $\Delta t$ of each free-flight step, the ``null-collision" method is utilized, which accounts for any changes in the electron's energy during the step. Once the step is completed, the electron's energy, direction, and position are updated, and the scattering process to occur is sampled based on the relative collision rates at the new energy $E_e^\prime$. The electron's energy and direction are then updated accordingly based on the type of collision that occurs. Overall, the microscopic tracking method provides a detailed and comprehensive view of the electron's movement and is useful for certain types of simulations.
 \subsubsection{Monte Carlo integration}
 The basic first-order equation of motion for a charged particle under the influence of electric and magnetic fields in a gaseous medium can be expressed as:
 \begin{eqnarray}\label{eqn:driftmotion}
 \dot{r}=v_d(E(r),B(r)),
 \end{eqnarray}
 where $v_d(E(r),B(r))$ is the drift velocity. The equation of drift motion (\ref{eqn:driftmotion}) can be numerically solved using two main methods: Runge-Kutta-Fehlberg integration and Monte-Carlo Integration. The latter approach, which we will focus on here, involves first calculating a step of length $\Delta s = v_d\,\Delta t$ in the direction of the drift velocity $v_d$ at the local electric and magnetic field. The user can then specify either the time step $\Delta t$ or the distance step $\Delta s$. 
 Afterwards, a stochastic displacement is chosen at random from three uncorrelated Gaussian distributions. The first component, which aligns with the velocity of the drift, has a standard deviation of $\sigma_L = \sqrt{D_L \Delta s}$, while the other two perpendicular components have a standard deviation of $\sigma_T = \sqrt{D_T \Delta s}$. The resulting vectors from these two displacements are then combined to compute the updated position. For the charge multiplication a modified version of Yale-Furry model is used in the class ``AvalancheMC", which is discussed in Chapter 2. In this thesis we have chosen this method for our simulation. 
 
\section{Calculation of the positions of electron and ion clusters}\label{sec:position}
\begin{figure}
\centering\subfloat[\label{fig:position of elc clouds}]{\includegraphics[width=0.65\textwidth]{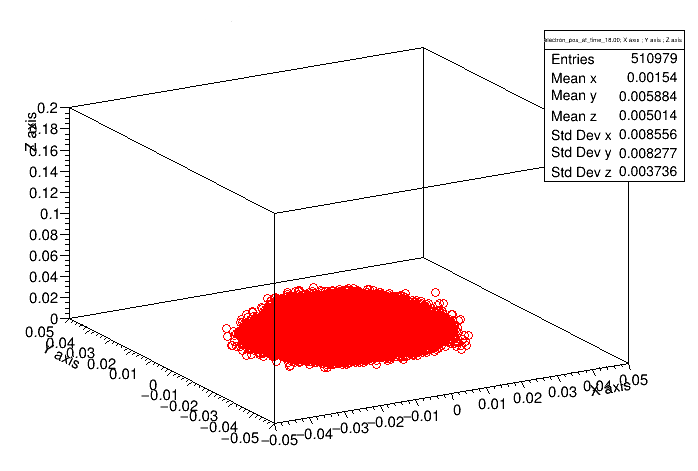}

}

\subfloat[\label{fig:picofring}]{\includegraphics[width=0.35\textwidth]{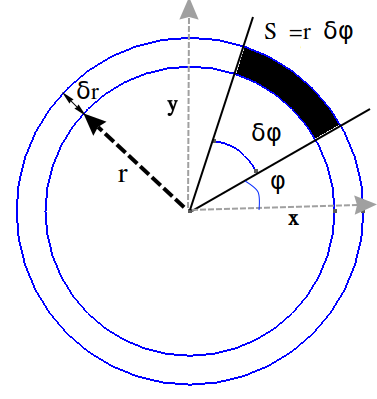}

}

\centering\subfloat[\label{fig:ringfieldpic}]{\includegraphics[width=0.5\textwidth]{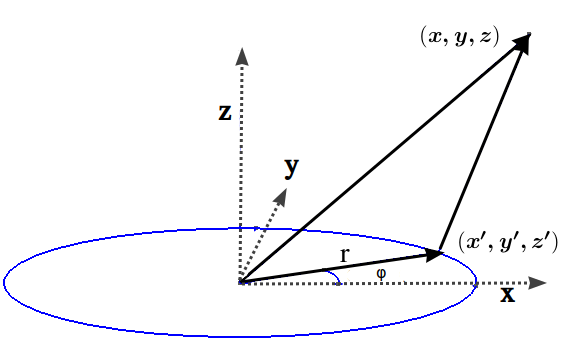}

}~~\subfloat[\label{fig:linefield}]{\includegraphics[width=0.45\textwidth]{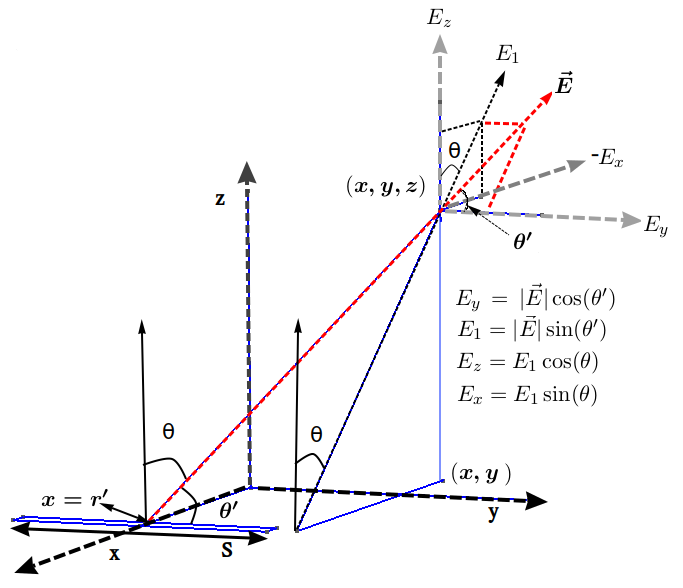}

}

\caption{(a) The position of the electron clouds at time 18 ns where the total number
of electron is 510979. (b) picture of a ring of width $\delta r$
(c) Computation of the electric field generated by a charged ring. (d) Components
of the electric field generated by a linear charge density \cite{Dey_2020}.}
\end{figure}

An avalanche has been simulated inside an RPC with an area of 30 cm $\times$ 30 cm and a 2 mm gas gap, starting from an electron created at the center of the gas gap (0,0,0) using the Garfield++ simulation tool. The gas mixture selected contains 97$\%$ of \ce{C2H2F4}, 2.5$\%$ of \ce{i-C4H10}, and 0.5\% of \ce{SF6}. A uniform electric field of $50$ kV/cm is applied perpendicular to the parallel plates of the detector (considered as the z-direction) to perform this simulation. The gas operates at a pressure of one atmosphere and a temperature of 293.15 K. Garfield++ keeps track of the drifting position and time of each primary and secondary electron and ion generated during the avalanche. A table of positions and corresponding times of those electrons and ions has been formed to calculate the interpolated positions of the electrons at a certain instant in time. The interpolated positions of the space charge cloud at 18 ns have been shown in Figure \ref{fig:position of elc clouds}.
\section{Calculation of the electric field due to the space charge distribution}

\label{subsec:algorithm_ring} 

\subsection{Ring approximation (method A) \label{subsubsec:ring-approximation}}
It is assumed that the avalanche charge distribution has rotational symmetry about the z-axis, as apparent from Figure \ref{fig:position of elc clouds}. Along the z-direction, the gas gap $g$ can be divided into $N_{z}$ steps with a step size of $\delta z=\frac{g}{N_{z}}$. The space charge region along the X-Y plane can also be divided into a number ($N_{r}$) of concentric charged rings centered at the z-axis and gradually increasing in radius $r$ (see Figure \ref{fig:picofring}). The size of the ring, $\delta r=\frac{r_{max}}{N_{r}}$ and $\delta z$, have been chosen according to the transverse and longitudinal spread of the avalanche, for example, $\delta r=\delta z=0.001,$cm and $r_{max}=0.045,$cm. Starting from a height $z=\bar{z}$ ($\bar{z}<g$), if any charge is present, the electric field due to all $N_{r}$ rings has been calculated by numerically integrating equations (\ref{eq:ring_field}) and (\ref{eq:p.fonte_cyl}), and then the value of $z$ is set to $z+\delta z$. This process is iterated until $z_{max}=g$. In this method, the whole region of space charge can be covered to calculate the electric field. The calculated components of the field due to all rings are summed to obtain the total electric field at any point.

\subsubsection{Components of electric field vector due to ring\label{subsec:Components-of-Electric}}

The X,Y and Z components of the electric field at any position (x,y,z)
due to a ring of radius \textbf{r} and of uniform charged density $\lambda$ (see Figure \ref{fig:ringfieldpic}),
centered at the z-axis can be written as follows \cite{Dey_2020},

\begin{equation}
\left.\begin{gathered}\begin{array}{c}
E_{x}^{A}=\frac{\lambda\,\boldsymbol{r}}{4\pi\epsilon_{0}}\int\limits _{0}^{2\pi}\frac{(x-\boldsymbol{r}\cos\left(\phi\right))}{\Delta}d\phi,\,E_{y}^{A}=\frac{\lambda\,\boldsymbol{r}}{4\pi\epsilon_{0}}\intop\limits _{0}^{2\pi}\frac{(x-\boldsymbol{r}\sin(\phi))}{\Delta}d\phi\\
E_{z}^{A}=\frac{\lambda\,\boldsymbol{r}}{4\pi\epsilon_{0}}\intop\limits _{0}^{2\pi}\frac{\,\,\,\,(\,z-\bar{z}\,)\,\,\,}{\,\Delta\,}d\phi
\end{array}\end{gathered}
\right\} \label{eq:ring_field}
\end{equation}

where, $\Delta=[(x-r\cos(\phi))^{2}+(y-r\sin(\phi))^{2}+(z-\bar{z})^{2}]^{\frac{3}{2}}$, $\bar{z}$ is the position of center of that ring along the z-axis, and $\phi$ is the angular displacement of an element of ring of length \textbf{r}$\delta\phi$ from the x-axis (see Figure \ref{fig:picofring}).
If $Q_{ring}$ is total charge of that ring then $\lambda=\frac{Q_{ring}}{2\pi\boldsymbol{r}}$.

\subsection{Straight-line approximation with uniform charge density (method B)\label{subsec:Straight-line-approximation-with}}

A ring can be equally segmented into a number of straight lines. If $r$ and $\delta r$ are the radius and thickness of that ring, respectively, then the length of an arc of any segmented element of that ring is $S = r \delta \phi$, where $\delta \phi$ is the angle in radians subtended to the center of the circle (see Figure \ref{fig:picofring}). $S$ can be approximated as a straight line of length $S$ and thickness $\delta r$ when $S/r \ll 1$.

It was discussed in Section \ref{subsubsec:ring-approximation} that the space charge region can be divided into a number of rings. As an extension of this algorithm, those rings are also split into a number of straight lines, where $\delta \phi$ is chosen by the user. Therefore, the electric field can be calculated for each charged straight line (for cases 1 and 2 in Section \ref{sec:Results-and-discussions}) corresponding to a ring using equations (\ref{eq:stline_eqn}).

Thus, after forming a ring by a number of straight lines, the same iteration process discussed in Section \ref{subsubsec:ring-approximation} can be followed to obtain electric fields for all charges. The components of fields of each charged line are added iteratively to obtain the total electric field. Thus, we can reproduce the results of the electric field of charged rings using the line charge approximation.
\subsubsection{Components of electric field vector due to a charged line\label{subsec:Components-of-Electric-1}}

Let us consider a straight line of constant charged density $\bar{\lambda}$
and of length $S$, aligned parallel to the y-axis at $x=\bar{r}$,
and $z=\bar{z}$. The X, Y, Z components of the electric
field at any position (x,y,z) due to this straight line can be written
as follows \cite{Dey_2020}(see Figure \ref{fig:linefield}),

\begin{equation}
\left.\begin{gathered}\begin{array}{c}
E_{x}^{B}=\frac{\bar{\lambda}(x-\bar{r})}{4\pi\epsilon_{0}P^{2}}\left[\frac{(y+\frac{S}{2})}{\sqrt{(y+\frac{S}{2})^{2}+P^{2}}}-\frac{(y-\frac{S}{2})}{\sqrt{(y-\frac{S}{2})^{2}+P^{2}}}\right],E_{y}^{B}=-\frac{\bar{\lambda}}{4\pi\epsilon_{0}}\left[\frac{1}{\sqrt{(y+\frac{S}{2})^{2}+P^{2}}}-\frac{1}{\sqrt{(y-\frac{S}{2})^{2}+P^{2}}}\right]\\
\\
E_{z}^{B}=\frac{\bar{\lambda}(z-\bar{z})}{4\pi\epsilon_{0}P^{2}}\left[\frac{(y+\frac{S}{2})}{\sqrt{(y+\frac{S}{2})^{2}+P^{2}}}-\frac{(y-\frac{S}{2})}{\sqrt{(y-\frac{S}{2})^{2}+P^{2}}}\right]
\end{array}\end{gathered}
\right\} \label{eq:stline_eqn}
\end{equation}

where, $P=\sqrt{(z-\bar{z})^{2}+(x-\bar{r})^{2}}$, and if $Q_{st}$
is the total charge of this straight line then, $\bar{\lambda}=\frac{Q_{st}}{S}$.

For a chosen value of $\delta\phi$ it can be said that the total number of straight lines needed to form a ring is $N_{st}$=$\frac{360}{\delta\phi}$. So $\delta\phi$ is the minimum rotation angle that is needed to move from one segment to the nearest subsequent one. Necessary coordinate transformations have been carried out to evaluate electric field in a consistent frame of reference. 

\subsection{Method available in literature (method C)\label{subsec:Method-available-in}}
The equation of the electric field at any point ($\rho,\alpha,z$) in a cylindrical coordinate system associated to a ring of uniform charged
density $\lambda$ and radius "$a$" centered at z-axis, can also be
found in ref. \cite{book-electro}(v. I pp. 176), (see Figure \ref{fig:ringfieldpic}),

\begin{equation}
\left.\begin{aligned}\vec{E}_{ring}^{C}(\rho,z,a) & =\frac{\lambda a}{\pi\epsilon_{0}}\left[\frac{1}{2\bar{r}_{1}\rho}\left(K_{1}(u)-\frac{(a^{2}-\rho^{2}+z^{2})K_{2}(u)}{\bar{r}_{1}^{2}(1-u^{2})}\right)\hat{\rho}+\frac{zK_{2}(u)}{\bar{r}_{1}^{3}(1-u^{2})}\hat{z}\right]=E_{r}^{C}\hat{\rho}+E_{z}^{C}\hat{z}\\
u & =\frac{2\sqrt{a\rho}}{\bar{r}_{1}},\,\bar{r}_{1}=\sqrt{(a+\rho)^{2}+z^{2}}
\end{aligned}
\right\} \label{eq:p.fonte_cyl}
\end{equation}

where $E_{r}^{C},E_{z}^{C}$ are the radial and z- components of the electric
field and $K_{1}(u)$ and $K_{2}(u)$ are the complete elliptic integrals
of the first and second kinds. Again, the alpha component of the electric field $E_{\alpha}^{C}$ has been considered as zero because of the axial symmetry. 
The components of the fields calculated in Cartesian co-ordinates (for
method A and B) have been converted into cylindrical coordinates by
using the ``Jacobi transformation'' to compare with the results of method
C. The required Jacobi matrix for this transformations is given below,

\begin{equation}
\begin{array}{c}
\begin{split}\left(\begin{array}{c}
E_{r}^{i}\\
E_{{\alpha}}^{i}\\
E_{z}^{i}
\end{array}\right) & =\left(\begin{array}{ccc}
\cos(\text{\ensuremath{{\alpha}}}) & \sin({\alpha}) & 0\\
-\sin({\alpha}) & \cos({\alpha}) & 0\\
0 & 0 & 1
\end{array}\right)\left(\begin{array}{c}
E_{x}^{i}\\
E_{y}^{i}\\
E_{z}^{i}
\end{array}\right), & {\alpha}=\tan^{-1}(\frac{y}{x})\end{split}
\end{array}
\end{equation}

$E_{r}^{i},E_{{\alpha}}^{i},E_{z}^{i}$ are the radial,
${\alpha}$ and z directional components of the electric fields at a position ($\rho,\alpha,z$) of any
charged ring or line. Where, $x=\rho \cos{(\alpha)}$, $y=\rho \sin{(\alpha)}$ and z=z and $i=A$,$B$ for method-A and method-B respectively.
These cylindrical form of the components for different methods can
be represented together as $E_{r}^{A,B,C},E_{{\alpha}}^{A,B,C},E_{z}^{A,B,C}$.
\section{Results and discussions \label{sec:Results-and-discussions}}

\begin{figure}
\centering\includegraphics[scale=0.30]{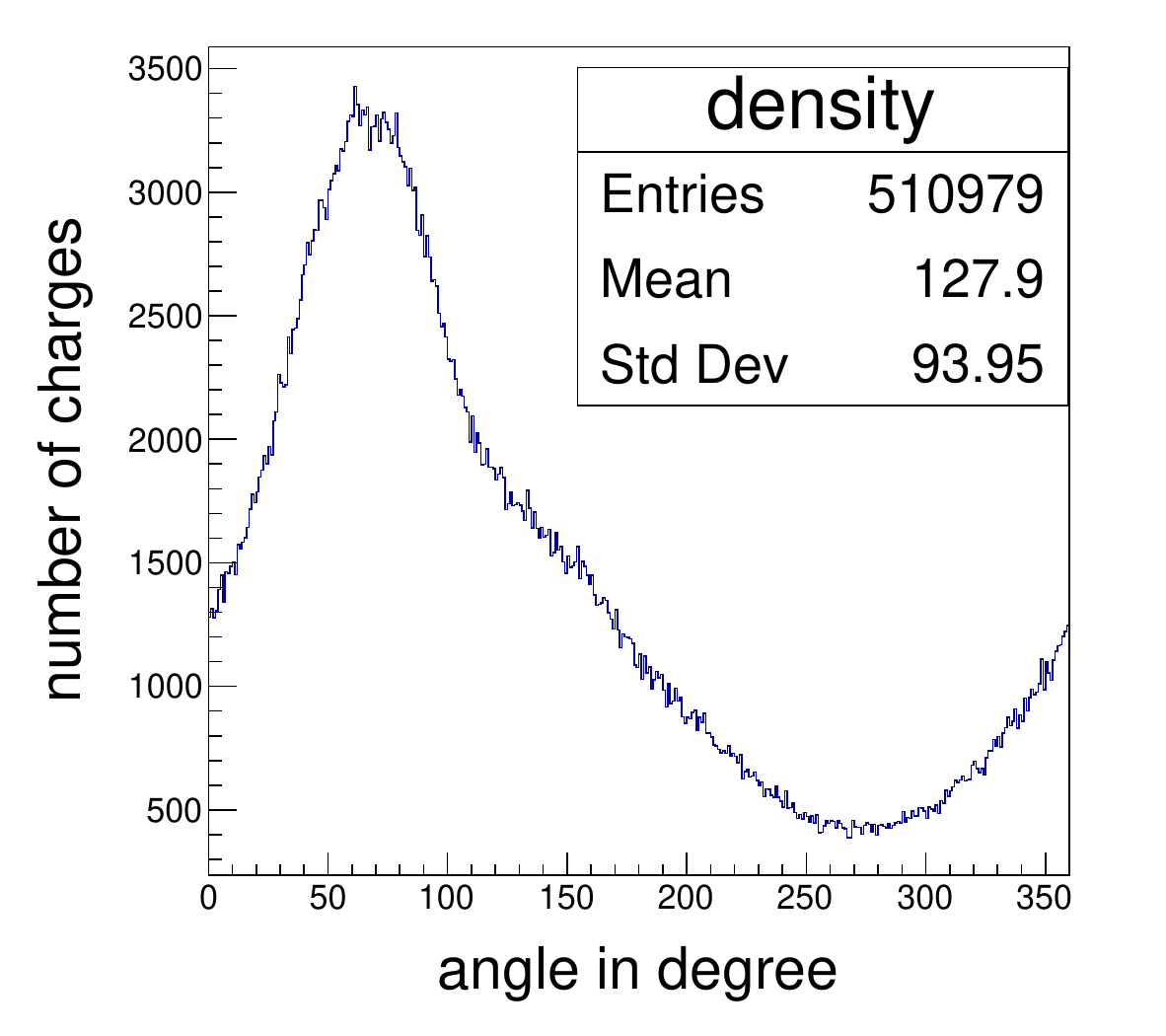}

\caption{\label{fig:angular-distribution-of-full-figure}Angular distribution
of charges at time 18 ns for case-2, \cite{Dey_2020}}
\end{figure}

The integrations of equations (\ref{eq:ring_field}) and (\ref{eq:p.fonte_cyl})
have been carried out numerically by using standard ``GSL-Integrator" from  ``GSL-library" available in "root 6.18/04" \cite{root-cern,BrunRoot}.
The absolute values of the calculated electric field components are set
to zero when they drop below a minimum value $\epsilon=10^{-8}$ V/cm,
because at this range the field due to the space charge is negligible
in comparison to the applied field. The results can be divided into two cases\textbf{:} 
\begin{description}
\item [{case-1:}] The charge density $\lambda$ has been considered constant
in equations (\ref{eq:ring_field}) and (\ref{eq:p.fonte_cyl}) of methods
A,C. In method-B the charge ($Q_{ring}$) of a ring is equally shared
between segmented lines from that ring, which is $Q_{st}=\lambda S=\frac{Q_{ring}S}{2\pi r}$
for a line. 
\item [{case-2:}] In this case, the conditions for equations (\ref{eq:ring_field})
and (\ref{eq:p.fonte_cyl}) remain the same as in the above \textbf{case-1}.
However, in the actual scenario the angular distribution of the total
$Q_{ring}$ charge may not be uniform over the ring. Therefore, the
amount of charge $Q_{st}$ will be different for each separate straight line of that ring. So charges are calculated separately for each line.
The Distribution of the charges which reside on each line at different angles
for all radii and heights, has been shown in Figure \ref{fig:angular-distribution-of-full-figure}. 
\end{description}

\subsection{Results of case-1\label{subsec:Results-of-case-1}}

\begin{figure}
\centering\subfloat[\label{fig:Ez-uniform-x0-y0}]{\includegraphics[scale=0.4]{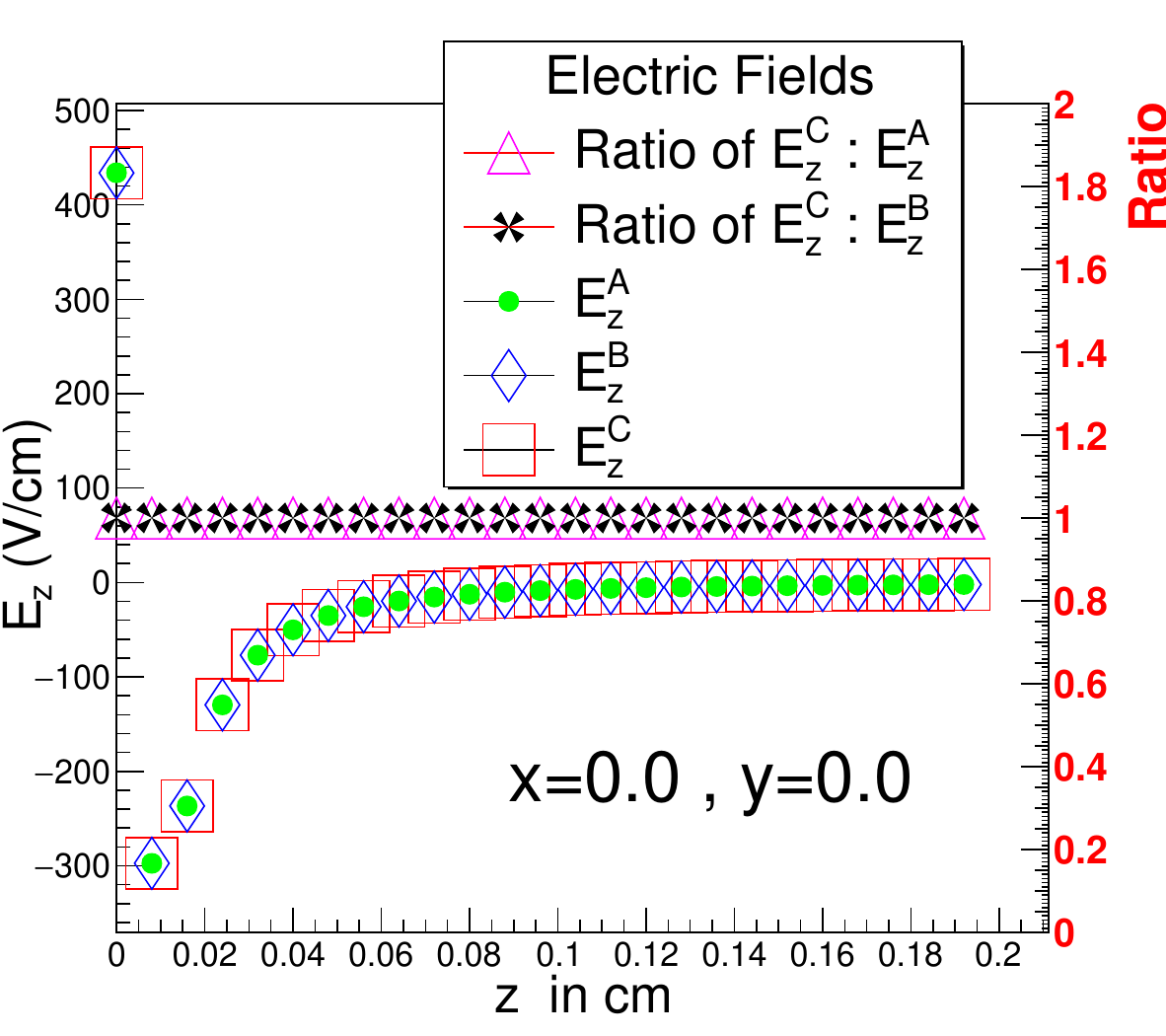}

}\subfloat[\label{fig:Er-uniform-x0-y0}]{\includegraphics[scale=0.4]{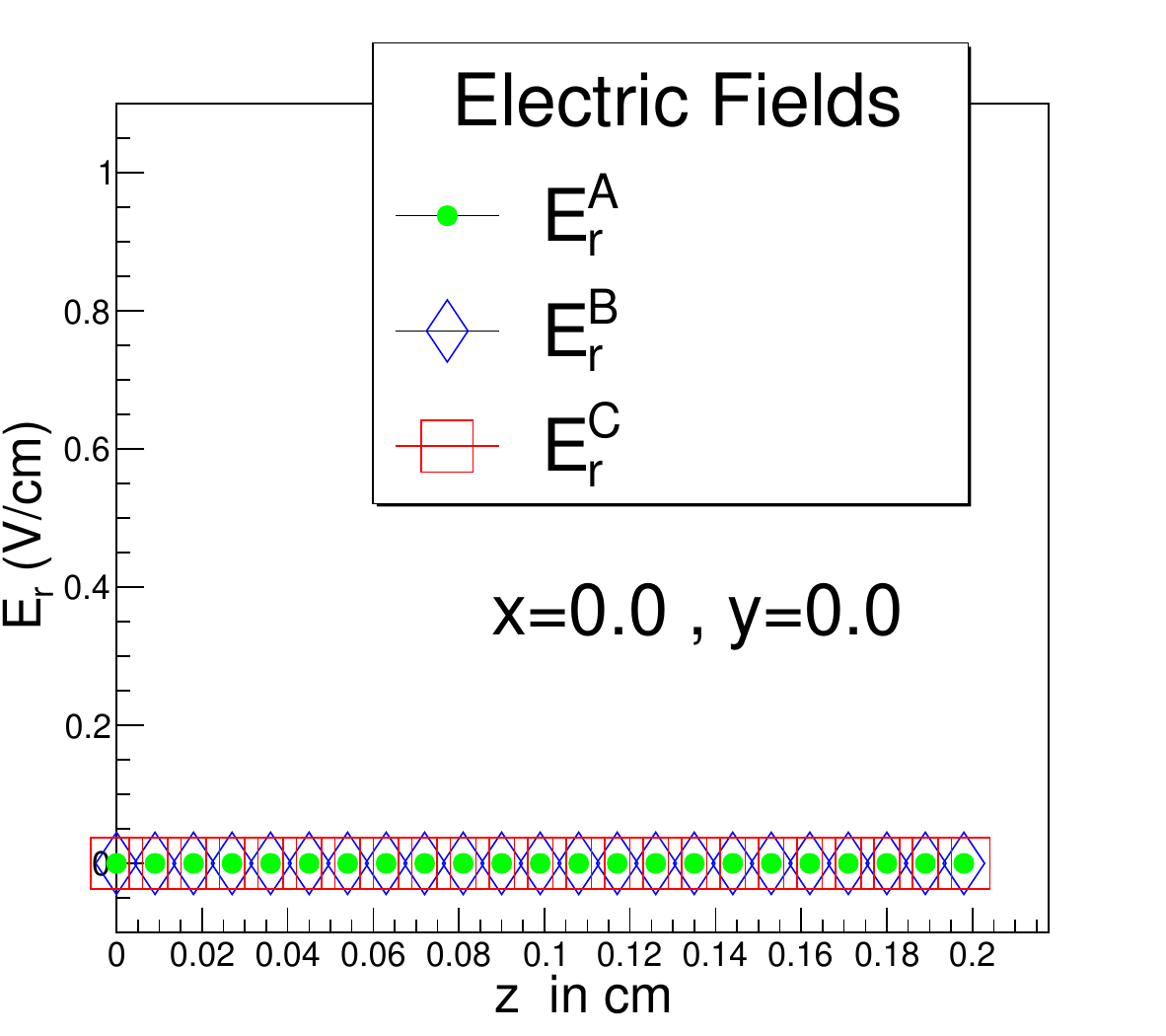}
}

\centering\subfloat[\label{fig:Ephi-uniform-x0-y0}]{\includegraphics[scale=0.35]{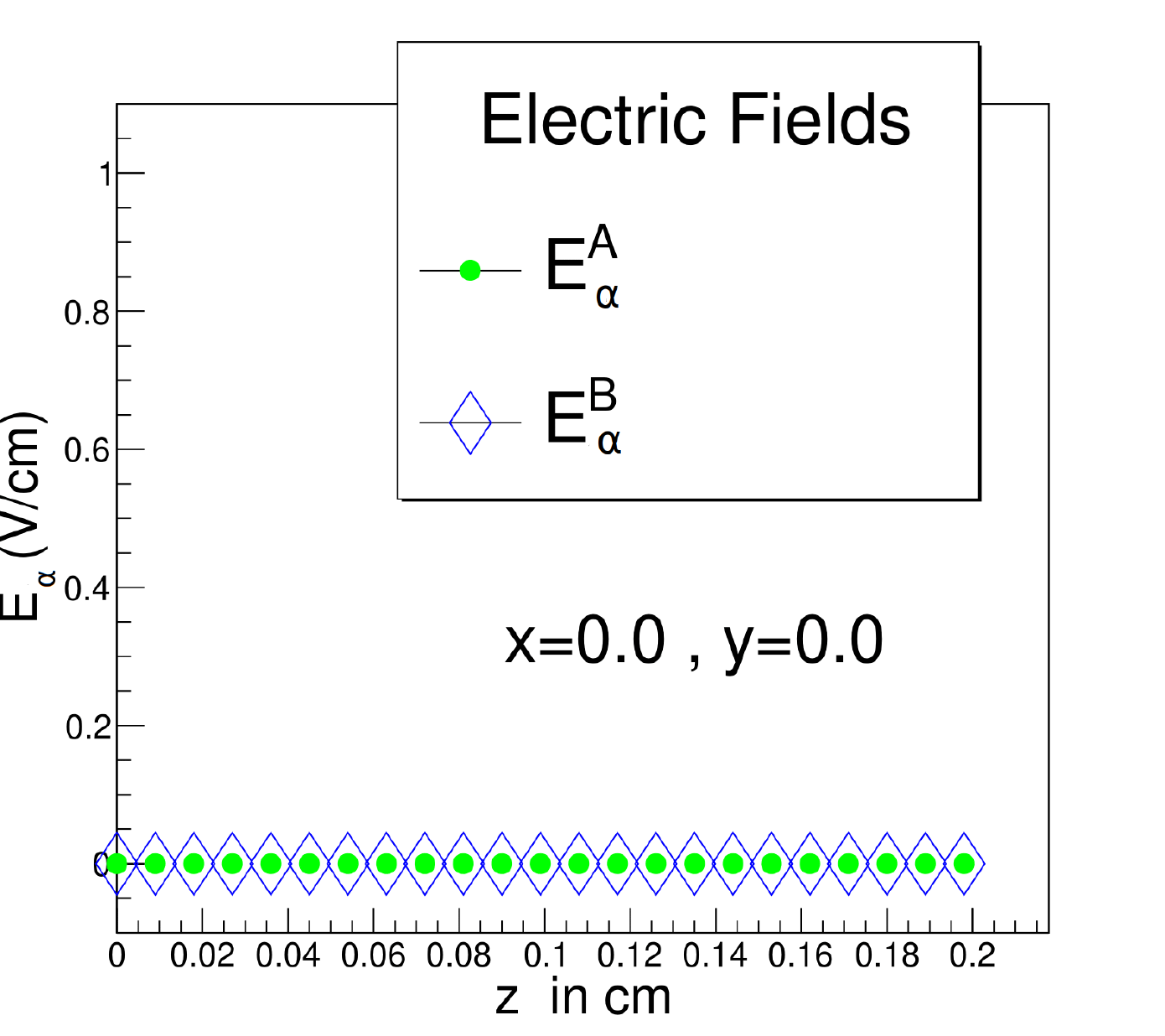}
}
\caption{\label{fig:uniform_density} Computation of electric field components
for case-1(section \ref{subsec:Results-of-case-1}) on the z-axis. (a) The
variation of $E_{z}^{A,B,C}$ on the z-axis has shown in the left-side
axis of Figures and right side axis contain the corresponding
ratios of the field components between $E_{z}^{C}$ and $E_{z}^{A,B}$,
(b) variation of $E_{r}^{A,B,C}$ on the z axis, (c) variation of
$E_{\alpha}^{A,B}$ on the z axis. \cite{Dey_2020}}
\end{figure}

It is well known to us that the electric field component $E_{z}$
is dominating over $E_{r}$ and $E_{{\alpha}}$ on the z-axis of
a uniformly charged ring. Because of the axial symmetry, the components
$E_{r}$ and $E_{{\alpha}}$ cancel out each other and become zero.
It can also be verified from Figures \ref{fig:Ez-uniform-x0-y0},
\ref{fig:Er-uniform-x0-y0} and \ref{fig:Ephi-uniform-x0-y0}, where the
components of electric field $E_{r}^{A,B,C},E_{{\alpha}}^{A,B},E_{z}^{A,B,C}$
have been plotted for three different methods A,B,C. Again, from the
same Figure \ref{fig:Ez-uniform-x0-y0} it is clear that the ratios $E_{z}^{C}/E_{z}^{A}=E_{z}^{C}/E_{z}^{B}=1$
on the z-axis. However, The values of $E_{r}^{A,B,C},E_{{\alpha}}^{A,B}$
are always zero on the z-axis, so the calculation of their ratios
is not possible . Hence, the ratio plot is not shown in Figures \ref{fig:Er-uniform-x0-y0}
and \ref{fig:Ephi-uniform-x0-y0}. 


\subsection{Results of case-2\label{subsec:Results-of-case-2}}
\begin{figure}
	\centering\subfloat[\label{fig:Ez-x=00003D00003D00003D0-y=00003D00003D00003D0-non-uni}]{\includegraphics[scale=0.4]{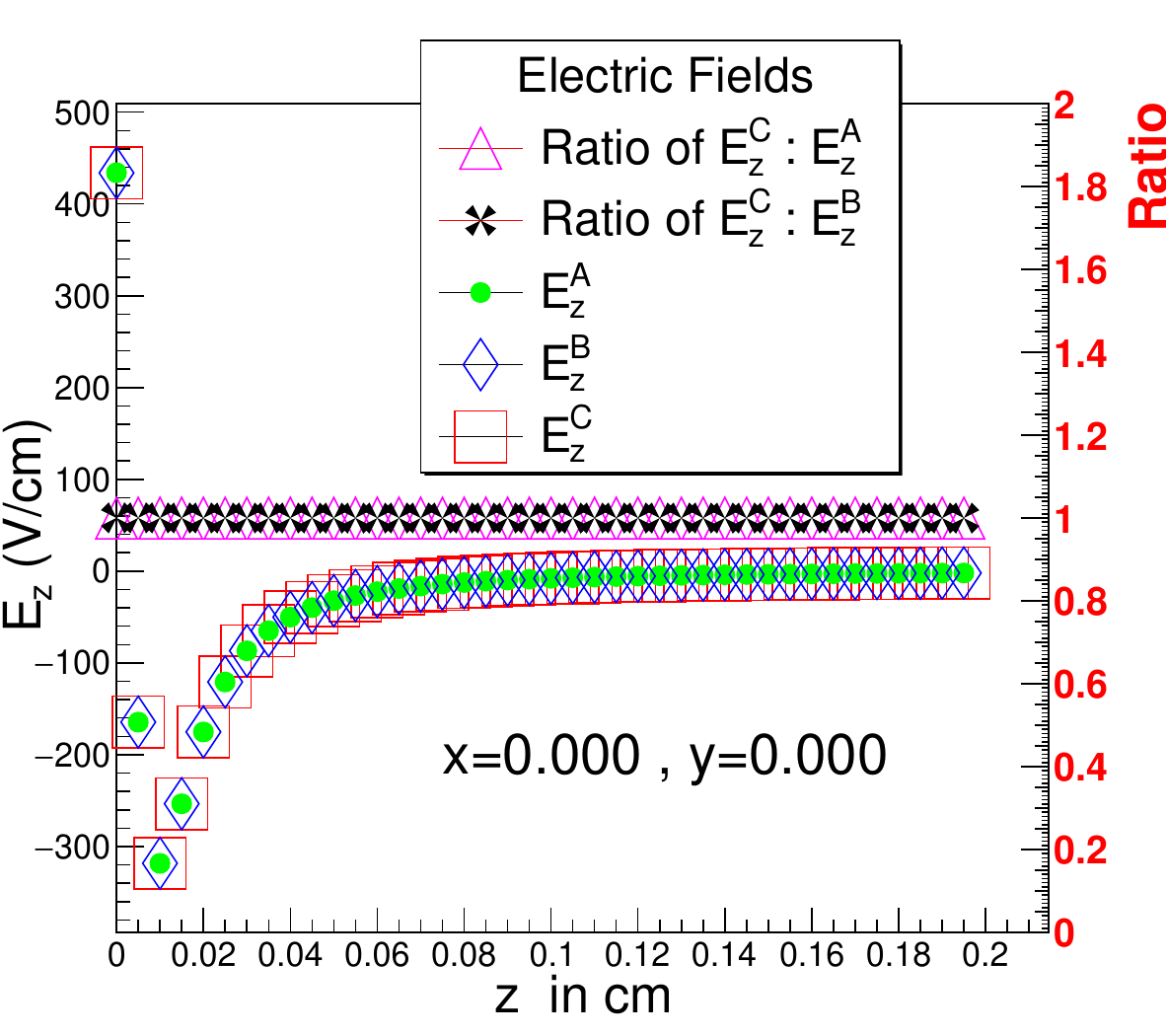}
	}\subfloat[\label{fig:Er-x=00003D00003D00003D0-y=00003D00003D00003D0-non-uni}]{\includegraphics[scale=0.4]{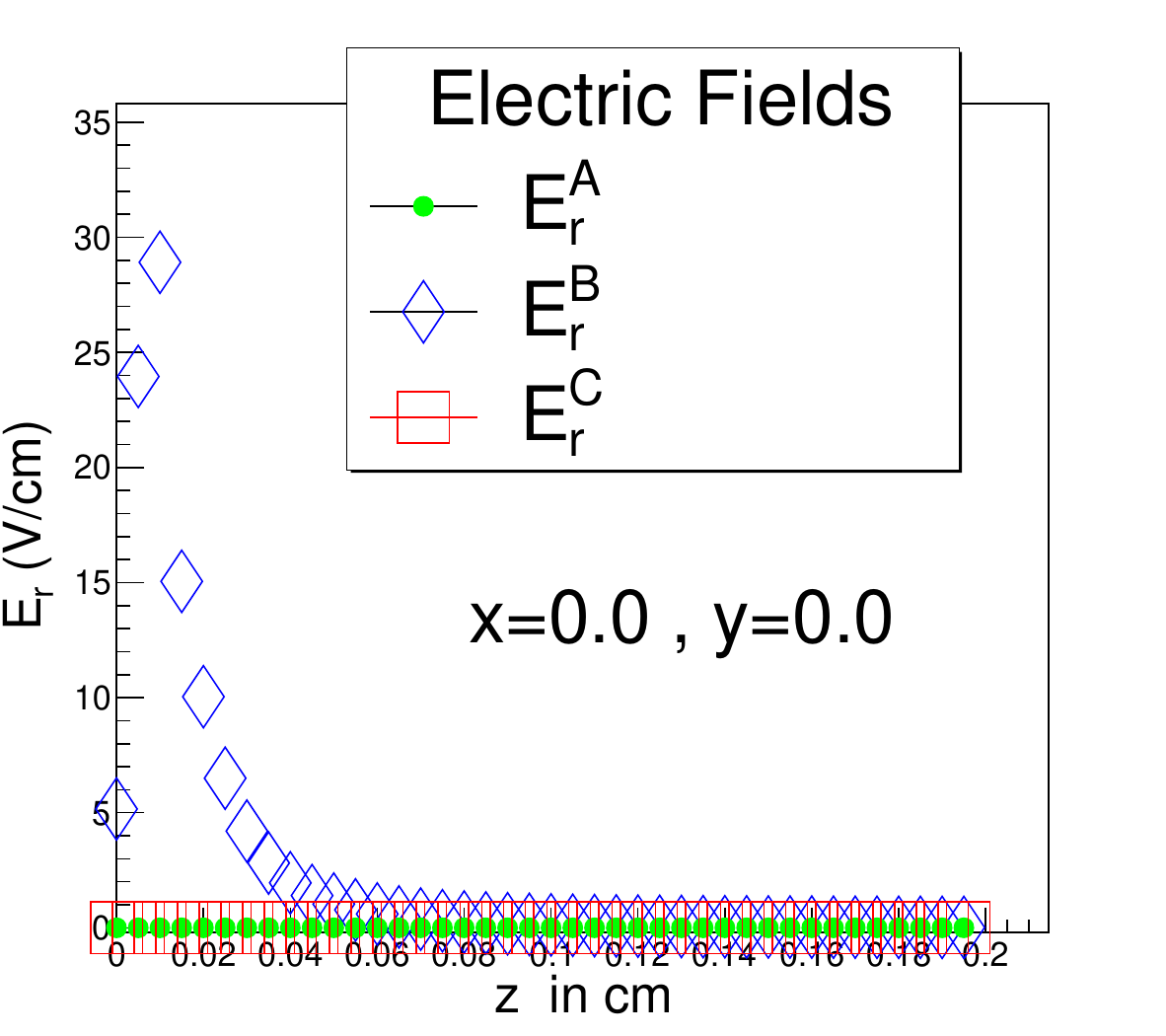}
	}
	
	\centering\subfloat[\label{fig:Ephi-x=00003D00003D00003D0-y=00003D00003D00003D0-non-uni}]{\includegraphics[scale=0.35]{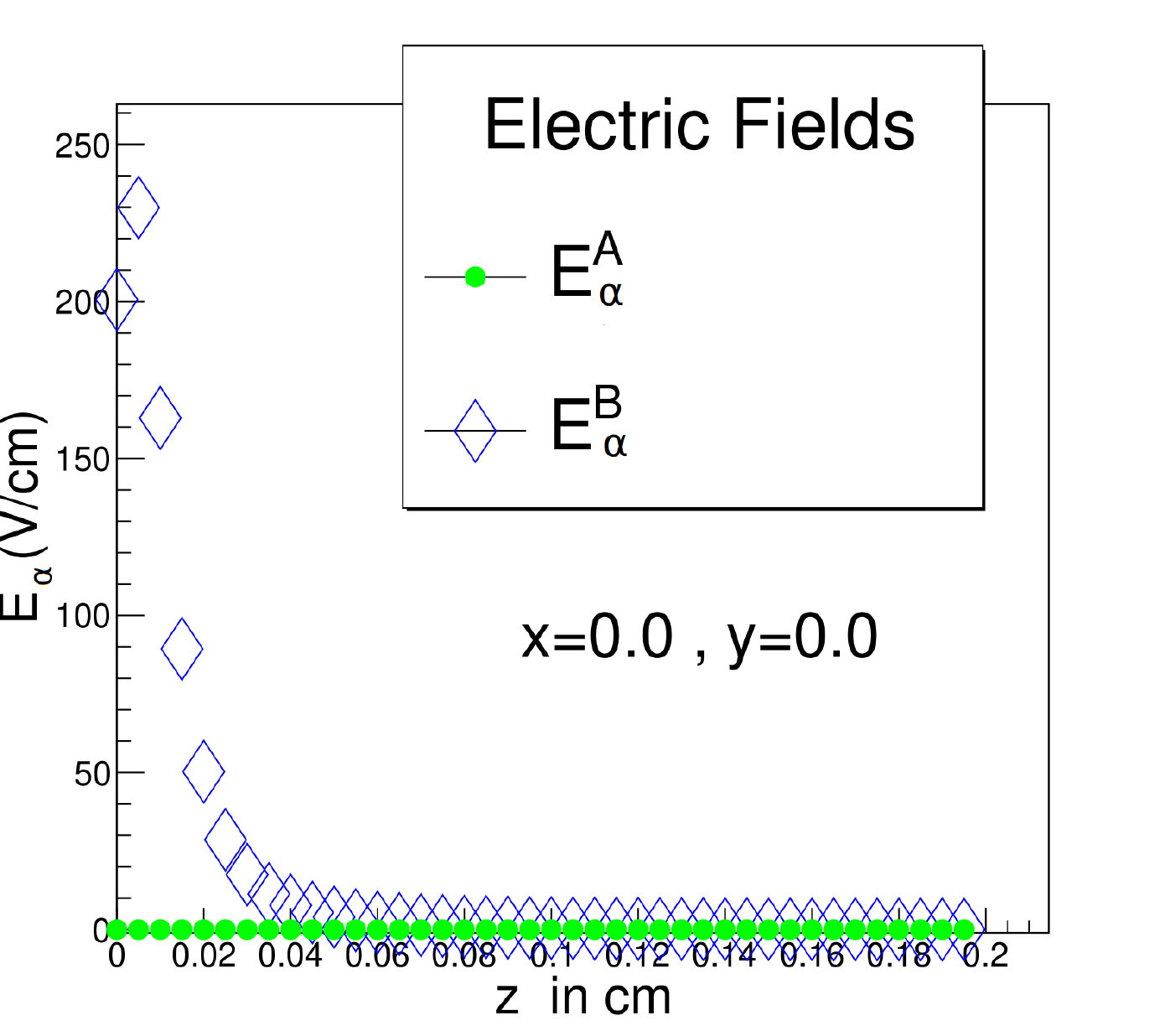}	}
	
	\caption{\label{fig:1-1} Computation of electric field components for case-2(section \ref{subsec:Results-of-case-2})
		on the z-axis. (a) The variation of $E_{z}^{A,B,C}$ on the z-axis
		has shown in the left-side axis of figures and right side axis
		contain the corresponding ratios of the field components between $E_{z}^{C}$and
		$E_{z}^{A,B}$, (b) variation of $E_{r}^{A,B,C}$ on the z axis, (c)
		variation of $E_{\alpha}^{A,B}$ on the z axis \cite{Dey_2020}.}
\end{figure}

The magnitudes of $E_{z}$ components for the three methods A, B, and C ($E_{z}^{A,B,C}$) are the same, as shown in Figure \ref{fig:Ez-x=00003D00003D00003D0-y=00003D00003D00003D0-non-uni}. However, there are discrepancies between the radial and ${\alpha}$ components of the electric field calculated in method B ($E_{r}^{B},E_{{\alpha}}^{B}$) and methods A and C ($E_{r}^{A,C},E_{{\alpha}}^{A}$), as shown in Figures \ref{fig:Er-x=00003D00003D00003D0-y=00003D00003D00003D0-non-uni} and \ref{fig:Ephi-x=00003D00003D00003D0-y=00003D00003D00003D0-non-uni}. The component $E_{\alpha}^{C}$ is always zero, so it is not shown in Figures \ref{fig:Ephi-uniform-x0-y0} and \ref{fig:Ephi-x=00003D00003D00003D0-y=00003D00003D00003D0-non-uni}.

The components $E_{r}^{A,C},E_{{\alpha}}^{A}$ still give the same result of zero along the z-axis, as in case-1 for methods A and C. However, the components $E_{r}^{B},E_{{\alpha}}^{B}$ show a non-zero value, especially near the charge distribution where the value is much higher than zero. These discrepancies can be explained by the angular distribution of charges shown in Figure \ref{fig:angular-distribution-of-full-figure}. It is clear from this figure that the angular distribution is not uniform; instead, most of the charges are within the angular range from $0$ to $150$ degrees.
\hspace{-1cm}
\subsubsection{Calculation of electric field including ion} 
\begin{figure}[H]
\centering\subfloat[\label{fig:Electrons-position-at3d}Electrons position at 19.06 ns]{\includegraphics[scale=0.28]{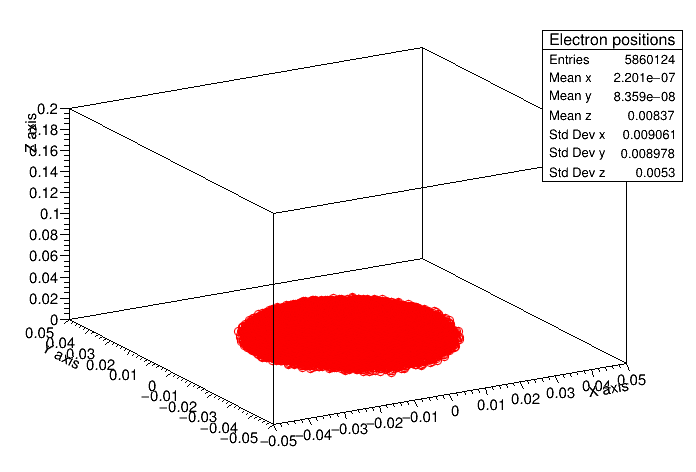}
	
}\subfloat[\label{fig:Ions-position-at3d}Ions position at 19.06 ns]{\includegraphics[scale=0.28]{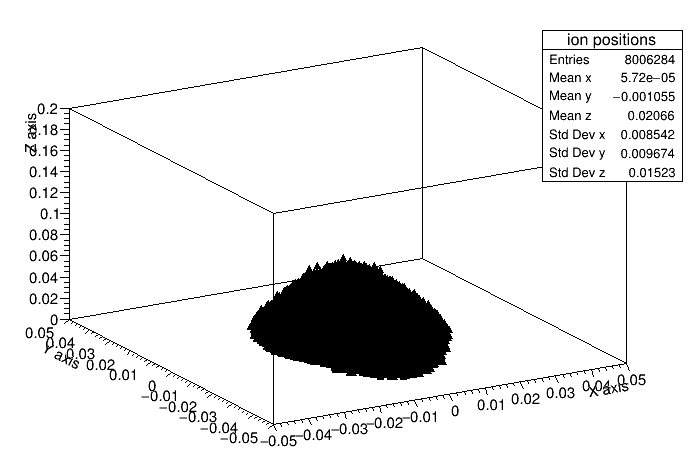}
	
}

\centering\subfloat[\label{fig:x-component-(Ex)-of3d}x-component (Ex) of space charge
field at different locations]{\includegraphics[scale=0.28]{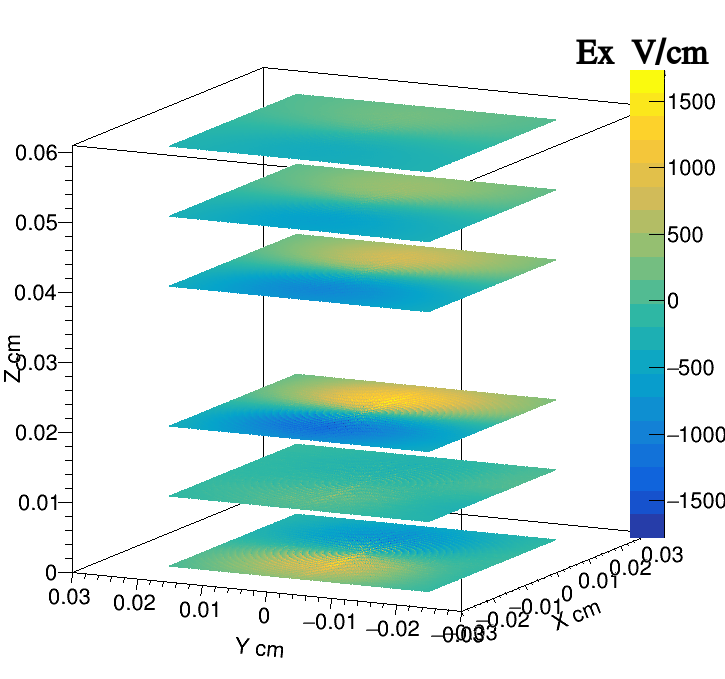}

}~~\subfloat[\label{fig:y-component-(Ey)-of3d}y-component (Ey) of space charge
field at different locations]{\includegraphics[scale=0.28]{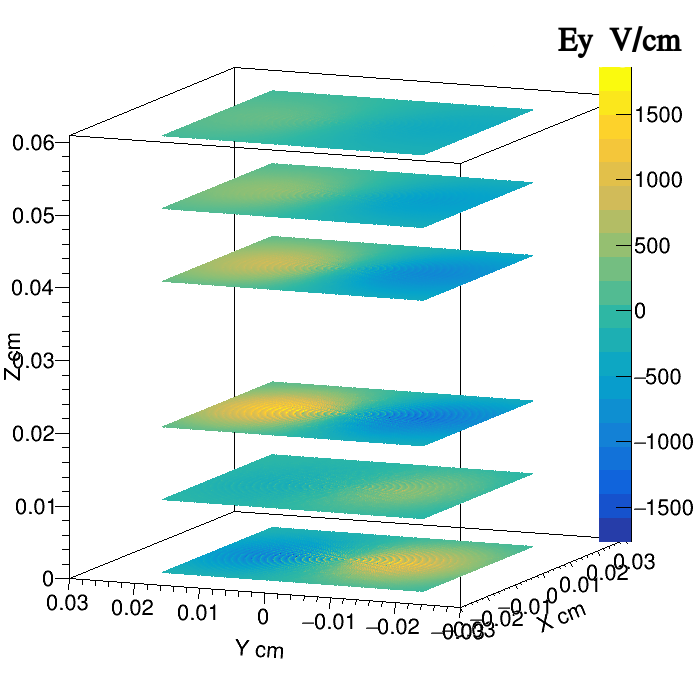}

}

\centering\subfloat[\label{fig:z-component-(Ez)-of3d}z-component (Ez) of space charge
field at different locations]{\includegraphics[scale=0.28]{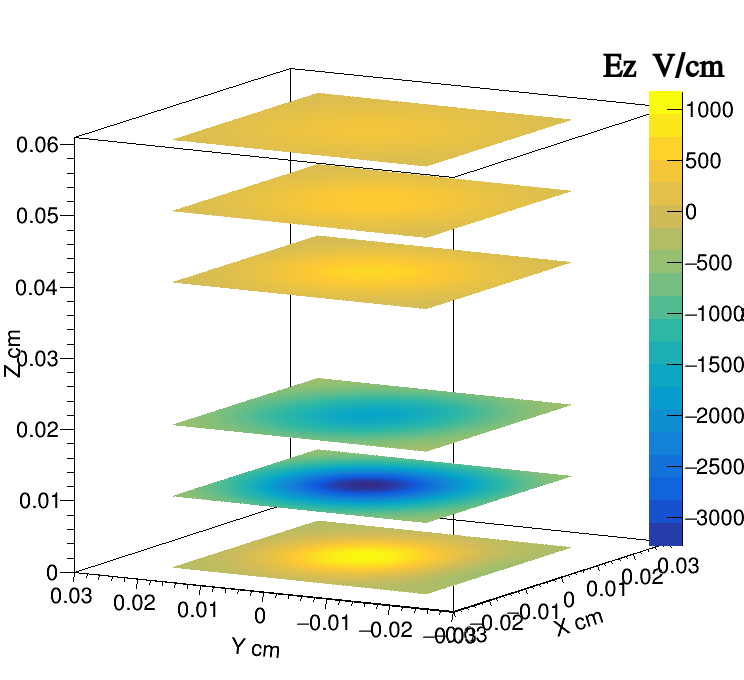}

}\subfloat[\label{fig:Magnitude-of-resultant3d}Magnitude of resultant ($E_{t}^{B}$)
space charge field at different locations]{\includegraphics[scale=0.28]{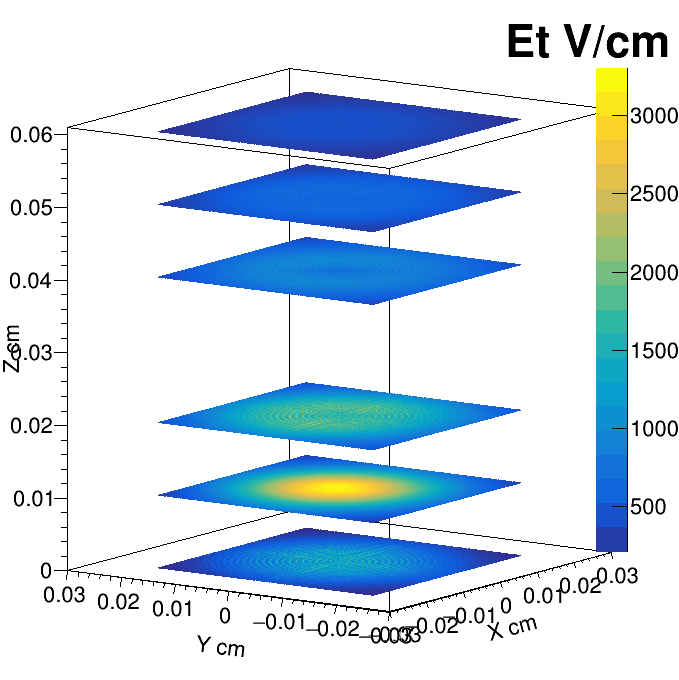}

}

\caption{Calculation of electric field due to ions and electrons\label{fig:3dField} \cite{Dey_2020}.}

\end{figure}

In this section, we have simulated the track of a muon with an energy of 2 GeV using Garfield++. The muon passed through the RPC and generated eight primary electrons inside the gas gap. Therefore, an avalanche was developed with those eight primary electrons under the same applied field, geometry, and gas mixture as discussed in section \ref{sec:position}. The positions of 5860124 electrons and 8006284 ions at 19.06 ns are shown in Figures \ref{fig:Electrons-position-at3d} and \ref{fig:Ions-position-at3d}, respectively. The electric field of the total space charge (ions+electrons) at the same instant of the avalanche was analyzed using 
\vspace{1cm}
method-B under the conditions of case 2. The $E_{x}^{B}$, $E_{y}^{B}$, and $E_{z}^{B}$ components of the field at different z-positions and parallel to the XY planes are shown in Figures \ref{fig:x-component-(Ex)-of3d}, \ref{fig:y-component-(Ey)-of3d}, \ref{fig:z-component-(Ez)-of3d}, and \ref{fig:Magnitude-of-resultant3d}. The color bar of figures shows the value of the electric field in V/cm unit. The maximum magnitude of the resultant ($E_{t}^{B}$) of the space charge field is 3423.25 V/cm, which is approximately 6.8\% of the applied field of 50 kV/cm.
\vspace{-2cm}
\section{Summary}
Initially, it was assumed that the electron cloud had some axial symmetry about the z-axis. Therefore, due to this symmetry, one could easily neglect the $E_{\alpha}$ component. However, from the analysis of an avalanche from a single primary electron, it was observed that the angular charge distribution is not uniform. Therefore, the $E_{\alpha}$ component plays a significant role in the development of the avalanche and can no longer be ignored. It has been previously discussed that the avalanche is generated from a single primary electron. However, in an actual event, the avalanche can be formed from several primaries. Therefore, the results for this case, along with the reflections of charges on the ground plates, need to be found, which is also a future interest.

The merit of using a straight-line approximation is that it produces similar results as uniformly charged rings and can easily be used when the charged density is nonuniform over the ring. Another remarkable feature is that the components of the field equations do not contain any elliptical integrals, so there is no need to worry about numerical integrations. Currently, Garfield++ does not have any module to generate an avalanche with the dynamic calculation of space charge field. In our local version of Garfield++, we have updated it with this feature using modules described in methods A and B.

\chapter{Numerical Study of Effects of Electrode Parameters and Image Charge on The Electric Field Configuration of RPCs}\label{ch4}
	\section{Introduction}
 The basic detector physics parameters of a Resistive Plate Chamber, such as ionization, multiplication, and signal induction, are functions of temperature, pressure, and electric field\footnote{The contents of this chapter are taken from the following publication by the author:
 	
 	\textbf{Numerical study of effects of electrode parameters and image charge on the electric field configuration of RPCs, T. Dey et al 2022 JINST 17 P04015 DOI 10.1088/1748-0221/17/04/P04015}}. Therefore, meticulous field measurements are always necessary. It is known that direct measurement of the electric field inside the RPC is difficult to achieve. However, an indirect measurement of the electric field in the gas discharge process using the Stark shift exists in the literature \cite{Cvetanovi__2015}. As a result, it becomes necessary to rely on analytical or numerical methods to estimate the electric field inside the RPC. An analytical approach to field calculation using the surface charge method (SCM) inside an RPC can be found in \cite{AMMOSOV1997217}. However, for solving realistic complex problems, it is preferable to use a numerical approach due to limitations in the analytical approach.
 
 In a single-gap RPC, in addition to the three primary layers (two electrodes and a gas gap), there are other dielectric layers and geometric non-uniformities such as side and button spacers. The electric field in such geometries, which have distinctly three-dimensional features, can be obtained using numerical solvers such as neBEM, COMSOL, among others \cite{MAJUMDAR2008346,MAJUMDAR2009719,comsol}.
RPC offers three basic operation modes, which are a) avalanche mode,
b) saturated avalanche mode, and c) streamer mode \cite{MOSHAII2012S168,Cardarelli1996AvalancheAS}. The charge accumulation near the electrode is high due to successive avalanche in high rate experiments.
As an effect, the distortion in the applied field due to the space charge effect is also significant \cite{Lippmann_1}. This phenomenon slows down the growth of avalanches generated from subsequent primaries, which reduces the particle detection efficiency \cite{rpc-book}. Therefore, limiting the average charge is an efficient way to work at a high rate\cite{Paolozzi:2012pt}. Indeed, in high rate experiments, it is preferable to work
in avalanche mode or low gain mode rather than the other two modes since gain, and consequently the generation of space charge, is larger in these other modes..

\begin{figure}
	\center{\includegraphics[scale=0.36]{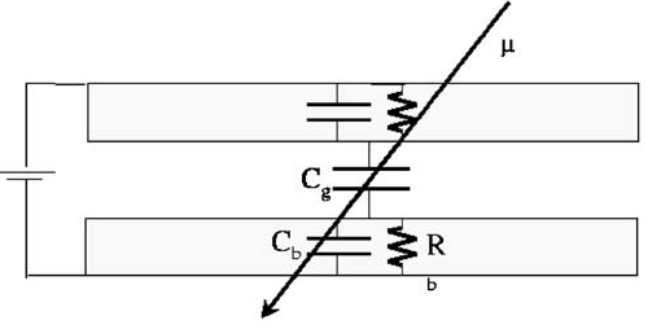}
		
	}
	
	\caption{\label{fig:equivalentRPC}Equivalent circuit of an RPC. \cite{ABBRESCIA20047}}
\end{figure}

The equivalent circuit of an RPC system has been shown in the figure \ref{fig:equivalentRPC}, where the gas gap is represented as the capacitor $C_{g}$ and and the resistive electrodes are represented as parallel combination of $C_{b}$ and R. The capacitor $C_{g}$ can be partially discharged when an ionising particle is passed through the gas gap and the effective voltage accross the gas gap reduced proportionally to the generated avalanche charge $q_{av}$ . Then, the voltage accross $C_{g}$ restore again with the help of external power supply. The behaviour of charging up process is exponential and the RC time constant $\tau_{g}$ of an RPC can be written as\cite{Gonzalez-Diaz:2006qno,ABBRESCIA20047}:
\begin{equation}\label{eq:RC_FORMULA}
	\tau_{g}=2R_b(2C_b+C_g)\\
	=\rho\epsilon_{0}(\epsilon_{r}+\frac{d}{g}),
\end{equation} where $\rho,\epsilon_{r},d$ are resistivity, relative permittivity, thickness of electrodes respectively and $g$, $\epsilon_{0}$ is the gas gap and free space permittivity. The bulk resistivity ($\rho$) of balekilte electrodes may vary between  $\backsim10^9-10^{11}\Omega\mbox{-}m$. Now let  the gas gap (g) and electrode thickness (d) is 2mm and relative permittivity ($\epsilon_r$) of bakelite is 5. Then, for $\rho\approx2\times10^{10}\Omega\mbox{-}m$ the typical value of the $\tau_g$ can be calculated using equation \ref{eq:RC_FORMULA} as follows:
\begin{equation}
	\tau_g=(2\times 10^{10})\times(8.85\times10^{-12})\times(5+1)=1.05\, sec.
\end{equation}
A low $\tau_{g}$ of electrodes can serve a good detection
rate. Hence, a search for optimized electrodes based on the parameters
$\rho,\epsilon_{r,}d,g$ is necessary \cite{Aielli_2016,Carboni:2003my}. A detailed simulation of the rate capability may help to optimize the same parameters. 
The accuracy of such simulation depends on the precise calculation of the dynamic space-charge electric field along with the two polarisation fields of electrodes a) polarisation due to space charge field and b) DC
polarisation (applied field). It is known that dielectric materials show a polarisation effect in the presence of the electric field. These polarisation fields can be calculated by solving the Poisson equation with proper boundary conditions or using the image method. A
Poisson equation solving approach can be found in \cite{HEUBRANDTNER,Lippmann:2003ar,Lippmann_1}. Alternatively the polarisation field due to space charge can be calculated using method of the image for such a simple geometric configuration, which  is a relatively easier and faster way. The basic geometry
of an RPC consists of five layers of dielectrics as shown in
figure \ref{fig:RPC_image_charge_formation}.
The formation of the image due to three layers of the dielectric can be found in the section 21. of chapter 6 of \cite{Weber-W}, where the charge resides inside the middle dielectric material
and the dielectric constant of outermost dielectrics is considered the same. The geometrical configuration used in the section 21. of chapter 6 of \cite{Weber-W} is nearly similar to that of an RPC. However, in that work the point charge location is not general, and the termination condition of infinite series of reflections for a general position of charge is also not discussed. A more detailed formation of infinite series of images of a point charge positioned in front of three dielectric layers has been discussed in \cite{Jomaa1983ElectricFD,multi-layer}(see figure \ref{fig:threelayer}). 

In this paper, in section \ref{sec:section2} we have discussed the variation of the applied DC field inside the gas-gap with the three RPC parameters i.e. $\epsilon_r,d$ and $g$. In section \ref{sec:section3}, we have first discussed the method of image for metal electrode and then generalized the method for an RPC of five dielectric layers  by considering that each dielectric electrode is different from the other. In this study the point charge can be located anywhere inside the middle dielectric layer (gas-gap). We also gave an instance of the
calculation of the 3D space-charge
field along with its polarisation effect on the electrodes of an RPC in section \ref{sec:Section4_avalancheCharge} for a specific avalanche charge distribution using the model described in \cite{Dey_2020} and compared the results with the results from neBEM field solver and other models from the literature. We have also discussed the behavior of the electrode polarisation field with important detector parameters $\epsilon_{r,}d$ and $g$.

\section{Variation of applied field inside the RPC due to different electrode parameters and gas gap} \label{sec:section2}
A numerical model of an RPC having sides 15 cm $\times$15 cm  and 10 micron thick of graphite layer has been designed in Garfield++. A DC voltage of $\pm$4.5KV is applied over the two graphite layers.  However, the parameters electrode thickness $d$, gas gap $g$, and relative permittivity $\epsilon_{r}$ are varied systematically, and the electric field calculated for different configurations using the neBEM solver. This study has been divided into three cases where at a time, we fixed two parameters and varied the third one.\\
Case 1:  $d$ and $g$ kept fixed and varied $\epsilon_{r}$.\\
Case 2:  $\epsilon_{r}$ and $g$ kept fixed and varied $d$.\\
Case 3:  $\epsilon_{r}$ and $d$ kept fixed and varied $g$.
\subsection{Case 1, variation of $\epsilon_{r}$}
\begin{figure}
	\center\subfloat[\label{electricField_diff_epr}]{\includegraphics[scale=0.36]{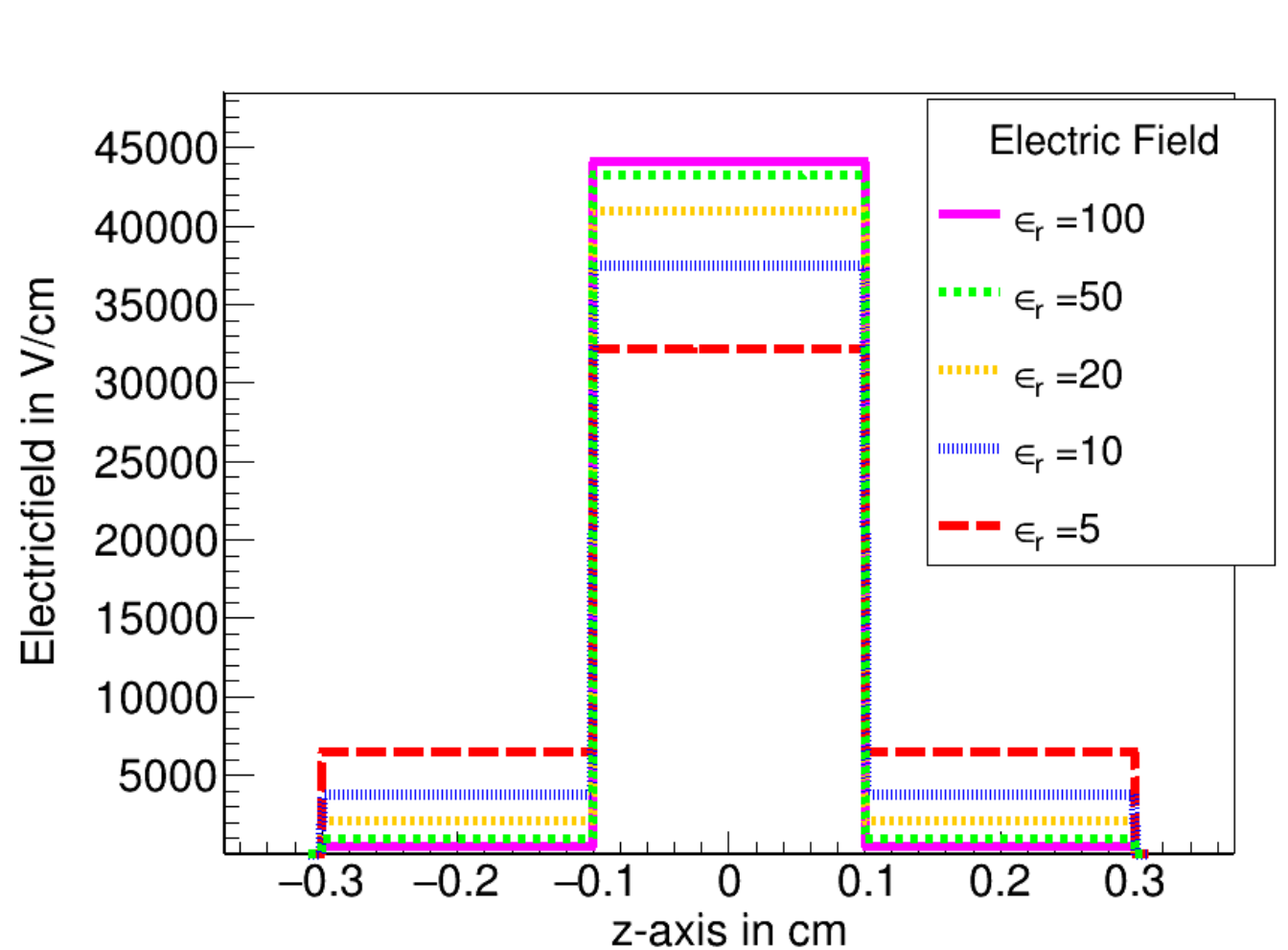}
		
	}
	
	\subfloat[\label{pot_diff_epr}]{\includegraphics[scale=0.36]{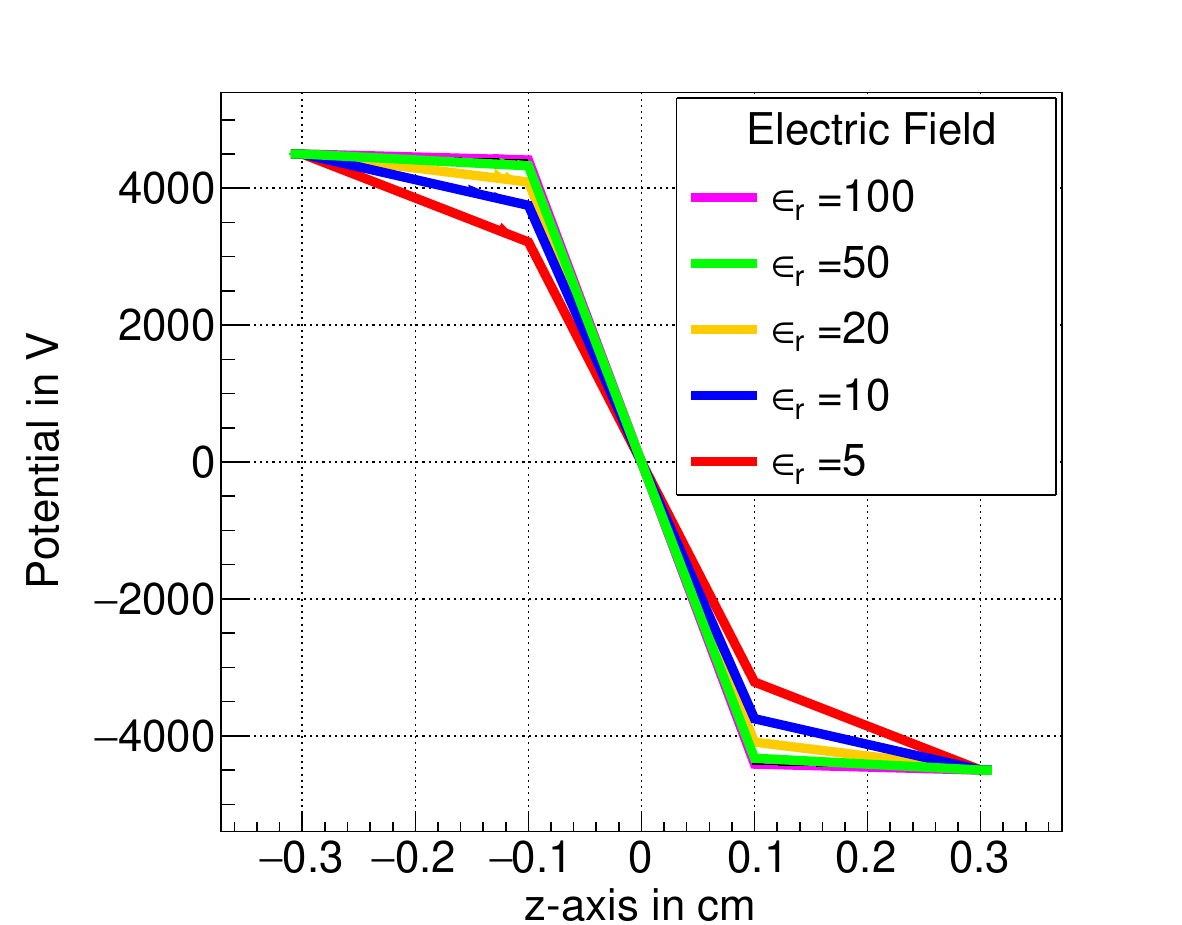}}\subfloat[\label{E_vs_epr}]{\includegraphics[scale=0.167]{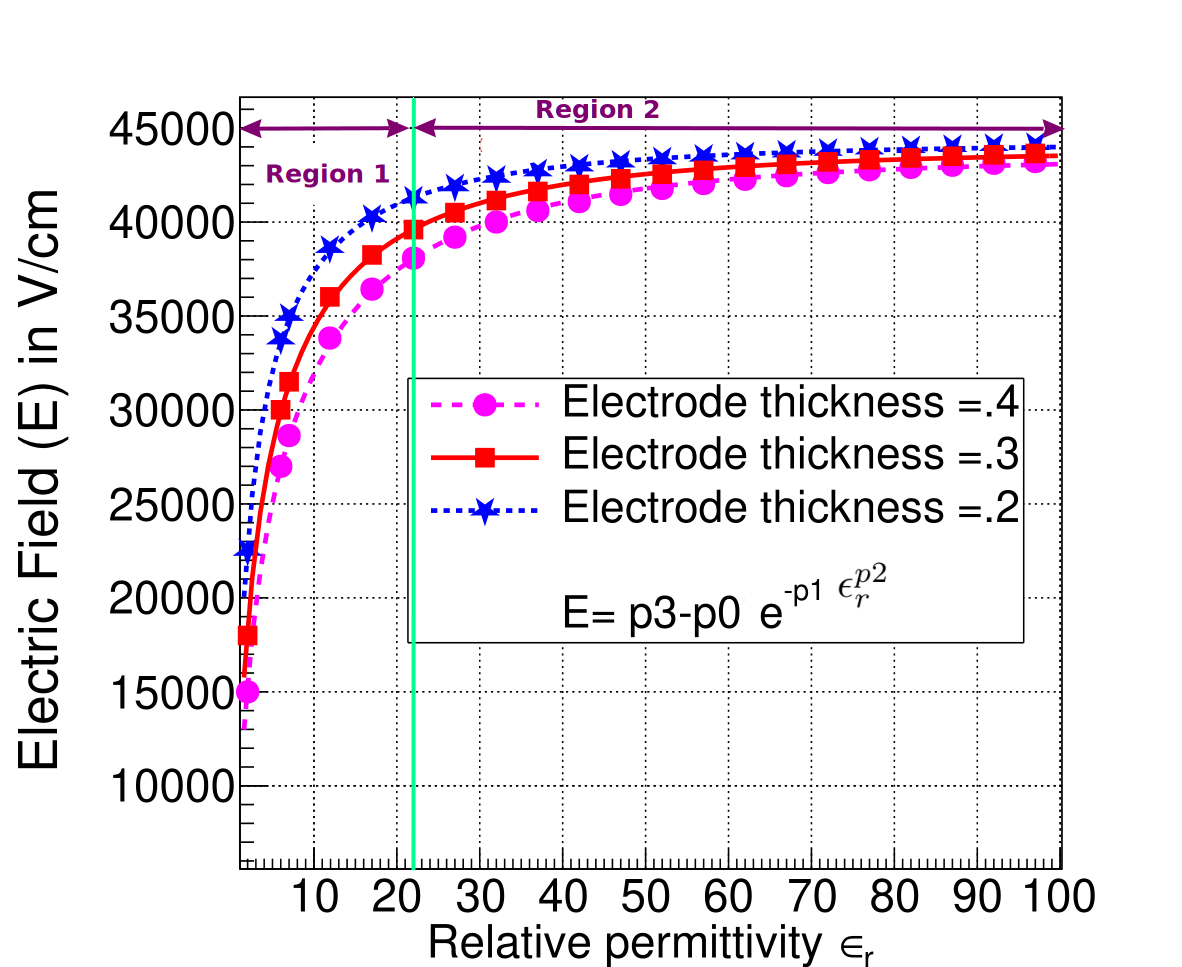}}
	
	\caption{\label{fig:electricField_diff_epsilon} (a) Applied electric field along the z-axis for different $\epsilon_{r}$ of electrodes. (b) Potential inside the RPC along the z-axis for different $\epsilon_{r}$. (c) Variation of electric field with $\epsilon_{r}$ at the middle of the gas gap. }
\end{figure}

It is known that the field inside the gas gap of the RPC is nearly constant at a fixed applied voltage. However, the field can be different for different electrode materials. The magnitude of the electric field and potential along the z-axis (perpendicular to the parallel plates) inside the RPC is shown in figures \ref{electricField_diff_epr} and \ref{pot_diff_epr} for different relative permittivity $\epsilon_{r}$ of electrodes. In figure \ref{electricField_diff_epr} the field is gradually increasing inside the gas gap (-0.1cm to 0.1cm) if we increase the value of permittivity $\epsilon_{r}$ but is gradually decreasing inside the dielectric electrodes (-0.3cm to -0.1cm and 0.1cm to 0.3cm) with the same $\epsilon_{r}$. Eventually, the slope of the variation of potential with z position is also dissimilar with different $\epsilon_{r}$ in all regions, which is shown in figure \ref{pot_diff_epr}. The variation of electric field with $\epsilon_{r}$ is visible more prominently in figure \ref{E_vs_epr}, where the electric field is calculated at the center of the detector. It is clear from the  figure  \ref{E_vs_epr} that the field initially grows and then saturates as $\epsilon_{r}$ is increased. We can roughly divide the figure \ref{E_vs_epr} based on electric field variations in two regions a) region 1 and b) region 2. Region 1 is from $\epsilon_{r}$=2 to 22, where the field is rapidly growing with $\epsilon_{r}$. Region 2 is from $\epsilon_{r}$=22 to 97, where the field tends to saturate on the increment of $\epsilon_{r}$. Now, the maximum percentage of change in the field on changing $\epsilon{_r}$ in both region 1 and 2 are tabulated below (Table \ref{table_diff_epsilon}) for three different thickness.
From Table \ref{table_diff_epsilon} it can be said that the percentage of change in the field is consistently falling off with the thickness of electrodes (d) in both regions 1 and 2. Hence, the dependence
of the electric field on  $\epsilon_{r}$ is low for the smaller thickness of the electrodes. 
\begin{table}[H]
	\center%
	\begin{tabular}{|c|c|c|}
		\hline 
		d (cm) & change of field in region 1 (\%)  & change of field in region 2 (\%)\tabularnewline
		\hline 
		\hline 
		0.4 & 153.8 & 13.5\tabularnewline
		\hline 
		0.3 & 120 & 10.22\tabularnewline
		\hline 
		0.2 & 83.33 & 6.88\tabularnewline
		\hline 
	\end{tabular}
	
	\caption{\label{table_diff_epsilon}Percentage of change in electric field for different thickness of electrodes.}
	
\end{table}
The data points of figure \ref{E_vs_epr} are fitted with the equation: 
\begin{equation}\label{eqn:2.1}
	f(\epsilon_{r})=p3-p0\, Exp(-p1\,\epsilon_{r}^{p2})
\end{equation}
to understand the functional behavior of the variation of the electric field with $\epsilon_{r}$ at a fixed electrode thickness and to allow quick interpolation at intermediate values of $\epsilon_{r}$. The fit results are tabulated below in Table \ref{table2_fitVAlue}.  From the dimensional analysis it is clear that the dimension of the parameters p3 and p0 should have the dimension of electric field, where p0 is normalisation constant and p3 can be taken as the offset. Since the term inside the exponential should be dimensionless, therefore the parameter p1 should be the function of $\epsilon{_r}$ and it can be called rate of reduction factor and p2 must be a constant. The other physical significance of these parameters are yet to be understood.

\begin{table}[H]
	\center%
	\begin{tabular}{|c|c|c|c|c|}
		\hline 
		d (cm) & p0 & p1 & p2 & p3\tabularnewline
		\hline 
		\hline 
		0.4 & $\begin{array}{c}
			60135\\
			\pm\\
			2189
		\end{array}$ & $\begin{array}{c}
			0.55\\
			\pm\\
			0.03
		\end{array}$ & $\begin{array}{c}
			0.5\\
			\pm\\
			0.02
		\end{array}$ & $\begin{array}{c}
			43511\\
			\pm\\
			129
		\end{array}$\tabularnewline
		\hline 
		0.3 & $\begin{array}{c}
			-68494\\
			\pm\\
			3396
		\end{array}$ & $\begin{array}{c}
			0.74\\
			\pm\\
			0.05
		\end{array}$ & $\begin{array}{c}
			0.43\\
			\pm\\
			0.02
		\end{array}$ & $\begin{array}{c}
			43858\\
			\pm\\
			110
		\end{array}$\tabularnewline
		\hline 
		0.2 & $\begin{array}{c}
			88207\\
			\pm\\
			6462
		\end{array}$ & $\begin{array}{c}
			1.1\\
			\pm\\
			0.07
		\end{array}$ & $\begin{array}{c}
			0.37\\
			\pm\\
			0.02
		\end{array}$ & $\begin{array}{c}
			44230\\
			\pm\\
			81
		\end{array}$\tabularnewline
		\hline 
	\end{tabular}
	
	\caption{\label{table2_fitVAlue}Fit parameters of the plots shown in figure \ref{E_vs_epr}.}
	
\end{table} 

\subsection{Case 2, variation of d}
\begin{figure}
	\center\subfloat[\label{E_Vs_thickness}]{\includegraphics[scale=0.3]{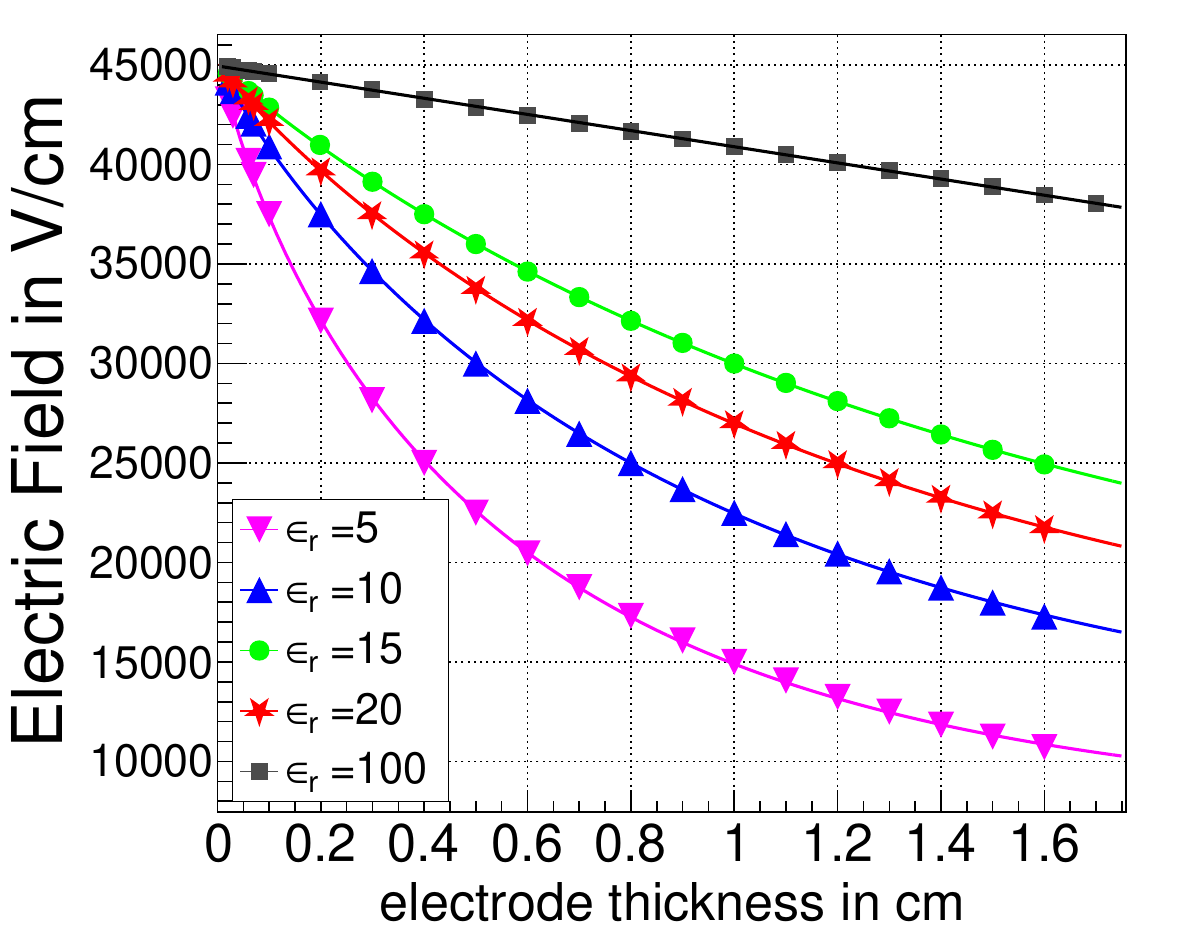}}\subfloat[\label{E_VS_gas_gap}]{\includegraphics[scale=0.3]{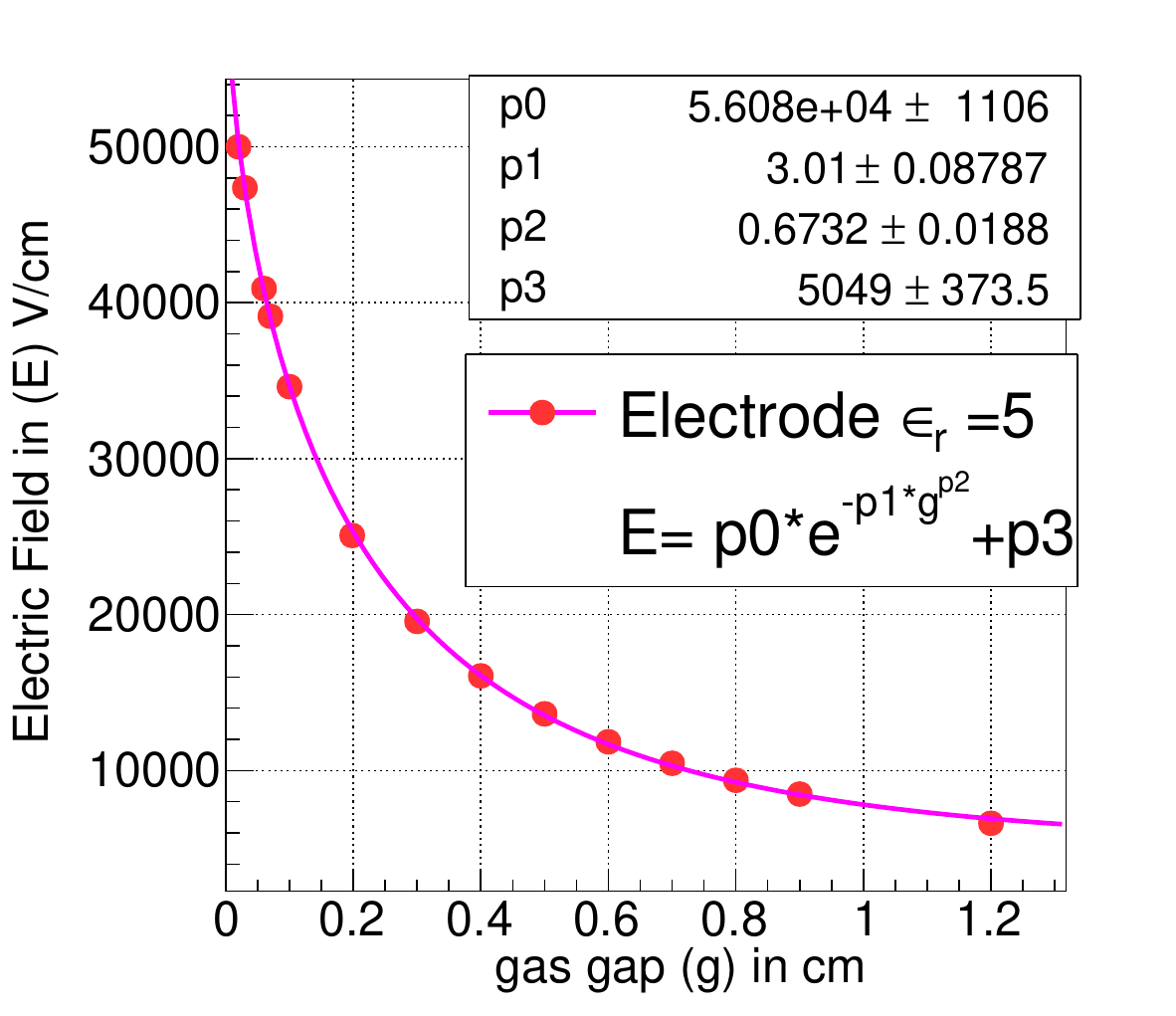}}
	
	\caption{\label{fig:reflection-1-1}(a) Variation of applied electric field with the electrode thickness for different $\epsilon_{r}$ of electrodes. (b) Variation of applied electric field with the gas gap.}
\end{figure}

As discussed earlier, electrode thickness is one of the most crucial factor when entering the high particle rate detection region. Therefore, the study of the dependence of electric fields on the same is necessary. The variation of the electric field inside the middle of the gas gap with electrode thickness has been shown in figure \ref{E_Vs_thickness}, where the electric field is decreasing with the increment of the thickness of electrodes. Each curve in figure \ref{E_Vs_thickness} has been shown for a fixed gas gap (g) of 2 mm and several permittivities ($\epsilon_{r}=5,10,15,20,100$), where it may be noted that an increase in $\epsilon_{r}$ leads to flatter curves. This is because the electrodes are shifting from the dielectric region to the conductive region. Hence, the dependence of the electric field on the electrode thickness for higher $\epsilon_{r}$ is getting small. The maximum percentage of the reduction of field on the variation of thickness from 0.02 cm to  1.6 cm  is shown in the table \ref{epsilon_vs_field_table_3} for several values of $\epsilon_{r}$, where the percentage of change in the field with respect to electrode thickness d is gradually reducing with the increament of $\epsilon_{r}$. Hence, it is expected that for the perfect conductor ($\epsilon_{r}\rightarrow\infty$), the curvature of figure \ref{E_Vs_thickness} will become flat and parallel to electrode thickness axis. So, the electric field will be independent of thickness of electrodes.   

\begin{table}
	\center%
	\begin{tabular}{|c|c|}
		\hline 
		$\epsilon_{r}$ & change in field with respect to d (\%)\tabularnewline
		\hline 
		\hline 
		5 & -75.24\%\tabularnewline
		\hline 
		10 & -60.81\%\tabularnewline
		\hline 
		15 & -51.06\%\tabularnewline
		\hline 
		20 & -44.05\%\tabularnewline
		\hline 
		100 & -15.29\%\tabularnewline
		\hline 
	\end{tabular}
	
	\caption{\label{epsilon_vs_field_table_3}Percentage of change in the electric field on variation of thickness from 0.02 cm to 1.6 cm for several permittivity of electrodes ($\epsilon_{r}$). }
	
\end{table}

\subsection{Case 3, variation of g}
It is expected that if we fix the electrode thickness and permittivity($\epsilon_{r}$), then the electric field will drop at any point with the increment of the gas gap. In figure \ref{E_VS_gas_gap} the variation of the electric field with the gas gap at the middle of the detector has been shown, where the electrode thickness and permittivity are fixed at d=0.2 cm and $\epsilon_{r}=5$, respectively. The figure \ref{E_VS_gas_gap} has been fitted with the equation: 
\begin{equation}\label{eqn:2.2}
	f(g)=p0\,exp(-p1\,g^{p2})+p3
\end{equation}
to understand the functional behaviour and convenient interpolation. The fit parameters  p0,p1,p2,p3 has been shown on the same figure \ref{E_VS_gas_gap}. As argued for the parameters of the equation \ref{eqn:2.1} here also the parameters p0 and p3 of the equation \ref{eqn:2.2} carry the dimension of electric field and p1 is the function of g and p2 a constant. The other physical significance of these parameters needs further study.

\section{Calculation of electrode polarisation fields}\label{sec:section3}
As discussed earlier the electric field inside an RPC is changing while an avalanche is generating. The reason of this change is the growing of space charges inside the RPC. The field due to those space charge can be divided in two sections, (a) field due to the space charges itself, (b) field due to polarisation of electrodes due to those space charges. The electrode polarisation field is calculated using method of image. In the following subsections we will discuss first the simplest case, charge particle inside the two metalic (no dielectric present) grounded plates and then the discussion will be continued to the case where dielectric layers will be present.
\subsection{Image charges of a point charge inside two grounded metallic conductor}

\begin{figure}
	\center\includegraphics[scale=0.5]{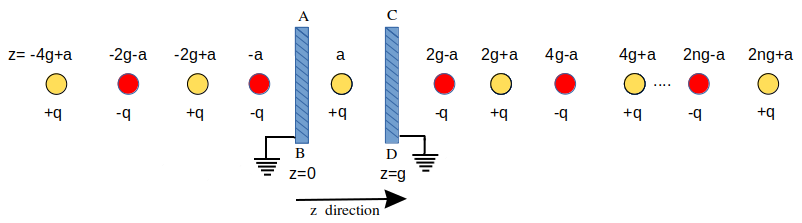}
	
	\caption{\label{fig:reflection}Formation of image charge due to a point charge between two metallic grounded conductors.}
\end{figure}
\begin{figure}
	
	
	
	\center\subfloat[\label{charge_vs_dist_field_D1}]{\includegraphics[scale=0.31]{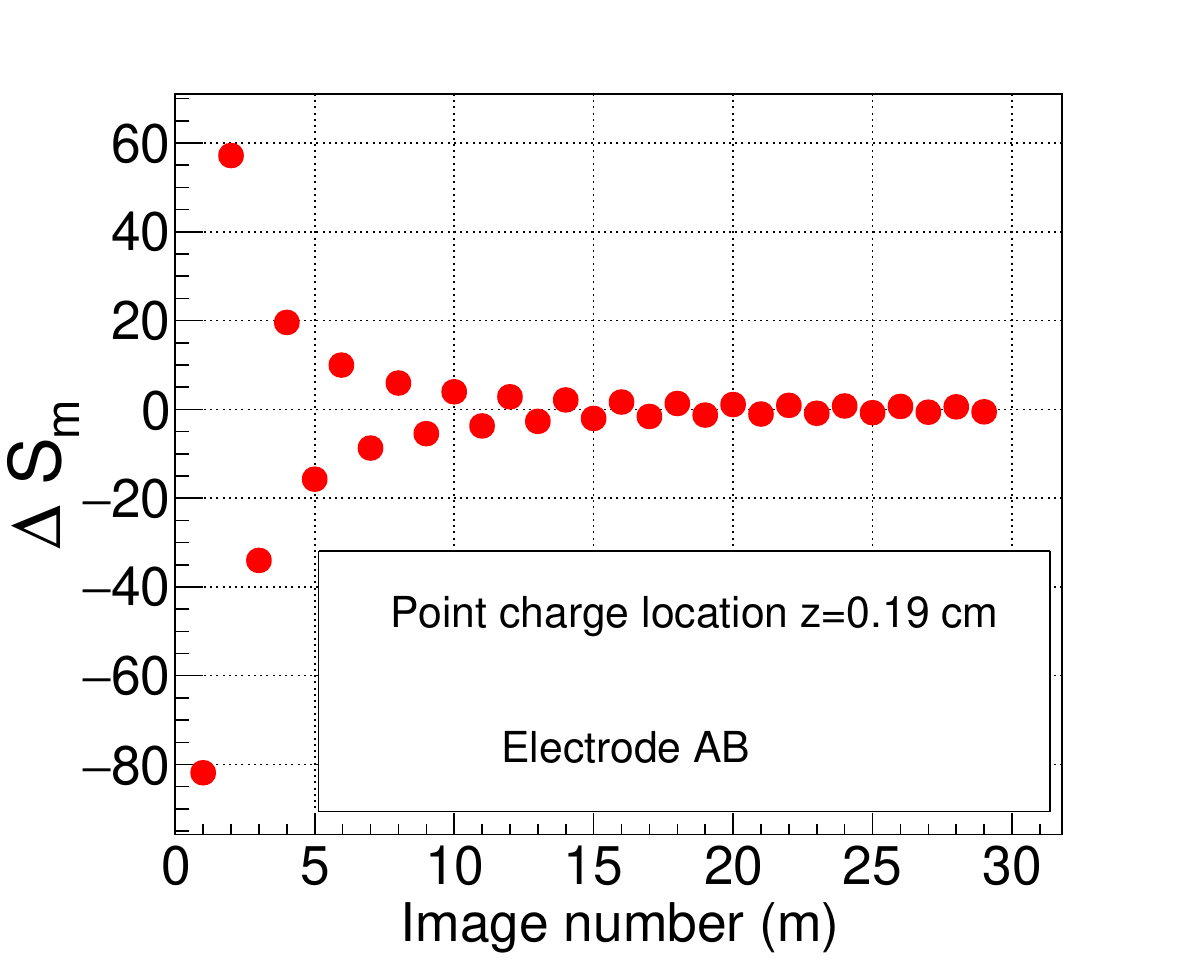}
		
	}\subfloat[\label{charge_vs_dist_field_D2}]{\includegraphics[scale=0.3]{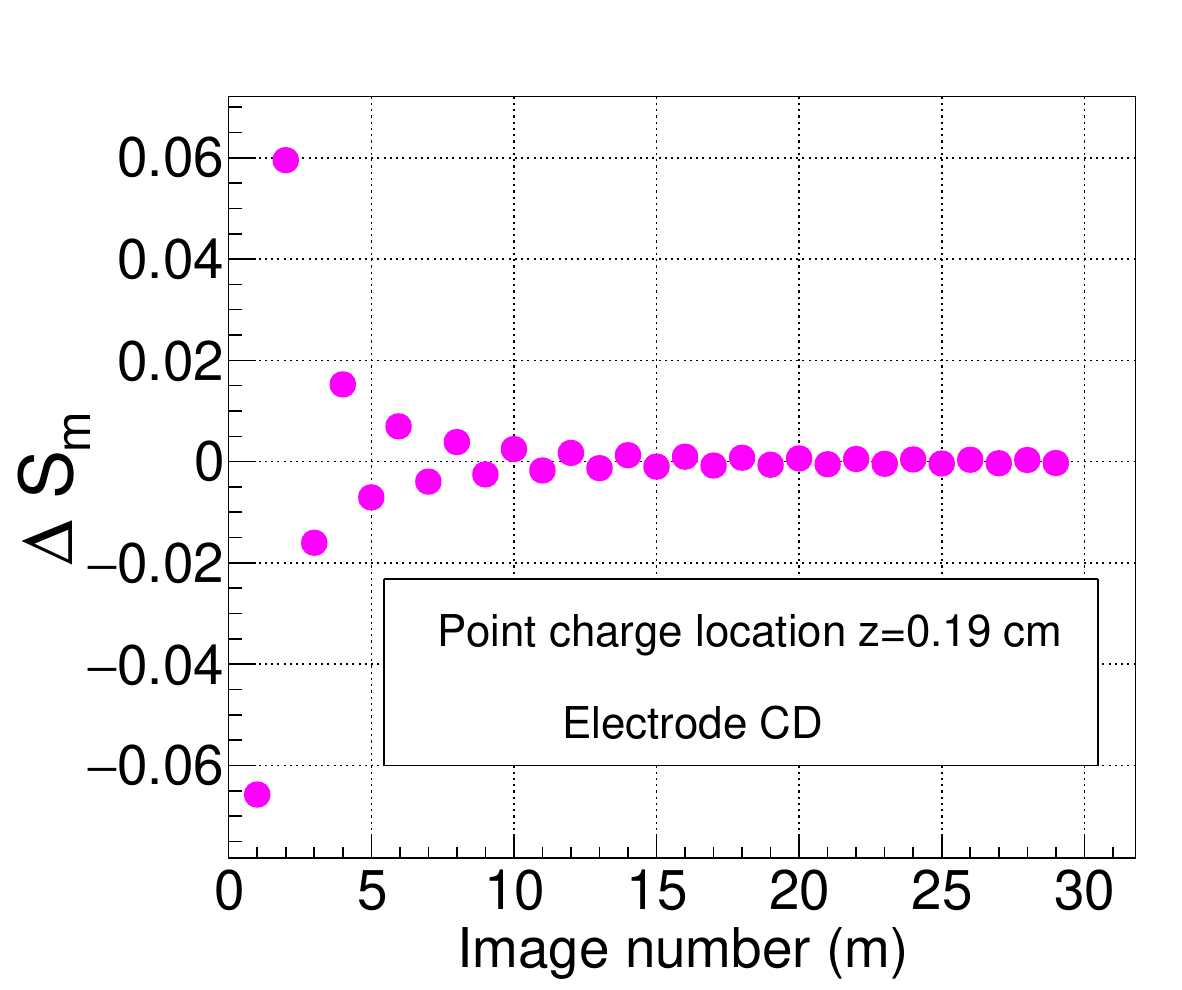}
	}
	
	\caption{(a) Percentage of contribution ($\Delta S_m$) of higher order image charges on the total electric field for electrode AB. (b) Percentage of contribution ($\Delta S_m$) of higher order image charges on the total electric field for electrode CD. }
	
\end{figure}

Let us consider a point charge between two metallic grounded conductors AB and CD, at a distance ``a" from the electrode AB (z=0, see figure \ref{fig:reflection}). Now, due to the planar geometry,
there will be an infinite number of reflections of the source
charge on both sides of the
grounded metallic electrodes, as shown in figure \ref{fig:reflection}. Then the electric field at any point between the electrodes is due to the sum of the fields of source charge and infinite image charges. The convergence of infinite series to calculate induced charge density and the total charge for such geometric configuration has been discussed in \cite{new_comb_image_plate}. Since in this work, our focus is on the calculation of electric field, we will find the convergence or termination of infinite series in terms of $E_{n}=\frac{q_{n}}{r_{n}^{2}}$, where $q_{n}=n^{th}$ image charge in electronic charge e unit and $r_{n}=\,\,$distance of $q_{n}$ from nearest electrode and $E_n$ will be different for different electrode. 
In figure \ref{fig:reflection} 
the magnitude and sign of a few image charges corresponding to the electrode AB and CD with their reflection number have been shown, where the magnitudes are same for all image charges but signs are alternating. Therefore, the sum of all n number of $E_n$ is $S_{m}=\sum\limits_{n=0}^m E_{n}$. If we multiply $E_{n}$ with coulomb constant and electron charge, then $S_{m}$ will give the total field due to m images on the surfaces of electrodes AB or CD. Hence, the percentage of change in $S_{m-1}$ on addition of one more image charge is:
\begin{equation}\label{eqn:DeltaSm}
	\Delta S_{m}=\frac{S_{m}-S_{m-1}}{S_{m-1}}\times 100
\end{equation}
where, m=1,2,3...., and $\Delta S_{m}$ gives the percentage of contribution of $m^{th}$ order image charge on field. From figure \ref{fig:reflection} we can see that the image charges are alternating between $\pm 1$. Hence, the sign of $\Delta S_{m}$ will also alternate and gradually converges to zero (see figure \ref{charge_vs_dist_field_D1}). For a=0.19, the charge is very close to the electrode CD. Therefore, the term $E_{0}$ (n=0) will be very large for electrode CD with respect to the all other $E_{n}$s (n=1,2,3..). Therefore, in $S_{m}$ the contribution from the higher-order $E_{n}$ (n=1,2,3..) will be small, following that $\Delta S_{m}$ (for m=1,2,3..) will be very small (see figure \ref{charge_vs_dist_field_D2}). On the other hand, the source charge is very far from the electrode AB. Therefore, few higher-order terms of $E_{n}$ (n=1,2,3..) can be comparable to $E_{0}$. 
This is the reason why $\Delta S_{m}$ is also higher for those few orders (see figure \ref{charge_vs_dist_field_D1}). However, the contribution of images of electrode AB on the entire field is smaller than that of electrode CD. Therefore, while calculating the image field for any electrode, one could apply a cut $\mid\Delta S_{m}\mid$ to terminate the series $S_{m}$ at certain order of reflection. For an example, if we cut $\mid\Delta S_{m}\mid\geq 20$ then from figure \ref{charge_vs_dist_field_D1} we can say that we will need to add four higher-order terms and one zeroth-order term to calculate the image field. Using the same cut value we could have only zeroth-order term for electrode CD (figure \ref{charge_vs_dist_field_D2}). Thus, this method opens the opportunity to select several images dynamically throughout the avalanche simulations inside the RPC.	
\\
Alternatively, it can be shown that the total charge induced on the electrodes due to the infinite images is finite. From figure \ref{fig:reflection} it is noted that the positive charges +q are at z=2ng+a and negative charges -q at z=2ng-a, where n runs from $-\infty$ to $\infty$. If we consider a ring of radius R centered at z-axis, along the line of image charges and width dR then at electrode AB (figure \ref{fig:reflection}) the surface-charge density $\sigma(R)$ can be represented as discussed in \cite{new_comb_image_plate}, which is as follows:
\begin{eqnarray}\label{eqn:surfaceChargeDensity}
	\sigma(R)&=&\frac{q}{4\pi} \sum_{n=-\infty}^{\infty}[-s(2ng+a,R)+s(2ng-a,R)]
\end{eqnarray} 
where
$s(z,R)=z(R^2+z^2)^{-3/2}$.
The convergence test of infinite series of equation \ref{eqn:surfaceChargeDensity} can be found in \cite{new_comb_image_plate}. 
The total surface charge can be found as follows \cite{new_comb_image_plate}:
\begin{equation}
	Q_{AB}=2\pi \int_{0}^{\infty} \sigma (R) RdR=-q\frac{g-a}{g}.
\end{equation} 
Similarly the total charge on the electrode CD can be calculated using the same method. The total surface chharge on CD is $Q_{CD}=-qa/g$. Therefore the total charge on both electrodes AB and CD is $Q=Q_{AB}+Q_{CD}=-q$.
\subsection{Formation of image charges in two layers of dielectric}

\begin{figure}
	\center\includegraphics[scale=0.5]{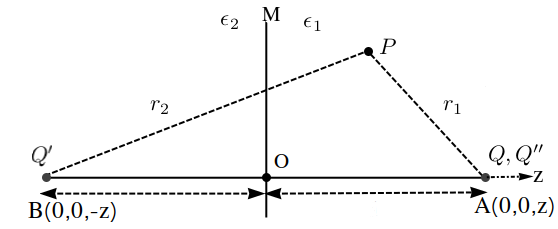}
	
	\caption{\label{fig:polarisation} Formation of image charge for two layers of dielectric case.}
\end{figure}

Let us consider a point charge $Q$ placed in a dielectric medium of permittivity $\epsilon_{1}$
at A(0,0,z) from the interface OM of two semi-infinite dielectric medium of permittivity $\epsilon_{1}$ and $\epsilon_{2}$, as shown in figure \ref{fig:polarisation}.
The field of $Q$ polarises the dielectric and the negative bound charges are
induced on the surface. The total field at any point P is the sum
of the field of bound charges and $Q$. Now to calculate potential
at P we imagine an image charge $Q^{\prime}$ in the dielectric
medium $\epsilon_{2}$ at position B(0,0,-z) away from the interfacing
surface of two dielectric medium (figure \ref{fig:polarisation}). Let
$\phi_{1},\phi_{2}$ denote potential in regions having dielectric permittivities $\epsilon_{1}$and
$\epsilon_{2}$. Now to satisfy boundary conditions 

\begin{equation}
	\begin{array}{c}
		\phi_{1}|_{z=0}=\phi_{2}|_{z=0}\\
		\epsilon_{1}\frac{\partial\phi_{1}}{\partial z}|_{z=0}=\epsilon_{2}\frac{\partial\phi_{2}}{\partial z}|_{z=0}
	\end{array}\label{eq:boundary_condition}
\end{equation}
$Q^{\prime}$should be $Q^{\prime}=-\alpha_{12}Q$
(where, $\alpha_{12}=\frac{\epsilon_{1}-\epsilon_{2}}{\epsilon_{1}+\epsilon_{2}}$).
Again to calculate potential at any point $P$ in the medium
$\epsilon_{2}$ another image charge $Q^{\prime\prime}$ can be
considered at the point A(0,0,z) in the medium $\epsilon_{1}$. Hence
to satisfy same boundary condition \ref{eq:boundary_condition} it
is found that the value of $Q^{\prime\prime}$ should be $Q^{\prime\prime}=\beta_{12}Q$
(where, $\beta_{12}=\frac{2\epsilon_{2}}{\epsilon_{2}+\epsilon_{1}}$).
The relation between permittivities can be written as follows to generalise the procedure for multilayered cases:  
\begin{eqnarray}
	\alpha_{mn}=\frac{\epsilon_{m}-\epsilon_{n}}{\epsilon_{m}+\epsilon_{n}}\\
	\beta_{mn}=\frac{2\epsilon_{n}}{\epsilon_{m}+\epsilon_{n}}
\end{eqnarray}
where, m=index of source charge and n=index of the reflecting medium. We call $\alpha_{mn}$ as reflection factor and $\beta_{mn}$ as equivalence factor as it defines equivalent charge \cite{Jomaa1983ElectricFD}.
\vspace{-0.6cm}
\subsection{Formation of image charges in three layers of dielectric}\label{section5}

\begin{figure}
	\center\includegraphics[scale=0.34]{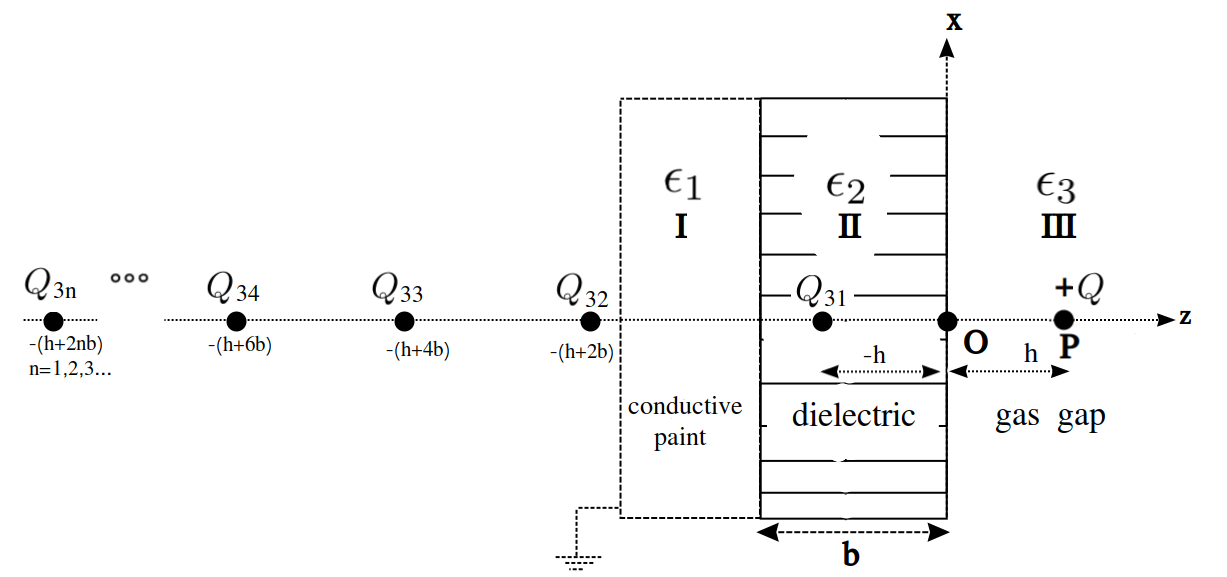}
	
	\caption{\label{fig:threelayer} Formation and positions of image charges due to three layers of dielectric. }
\end{figure}
\begin{figure}
	\center\subfloat[\label{three_dielectric_charge}]{\includegraphics[scale=0.32]{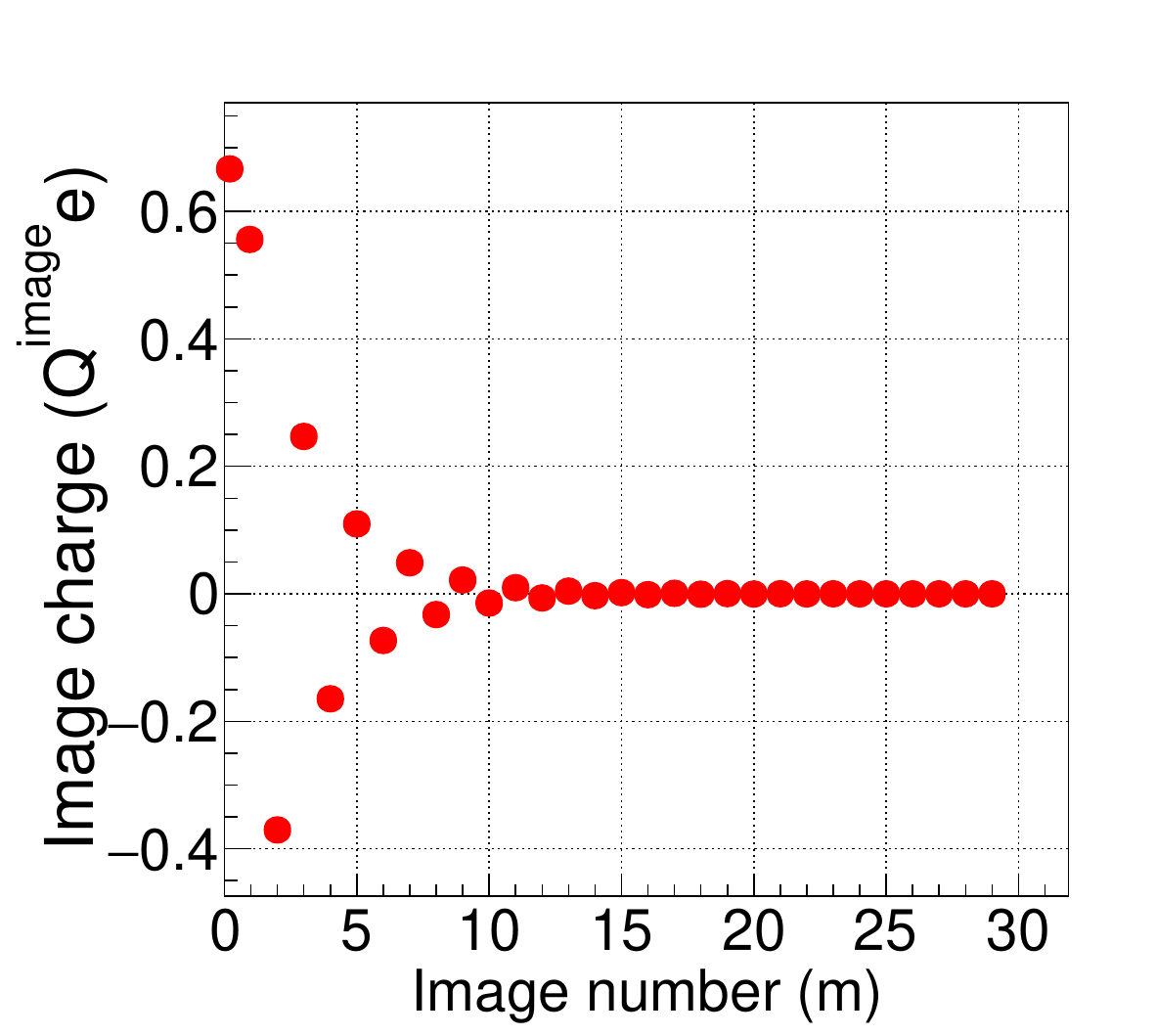}
		
	}\subfloat[\label{three_dielectric_field}]{\includegraphics[scale=0.32]{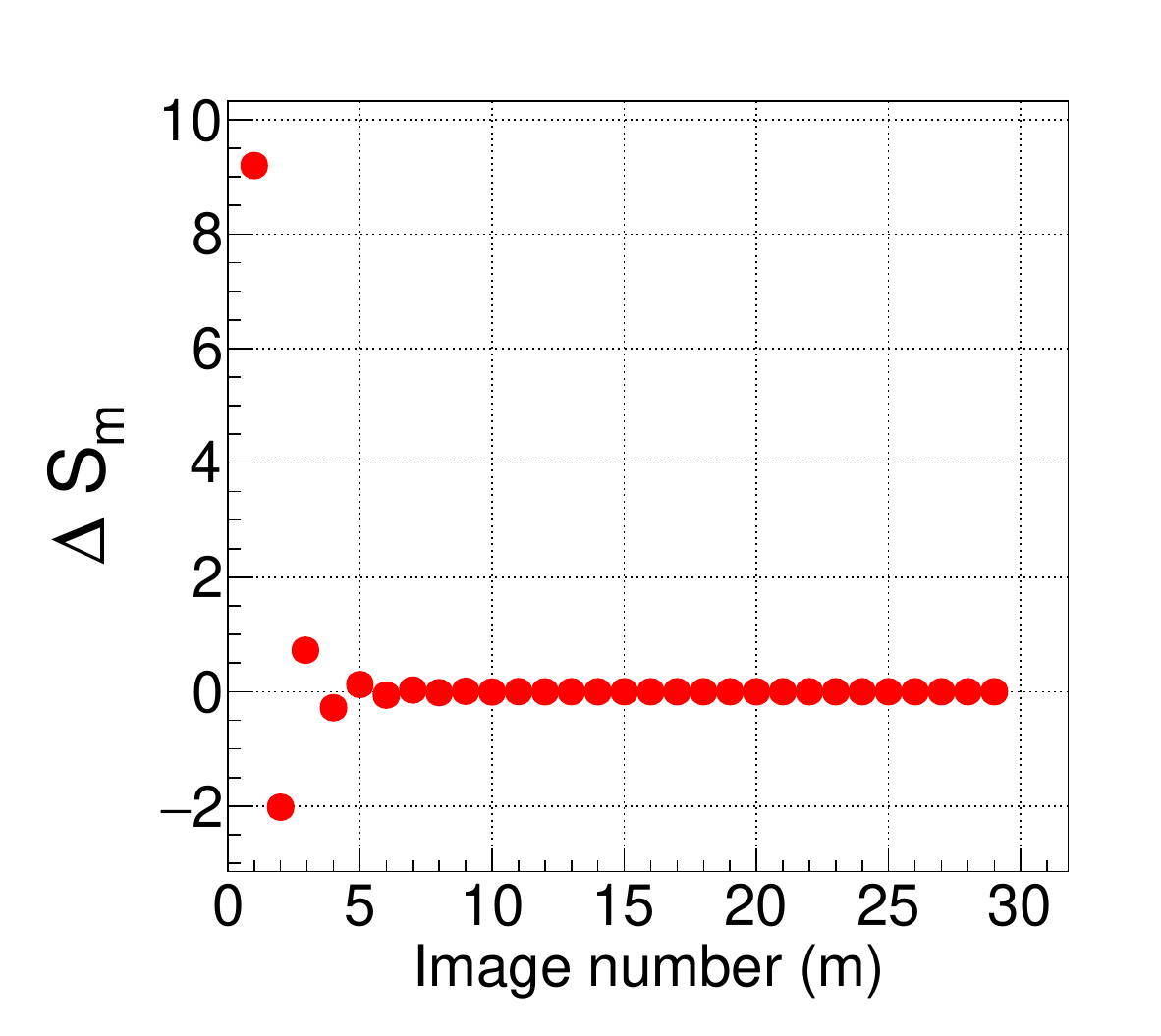}
	}
	
	\caption{\label{fig:charge_intensity-1}(a) Variation of image charges with order of reflection for three layers of dielectric case. (b) Percentage of contribution of higher order image charges on the electric field for three layers of dielectric case.}
\end{figure}

\vspace{-0.3cm}
If we extend from two layers to three layers, it can be found that due to two boundary interfaces between medium I and II and between II and III see (figure \ref{fig:threelayer}), an infinite series of image charges is induced\cite{Jomaa1983ElectricFD} due to a source charge +Q placed at a distance h from the interface of dielectric II and III. The series formed by a set of infinite of image charges shown in figure \ref{fig:threelayer} is named as $M3\footnote[1]{The name $M3$ has been chosen since the images is formed in three layerd medium}$  which can be written as follows:
\begin{align}
	M3=\left\lbrace   Q_{31},Q_{32},Q_{33}...Q_{3n}\right\rbrace,
\end{align}
where the $n^{th}$ image charge can be written as $Q_{in}$.
The first index $i$ represents the medium in which we calculate the electric field, and the second index $n$ represents the order of reflection or image charge. Let us consider the thickness of the medium II is b. The positions and charges of the images are shown in the table \ref{tab:threelayer_image}. Here, all distances are measured from the boundary interface between medium III and II. Again to calculate the contribution of higher-order reflections $\Delta S_m$ using equation \ref{eqn:DeltaSm} ( m=2,3..) define $E_{n}=\frac{Q_{in}}{r_n^2}$ and $S_m=\sum_{n=0}^{m}E_n$, where $Q_{in}$ is the $n^{th}$ order image charge and $r_n=-(h+2(n-1)b)$ (for n=1,2,3...) is the distance of $Q_{in}$ from the interface of medium III and II. In figure \ref{three_dielectric_charge} the variation of image charges (for h=0.1mm and b=2mm) with its order of reflection has been shown, where it is clear that the image charges change their sign, and magnitudes gradually reduce with the order. From figure \ref{three_dielectric_field} we can also conclude that the percentage of contribution $\Delta S_m$ of higher-order $m^{th}$ (m=2,3,4..) reflection is also gradually decreasing to zero.  After seven to eight order the value drops from 9.2\% to 0.008\%. 

\begin{table}
	\center%
	\begin{tabular}{|c|c|c|c|}
		\hline 
		Order(n) &  & Charge & Position\tabularnewline
		\hline 
		\hline 
		$1^{st}$ & $Q_{31}$ & $\alpha_{32}Q$ & -h\tabularnewline
		\hline 
		$2^{nd}$ & $Q_{32}$ & $(1-\alpha_{32}^{2})\alpha_{21}Q$ & -(h+2b)\tabularnewline
		\hline 
		$3^{rd}$ & $Q_{33}$ & $(1-\alpha_{32}^{2})(-\alpha_{32})\alpha_{21}^{2}Q$ & -(h+4b)\tabularnewline
		\hline 
		&  & $\begin{array}{c}
			.\\
			.\\
			.
		\end{array}$ & \tabularnewline
		\hline 
		$n^{th}$ & $Q_{3n}$ & $(1-\alpha_{32}^{2})(-\alpha_{32})^{n-2}(\alpha_{21})^{n-1}Q$ & -(h+2(n-1)b)\tabularnewline
		\hline 
	\end{tabular}
	
	\caption{\label{tab:threelayer_image}Magnitudes and locations of image charges for the three layer case.}
\end{table}


\par In our calculation we have considerd that medium-I is a conductor so $\epsilon_{1}$ is considered as infinity.Hence, $\alpha_{21}$ can  be aproximated as below:
\begin{eqnarray}\label{eqn:alpha21}
	\alpha_{21}&=&\frac{\epsilon_{2}-\epsilon_{1}}{\epsilon_{2}-\epsilon_{1}}\\\nonumber 
	&=&\frac{\frac{\epsilon_{2}}{\epsilon_{1}}-1}{\frac{\epsilon_{2}}{\epsilon_{1}}+1}\\\nonumber
\end{eqnarray}
As $\epsilon_{1}\rightarrow\infty$, so from equation \ref{eqn:alpha21} we can write $\alpha_{21}=-1$.
\vspace{-1.5cm}
The sum of all image charges from table	\ref{tab:threelayer_image} can be written as follows:
\begin{eqnarray}
	Q_{image}^{Total}&=&Q\left[\alpha_{32}+(1-\alpha_{32}^2)\sum_{n=2}^{\infty}\alpha_{21}^{n-1}(-\alpha_{32})^{n-2} \right]\\\nonumber
	&=&Q\left[\alpha_{32}+\frac{(1-\alpha_{32}^2)}{\alpha_{21}\alpha_{32}^{2}}\sum_{n=2}^{\infty}\alpha_{21}^{n}(-\alpha_{32})^{n}\right]\\\nonumber
	&=&Q\left[\alpha_{32}+\frac{(1-\alpha_{32}^2)}{\alpha_{21}\alpha_{32}^{2}}\Bigg\{\sum_{n=0}^{\infty}[\alpha_{21}^{n}(-\alpha_{32})^{n}]-(1-\alpha_{21}\alpha_{32})\Bigg\}\right]\\\nonumber
	&=&Q\left[\alpha_{32}-\frac{(1-\alpha_{32}^2)}{\alpha_{32}^{2}}\Bigg\{\sum_{n=0}^{\infty}[(-1)^{2n}\alpha_{32}^{n}]-(1+\alpha_{32})\Bigg\}\right],(As\; \alpha_{21}=-1)\\ \nonumber
	&=&Q\left[\alpha_{32}-\frac{(1-\alpha_{32}^2)}{\alpha_{32}^{2}}\Bigg\{\frac{1}{1-\alpha_{32}}-(1+\alpha_{32})\Bigg\}\right],(As\; \sum_{n=0}^{\infty}\alpha_{32}^{n}=\frac{1}{1-\alpha_{32}},\,and \mid\alpha_{32}\mid<1)\\ \nonumber
	&=&-Q
\end{eqnarray}

\subsection{Formation of image charges in RPC} \label{sec:sec_6}
\begin{figure}
	\center\includegraphics[scale=0.27]{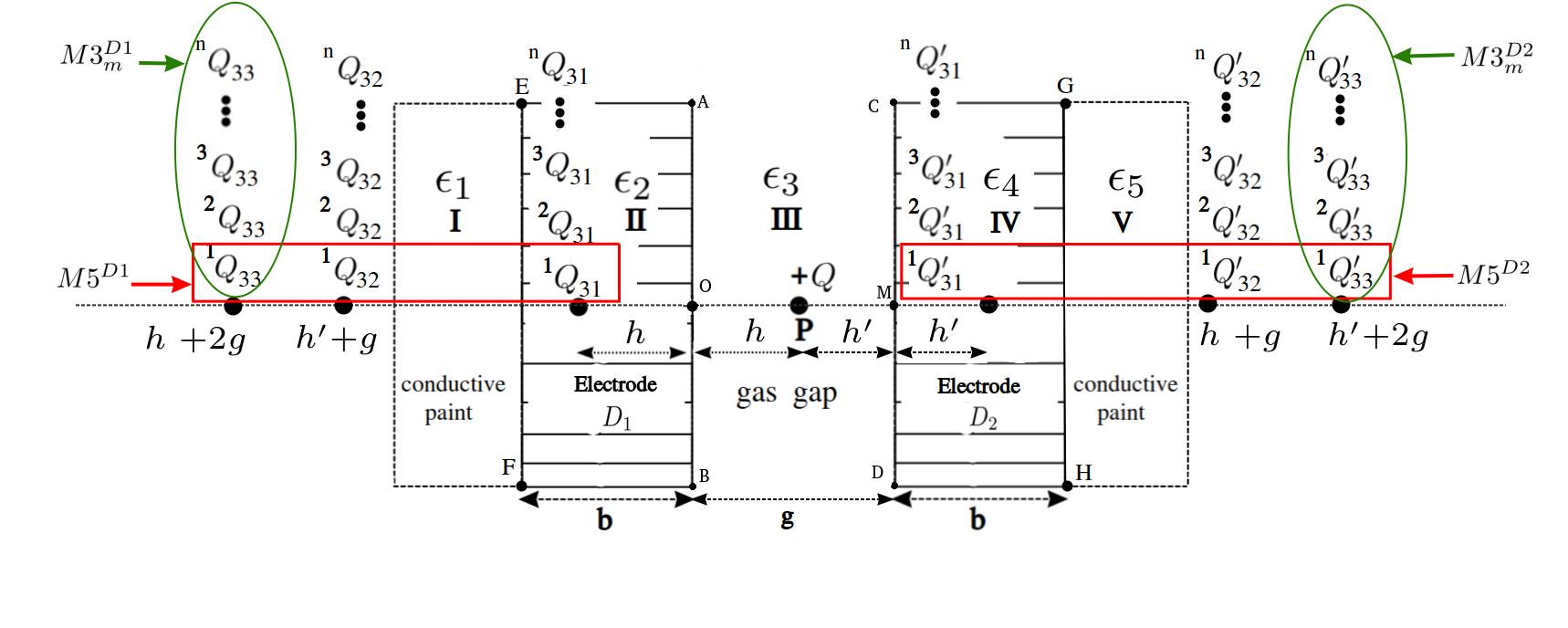}
	
	\caption{\label{fig:RPC_image_charge_formation}Formation of image charges in an  RPC with positions of $1^{st}$ order $M3$ image charges.}
\end{figure}

It is known that RPC contains two dielectric electrodes along with a layer of conductive graphite paint (see figure \ref{fig:RPC_image_charge_formation}). For now, we are considering graphite paint as a perfect conductor. Due to the presence of dielectric, the formation of the image will be different from the metal electrode case, as described below.

\subsubsection{Formation of $M3$ series for electrodes D1 and D2}\label{subsec:6.1}	
Let us consider a charge Q located inside the gas gap (medium III) at a distance of $h$ from the left electrode D1 and $h^{\prime}$ from the inner surface of the right electrode D2 (see figure \ref{fig:RPC_image_charge_formation}). To calculate the position of image charges, we divide the whole system of figure \ref{fig:RPC_image_charge_formation}, such that there are three layers of the medium on the left side (I, II, III) and three on the right (III, IV, V). In section \ref{section5} for the three-layer case (see figure \ref{fig:threelayer}), we have seen a series $M3$ of infinite image charges formed due to a single point charge. Similarly, here also, we will get two similar kind $M3$ series for three layers of medium-I, II, III, on left of D1 and for medium-III, IV, V on right of D2 (see figure \ref{fig:RPC_image_charge_formation}). Since the left and right $M3$ series of the figure \ref{fig:RPC_image_charge_formation} is corresponding to electrodes D1 and D2 respectively ; hence the left and right $M3$ series can be written as follows:
\begin{align}
	M3^{D1}=\left\lbrace   ~^1\!Q_{31},~^2\!Q_{31},~^3\!Q_{31}...~^n\!Q_{31}\right\rbrace \\ 
	M3^{D2}=\left\lbrace ~^1\!Q_{31}^\prime,~^2\!Q_{31}^\prime,~^3\!Q_{31}^\prime...~^n\!Q_{31}^\prime\right\rbrace
\end{align}
where the  index at superscript ($n$)  represents the order of reflection and the prime symbol ($~^1\!Q_{31}^\prime$) on image charges of D2  has been used to separate from image charges of D1. The significance of all indices at the subscript will be discussed in next section. As discussed in section \ref{section5} the value of $\Delta S_{m}$ can be used to decide the significant number of terms of series $M3^{D1,D2}$.
\subsubsection{Formation of image charges in five layers of dielectric}\label{subsec:6.2}	
If we consider 1st order reflection $ ~^1\!Q_{31} $ as a source charge for electrode D2, it will generate another $M3^{D2}$ series of second order at D2, which is $\left\lbrace ~^1\!Q_{32}^\prime,~^2\!Q_{32}^\prime,~^3\!Q_{32}^\prime...~^n\!Q_{32}^\prime\right\rbrace$.  Similarly, if we take $~^1\!Q_{31}^\prime$  as a source charge for D1, it will generate another  $M3^{D1}$ series at electrode D1, which is $\left\lbrace   ~^1\!Q_{32},~^2\!Q_{32},~^3\!Q_{32}...~^n\!Q_{32}\right\rbrace $. This process will continue infinitely and generate two series of infinite images for electrodes D1 and D2 . As the formation of these series is corresponding to five layers of dielectric medium we named this $M5$. The mathematical representation of $M5$ for D1 and D2 is given below,
\begin{align}
	M5^{D1}=\left\lbrace \left\lbrace ~^1\!Q_{31}..~^n\!Q_{31} \right\rbrace , \left\lbrace ~^1\!Q_{32}..~^n\!Q_{32} \right\rbrace,\left\lbrace ~^1\!Q_{33}..~^n\!Q_{33} \right\rbrace ...\left\lbrace ~^1\!Q_{3m}..~^n\!Q_{3m} \right\rbrace  \right\rbrace \\
	M5^{D2}=\left\lbrace \left\lbrace ~^1\!Q_{31}^\prime..~^n\!Q_{31}^\prime \right\rbrace  , \left\lbrace  ~^1\!Q_{32}^\prime..~^n\!Q_{32}^\prime \right\rbrace , \left\lbrace  ~^1\!Q_{33}^\prime..~^n\!Q_{33}^\prime \right\rbrace ...\left\lbrace  ~^1\!Q_{3m}^\prime..~^n\!Q_{3m}^\prime \right\rbrace  \right\rbrace
\end{align} 
where $M5^{D1}$, $M5^{D2}$ stands for the electrode D1, D2 respectively,  m is the order of the  $M5^{D1,D2}$ series and the index 3 of the charges ($~^n\!Q_{3m},~^n\!Q_{3m}^\prime$) is corresponding to the medium at which we are calculating the electric field. 
Alternatively, in a composite form for D1 and D2 we can write,
\begin{equation}
	M5^{D1,D2}=\left\lbrace M3^{D1,D2}_{1},M3^{D1,D2}_{2},M3^{D1,D2}_{3}...M3^{D1,D2}_{m} \right\rbrace
\end{equation}
where, $M3_m^{D1,D2}=$ $\left\lbrace ~^1\!Q_{3m}..~^n\!Q_{3m} \right\rbrace$ or $\left\lbrace  ~^1\!Q_{3m}^\prime..~^n\!Q_{3m}^\prime \right\rbrace$, and it represents $m^{th}$ order of  $M3^{D1,D2}$ series for D1 or D2. Since the higher-order reflections $~^2\!Q_{3m},~^3\!Q_{3m}...~^n\!Q_{3m}$ or $~^2\!Q_{3m}^\prime,~^3\!Q_{3m}^\prime...~^n\!Q_{3m}^\prime$ etc. are at considerably far from the electrode D1 and D2, so their image charge will be generated even further from both the electrodes. Hence the effect of images of them in the field calculations is negligible
and we only consider the images of $~^1\!Q_{3m}$ and $~^1\!Q_{3m}^\prime$. Therefore now the $M5^{D1,D2}$ becomes:
\begin{align}
	M5^{D1}=\left\lbrace  ~^1\!Q_{31}  ,  ~^1\!Q_{32} , ~^1\!Q_{33} ... ~^1\!Q_{3m} \right\rbrace \\
	M5^{D2}=\left\lbrace ~^1\!Q_{31}^\prime   , ~^1\!Q_{32}^\prime  ,  ~^1\!Q_{33}^\prime  ...  ~^1\!Q_{3m}^\prime \right\rbrace.
\end{align} 

Indeed, one can include some of the higher-order reflections in the field calculation according to the need of precision but not their images.

\begin{figure}[H]
	\center{\includegraphics[scale=0.5]{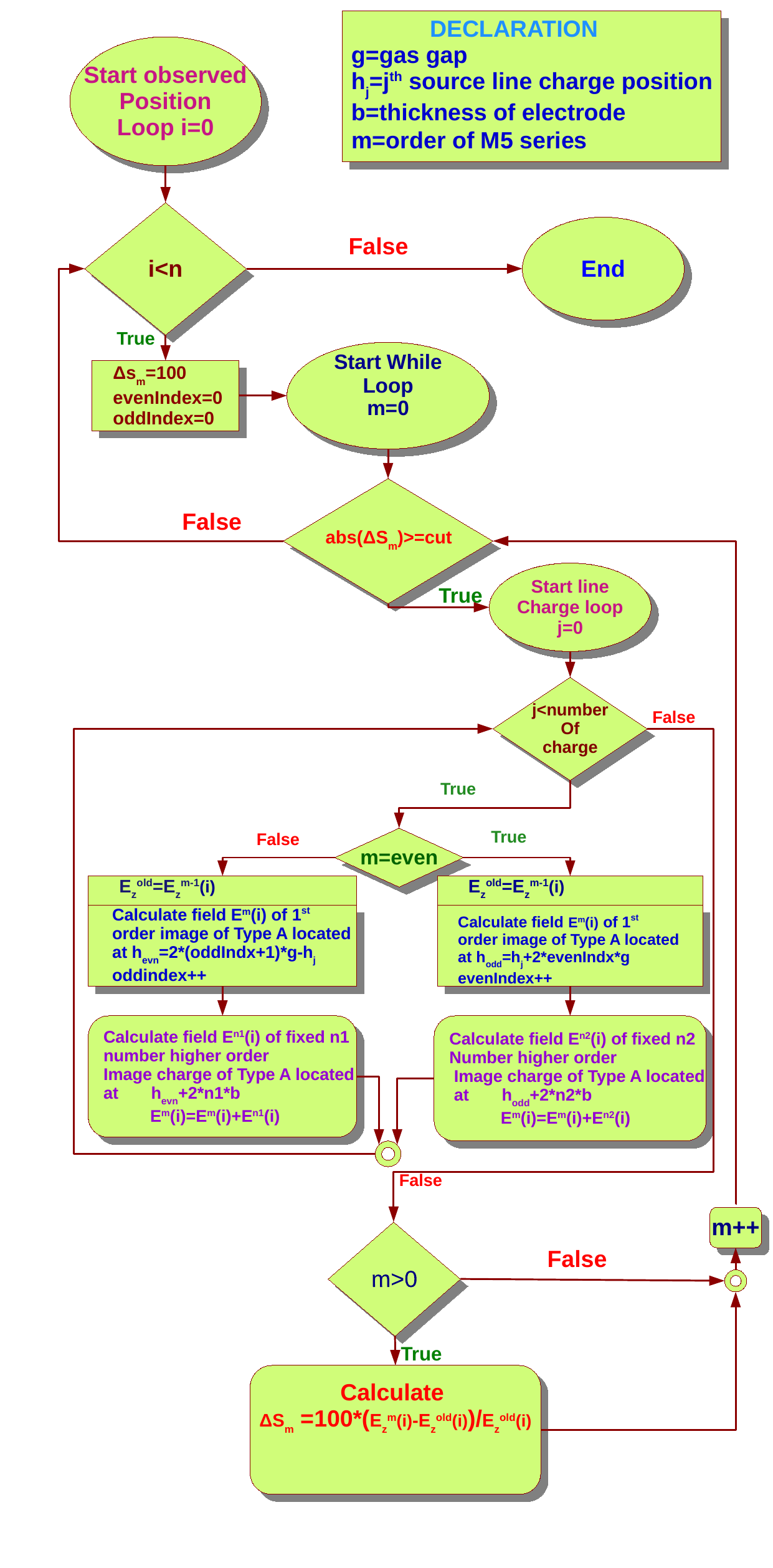}
		
	}
	\caption{Algorithm to generate image charge\label{fig:algo}}
\end{figure}

The magnitude and location of all charges in $M5^{D1,D2}$ series have been shown in table \ref{tab:genaration of image RPC}. The arrow's tail in table \ref{tab:genaration of image RPC} denotes the source charge, and the head indicates the image charge corresponding to the source. $\alpha_{32}$ and $\alpha_{34}$ is the reflection factor of D1 and D2, respectively.
The series shows in table \ref{tab:genaration of image RPC}  can be divided into even and odd series on the basis of their order of reflection for ease of calculations, which is represented in table \ref{tab:even_odd_series_table6}. It is noted that the $h$ and $h^{\prime}$ are just the magnitude of the distance from the inner surface of electrodes D1 and D2. Hence, to find the real and image charge locations about a fixed origin, one will need to use suitable co-ordinate transformations. The algorithm of generation of image charge for D1 electrode has been discussed in Figure \ref{fig:algo}. The same method can be applied to generate the image for the D2 electrode.		
\begin{table}
	\center\includegraphics[scale=0.35]{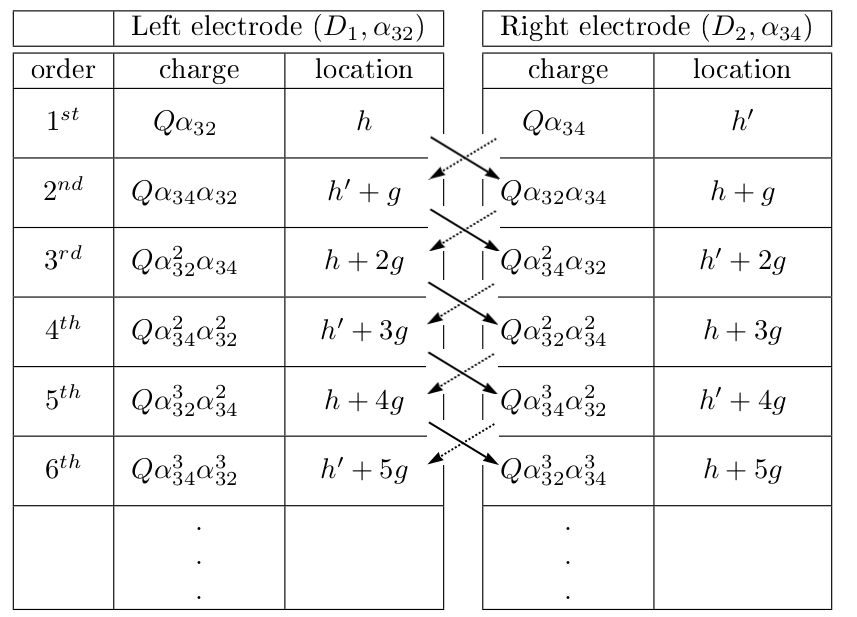}
	
	\caption{\label{tab:genaration of image RPC}Locations of image charges of $M5^{D1,D2}$, where the direction of arrow denotes source to image charge.}
\end{table}
\begin{table}
	\center%
	\begin{tabular}{|>{\centering}p{2cm}|>{\centering}p{2cm}|>{\centering}p{2cm}|}
		\hline 
		\multicolumn{3}{|c|}{Odd Series $(Qp=Q \alpha_{32})$}\tabularnewline
		\hline 
		\hline 
		Name & Charge & Location\tabularnewline
		\hline 
		$Q_{11}$$\begin{array}{c}
			\\
			\\
		\end{array}$ & $Qp$ & $h$\tabularnewline
		\hline 
		$Q_{13}$$\begin{array}{c}
			\\
			\\
		\end{array}$ & $Qp\,\alpha_{32}\alpha_{34}$ & $h+2g$\tabularnewline
		\hline 
		$Q_{15}$$\begin{array}{c}
			\\
			\\
		\end{array}$ & $Qp\,\alpha_{32}^{2}\alpha_{34}^{2}$ & $h+4g$\tabularnewline
		\hline 
		& $\begin{array}{c}
			.\\
			.
		\end{array}$ & \tabularnewline
		\hline 
		$Q_{1(2n_{1}+1)}$$\begin{array}{c}
			\\
			\\
		\end{array}$ & $Qp\,\alpha_{32}^{n_{1}}\alpha_{34}^{n_{1}}$ & $h+2n_{1}g$\tabularnewline
		\hline 
	\end{tabular}~~%
	\begin{tabular}{|>{\centering}p{2cm}|>{\centering}p{2.5cm}|>{\centering}p{2.6cm}|}
		\hline 
		\multicolumn{3}{|c|}{Even Series $(Qp=Q \alpha_{32})$}\tabularnewline
		\hline 
		\hline 
		Name & Charge & Location\tabularnewline
		\hline 
		$Q_{12}$$\begin{array}{c}
			\\
			\\
		\end{array}$ & $Qp\,\alpha_{34}$ & $2g-h$\tabularnewline
		\hline 
		$Q_{14}$$\begin{array}{c}
			\\
			\\
		\end{array}$ & $Qp\,\alpha_{34}^{2}\alpha_{32}$ & $4g-h$\tabularnewline
		\hline 
		$Q_{16}$$\begin{array}{c}
			\\
			\\
		\end{array}$ & $Qp\,\alpha_{34}^{3}\alpha_{32}^{2}$ & $6g-h$\tabularnewline
		\hline 
		& $\begin{array}{c}
			.\\
			.
		\end{array}$ & \tabularnewline
		\hline 
		$Q_{1(2n_{2})}$$\begin{array}{c}
			\\
			\\
		\end{array}$ & $Qp\,\alpha_{34}^{n_{2}+1}\alpha_{32}^{n_{2}}$ & $2(n_{2}+1)g-h$\tabularnewline
		\hline 
	\end{tabular}
	
	\caption{\label{tab:even_odd_series_table6}Division of table \ref{tab:genaration of image RPC} into odd and even series.}
\end{table}
\subsubsection{ Image charge selection criteria for an RPC }\label{subsec:6.3}
Since the images are symmetric for electrode D1 and D2, we will show results for D1 only. Unlike metal electrodes, in the case of an RPC, $m^{th}$ order of reflection is a sum of n number of image charges of  $M3^{D1}_m$ series(see figure \ref{fig:RPC_image_charge_formation}). Therefore, we can define $ E_{j}$ as, $E_{j}=\sum\limits_{i=1}^{n}\frac{^i\!Q_{3j}}{r_{i}^{2}}$, where $r_{i}$= position of the image charge $^i\!Q_{3j}$ measured from the corresponding electrode and n is the order of  $M3^{D1}_m$ series. Since we have neglected the higher order terms of $M3_m^{D1}$; hence in this case the value of n is 1. If m is the order of the  $M5^{D1}$, till which effects are considered, then we can write ${S}_{m}=\sum\limits_{j=1}^{m}{E}_{j}$. Therefore, the value of $\Delta S_{m}$ can be calculated using equation \ref{eqn:DeltaSm}, where m = 2,3,4....
\begin{figure}
	\center\subfloat[\label{RPCcharge_vs_dist_D1}]{\includegraphics[scale=0.3]{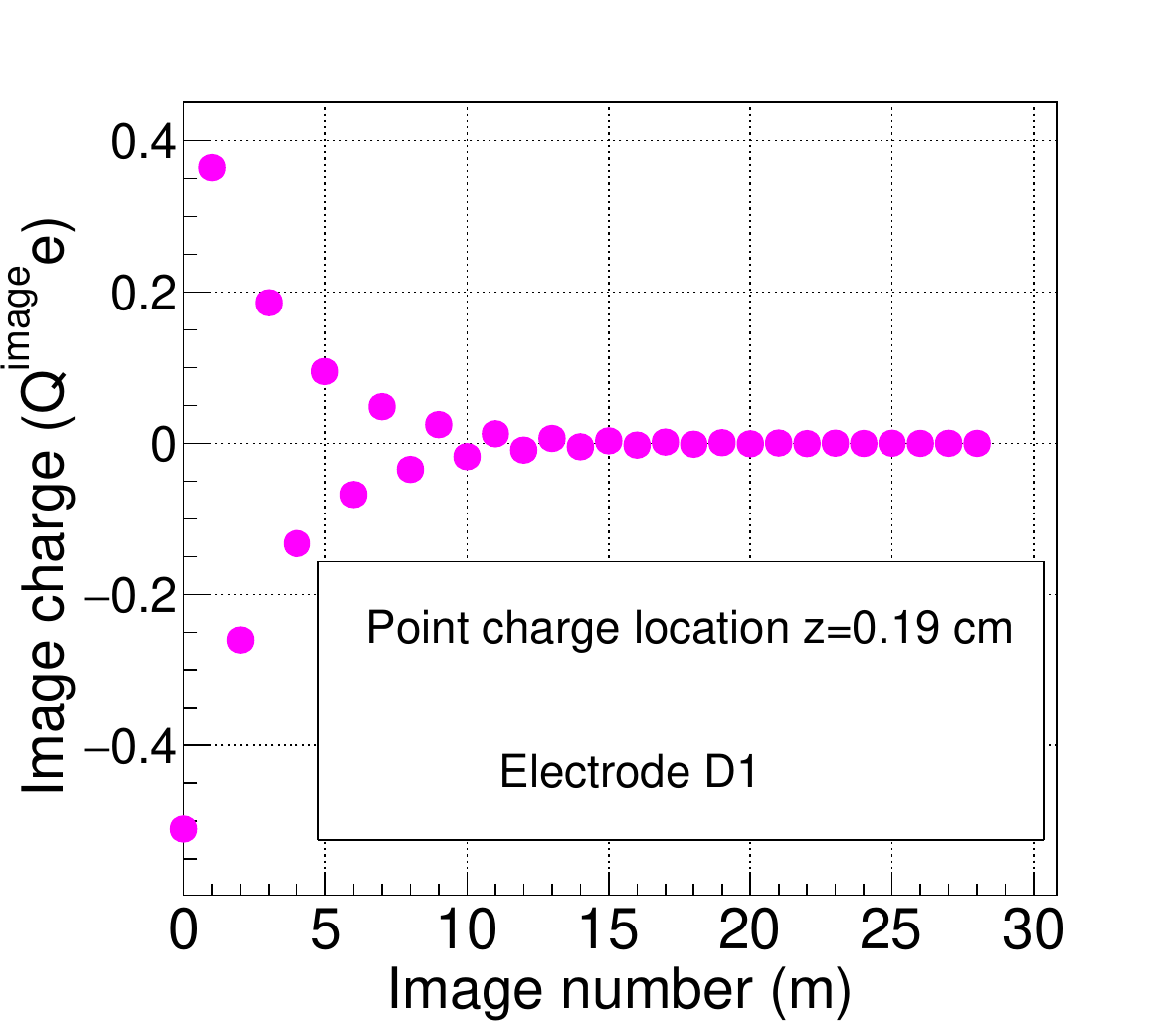}
		
	}\subfloat[\label{RPCcharge_vs_dist_D2}]{\includegraphics[scale=0.32]{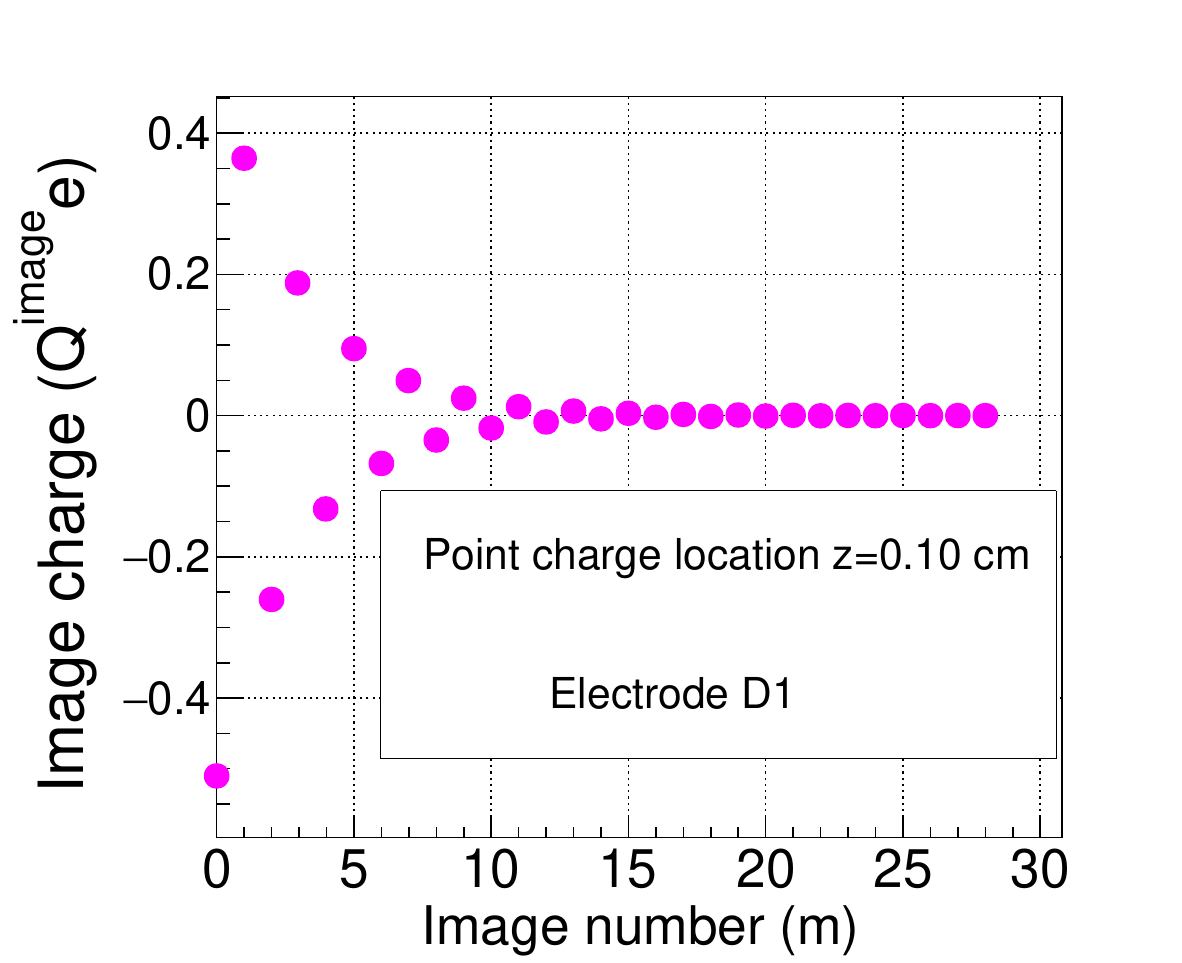}
		
	}
	
	\center\subfloat[\label{RPCcharge_vs_dist_field_D1}]{\includegraphics[scale=0.31]{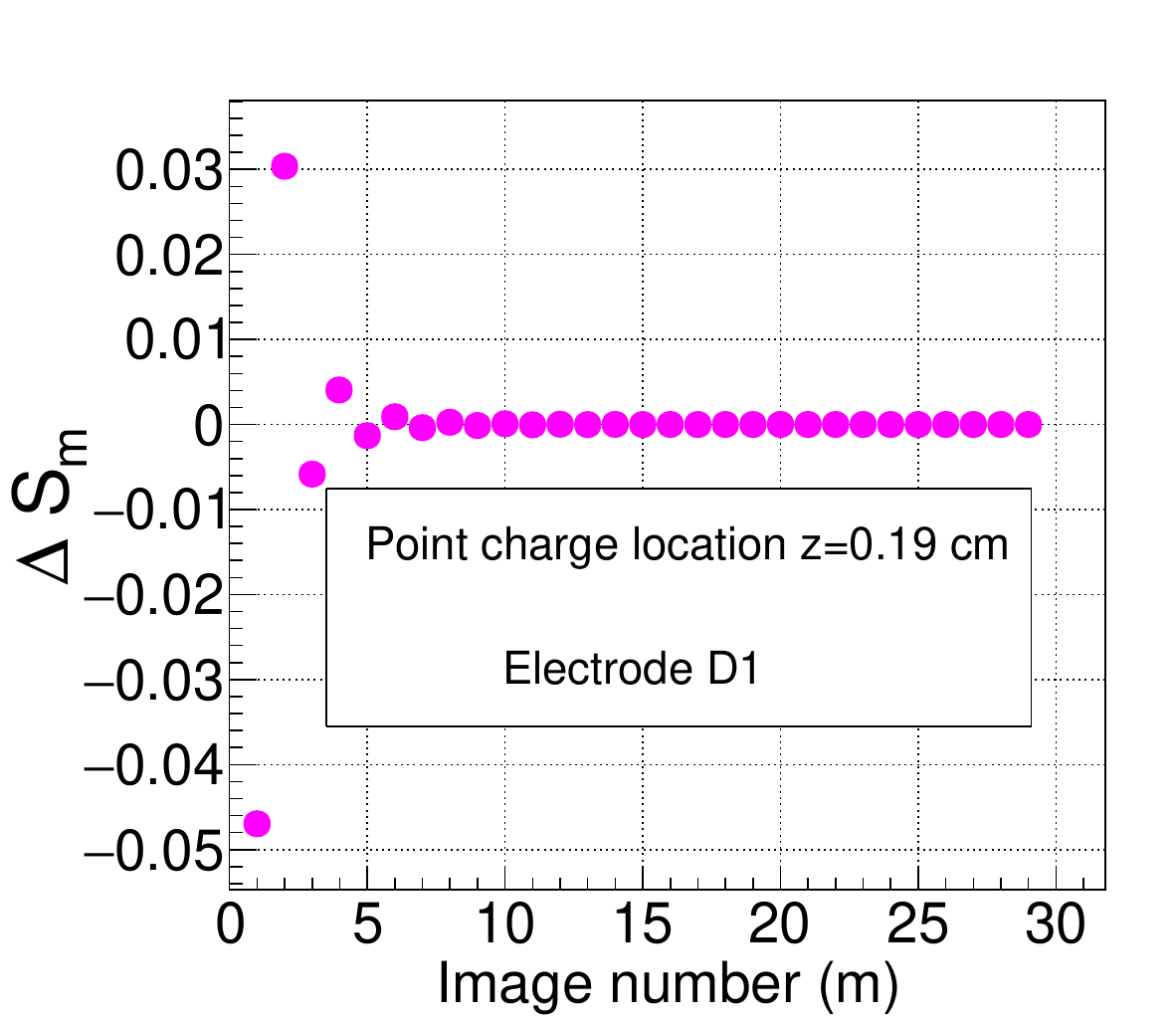}
		
	}\subfloat[\label{RPCcharge_vs_dist_field_D2}]{\includegraphics[scale=0.32]{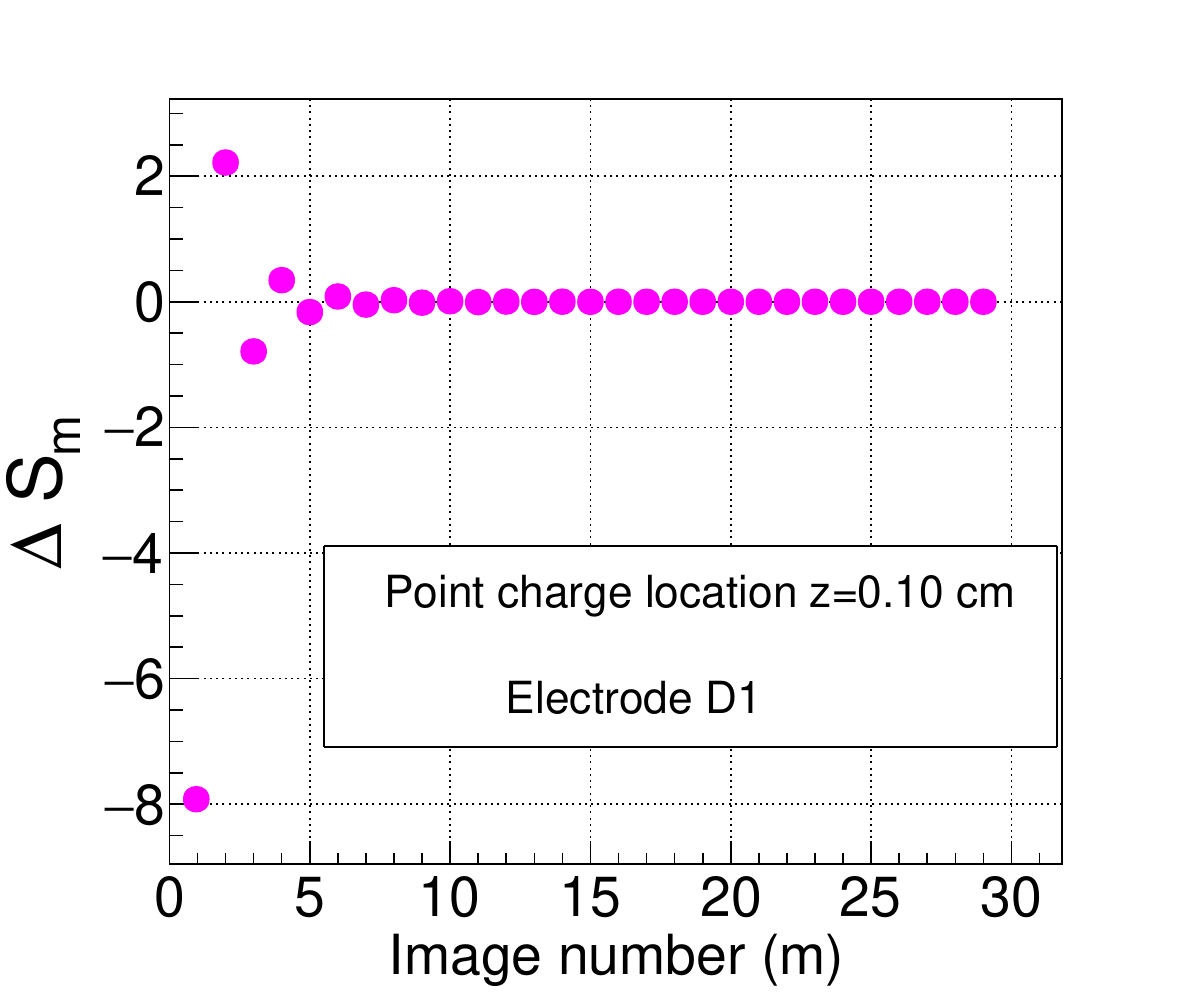}
	}
	
	\caption{(i) Variation of image charge with the order of reflection for electrode D1 of an RPC,when the point charge location at (a) z=0.019 cm and (b) z=0.1 cm. (ii) Percentage of contribution of
		higher order image charges on the electric field, when the point charge is located at (c) z=0.019 cm and (d) z=0.1 cm. }
	
\end{figure}
Let us now consider electrode thickness b = 2 mm and gas-gap g = 2 mm. The magnitude and sign of the image charges of $M5^{D1}$ series due to a point source charge located at two different positions a) near to the electrode D1 (0,0,0.19) and b) middle of the gas-gap (0,0,0.1), has been shown in figures \ref{RPCcharge_vs_dist_D1} and \ref{RPCcharge_vs_dist_D2}, where the origin is at the inner surface of electrode D2. The sign of image charges alternate between plus and minus and their magnitude gradually decrease with order increment. The contribution in field evaluation of few terms from the  $M5^{D1}$ series for two different locations of the source charge has been shown in figures \ref{RPCcharge_vs_dist_field_D1} and \ref{RPCcharge_vs_dist_field_D2}. 
It is found that the maximum contribution of the higher order term of  $M5^{D1}$ series is nearly 8$\%$ and continuously converges to zero as order increases.\\
The sum of image charges of $M5^{D1}$ series can be calculated from the table \ref{tab:even_odd_series_table6} as follows:
\vspace{-0.6cm}
\begin{eqnarray}\label{eqn:imageChargeD1}
	Q_{Total}^{D1}&=&(Q_{11}+Q_{13}+Q_{15}...) +(Q_{12}+Q_{14}+Q_{16}...)\\\nonumber
	&=&Q\,\alpha_{32}\sum_{n_1=0}^{\infty}[(\alpha_{32}\,\alpha_{34})^{n_{1}}]+Q \alpha_{32}\,\alpha_{34}\sum_{n_2=0}^{\infty}[(\alpha_{32}\,\alpha_{34})^{n_2}]\\\nonumber
	&=&\frac{Q\,\alpha_{32}}{1-\alpha_{32}\,\alpha_{34}}+\frac{Q\,\alpha_{32}\,\alpha_{34}}{1-\alpha_{32}\,\alpha_{34}},(As,\; \sum_{n=0}^{\infty}(\alpha_{32}\,\alpha_{34})^{n}=\frac{1}{1-\alpha_{34}\,\alpha_{32}}, and \, \mid\alpha_{32}\,\alpha_{34}\mid<1)\\\nonumber
	&=&Q\alpha_{32}\frac{1+\alpha_{34}}{1-\alpha_{32}\,\alpha_{34}}
\end{eqnarray}  
Let's consider that the permittivity of D1 and D2 is the same. Hence, $\alpha_{32}=\alpha_{34}$  and so from the equation \ref{eqn:imageChargeD1} we can write:
\begin{eqnarray}
	Q_{Total}^{D1}=\frac{Q\,\alpha_{32}}{1-\alpha_{32}}.
\end{eqnarray}
Similarly we can find the total image charges for $D2$ electrodes.

\section{Calculation of the space charge field  and image field of avalanche inside an RPC} \label{sec:Section4_avalancheCharge}
\begin{figure}
	\center\subfloat[\label{avalanche_electron_dist}]{\includegraphics[scale=0.36]{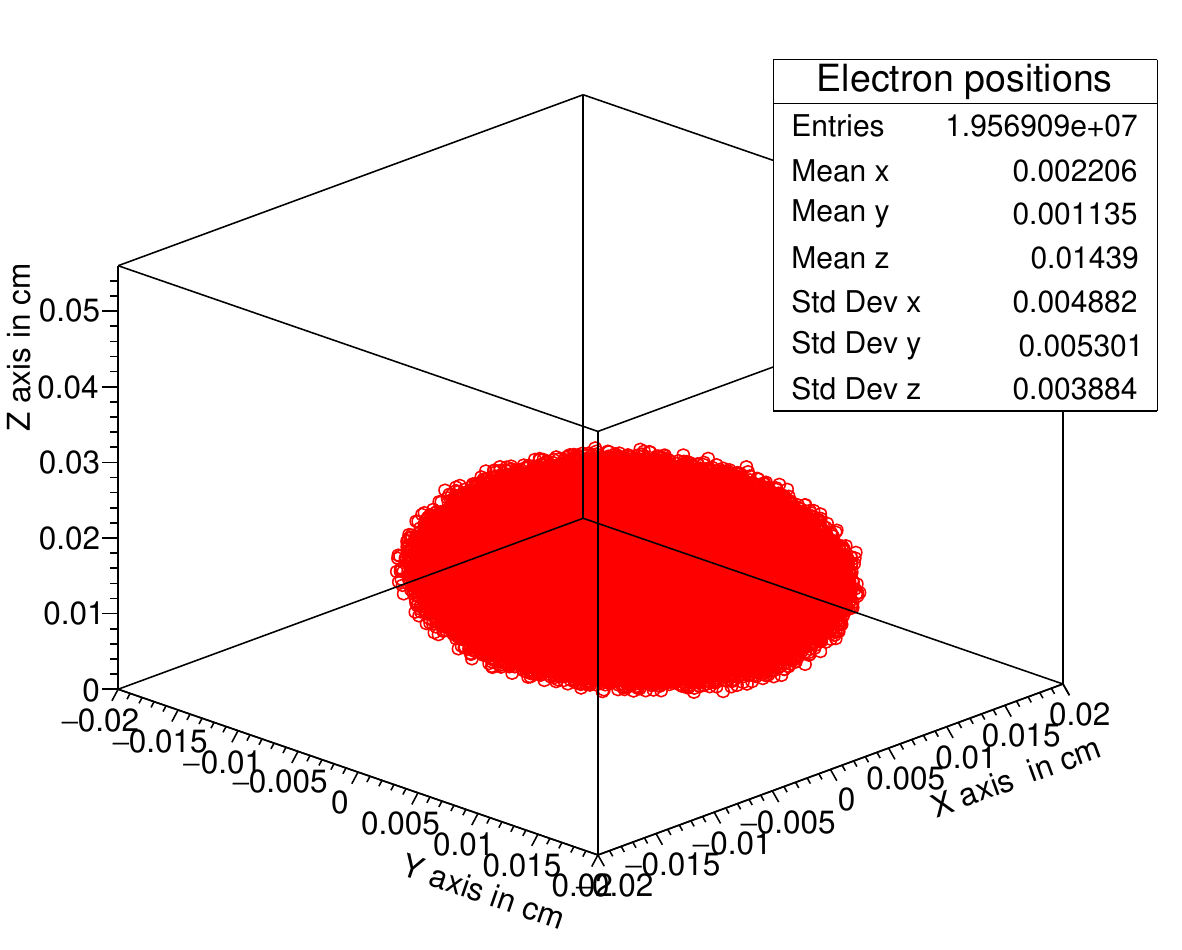}
		
	}\subfloat[\label{avalanche_ion_dist}]{\includegraphics[scale=0.36]{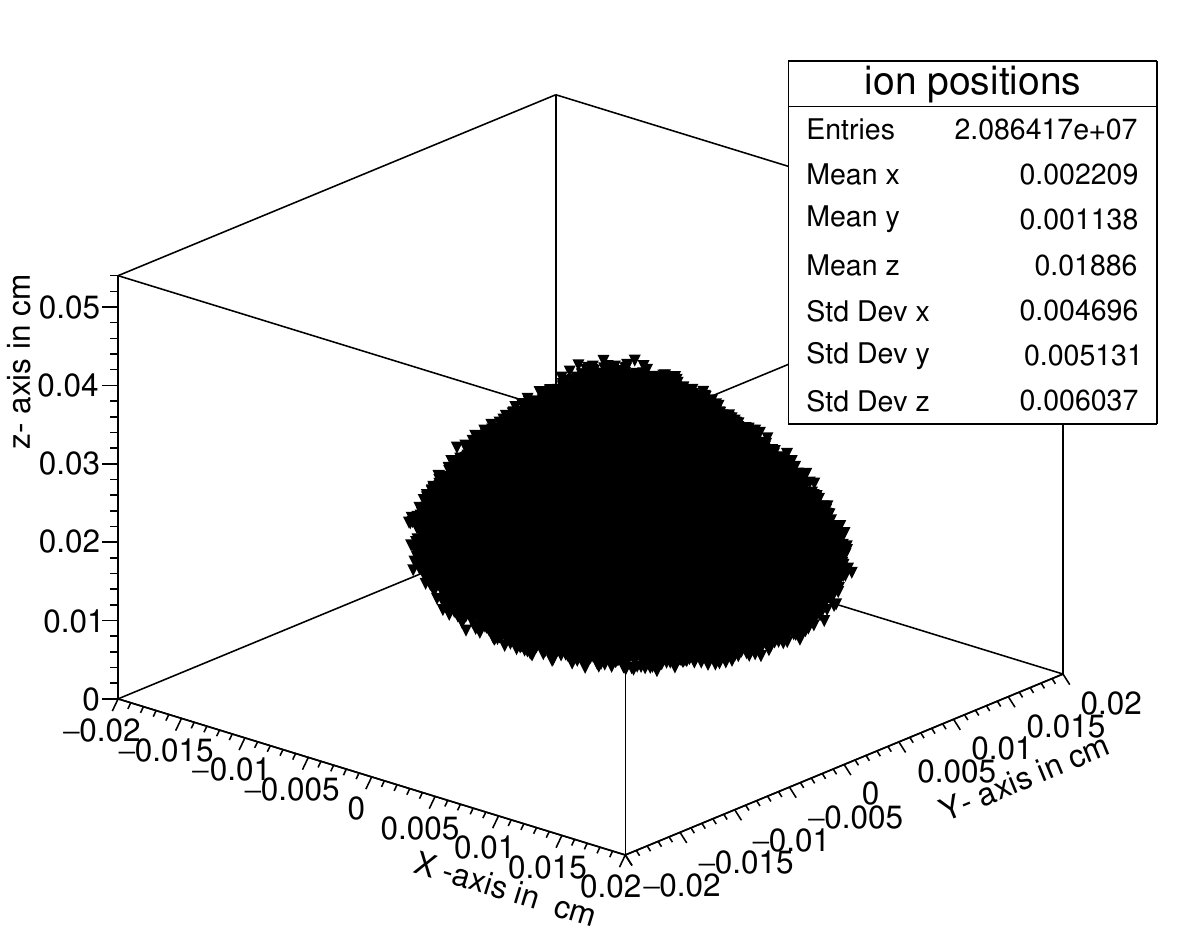}
		
	}

	\caption{(a) Simulated avalanche electron distribution at certain instant of time. (b) Simulated avalanche ion distribution at a certain instant of time. }
\end{figure}
It is known that the generation of space charge inside the RPC is high when the incoming particle rate is high. This condition also can be achieved with a single primary electron if the applied field is sufficiently high for a certain gas mixture. An avalanche charge distribution at an instant of time has been simulated from a single primary electron inside an RPC using Garfield++, where the geometry and applied voltage discussed in section \ref{sec:section2} is used and the initial position of electron has been chosen at the center of the RPC. The thickness of electrode and gas gap is fixed at 2 mm. To introduce drift motion in the simulation the drift velocity have been calculated using MAGBOLTZ \cite{BIAGI1989716} and for the thermal diffusive motion we have considered that the diffusion is described as gaussian distribution. It is known that under the electric field the gaussian diffusion becomes anisotropic and so the distribution becomes \cite{LippmanThesis}:
\begin{eqnarray}
	\phi_L(z,l)&=&\frac{1}{\sqrt{2\pi l}D_L}exp\Big(-\frac{(z-z_0)^2)}{2D_L^2 l}\Big)\\
	\phi_T(r,l)&=&\frac{1}{D^2_T l}exp \Big(-\frac{(r-r_0)^2}{2 D^2_T l}\Big),
\end{eqnarray}

where,$\phi_L$ and  $\phi_T$ is longitudinal and transverse gaussian distributions, $D_L$ and $D_T$ is longitudinal and transvers diffusion constants which are calculated using MAGBOLTZ \cite{BIAGI1989716}, $z_0$ and $r_0$ is the position of center of mass of the distribution, $l$ is the drifted distance at time t, r and z are the position of electron in cylindrical co-ordinate system. When the number of electrons and ions became of the order of $\approx 10^7$ the simulation is stopped and the position of electrons and ions are stored (see figures \ref{avalanche_electron_dist} and \ref{avalanche_ion_dist}) to calculate the electric field.  \\
In the following subsections the calculation of space charge field
along with image charge field has been done using three models,
(a) charged ring, (b) line charge, and (c) neBEM and the comparison between them also been discussed.

\subsection{Field of a uniformly charged ring}
The electric field of the charge distribution shown in figures \ref{avalanche_electron_dist} and \ref{avalanche_ion_dist} can be calculated by modeling the charge region as a number of concentric rings. The field at any point $(r,\phi,z)$ due to a uniformly charge ring of radius $r^\prime$ located at $z^\prime$ can be expressed as \cite{Lippmann_1,LippmanThesis}, 
\begin{subequations}
	\begin{align}
		E_{r}^{ring}(r,z,r^{\prime},z^{\prime})&\approx \frac{Q}{2\pi\epsilon_{0}}\frac{1}{ra^2b}\times
		\left[ c^{2}E\left(\frac{-4rr^{\prime}}{b^{2}}  \right) +a^{2}K\left(\frac{-4rr^{\prime}}{b^{2}}  \right)\right]\label{eqn:Er_lip}  \\
		E_{\phi}^{ring}(r,z,r^{\prime},z^{\prime})&=0\label{eqn:Ephi_lip} \\
		E_{z}^{ring}(r,z,r^{\prime},z^{\prime})&\approx\frac{Q}{\pi \epsilon_{0}}\frac{(z-z^\prime)}{a^2b}E\left(\frac{-4rr^{\prime}}{b^{2}}\right)\label{eqn:Ez_lip}
	\end{align}
\end{subequations}
where,
\begin{subequations}
	\begin{align}
		a^2&=(r+r^\prime)^2+(z-z^\prime)^2\\
		b^2&=(r-r^\prime)^2+(z-z^\prime)^2\\
		c^2&=r-(r^\prime)^2+(z-z^\prime)^2
	\end{align}
\end{subequations}
and
\begin{align}
	K(x)&=\int\limits_{0}^{\frac{\pi}{2}}\frac{1}{\sqrt{1-x \sin^2(\zeta)}}d\zeta\\
	E(x)&=\int\limits_{0}^{\frac{\pi}{2}}\sqrt{1-x \sin^2(\zeta)}\,\,d\zeta
\end{align}
$K(x)$ and $E(x)$ represents the first and second kind elliptic integrals respectively.	The steps of finding charge inside the rings are also discussed in \cite{Lippmann_1}. The field of  mirror charged rings located at $(r^\prime,\phi^\prime,2g-z^\prime)$ and $(r^\prime,\phi^\prime,-z^\prime)$ can be found by replacing $z^\prime$ with $2g-z^\prime$ and $-z^\prime$ in equations \ref{eqn:Er_lip} and \ref{eqn:Ez_lip}.

\subsection{Field due to a single line charge}
\vspace{-0.5cm}
In the above ring approximation, due to the rotational symmetry of the avalanche charged region, the $\phi$ directional field $E_{\phi}$ is considered zero. However, depending on the experimental situation, the charge region may not be properly rotationally symmetric \cite{Dey_2020}. So the approximation of uniform ring is not always good enough. Hence, in those cases, one can divide the ring into several uniformly charged straight lines along the periphery where each can carry a different charge. Therefore, the sum of all lines over a ring together can be represented as a non-uniform charged ring. The method of division of rings in several lines is discussed in \cite{Dey_2020}.  

The electric field at any position (x,y,z) due to a line of uniform charged density $\bar{\lambda}$ and length S, located at $x^\prime=\bar{r}$,$z^\prime=\bar{z}$ and parallel to y-axis can be expressed as follows \cite{Dey_2020},
\begin{subequations}
	\begin{align}
		E_{x}^{line}&=\frac{\bar{\lambda}(x-\bar{r})}{4\pi\epsilon_{0}P^{2}}\left[\frac{(y+\frac{S}{2})}{\sqrt{(y+\frac{S}{2})^{2}+P^{2}}}-\frac{(y-\frac{S}{2})}{\sqrt{(y-\frac{S}{2})^{2}+P^{2}}}\right]\label{eqn:ex_line}\\
		E_{y}^{line}&=-\frac{\bar{\lambda}}{4\pi\epsilon_{0}}\left[\frac{1}{\sqrt{(y+\frac{S}{2})^{2}+P^{2}}}-\frac{1}{\sqrt{(y-\frac{S}{2})^{2}+P^{2}}}\right]\label{eqn:ey_line}\\
		E_{z}^{line}&=\frac{\bar{\lambda}(z-\bar{z})}{4\pi\epsilon_{0}P^{2}}\left[\frac{(y+\frac{S}{2})}{\sqrt{(y+\frac{S}{2})^{2}+P^{2}}}-\frac{(y-\frac{S}{2})}{\sqrt{(y-\frac{S}{2})^{2}+P^{2}}}\right]\label{eqn:ez_line}
	\end{align}
\end{subequations}
where $P=\sqrt{(z-\bar{z})^{2}+(x-\bar{r})^{2}}$, and if $Q_{st}$
is the total charge of this straight line then, $\bar{\lambda}=\frac{Q_{st}}{S}$. The field of a mirror line charge at any point inside the gas gap g  can be found by replacing $\bar{z}$ with the position of the mirror line from the respective electrode in equations \ref{eqn:ex_line},\ref{eqn:ey_line} and \ref{eqn:ez_line}. The positions of mirror lines can be found from table \ref{tab:threelayer_image} and \ref{tab:genaration of image RPC}.


\subsection{Comparison of Z-directional field $E_{z}$}
\subsubsection{  Ring and line approximation}
The combined z-directional field of avalanche electron and ion distribution (see figures \ref{avalanche_electron_dist} and \ref{avalanche_ion_dist}) has been calculated using equations \ref{eqn:Ez_lip} and \ref{eqn:ez_line}, where both radial and z-directional thickness of each ring is 0.001 cm. As discussed in the \cite{Dey_2020}, to segment a ring in several lines, one needs to assign one more parameter $\delta\phi$, taken as 1 degree in this calculation, where $\delta\phi$ is the angle, subtended to the center of the circular ring.  Here it is considered that the charge enclosed by a ring is uniformly distributed over the ring. Therefore after segmentation, each line will carry the same amount of charge, which is similar to case-1 in \cite{Dey_2020}. 
\begin{figure}
	\center\subfloat[\label{fig:ezfieldlip}]{\includegraphics[scale=0.4]{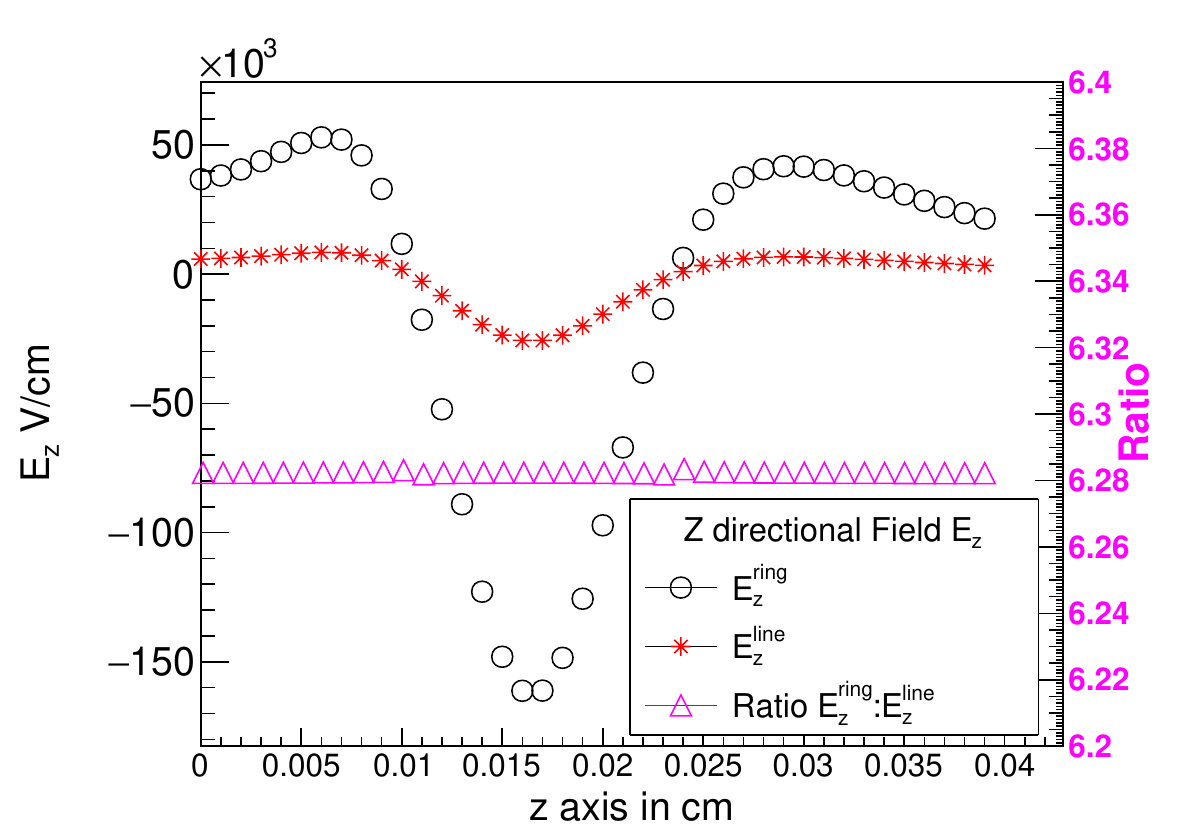}
		
	}\subfloat[\label{fig:ezfieldlipCorrected}]{\includegraphics[scale=0.4]{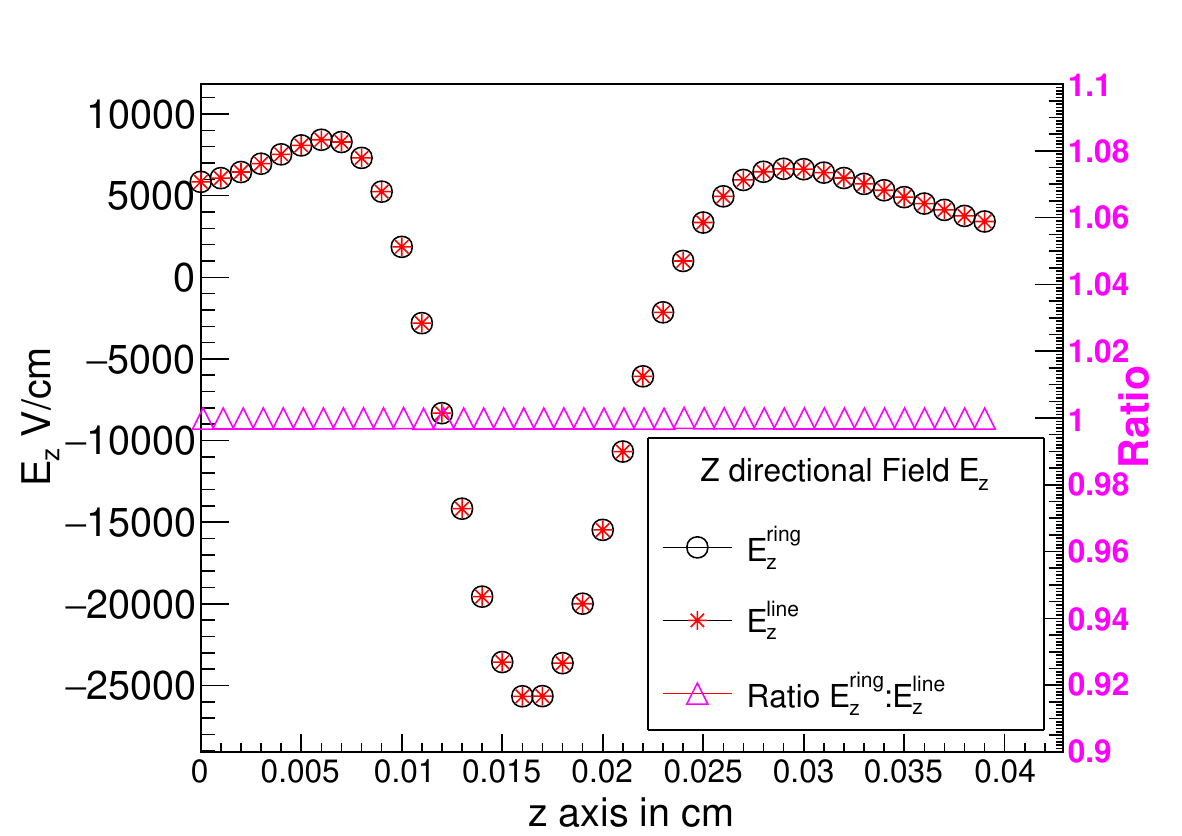}
		
	}
	
	\caption{(a) Comparison between ring and line z-directional field (source+image) before $2\pi$ division. (b) Comparison between ring and line z-directional field (source+image) after $2\pi$ division.}
\end{figure}
\begin{figure}
	\center\subfloat[ \label{fig:ezfieldImagelip}]{\includegraphics[scale=0.4]{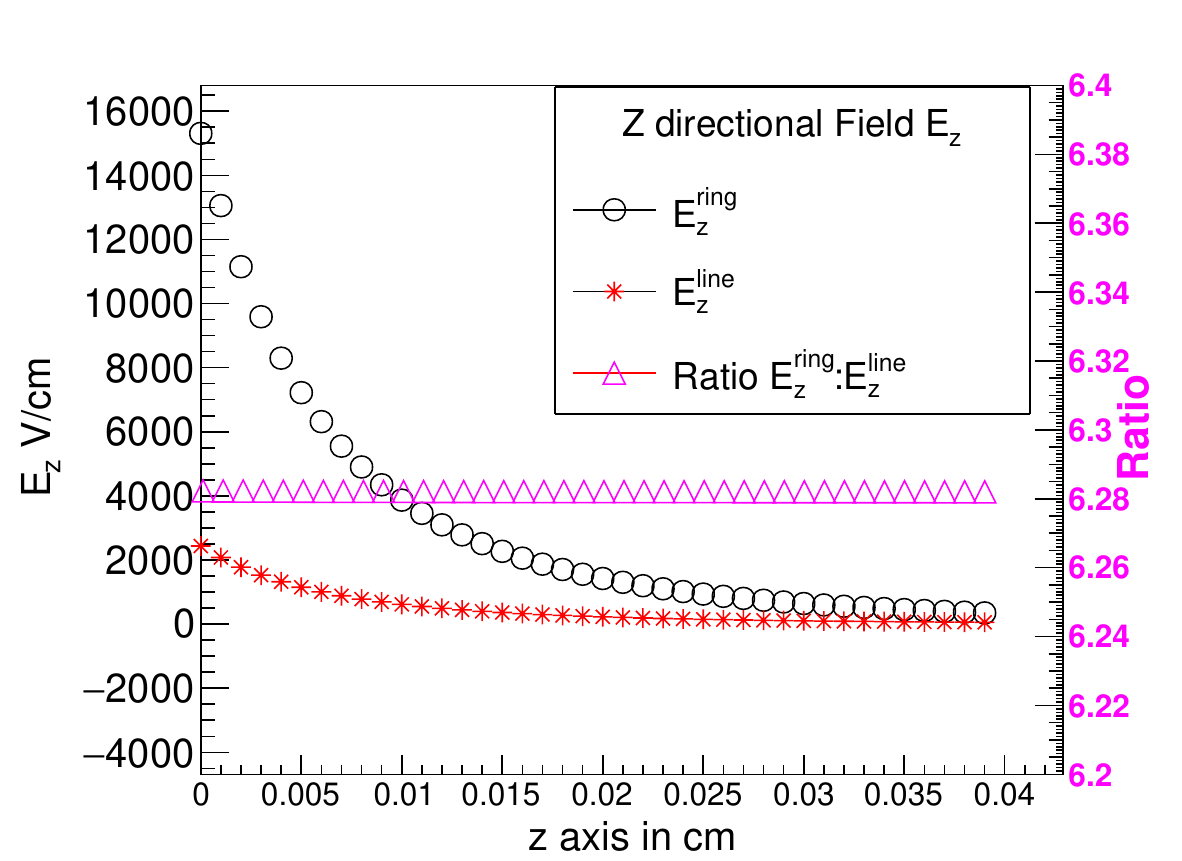}
	}\subfloat[ \label{fig:ezfieldImageLipCorrected}]{\includegraphics[scale=0.4]{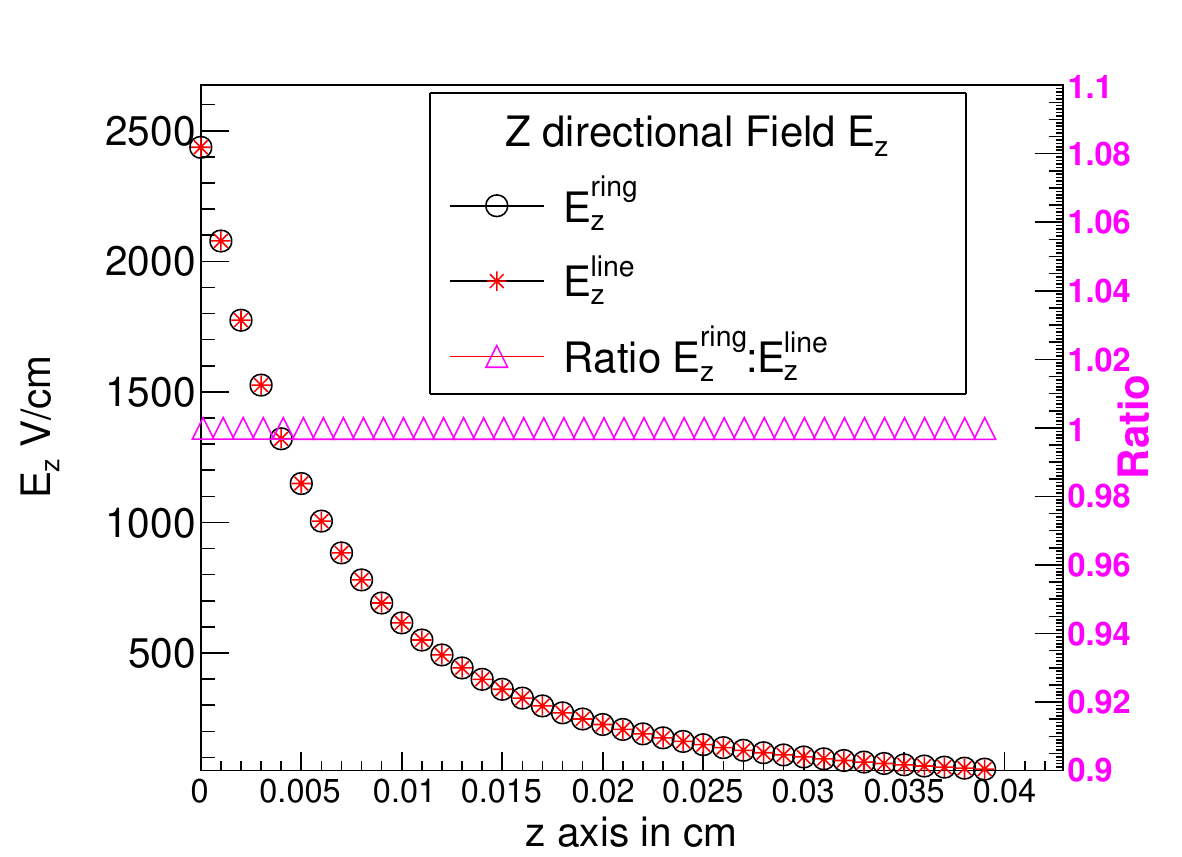}
		
	}
	
	\caption{(a)
		Comparision between ring and line z-directional image field before $2\pi$ division. (b) Comparision between ring and line z-directional image field after $2\pi$ division.}
\end{figure}
The variation of total (source+image) and only image z-directional field along the z-axis inside the gas gap for both ring and line approximation with their ratios at each point $(E_{z}^{ring}:E_{z}^{line})$ have been shown in figures \ref{fig:ezfieldlip} and \ref{fig:ezfieldImagelip} respectively. It is found that in both cases the ratio at each point is $\frac{E_{z}^{ring}}{E_{z}^{line}}\approx6.28\approx\,2\pi$ approximately. Therefore the field values are large for ring approximation in figures \ref{fig:ezfieldlip} and \ref{fig:ezfieldImagelip}. If we divide the values of $E_{z}^{ring}$ with $2\pi$ then the field values match exactly for both ring and line approximations, and the ratio becomes $\frac{E_{z}^{ring}}{E_{z}^{line}}\approx1$ (see figure \ref{fig:ezfieldlipCorrected} and \ref{fig:ezfieldImageLipCorrected}).
\subsubsection{ neBEM and line approximation}
\begin{figure}
	\center\subfloat[\label{fig:neBEM_no_image} ]{\includegraphics[scale=0.4]{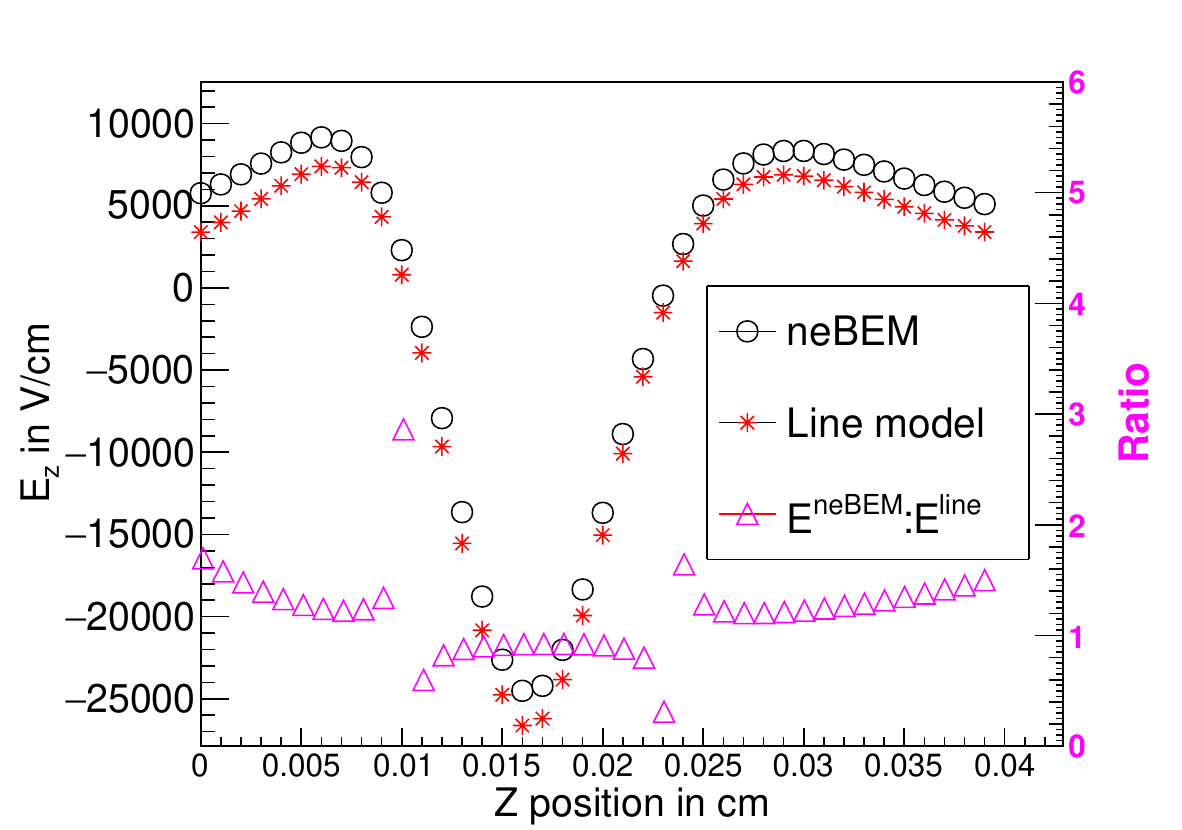}
		
	}\subfloat[\label{fig:neBEM_with_image} ]{\includegraphics[scale=0.4]{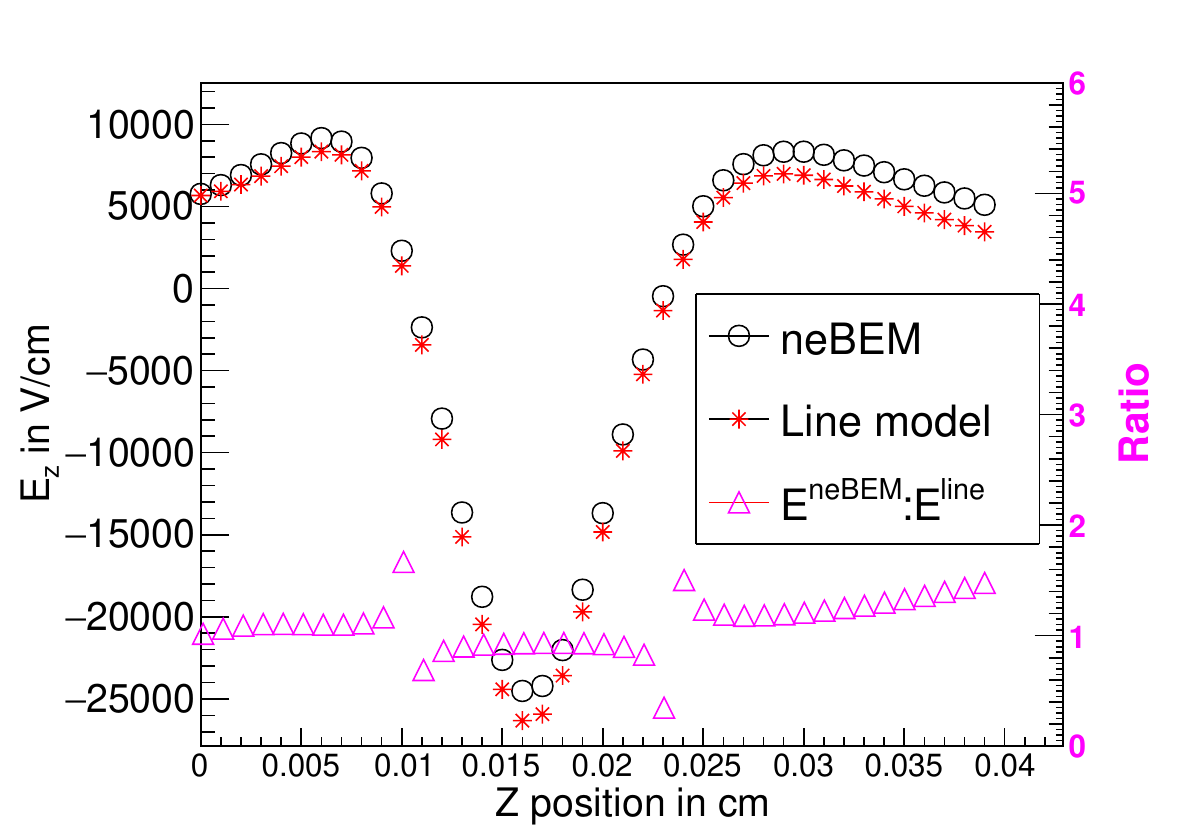}
		
	}
	
	\caption{(a) Comparision of  z-directional field without considering effects due to image charges with neBEM. (b) Comparision of total z-directional field (source+image) with neBEM. }
\end{figure}

In figures \ref{fig:neBEM_no_image} and \ref{fig:neBEM_with_image}, we have compared the results obtained using the proposed method with those obtained using the numerical solver neBEM. In the former figure, line model without image charge has been used to estimate the field. The estimated fields are different from those obtained using neBEM, the larger deviations (ratio between the estimates is $\approx
1.75$) being close to the RPC surface where the effect of image charges are expected to be important. Beyond this region, the comparison is reasonable-the ratio between the estimated values varying from 0.8 to 1.25. The ratio is much larger close to regions where the field values themselves are close to zero and can be safely ignored. The estimates from the line model with image charge have been compared to the same neBEM estimates in the latter figure. Here, remarkable improvement is observed close to the RPC surface, the ratio between the two estimates being close to 1. In the rest of the domain, the quality of agreement remains unchanged. This comparison clearly shows the efficacy of the proposed approach to incorporate effects of image charge using the line model.    
\vfill
\subsection{Selection of number of image charge for avalanche charge distribution} \label{Subsec:Selection_image_avCharge}
\begin{figure}
	\center\subfloat[ \label{fig:deltaSm_avalanche_chargeD1}]{\includegraphics[scale=0.38]{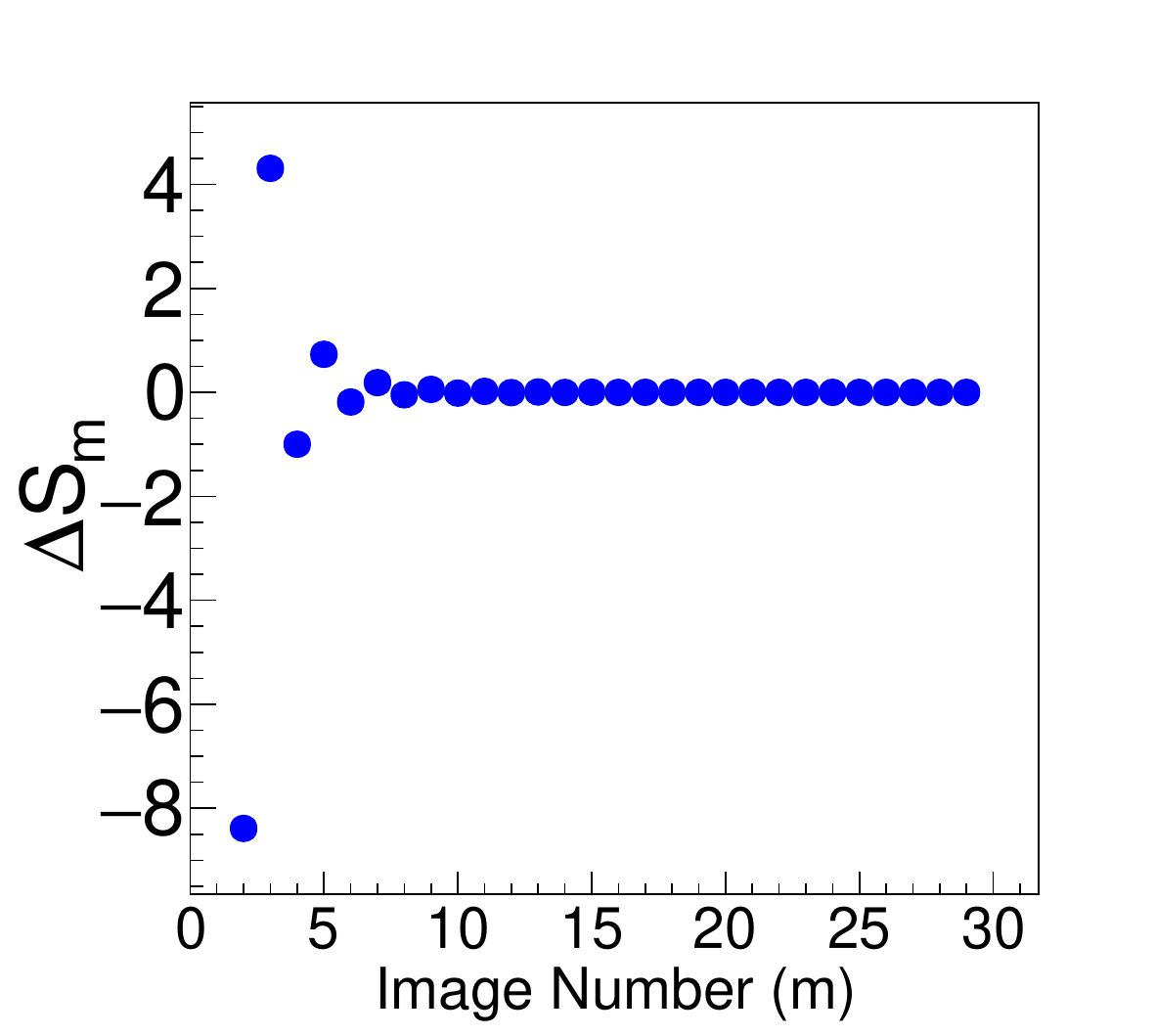}
		
	}\subfloat[ \label{fig:deltaSm_avalanche_chargeD2}]{\includegraphics[scale=0.38]{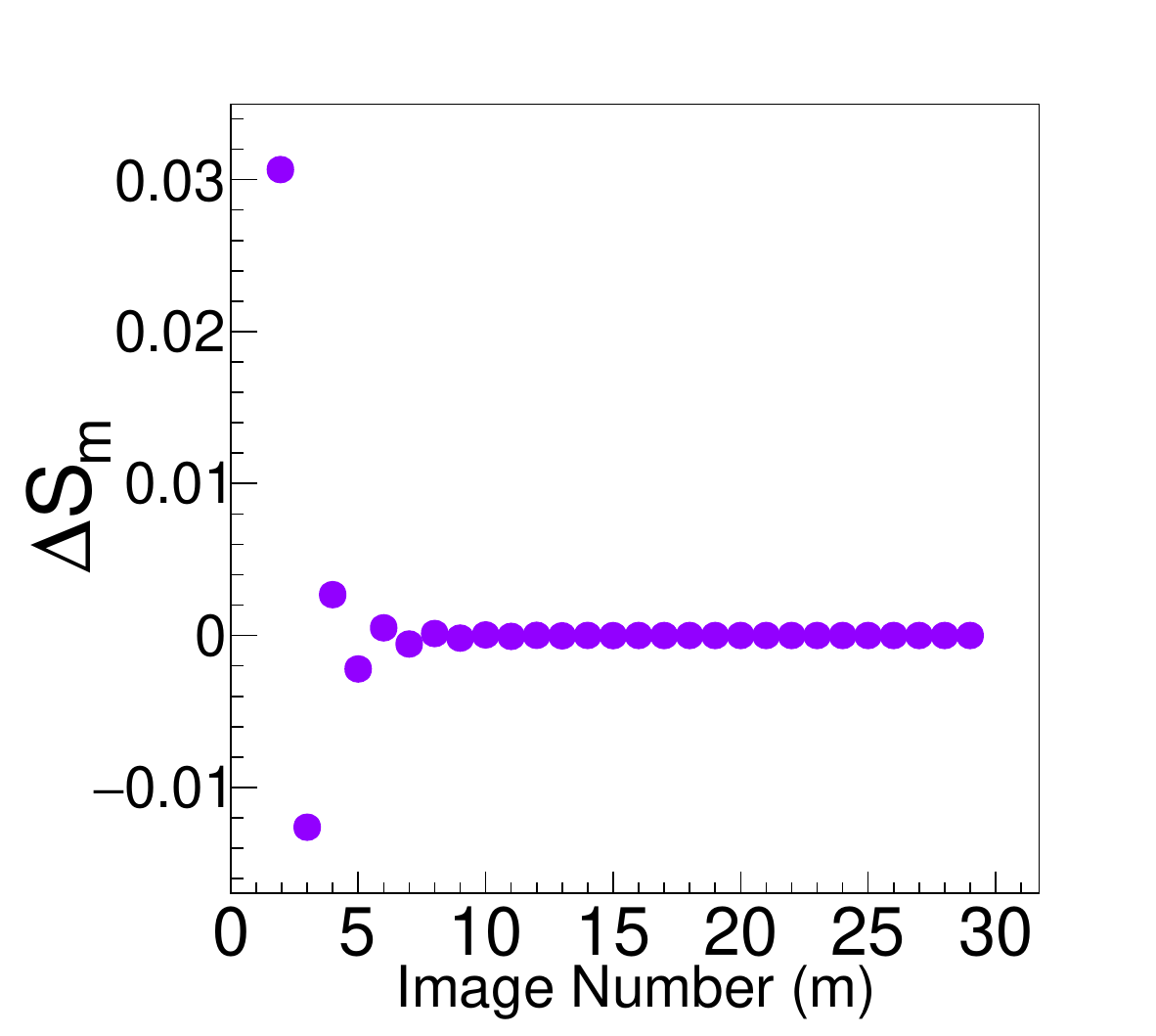}
		
	}
	
	\caption{ (a) Percentage of contribution of higher order avalanche image charges on the total electric field
		for dielectric electrode D1 of figure \ref{fig:RPC_image_charge_formation}. (b) Percentage of contribution of higher order avalanche image charges on the total electric
		field for dielectric electrode D2 of figure \ref{fig:RPC_image_charge_formation}.
	}
\end{figure}

In subsection (\ref{subsec:6.3}), we have discussed the selection criteria of the number of images due to a single point charge inside RPC. In this section, we will discuss the same for the avalanche charge distribution. The value of $\Delta S_{m}$ for both electrodes D1 and D2 has been shown in figures \ref{fig:deltaSm_avalanche_chargeD1} and  \ref{fig:deltaSm_avalanche_chargeD2}, where the results are more or less similar with single point charge case. Since the charge cluster is near electrode D2, the first-order  term $Q_{31}^\prime$ of $M5^{D2}$ has more contribution on the field than the other higher-order terms. Hence, the maximum contribution of higher-order terms or the absolute maximum value of  $\Delta S_{m}$ ($\approx0.03\%$) is very small.
On the other hand, electrode D1 is far from the charge cluster. Hence, the first order term $Q_{31}$ of $M5^{D1}$ is itself a very weak contributor and so the other corresponding higher orders.  Therefore, the maximum absolute value of $\Delta S_{m}$ ( $\approx 8\%$) for D1 is found to be larger than maximum absolute value $\Delta S_{m}$ for D2 (m=2,3,4..). However, the strength of the field is significantly less for the images of electrode D1.	

\subsection{Variation of image field with electrode material}\label{sec:varImFieldWPer}

The variation of the applied electric field with the relative permittivity of the electrode has been discussed in section \ref{sec:section2} (case 1). We have seen in sections \ref{section5} and \ref{sec:sec_6} that the contribution of the higher-order image charges are very small and from the table \ref{tab:threelayer_image} and \ref{tab:genaration of image RPC} it is confirmed that the position of image charges are depending on the electrode thickness and gas-gap. Therefore, it can be concluded that the dependence of image charge field on the electrode thickness and gas gap is also less. Thus, we will only discuss the variation of the image field on $\epsilon_{r}$. As higher order terms are less significant, we neglect them from $M5^{D1,D2}$, and then we are left with only a single image charge $Q\,\alpha_{32}$ for D1 and $Q\,\alpha_{34}$ for D2.  Therefore,  the image field at any point (x,y,z) inside the gas gap can be expressed as:

\begin{align}
	E^{image}_{z}=E_{z}^{D1}(Q\,\alpha_{32},h,g,x,y,z,\bar{r},\bar{z})+E_{z}^{D2}(Q\,\alpha_{34},h^\prime,g,x,y,z,,\bar{r},\bar{z}).
\end{align}

\begin{figure}
	\center{\includegraphics[scale=0.4]{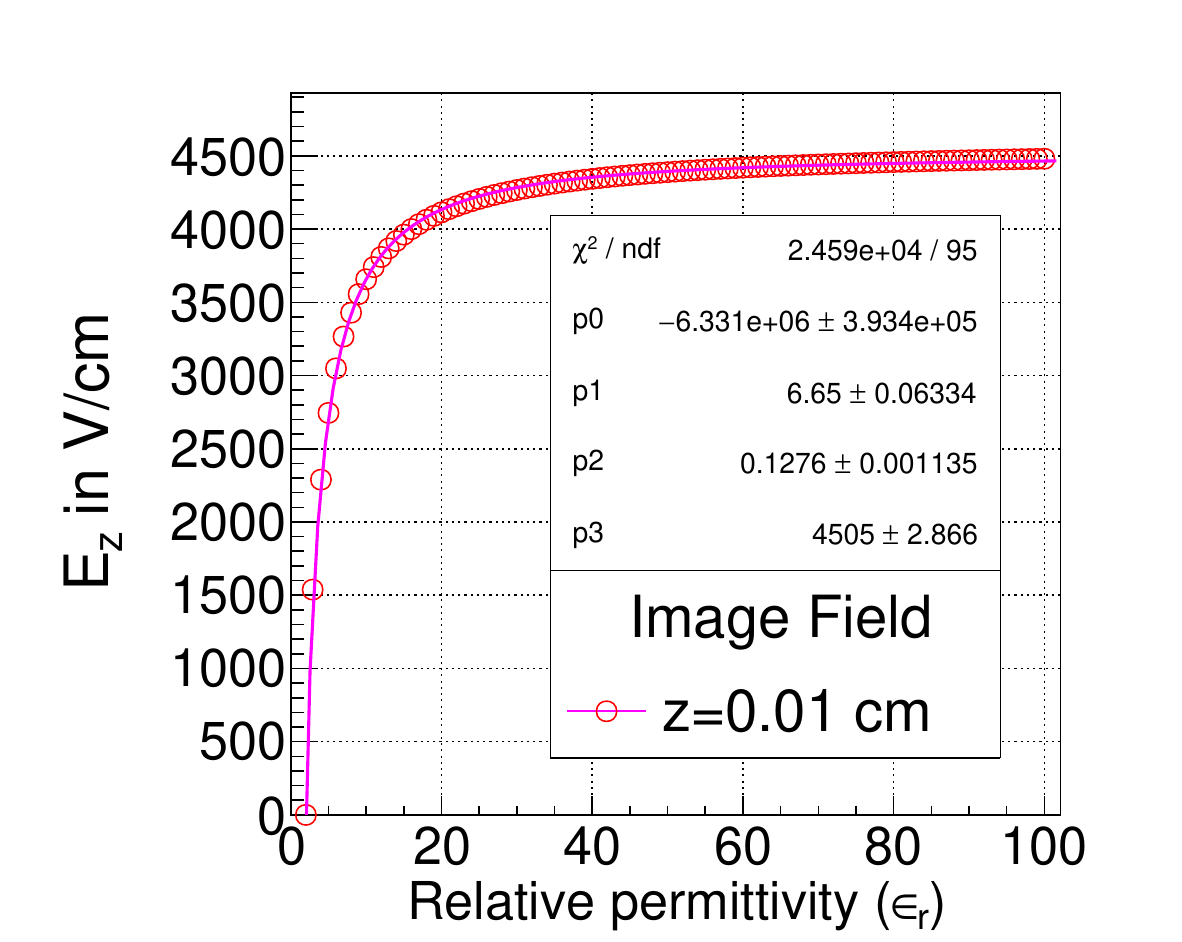}
		
	}
	
	\caption{\label{fig:Ez_vs_epr_image}Variation of avalanche image field at z=0.01 cm with relative permittivity of electrodes ($\epsilon_{r}$) of an RPC }
\end{figure}

Here we are using line equation \ref{eqn:ez_line} to calculate both $E_z^{D1,D2}$. As $\alpha_{32}=\alpha_{34}$ we can write,  

\begin{equation}
	\begin{split}
		E^{image}_{z}&=\alpha_{32}(E_{z}^{D1}(Q,h,g,x,y,z,,\bar{r},\bar{z})+E_{z}^{D2}(Q,h^\prime,g,x,y,z,,\bar{r},\bar{z})).
	\end{split}
\end{equation}
As medium III is gas so $\epsilon_{3}=\epsilon_{0}$ then $\alpha_{32}=\frac{1-\epsilon_{r}}{1+\epsilon_{r}
}$, where  $\epsilon_{r}$=$\frac{\epsilon_{2}}{\epsilon_{0}}$ and $\epsilon_{0}$=permittivity of gas/air. The variation of $E_z^{image}$ with relative permittivity $\epsilon_{r}$ of electrode has been shown in figure \ref{fig:Ez_vs_epr_image} at z=0.01cm, where it is found that on an increment of $\epsilon_{r}$, the field value also increases first, then it starts showing the saturation after $\epsilon_{r}>20$. The data points of figure \ref{fig:Ez_vs_epr_image} are fitted with the equation:
\begin{equation}\label{eqn:4.8}
	f(\epsilon_{r})=p0\,exp(-p1\,\epsilon_{r}^{p2})+p3.
\end{equation}
The fitted values of the parameters p0,p1,p2,p3 have been shown in same figure.
Again like equation \ref{eqn:2.1} the parameters p0 and p3 of equation \ref{eqn:4.8}  has the dimension of the electric field and p1 is the function of $\epsilon_r$ and p2 must be constant. Other physical significance of the same parameters are yet to be understood, but the functional form allows interpolation for arbitrary values of $\epsilon_{r}$.

\section{Summary}
We have discussed the dependence of the applied electric field inside an RPC with electrode parameters such as permittivity, thickness of electrodes, and gas gap. It is seen that at fixed applied voltage, electrode thickness, and gas gap, the electric field increases rapidly with permittivity and starts to saturate approximately after the value of 20. Also, the variation of electric field with the permittivity of the electrode decreases for smaller thickness of the electrodes.  
On the other hand, the electric field reduces with the increment of electrode thickness when the permittivity and gas gap is fixed. The electric field inside the middle of the gas gap diminishes on the increment of the gas gap, as expected.
\par In sections \ref{sec:section3} and \ref{sec:Section4_avalancheCharge}, the calculation of the total field (source + image) of charges at arbitrary locations in an RPC has been discussed. The electric field caused by the space charge-induced dipoles on the electrodes has been calculated using the method of images. This enables us to compute the total field-induced due to the presence of an avalanche. We have also given an example of electric field calculation of an avalanche charge distribution where we have used a straight-line model.  The validity of our straight-line model with existing models in the literature has also been checked, and a very good agreement is observed. 
\par It is clear that the number of space charges of a growing avalanche is a stochastic process, and the image charge field depends on the number of space charges and their distances from the electrodes. Therefore, at every step of a growing avalanche, the number and distance of space charges and the images of them will change. Hence,  we have proposed a technique to dynamically choose reflections of the image charges while the avalanche is growing.   
Another important issue is the variation of the image field with permittivity of the electrodes, which is discussed in subsection \ref{sec:varImFieldWPer}. The image field is increased with permittivity, but approximately after $\epsilon_{r}>20$, the field value attains saturation.
\par The advantage of using the line model is that it removes the constraint of rotational symmetry of the avalanche charge region. Since, the electric field equations are analytical and does not include any numerical integration, the method is fast and useful while simulating an avalanche or avalanche to streamer transition. In the case of streamers, the charge can be distributed over the full gas gap. Hence, the variation of space charge electric field and image field, will need to be further investigated. We hope to incorporate the proposed model in Garfield++ in the near future.



\section{Algorithm}\label{appendix_ch5}

\chapter{Parallelization of Garfield++ and neBEM to Simulate Space Charge Effects in RPCs}\label{chp:chap_fast_aval}
	\section{Introduction}\label{sec:1_intro}
The development of gaseous ionization detectors such as Gas Electron Multipliers (GEM) \cite{SAULI20162}, Resistive Plate Chambers (RPC) \cite{cardeli-1, cardeli-2}, and Time Projection Chambers (TPC) \cite{Hilke:2010zz} has been crucial for progress in various fields, including high energy physics, astronomy, and medical physics\footnote{The contents of this chapter are taken from the following arxiv publication by the author:
	
	\textbf{Parallelization of Garfield++ and neBEM to simulate
		space-charge effects in RPCs, T. Dey, et al., Computer Physics Communications, Volume
		294, 2024, ,https://doi.org/10.1016/j.cpc.2023.108944}}.
Although the geometry of an RPC is simpler than that of other gaseous detectors, the physics processes associated with it (such as primary ionization, charge transport, electron multiplication, and signal formation) are just as complex. Despite this complexity, RPCs are considered reliable enough to be regularly used as timing, triggering, and tracking devices in numerous experiments \cite{Goswami_2017, Kumari_2020, Collaboration_2012, MONDAL2021166042}. To adequately explain experimental results, a precise simulation model is needed. A one-dimensional hydrodynamic model can provide quick results in less computational time \cite{MOSHAII2012S168}. This model works well when the number of space charges is significant, as the collective behavior of space charges can be considered as fluid-like \cite{Rout_2021}. Another option is a detailed Monte Carlo simulation particle model, in which particles are tracked individually \cite{LippmanThesis, Lippmann_1, Lippmann:2003ar}. However, this model can be computationally expensive due to the large number of avalanche charges it simulates
\par Garfield++ ~\cite{Garfield,Veenhof:1993hz,VEENHOF1998726} is a C++-based simulation tool to simulate ionization detectors.
It has its own geometry tools, which can be used to make simple virtual detectors.
However, for complex geometries, one can borrow field maps and geometry files from external field solvers COMSOL ~\cite{comsol}, neBEM ~\cite{MAJUMDAR2008346,MAJUMDAR2009719} etc. There are several methods to generate avalanches inside gaseous detectors in Garfield++, for example, microscopic tracking, Monte Carlo tracking, etc., that are based on particle models. 
These tracking methods are very detailed but slow in a computational sense because all codes run serially.
One more drawback of tracking methods in Garfield++ is the absence of the dynamic space charge effect, which is crucial when the number of space charges becomes significantly large ($\geq10^6$) so as to modify the applied field.
\par To address the above issues of Garfield++, we have developed and implemented an algorithm and introduced an extra class, pAvalancheMC, to generate avalanches with multithreading techniques using OpenMP \cite{OpenMp}.
In this class, we have incorporated the space charge effect using the line charge model discussed in \cite{Dey_2020,Dey_2022}.
Another approach to parallelize Garfield++ using MPI can be found in \cite{pGarfield1, BOUHALI201892}, named pGARFIELD. However, pGARFIELD does not include the space charge effect.

In section \ref{sec:2_randomNumber}, we describe the method of generating uniform and uncorrelated parallel random numbers using the TRandom3 class of ROOT~\cite{root-cern}, which is the heart of the Monte Carlo simulations. Similarly, the use of OpenMP and FastVol method of neBEM \cite{MAJUMDAR2008346,MAJUMDAR2009719} to accelerate electric field computation is described in section \ref{sec:3_neBEM}.

Section \ref{sec:4_stepsOfAvalanche} describes the steps or algorithm of generating Monte Carlo avalanche, including the dynamic space charge effect. In section \ref{sec:5_instanceAvalancheConst}, we provide an example of an avalanche in an RPC generated through pAvalancheMC.

In section \ref{sec:6_speedUp}, we discuss the performance of the multithreaded code. Section \ref{sec:7_neBEM} compares the induced charge distribution of a timing RPC for three different voltages (1720 V, 1730 V, and 1735 V) with space charge effect calculated using neBEM and the proposed version of Garfield++.

 In Section \ref{sec:field}, the applied electric field configuration of RPCs was discussed. The number of primary electron distributions inside the gas gap of two RPCs was discussed in Section \ref{sec:track}. Sections \ref{sec:indCh} and \ref{sec:risetime} presented a comparison of induced charge and signal rise time for different geometrically configured RPCs at the same applied field.
\section{Uncorrelated and thread safe parallel Random number generation using Trandom3 and OpenMP} \label{sec:2_randomNumber}
\begin{figure}
	\center\subfloat[\label{fig:2d_corr_th0,2&0,1}]{\includegraphics[width=0.7\linewidth]{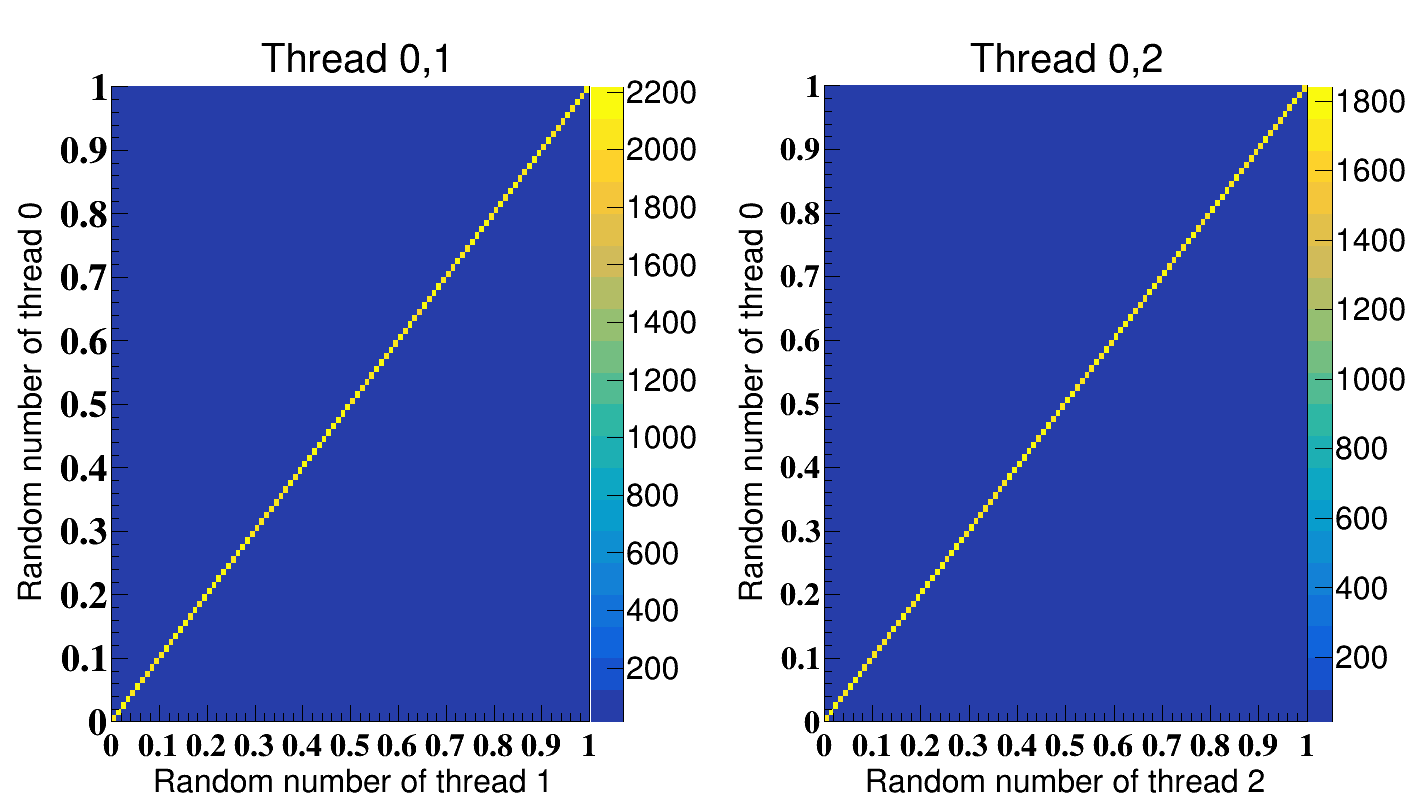}}
	\center\subfloat[\label{fig:3d_corrth0,1,2}]{\includegraphics[width=0.6\linewidth]{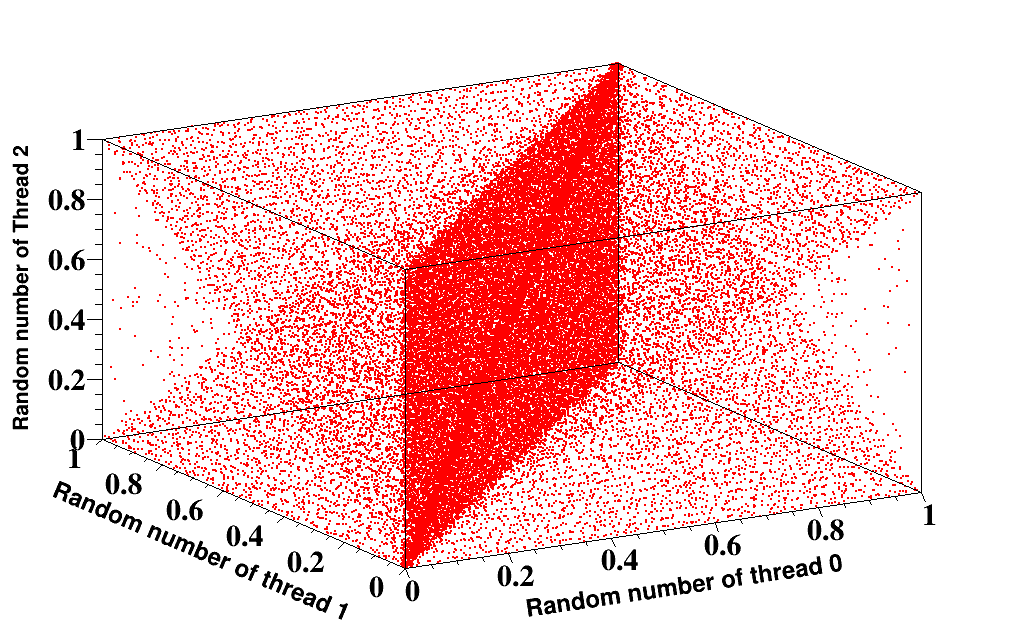}}
	\caption{Parallel random number generation with the same TRandom3 object and the same seed used by all threads,
		(a) 2D Correlation between random numbers generated by threads 0,1 (left) and threads 0,2 (right) (b) 3D correlation between random numbers generated by threads 0,1,2. \label{fig:Correlation_radnds}}
	
\end{figure}
\begin{figure}
	\center\subfloat[\label{fig:2d_uncorr_th0,2&0,1}]{\includegraphics[width=0.7\linewidth]{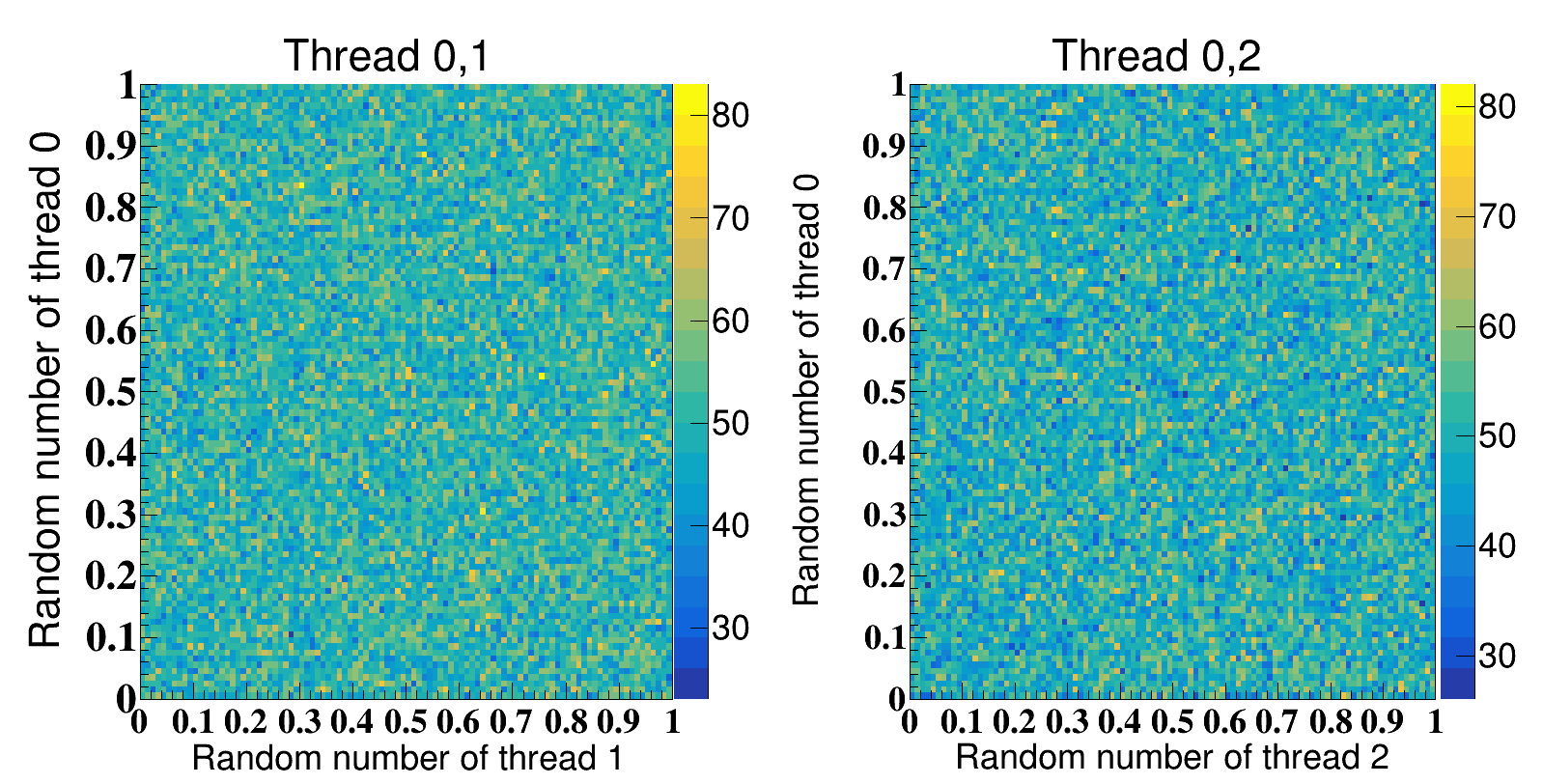}}	
	\center\subfloat[\label{fig:unCorrelation_radnds}]{\includegraphics[width=0.6\linewidth]{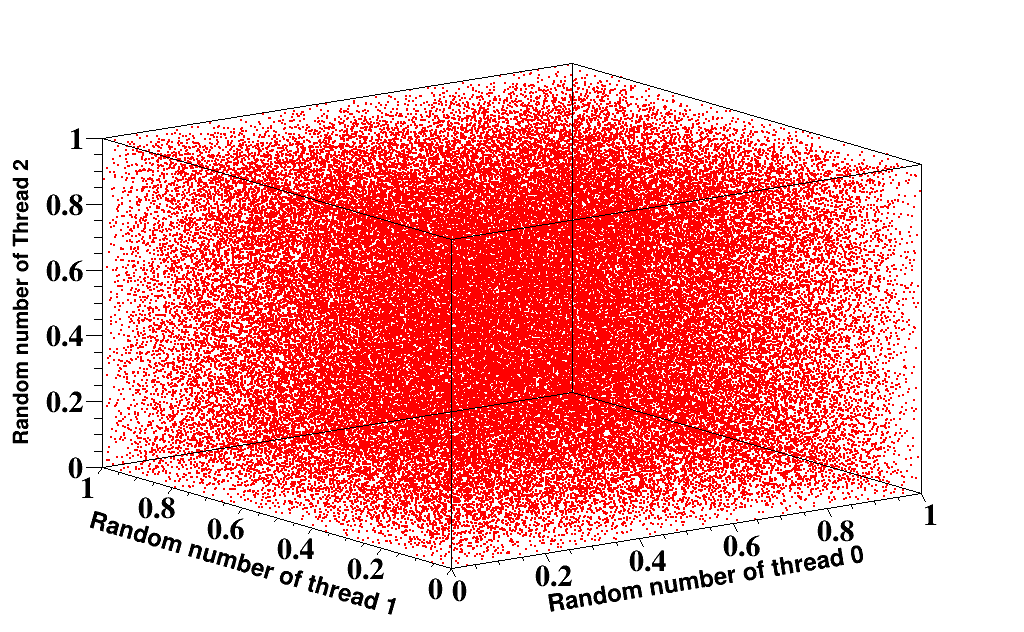}}
	
	\caption{Parallel random number generation with different TRandom3 object and seed used by different threads, (a) 2D Correlation between random numbers generated by threads of IDs 0,1 (left) and threads 0,2 (right) (b) 3D correlation between random numbers generated by threads of IDs 0,1,2.}
	
\end{figure}

We know that the outcome of any Monte Carlo simulation depends on the quality of the generated random numbers. It is important that the generated random numbers are uncorrelated, and the corresponding random engine has a very long period. TRandom3 is a class of ROOT~\cite{BrunRoot} that can be used to generate uniform random numbers. It is based on the Mersenne Twister (MT) algorithm and provides a period of $2^{19937}-1$ and 623-dimensional equidistribution up to 32-bit accuracy~\cite{MersenneTwister}. We have chosen TRandom3 as the random engine in our simulation code. The random numbers generated serially using TRandom3 are uncorrelated.
However, we have observed that the former engine leaves some correlation between random numbers generated in parallel by different threads (see Figure~\ref{fig:Correlation_radnds}) when all threads use the same random engine and seed. The 2D and 3D correlation between random numbers corresponds to threads with IDs 0, 1, and 2, generated from the same object of the random engine. This effect is shown in Figure~\ref{fig:2d_corr_th0,2&0,1} and Figure~\ref{fig:3d_corrth0,1,2}, respectively, where it can be seen that the random numbers are strongly correlated and mostly lie on three planes. To solve this problem, objects of the TRandom3 class have been assigned for three threads with three different seed values. As a result, the random numbers corresponding to each thread become uncorrelated, as shown in Figure~\ref{fig:2d_uncorr_th0,2&0,1}. Furthermore, from the 3D correlation plot among random numbers corresponding to thread IDs 0, 1, and 2 (see Figure~\ref{fig:unCorrelation_radnds}), it can be inferred that they are uniformly distributed throughout the volume.

The method outlined above is implemented in the pAvalancheMC class. Using the function SetNumberOfThreads(n), one can specify the number of threads used for random number generation. The different objects of the TRandom3 class for different threads can be assigned using the same function. Using a unique thread ID created by OpenMP, all threads can generate random numbers from their corresponding objects. An example C++ code for the function SetNumberOfThreads(n) is provided below:
\vspace{3cm}
\begin{lstlisting}%[frame=none]

vector<TRandom3 *>rnd;
void SetNumberOfThreads(int nThread)
{
	rnd.resize(nThread);
	for(int ThreadId=0; ThreadId<nThread; ThreadId++)
		rnd[ThreadId]= new TRandom3(0);
}
\end{lstlisting}


\section{Acceleated electric field solution using OpenMP and FastVol}
\label{sec:3_neBEM}
The Open Multi-Processing (OpenMP) is an Application Programming Interface (API) that supports multi-platform shared memory multiprocessor programming in C, C++ and FORTRAN on most processor architectures and operating systems \cite{OpenMp}.
It consists of a set of compiler directives, library routines and environment variables that influence runtime behavior and uses a simple, scalable API for developing parallel applications on platforms ranging from the standard desktop to supercomputers.
We have successfully implemented OpenMP for the neBEM field solver \cite{MAJUMDAR2009719}.
The parallelization has been implemented in several sub-functions of the toolkit, such as computation of the influence coefficient matrix, matrix inversion and evaluation of the field and potential at desired locations.
These routines are computation intensive since there can be thousands of elements where the
charge densities need to be evaluated and /or influence due to all these elements need to be taken
care of.
The matter is even more complicated through the use of repetition of the basic structure
in order to conform to the real geometry of a detector.
This has proved to be very important in improving the computational efficiency of the solver.
We have tested these implementations on upto 24 cores.
The observed reduction in the computational time has been found to be significant while the precision of the solution has been found to be preserved.
\par Even after adopting OpenMP while solving for the charge distribution on all the material interfaces of a given device, the time to estimate potential and field for a complex device
can become prohibitive.
This is especially true if the device (not usually necessary for an RPC) is composed of hundreds of primitives, thousands of elements and several tens of repetitions.
Reduction of time taken to estimate the electrostatic properties becomes increasingly important when complex processes such as Avalanche, Monte-Carlo tracking and Micro-Tracking are being modelled using Garfield++.
In order to model these phenomena within a reasonable span of time, we have implemented the concept of using pre-computed values of potential and field at large number of nodal points in a set of suitable volumes.
These volumes are chosen such that they can be repeated to represent any region of a given device and 	simple trilinear interpolation is used to find the properties at non-nodal points.
The associated volume is named as the Fast Volume (FastVol) which may be staggered, if necessary.
In order to preserve the accuracy despite the use of trilinear interpolation, it is natural that the nodes should be chosen such that they are sparse in regions where the potential and fields are changing slowly and closely packed where these properties are changing fast.

\section{Steps of the avalanche simulation with space charge effect in Garfield++} \label{sec:4_stepsOfAvalanche}

This section will discuss techniques for implementing the space charge effect with multithreading in the newly introduced pAvalancheMC class of Garfield++. In further discussions, we will use the short name aE for the applied field, sE for the space charge field, and tE for the sum of applied and space charge fields. The necessary steps to simulate an avalanche are given below:
\begin{enumerate}[(i)]
	\item Selection of primaries to start avalanche according to their generation time.
	\item Parallel calculations of drift line of avalanche charges.
	\item Calculation of Gain in each step.
	\item Generation of cylindrical grid of space charge region.
	\item Calculation of dynamic space charge field.
\end{enumerate}

\subsection{Selection of primaries to start avalanche according to their generation time}
When a charged particle passes through a gas detector, it can interact with gas molecules and leave some primary ionizing particles along its path. A track of a charged particle through the gas gap and corresponding primary ionizations can be simulated using the class TrackHeed of Garfield++. The locations and time of generation of primary clusters are then stored in an array. The primaries (here electrons) can be generated at different instants of time along the track. Hence, the starting time of the avalanche generation from different primaries can be different. Therefore, if we arrange primaries in  an ascending order of the generation time, then depending on the simulation step $\delta t$, they will start the avalanche process. For example, if t is the generation of time of any primary electron, it will start an avalanche when $n\delta t\geq t$, where n is the number of simulation steps that have elapsed. 
\subsection{Parallel calculations of drift line of avalanche charges}
The AvalancheMC class in Garfield++ is capable of simulating the drift path of charges in a serial manner. However, it can become slow when dealing with a large number of charges. To address this issue, we have utilized OpenMP to divide the secondary particles among available threads and propagate them in the gaseous medium simultaneously at each time step to calculate further drift points. The diffusion component of pAvalancheMC remains the same as that in the AvalancheMC class, where Gaussian and anisotropic thermal diffusion is considered when an electric field is present \cite{LippmanThesis}. Therefore, the diffusion occurs in both longitudinal and transverse directions. The drift velocity ($v_D(tE)$), longitudinal ($D_L(tE)$) and transverse diffusion ($D_T(tE)$) constants have been calculated using MAGBOLTZ \cite{BIAGI1989716,BIAGI1999234} for the total electric field tE. The variance of the longitudinal Gaussian distribution is given by $\sigma_L=D_L\sqrt{v_D\delta t}$, and the variance of the transverse diffusion is given by $\sigma_T=D_T\sqrt{v_D\delta t}$.
\subsection{Calculation of Gain in each step}
The pAvalancheMC class calculates gain using the same method as the AvalancheMC class, which employs a modified version of the Yule-Furry model described in Section \ref{sec:ch2yuleFurry} of Chapter \ref{ch2}.
\subsection{Calculation of dynamic space charge field}
The space charge effect is turned on when the electron gain crosses a threshold. This threshold is determined by the user and, for the computations presented in this paper, the value has been set to $10^4$. The generation of grid elements and finding locations of the space charges in the grids have been divided into four steps. The steps are as follows:\\

\begin{figure}
	\center\subfloat[\label{fig:zGrid}]{\includegraphics[scale=0.4]{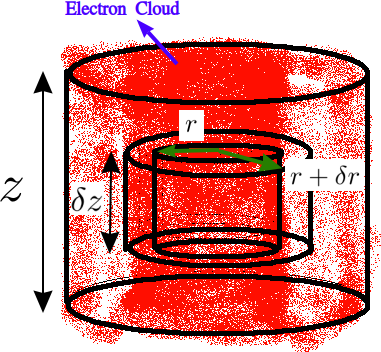}}\subfloat[\label{fig:rGrid}]{\includegraphics[scale=0.5]{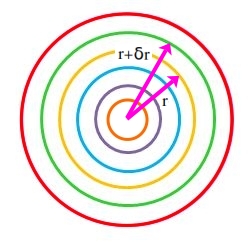}}
	
	\center\subfloat[\label{fig:cGrid}]{\includegraphics[scale=0.1]{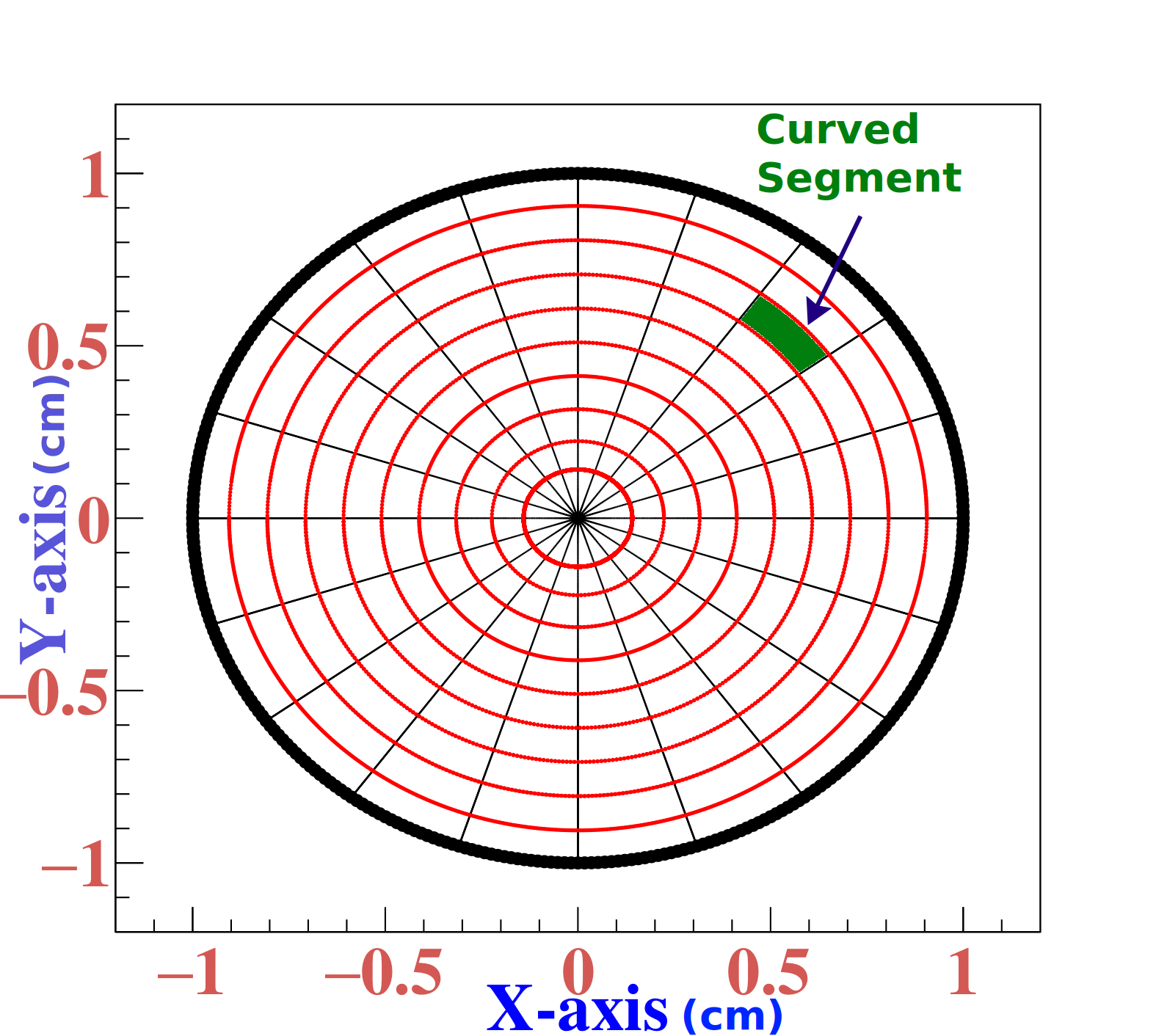}}\subfloat[\label{fig:sGrid}]{\includegraphics[scale=0.35]{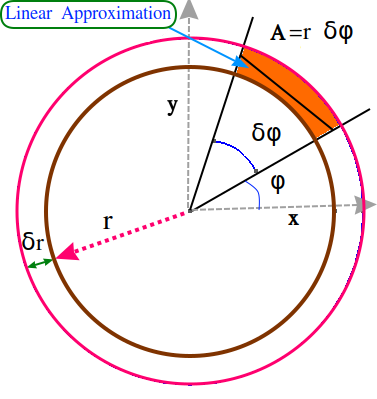}}
	\caption{Steps to generate cylindrical grid (a) Dividing z space by $\delta z$ element (b) radial space divided into several co-centric rings of increasing radius r \& thickness $\delta r$, (c) Divide rings into several small segments of curvatures, (d) Small $\delta \phi$  so that curvature becomes a line.}
\end{figure}

Step  1: Let us consider at a particluar time step of a growing avalanche, the space charges are distributed as shown in Figure \ref{fig:zGrid}. To calculate the field for this charge distribution, the z-directional space is divided into the ``S" number of $\delta z$ elements as shown in Figure \ref{fig:zGrid}. 

Step  2: For a particular z position, the avalanche region can be divided into several co-centric ``R" number of rings each having a thickness $\delta r$ as shown in Figures \ref{fig:rGrid}. For all z, the thickness $\delta z$ is considered as small as 0.001 cm.

Step  3: In the third step, each ring can be divided into ``L" curved segments (see Figure \ref{fig:cGrid} ). Hence, each curved segment will lie between angles of $\phi$ and $\phi+\delta \phi$, where $\phi$ is the azimuthal angle subtended to the center of the circle or ring (see Figure \ref{fig:sGrid}). Therefore, the $\phi$ directional space will also be divided into L number of $\delta \phi$ segments, so that $L\delta \phi=360$\textdegree \space(for more details see ~\cite{Dey_2020}).

Step  4: In this calculation, $\delta \phi$ is considered very small ($\approx 1$\textdegree) so that each curved segment can be considered as a straight line of length $r\delta \phi$ (see Figure \ref{fig:sGrid}).
The entire avalanche volume is segmented into small volumes, as described above. The charges within a specific volume are represented by a line charge with a constant charge density situated within that volume. The total number of grid elements will be $T=S\times R\times L$. Consider an electron or ion situated at ($r,\phi,z$) in a cylindrical coordinate system. Since a location ($r,\phi,z$) corresponds to one elemental volume of the grid, we need to find the serial number of the grid element to calculate the space-charge field. Suppose the electron at ($r,\phi,z$) corresponds to the $n^{th}$ ring, $m^{th}$ line, and $l^{th}$ z-segment, then the corresponding serial number of the elemental volume of the grid (G) is,
\begin{equation}
G=l\times R \times L+n \times L+m.
\end{equation}     
\par The space charge field at any grid element of number G is the sum of the electric fields due to other grid elements containing non-zero charges. The electric field is calculated in those grids where at least one electron is found. The space charge field is considered to be equal for all charges sharing the same grid element of number G.
The total field at any location ($r,\phi,z$) is the resultant of the applied field and space-charge field. Since grid elements are considered as line segments, the line charge formula has been used to calculate the space-charge field, as discussed in ~\cite{Dey_2020,Dey_2022}. 

\section{Instance of a simulated avalanche inside an RPC with space charge effect}\label{sec:5_instanceAvalancheConst}

\begin{figure}
	\center\includegraphics[width=0.5\linewidth]{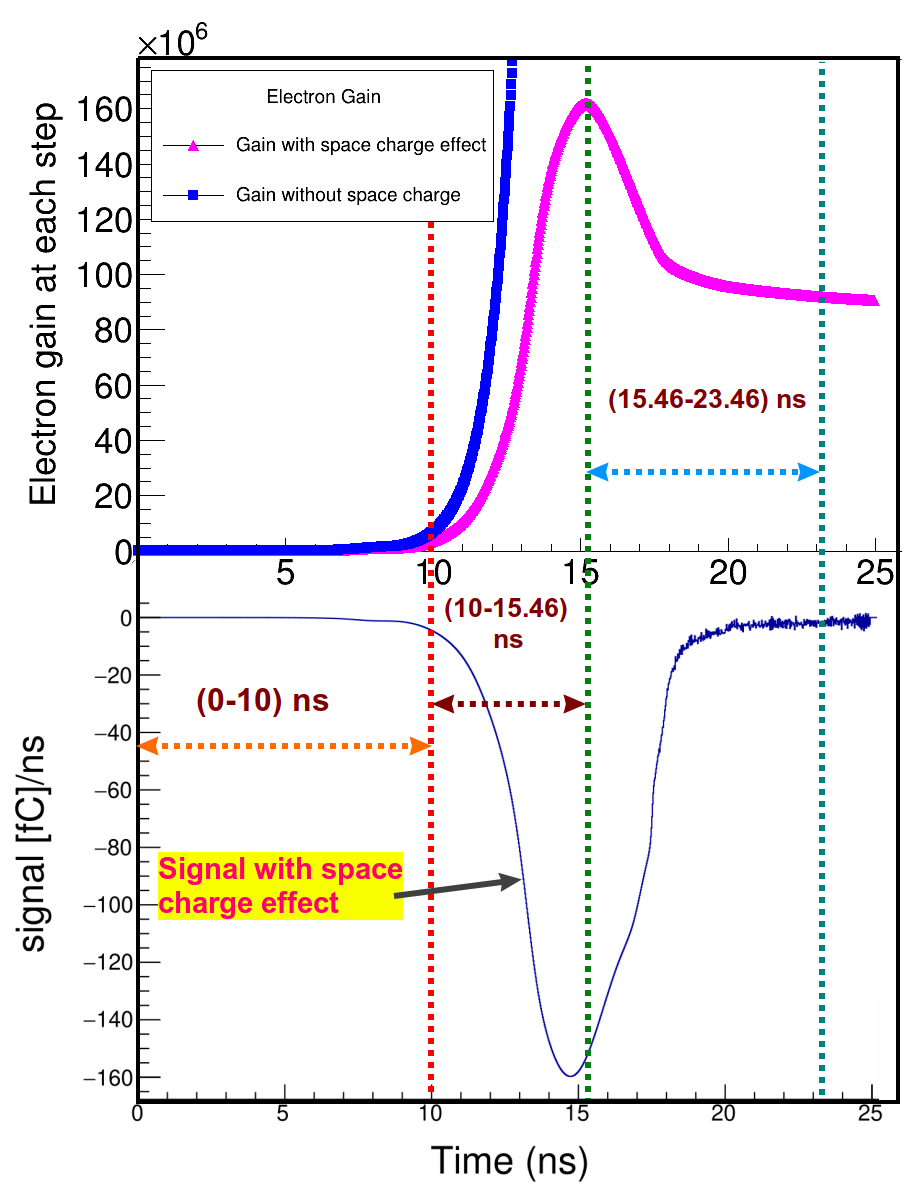}
	
	\caption{Variation of electron gain and signal with time}\label{fig:gain}
	
\end{figure}	
This section will discuss an example of an avalanche simulated in an RPC using pAvalancheMC with and without the space charge effect. A gas mixture of 70\% of Ar and 30\% of $\ce{CO_2}$ has been used. The electrodes of the RPC are made of a 2 mm thick Bakelite with an area of $30\times 30~cm^2$. The gas gap is also taken as 2 mm. A constant electric field (aE) of 23.5 kV/cm has been applied across the gas gap with the help of ComponentConstant class of Garfield++. The step of the simulation is taken as 20 ps. Muon track and primary ionisations inside the gas gap are generated by using the HEED.
\par In Figure \ref{fig:gain}, the electron gain at each step of time with and without the space charge effect is compared. The blue and pink curves specify the electron gain without and with the space charge effect, respectively. It is clear from the prior figure that till 10 ns, both curves almost overlap each other. After 10 ns, the blue curve continuously grows at a higher rate and never reaches saturation. On the other hand, the pink curve grows slowly, reaches a peak at around 15.46 ns, and then starts showing a saturation effect approximately after  17.96 ns.
\par  We can divide the pink curve into three regions a) before 10 ns in which space charge effects are negligible, b) From 10 ns to 15.46 ns, when gain increases rapidly, and c) after 15.46 ns, during which gain drops and maintains a saturated value. 
\par 
In order to understand the transport of charged particles through the gas volume, the distribution of z-positions vs. radial positions of the electrons at different instants of avalanche growth have been analyzed (e.g., Figure \ref{fig:4.5ns}). The following discussions contain the results of the analysis. 

\subsection{ From 0 ns to 10 ns, overlap region of gain}
\begin{figure}
	\center\subfloat[\label{fig:0ns}]{\includegraphics[scale=0.17]{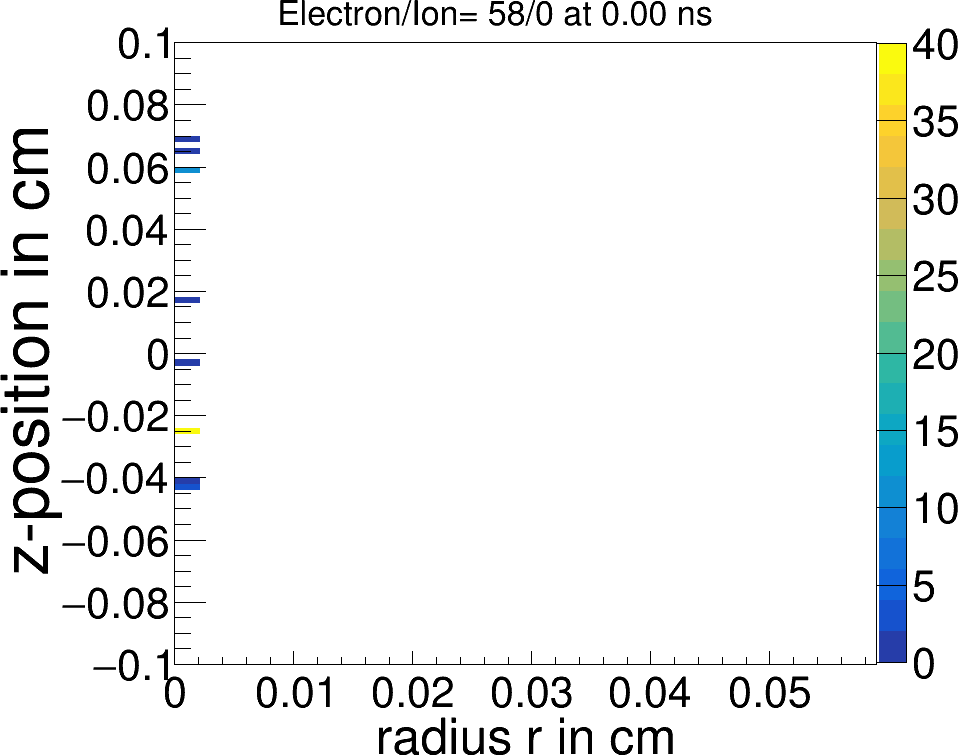}
		
	}\subfloat[\label{fig:4.5ns}]{\includegraphics[scale=0.17]{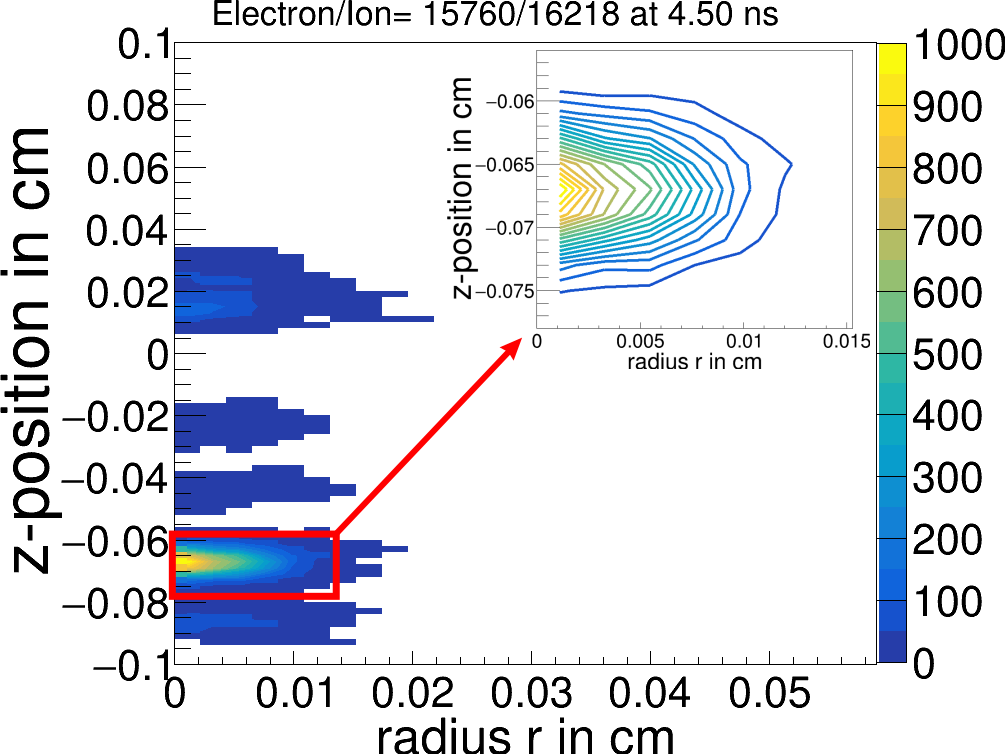}
		
	}
	
	\center\subfloat[\label{fig:7.46ns}]{\includegraphics[scale=0.17]{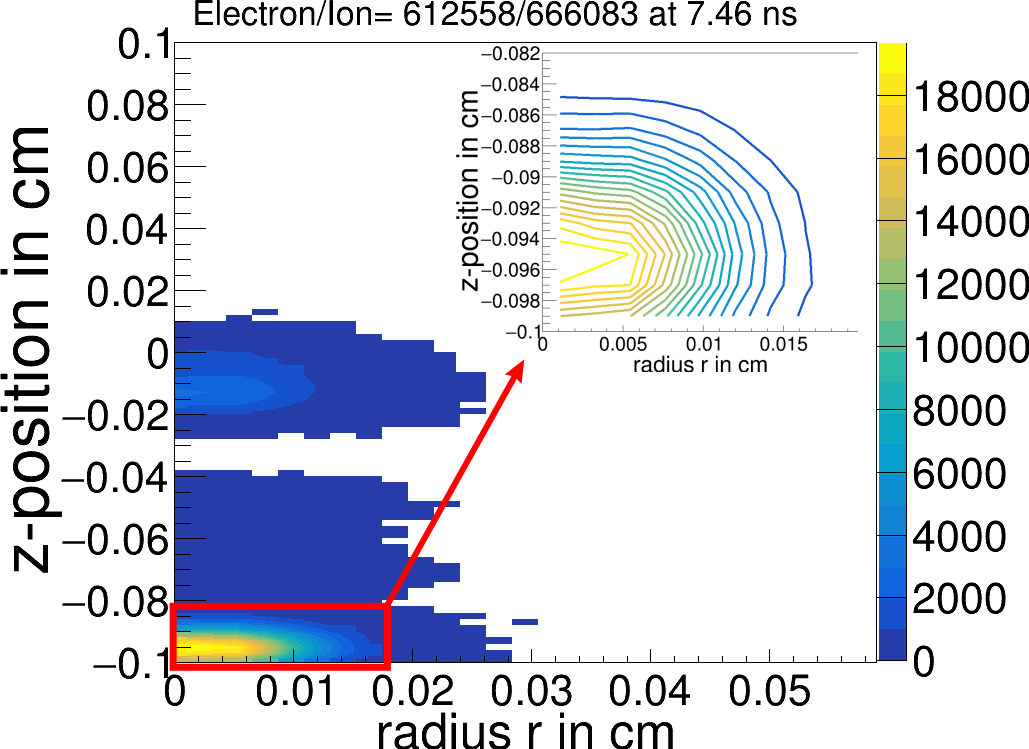}
		
	}\subfloat[\label{fig:10.46}]{\includegraphics[scale=0.17]{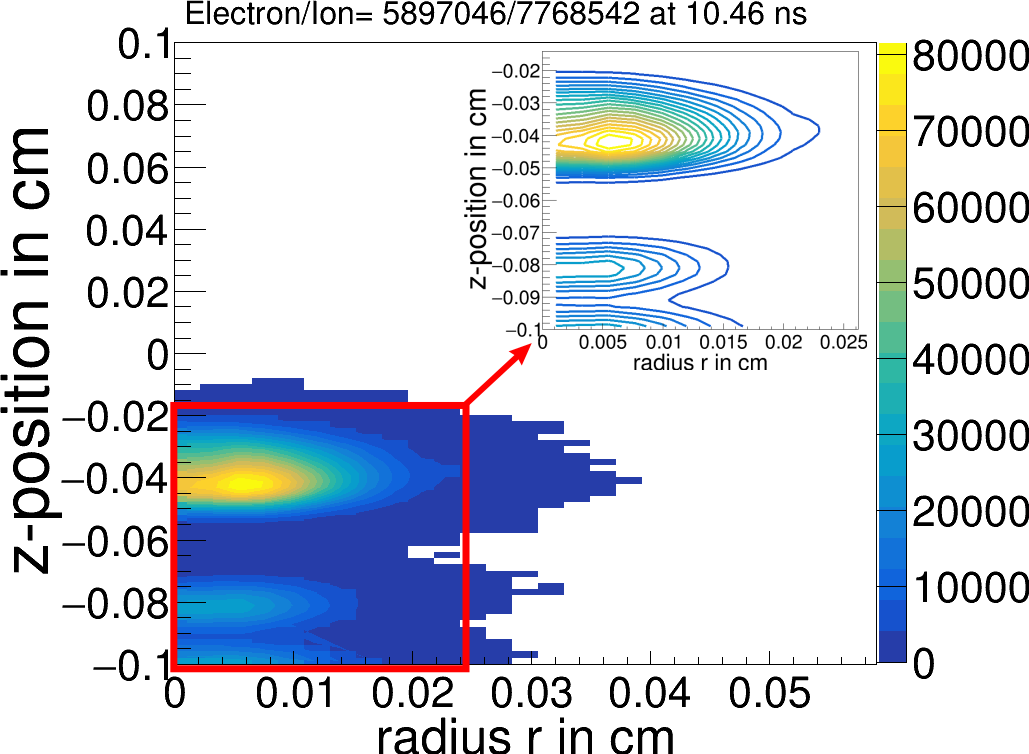}}
	
	\caption{Location of electrons in z-r plane at time (a) 0. ns, (b) 4.5 ns, (c) 7.46 ns and (d) 10.46 ns.}\label{fig:trk_beforesp}
	
\end{figure}
\begin{enumerate}[i.]
	\item At time 0 ns (see Figure \ref{fig:0ns}) there is only total 58 electrons  distributed in 7 clusters. It is noted that z=0.1 is the anode plane and z=-0.1 is the cathode plane. The color bar represents the number of electrons in a ring of radius r and thickness $\delta r$.
	\item In Figure \ref{fig:4.5ns}, it is shown that at time 4.5 ns electrons have already started participitating in the ionization process, and a total of 15760 electrons and 16218 ions are generated. In the former figure, two clusters near the cathode plane at z=-0.1 merge and form a big cluster. Therefore, the total number of electron clusters at this stage is four. From the color of the electron density contours of the prior figure, it can be said that the location of maximum charge density, or let us say ``charge center," is residing in the middle of the distribution (at radius r=0). Therefore it can be concluded that the electron distribution is roughly symmetric. 
	\item In Figure \ref{fig:7.46ns} at time 7.46 ns, there are only two electron clusters. The total number of electrons and ions is 612558 and 666083, respectively. Also, the charged center of the merged distribution has reached the anode, and the electron density contour is smooth and symmetric.
	
\end{enumerate}
\begin{figure}
	\center\subfloat[\label{fig:10.46ns_z}]{\includegraphics[scale=0.19]{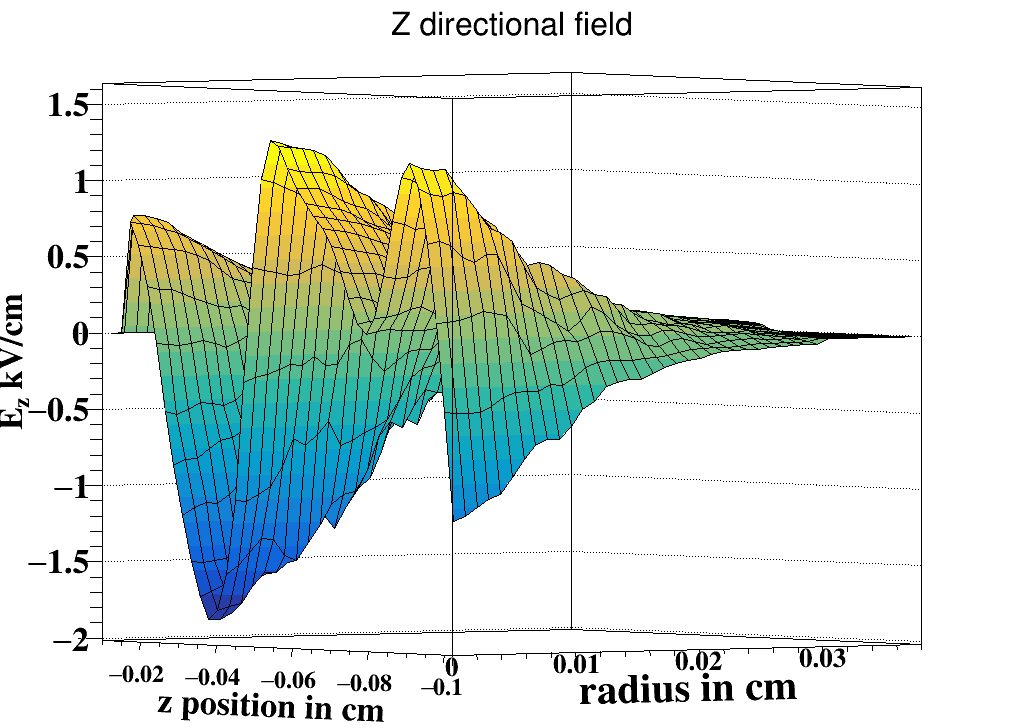}}\subfloat[\label{fig:10.46ns_rad}]{\includegraphics[scale=0.19]{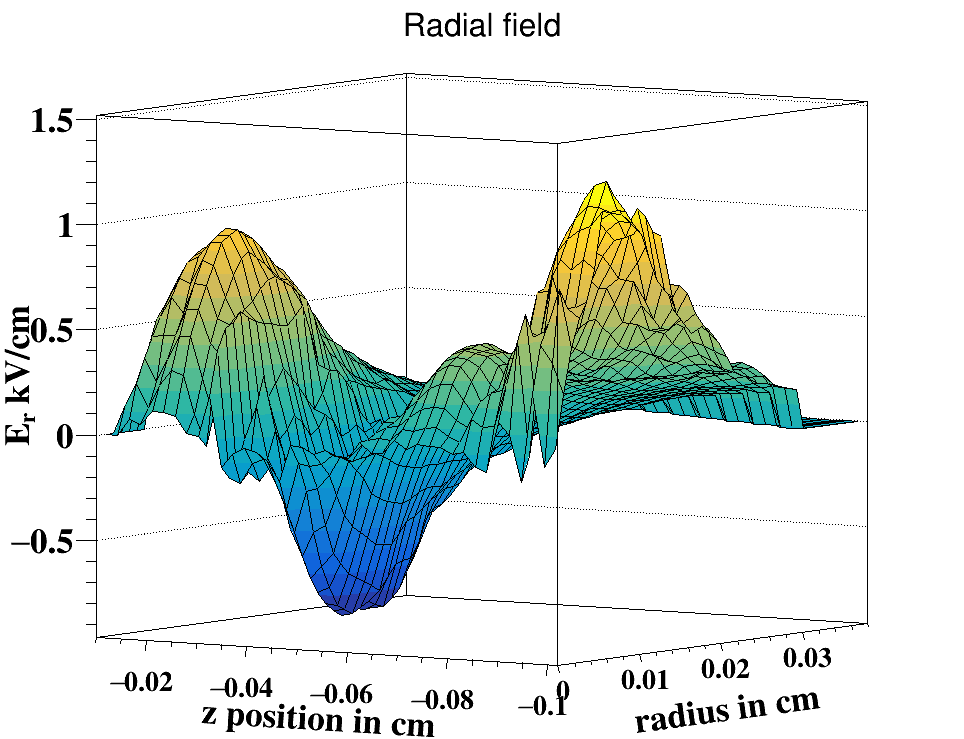}}
	
	\center\subfloat[\label{fig:10.46ns_phi}]{\includegraphics[scale=0.25]{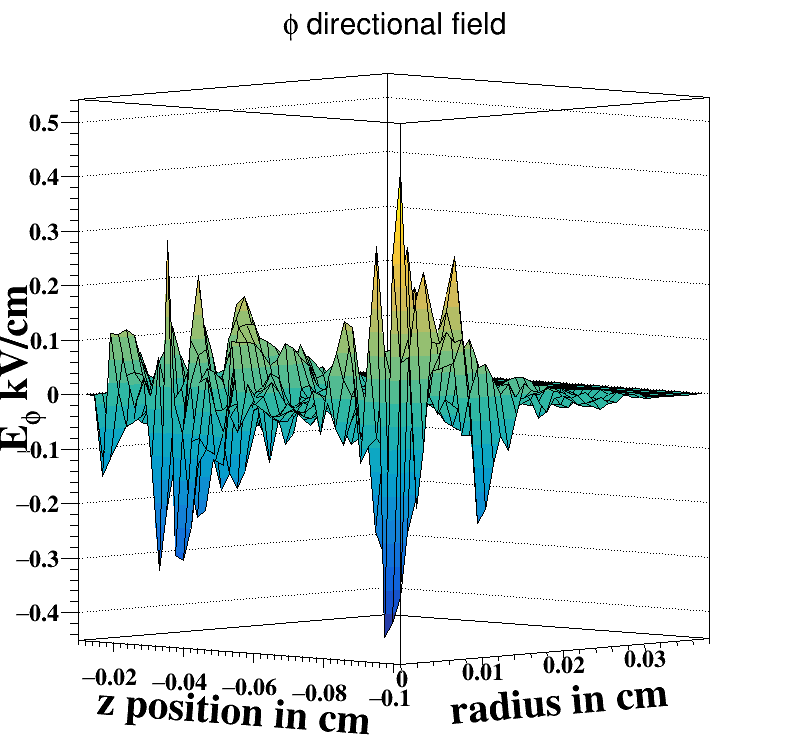}	
		
	}~\subfloat[\label{fig:10.46nsE}]{\includegraphics[scale=0.27]{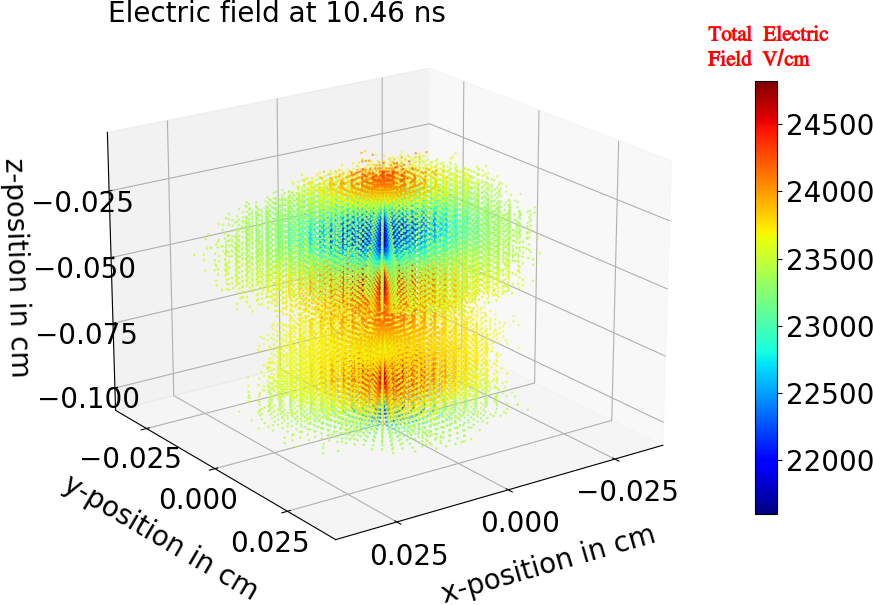}}
	
	\caption{(a) z- component of space charge field at 10.46 ns, (b) Radial- component of space charge field at 10.46 ns, (c) $\phi$- component of space charge field at 10.46 ns, (d) Shape of the electron cloud and electric field magnitude (tE) at different grid locations at time 10.46}
\end{figure}
\subsection{From 10 ns to 15.46 ns, till peak region }

\begin{enumerate}[i.]
	\item At 10.46 ns, the blue and pink curves of Figure \ref{fig:gain} start to separate, which signifies the relative importance of the space charge effect. There are 5897046 electrons and 7768542 ions (see Figure \ref{fig:10.46}). At this stage, the last two clusters also merge and form a single big cluster. A very close look at Figure \ref{fig:10.46} shows that the charge center of the last cluster from the anode has not remained at r=0; rather, it shifted to the right. This is because z-directional space charge field (sE$_z$) at z=-0.04 and r=0 is more negative (opposite to the applied field) in comparison to other regions (see Figure \ref{fig:10.46ns_z}). Therefore, the total field (tE) and the ionization probability at the center are also less compared to the other regions. The radial component of the space charge field (sE$_r$) plays an important role in spreading the avalanche along the transverse direction. The radial field at 10.46 ns at different z and radius r has been shown in Figure \ref{fig:10.46ns_rad}. The radial field sE$_r$ of ions is positive, and of electrons is negative. A positive sign indicates an outward radial field from the center, and a negative sign signifies an inward radial field to the center. It is noted that near the region z=-0.04 and r=0.005
	(see Figure \ref{fig:10.46ns_rad})the radial field is negative, meaning the number of electrons dominates the number of ions at those places. The phi (sE$_\phi$) component of the field measures the axial symmetry of the avalanche charge distribution. The charge distribution will be axially symmetric when the value of sE$_\phi$ is close to zero. A nonzero value of sE$_{\phi}$ signifies axial asymmetry. In Figure \ref{fig:10.46ns_phi} the variation of sE$_\phi$ with z position and radius r has been shown. At 10.46 ns,  sE$_\phi$ is not as strong as like sE$_r$ and sE$_z$ (maximum sE$_\phi$ is only 26\% of sE$_z$). But, still, sE$_\phi$ can act as a perturbation to the applied field.
	
	The shape of the electron cloud corresponding to time 10.46 ns is shown in Figure \ref{fig:10.46nsE}, and the magnitude of the total field (tE) in V/cm at different grid locations are shown in the color bar of the same figure. The maximum increment and decrement of the magnitude of the total field from the initially applied field are 5.62\% and 8.0\%, respectively.

	\begin{figure}
		\center\subfloat[\label{fig:12.46}]{\includegraphics[scale=0.17]{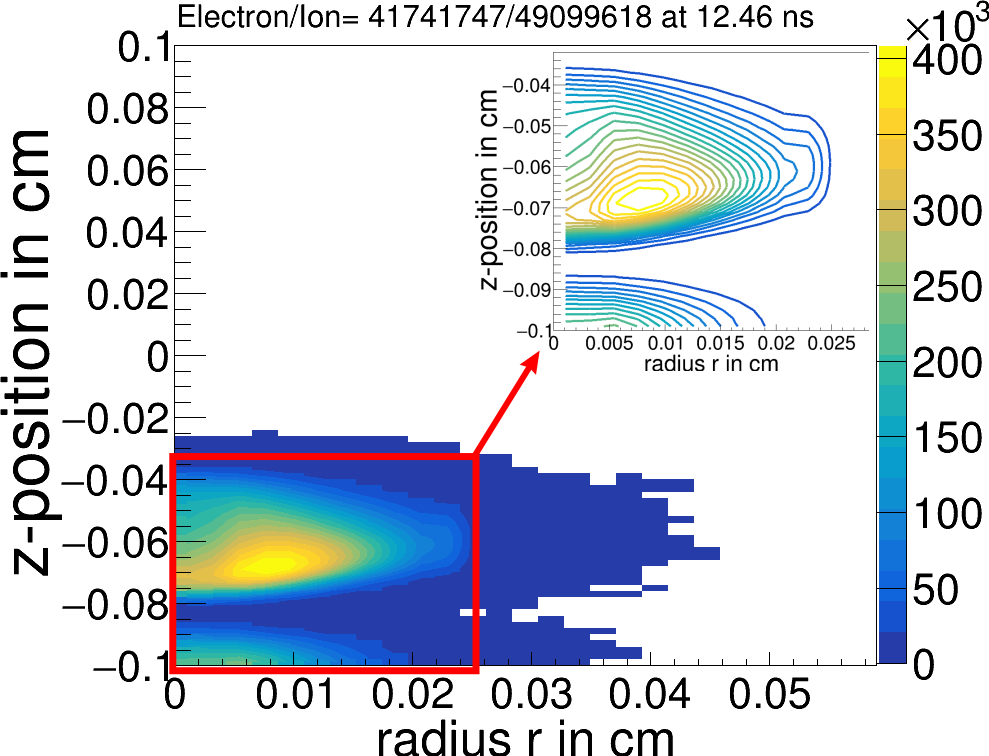}
			
		}~~\subfloat[\label{fig:12.96}]{\includegraphics[scale=0.17]{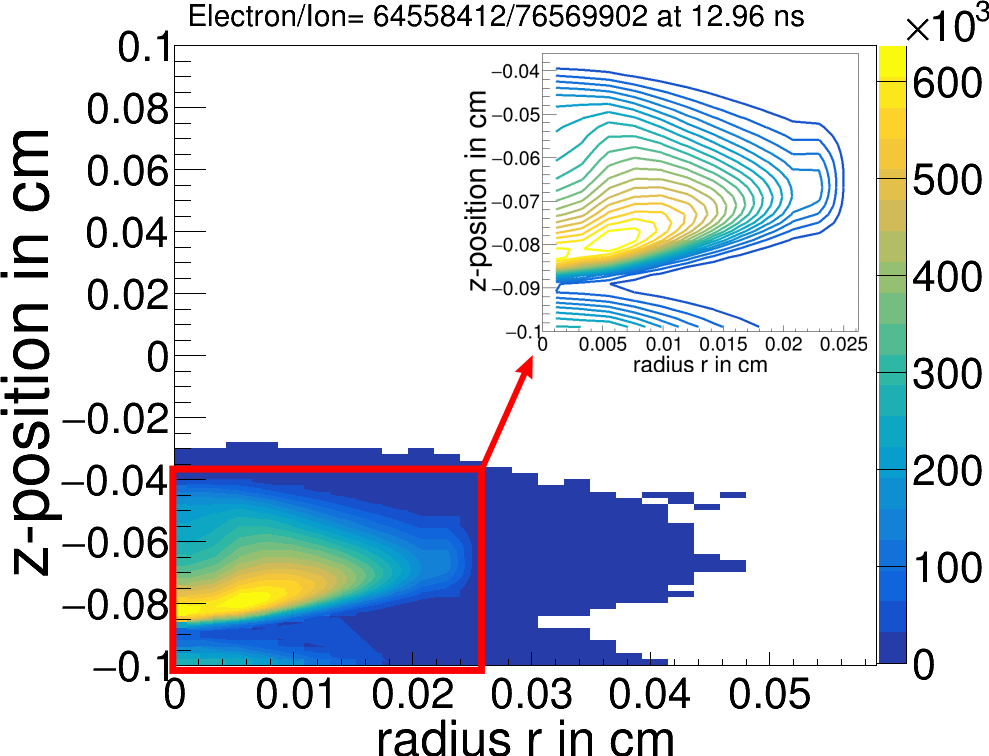}}
		
		\center\subfloat[\label{fig:13.46}]{\includegraphics[scale=0.17]{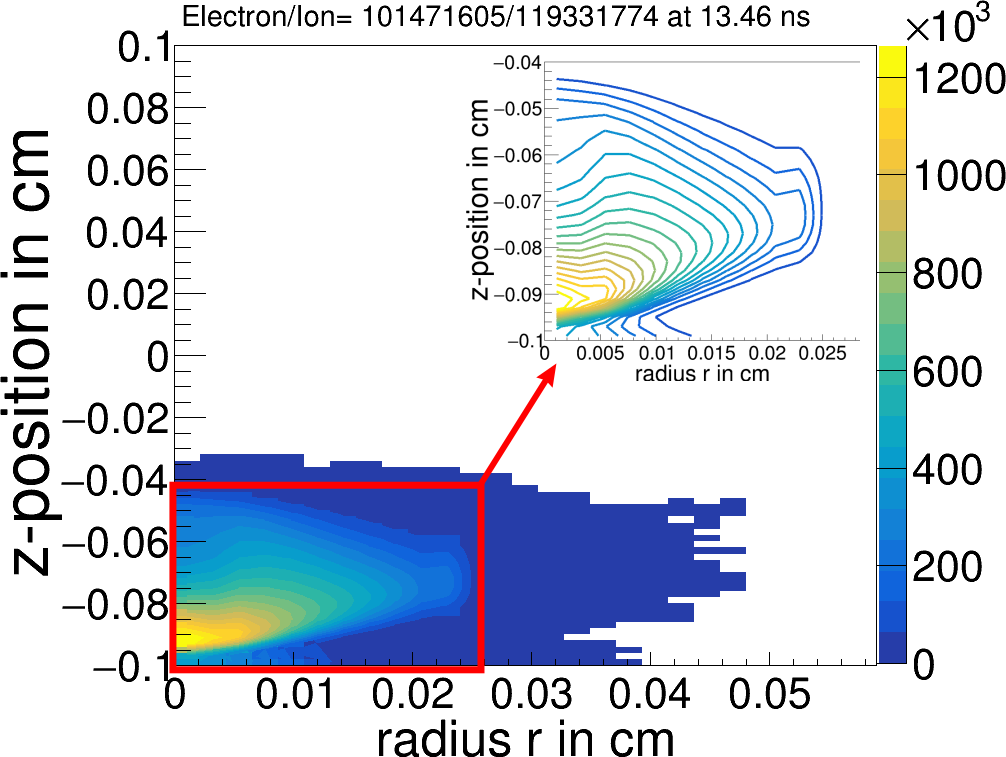}
		}
		\caption{(a) Location of electrons in z-r plane at time 12.46 ns, (b) Location of electrons in z-r plane at time 12.96 ns, (c) Location of electrons in z-r plane at time 13.46 ns.}
	\end{figure}
	\begin{figure}
		\center\subfloat[\label{fig:12.46ns_z}]{\includegraphics[scale=0.19]{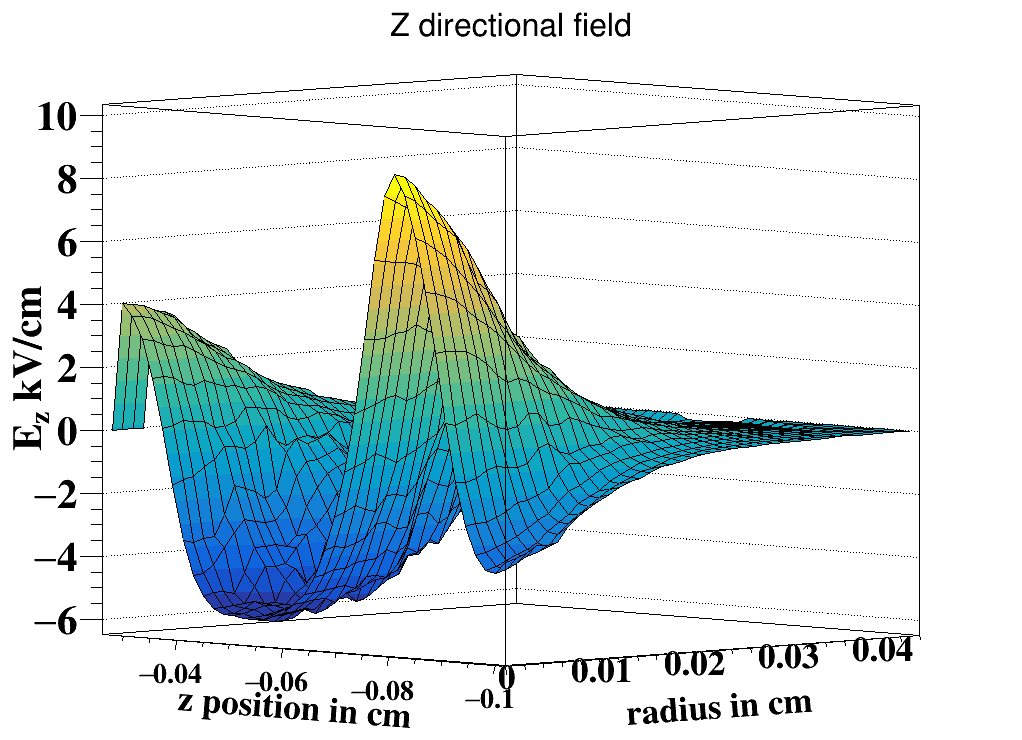}
		}\subfloat[\label{fig:12.46ns_rad}]{\includegraphics[scale=0.19]{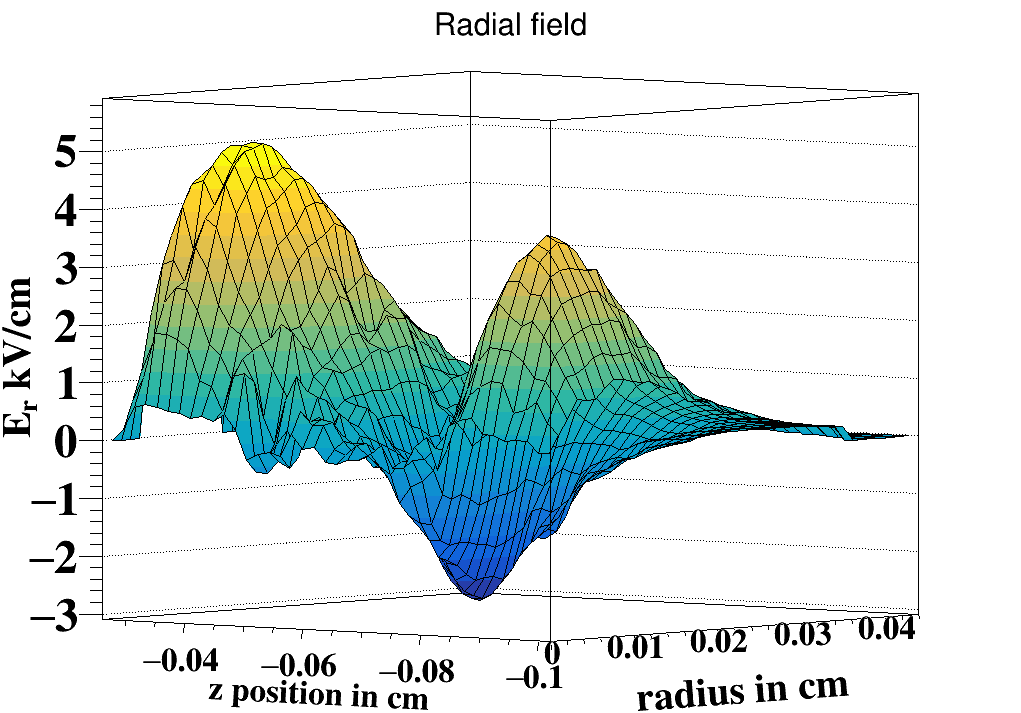}}
		
		\center\subfloat[\label{fig:12.46ns_phi}]{\includegraphics[scale=0.22]{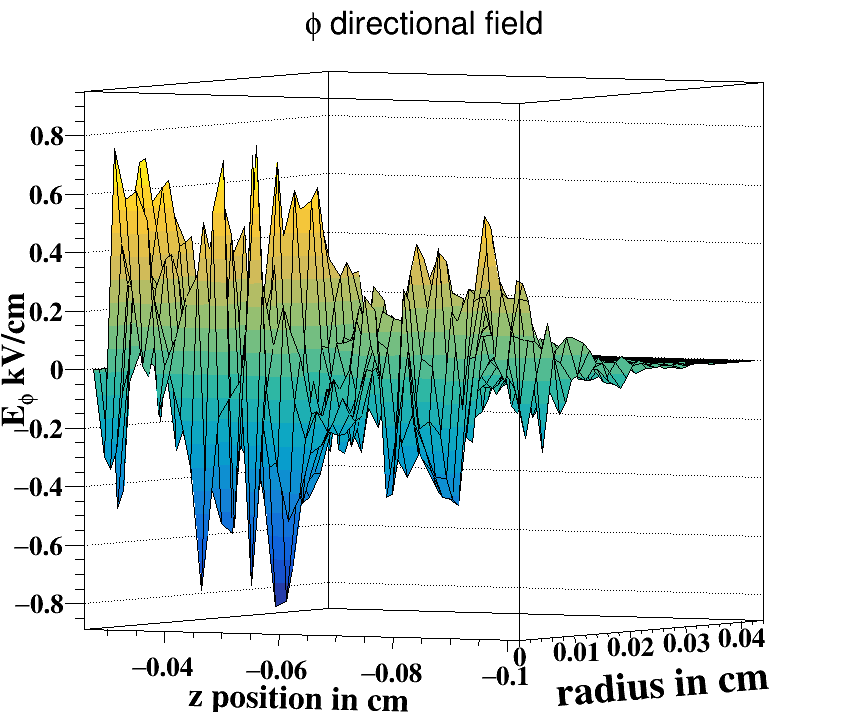}	
			
		}
		
		\caption{(a) z- component of space charge field at 12.46 ns, (b) radial- component of space charge field at 12.46 ns, (c) $\phi$- component of space charge field at 12.46 ns}
	\end{figure}

	\item From the collective view of Figures \ref{fig:12.46},\ref{fig:12.96} and \ref{fig:13.46} it can be said that from 12.46 ns to 13.46 ns the  charge center again gradually  move back to the center r=0. The z directional space charge field (sE$_{z}$) at 12.46 ns near z=-0.08 and r=0 becomes positive and higher than other regions (see Figure \ref{fig:12.46ns_z}). Therefore the ionization rate becomes high at that location. This is the reason the charge center again moves towards r=0. In Figure \ref{fig:12.46ns_rad}, the radial field at a different location at time 12.46 ns has been shown. At time 12.46 ns, the radial field increases significantly (absolute maximum along positive and negative directions around 5 kV/cm and 3 kV/cm respectively) at some places, which are reflected in the transverse or radial spread ($r_{max}=0.044$ cm)  of the avalanche (see Figure \ref{fig:12.46}). The variation of sE$_\phi$ at different z location and radius has been shown in Figure \ref{fig:12.46ns_phi}. The maximum magnitude of sE$_\phi$ at time 12.46 ns is 0.8 kV/cm, which is double the value of sE$_\phi$ at 10.46 ns. 
	\begin{figure}
		\center\subfloat[\label{fig:12.96elcden}]{\includegraphics[scale=0.23]{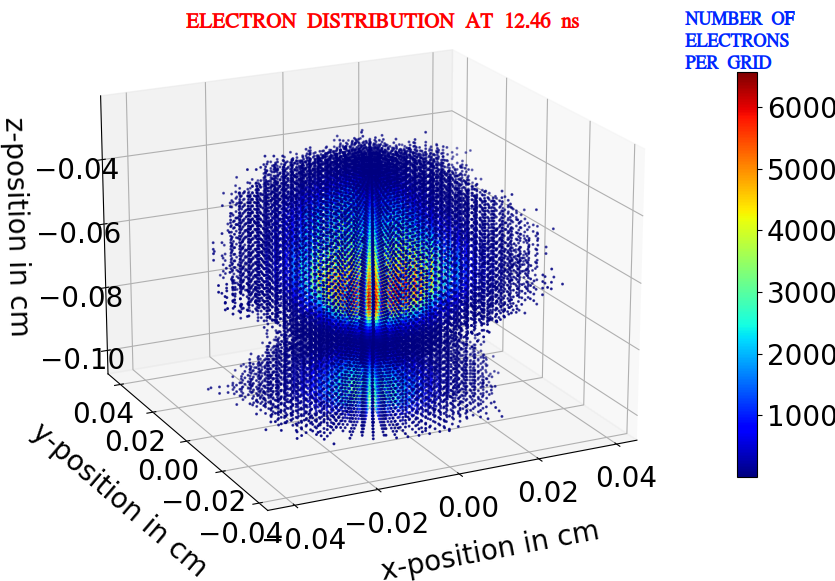}
			
		}\subfloat[\label{fig:12.96ionden}]{\includegraphics[scale=0.24]{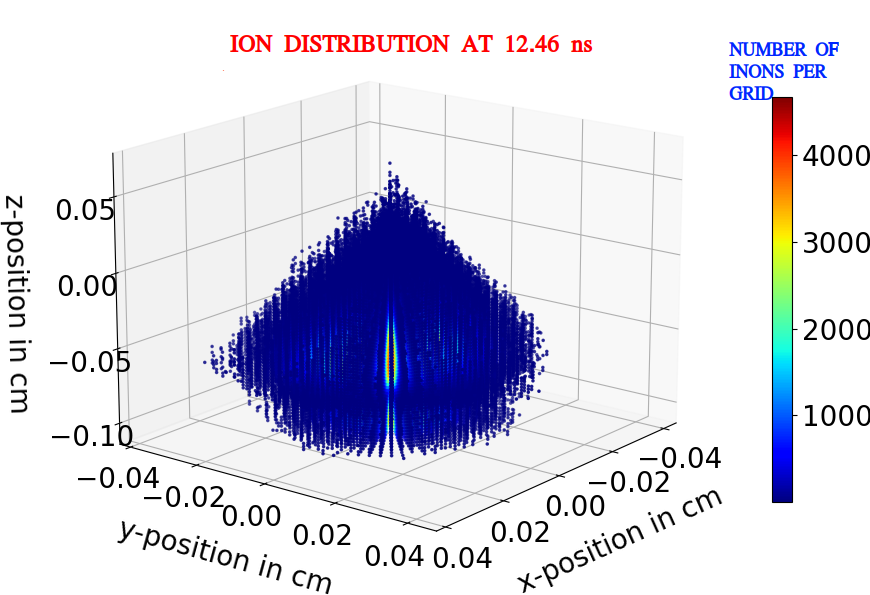}
			
		}
		\caption{(a) Shape of the electron cloud and number of electrons at different grid elements
			(color bar) at time 12.46 ns, (b) Shape of the ion cloud and number of ions at different
			grid elements (color bar) at time 12.46 ns.}
		
	\end{figure}	
	
	\par It is clear from Figure \ref{fig:12.96elcden} (color bar represents the number of electrons per grid) that the range of electron cloud along the z direction is -0.1 cm to -0.04 cm.
	As shown in Figure \ref{fig:12.96ionden} (color bar represents the number of ions per grid), the range of ion cloud is -0.1 cm to 0.05 cm. Therefore, along the z-axis, ions are on both sides of the electron cloud. Hence, at the tip and tail of the electron cloud, ions attract the nearest electrons toward them. As a result, the shape of the density contours of the charge distribution becomes asymmetric along the z-axis.
	\begin{figure}
		\center\subfloat[\label{fig:12.46nsE}]{\includegraphics[scale=0.23]{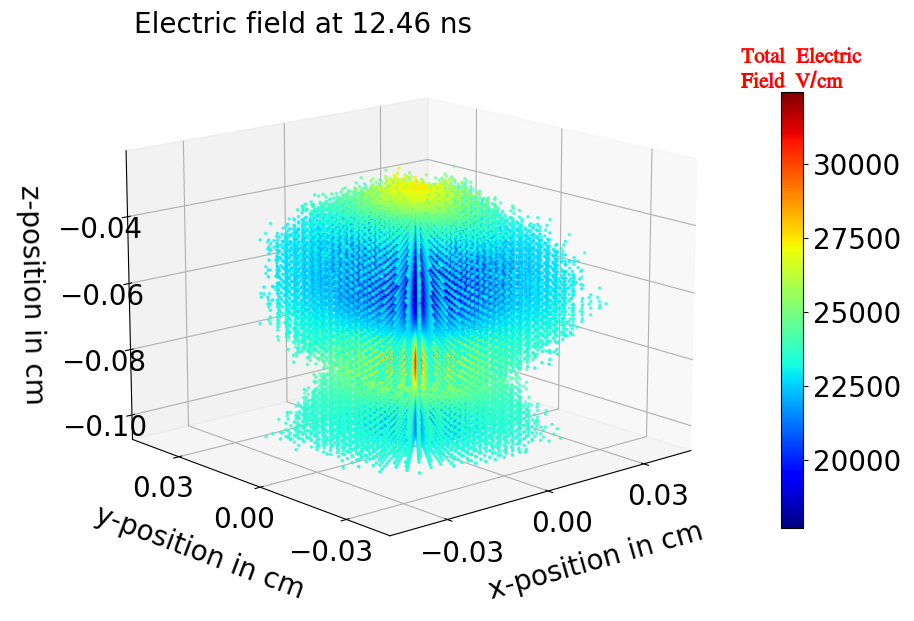}
			
		}\subfloat[\label{fig:12.96nsE}]{\includegraphics[scale=0.23]{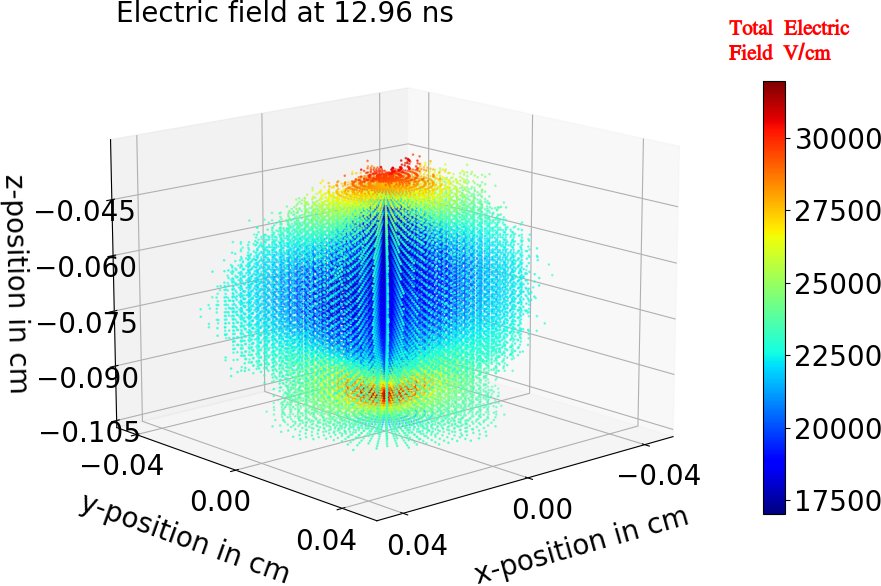}
			
		}
		
		\center\subfloat[\label{fig:13.46nsE}]{\includegraphics[scale=0.23]{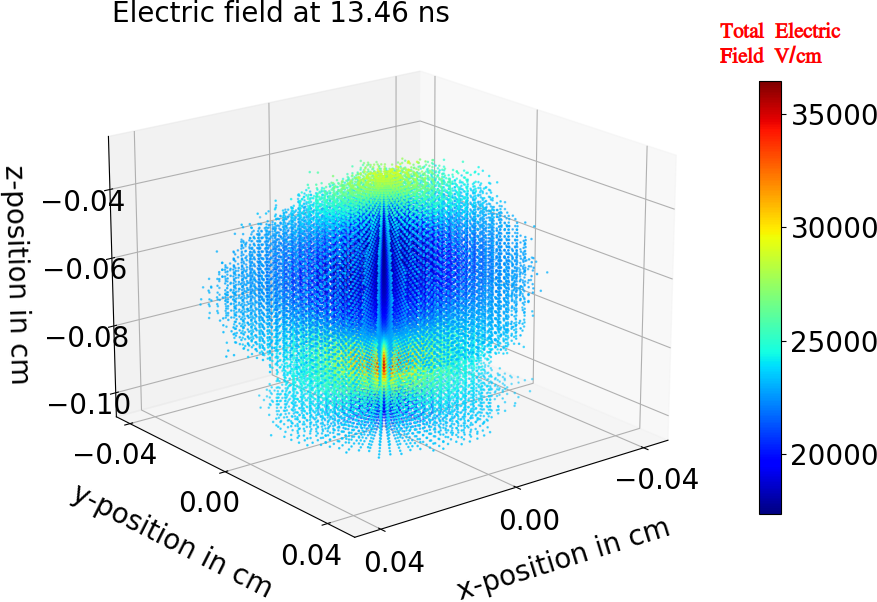}
		}
		\caption{(a) Shape of the electron cloud and electric field magnitude (tE) at different grid elements (color bar) at time 12.46 ns, (b) Shape of the electron cloud and electric field magnitude (tE) at different grid elements (color bar) at time 12.96 ns, (d) Shape of the electron cloud and electric field magnitude (tE) at different grid elements (color bar) at time 13.46 ns.}\label{fig:trk_aftersp1}
		
	\end{figure}
	
	\par In Figures \ref{fig:12.46nsE}, \ref{fig:12.96nsE} and \ref{fig:13.46nsE} the shape of the electron cloud and magnitude of total field at different places of the cloud and at different times 12.46 ns,12.96 ns and 13.46 ns have been shown. It is found from the former figures that though the electron density at different places inside the cloud is changing with time but the overall shape of the distribution is not changed much. In this period of transition, the maximum percentage of increment and decrement of the total electric field from the applied field at 13.46 ns is 36.02\% and 27.5\% respectively. 
\end{enumerate}

\subsection{ After 15.46 ns, peak to saturated region}

\begin{figure}
	\center\subfloat[\label{fig:15.46}]{\includegraphics[scale=0.17]{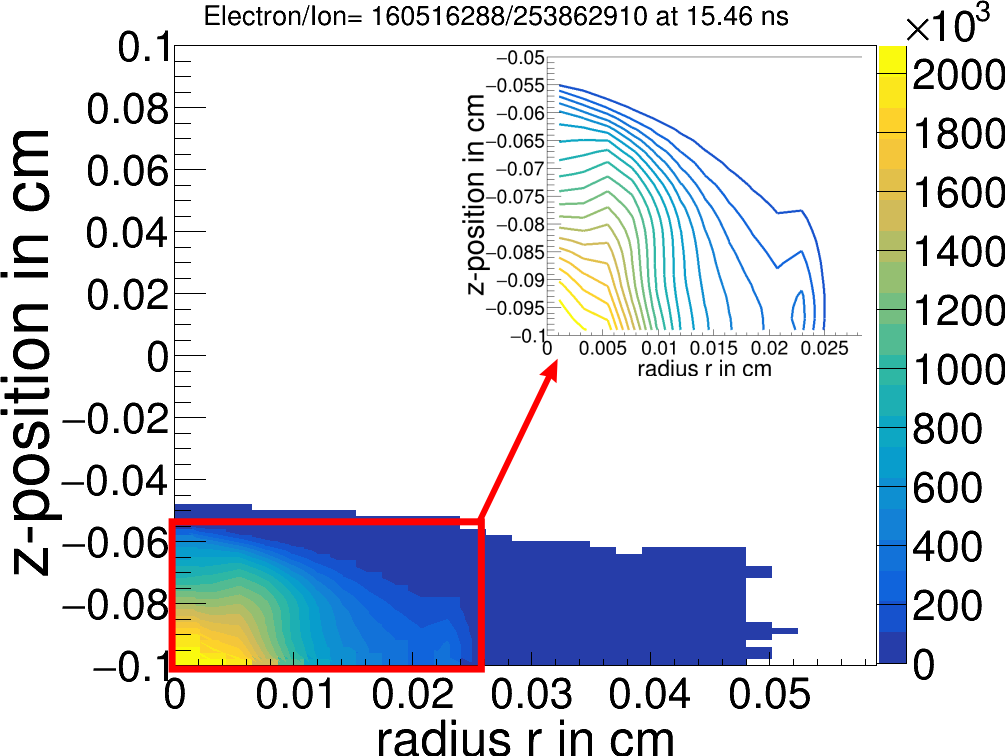}
		
	}~~\subfloat[\label{fig:15.46nsE}]{\includegraphics[scale=0.23]{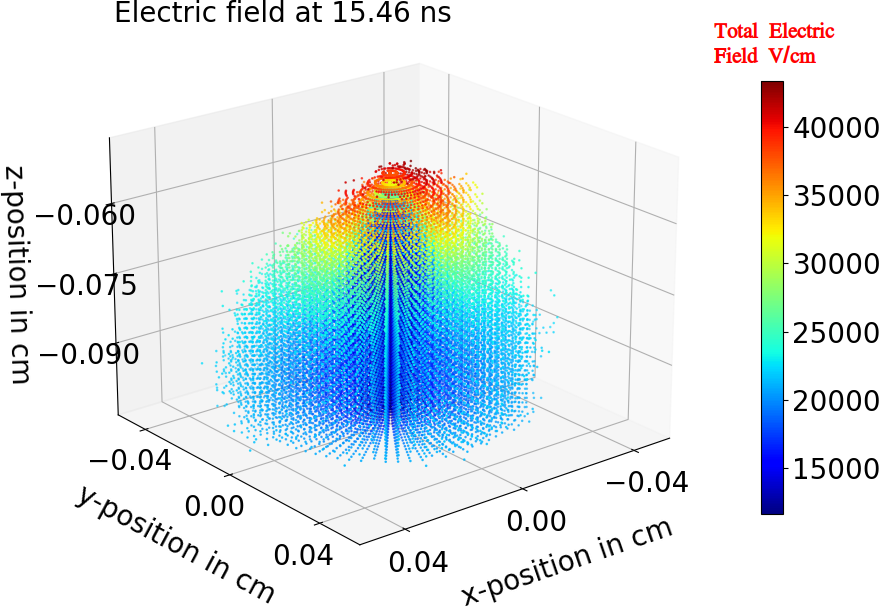}
	}

	\center\subfloat[\label{fig:16.96}]{\includegraphics[scale=0.17]{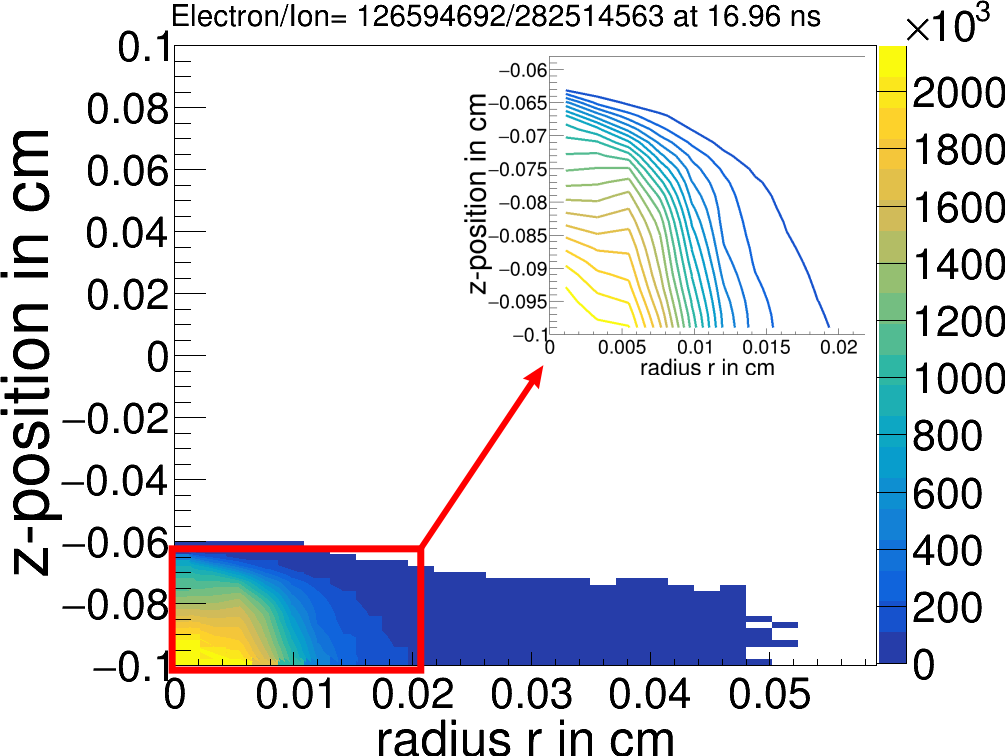}
		
	}~~\subfloat[\label{fig:16.96nsE}]{\includegraphics[scale=0.23]{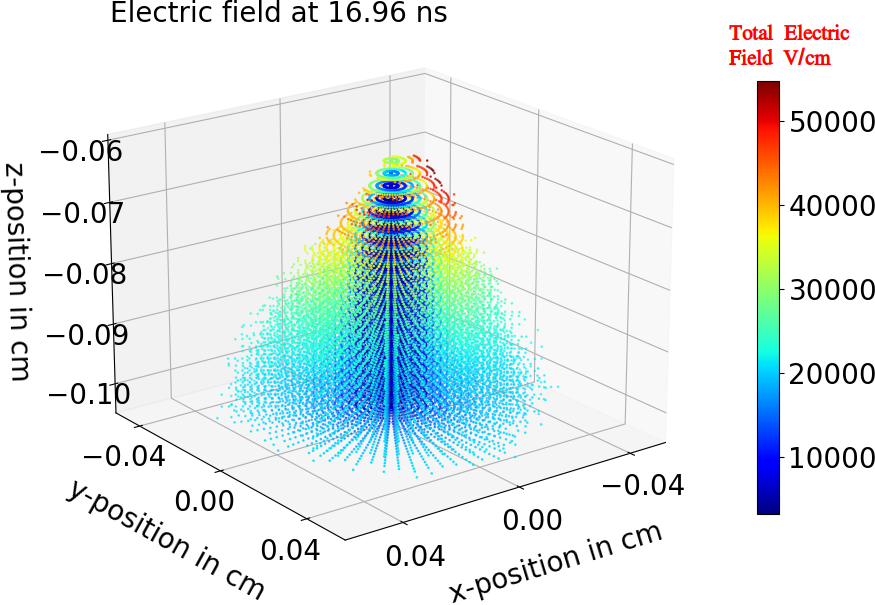}
		
	}
	\caption{(a) Location of electrons in z-r plane at time 15.46 ns, (b) Shape of the electron cloud and electric field magnitude (tE) at different grid elements (color bar) at time 15.46 ns, (c) Location of electrons in z-r plane at time 16.96 ns, (d) Shape of the electron cloud and electric field magnitude (tE) at different grid elements (color bar) at time 16.96 ns.}\label{fig:trk_aftersp2}
	
\end{figure}
\begin{figure}
	\center\subfloat[\label{fig:15.46ns_z}]{\includegraphics[scale=0.19]{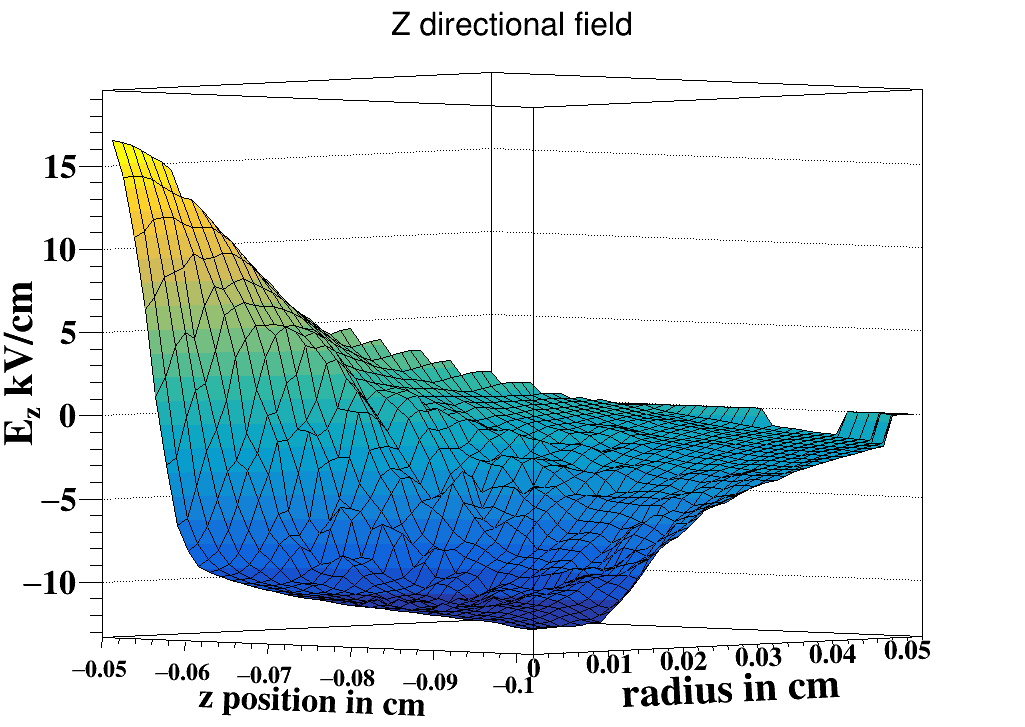}
	}\subfloat[\label{fig:15.46ns_rad}]{\includegraphics[scale=0.19]{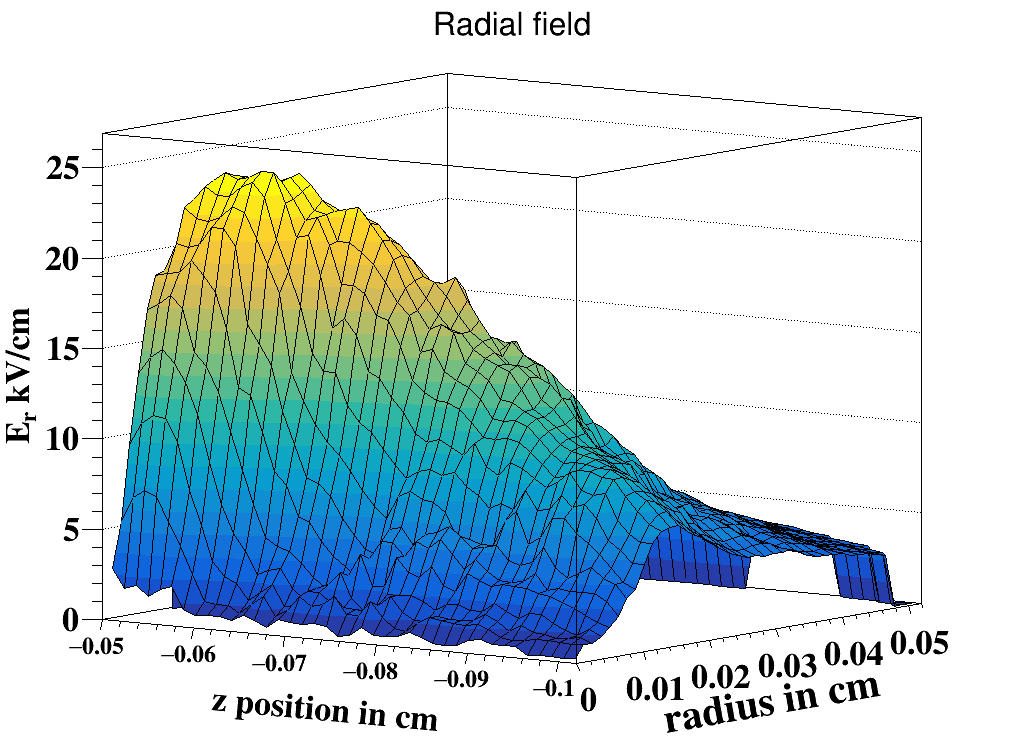}}
	
	\center\subfloat[\label{fig:15.46ns_phi}]{\includegraphics[scale=0.22]{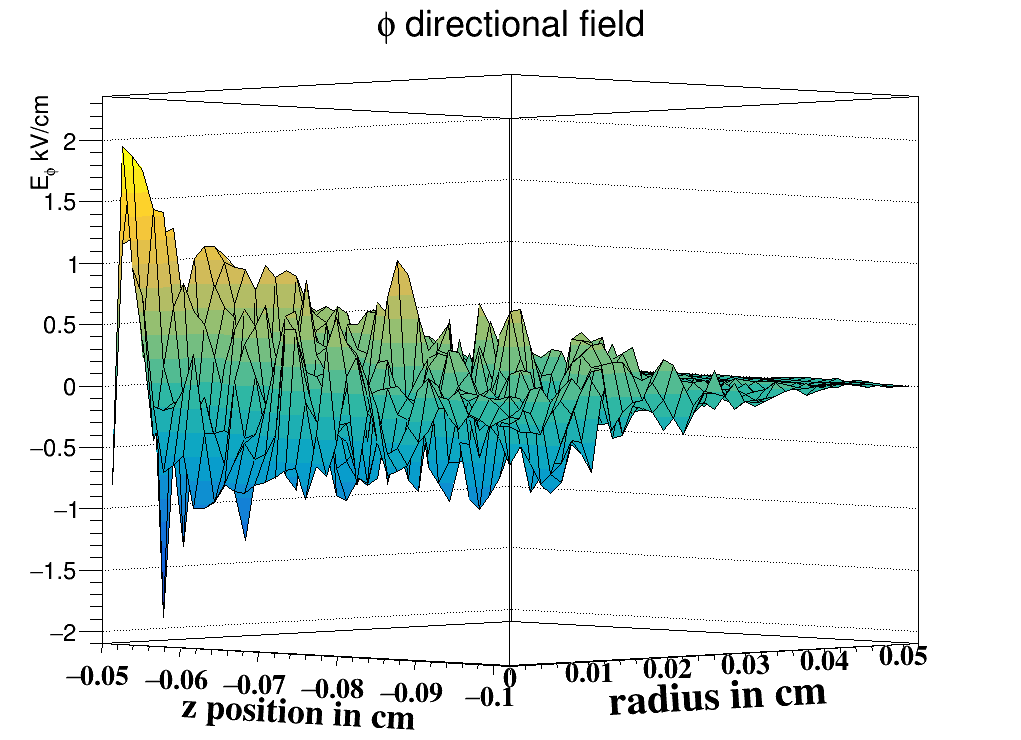}	
		
	}
	
	\caption{(a) z- component of space charge field at 15.46 ns, (b) radial- component of space charge field at 15.46 ns, (c) $\phi$- component of space charge field at 15.46 ns}
\end{figure}

\begin{figure}
	\center\subfloat[\label{fig:18.46}]{\includegraphics[scale=0.17]{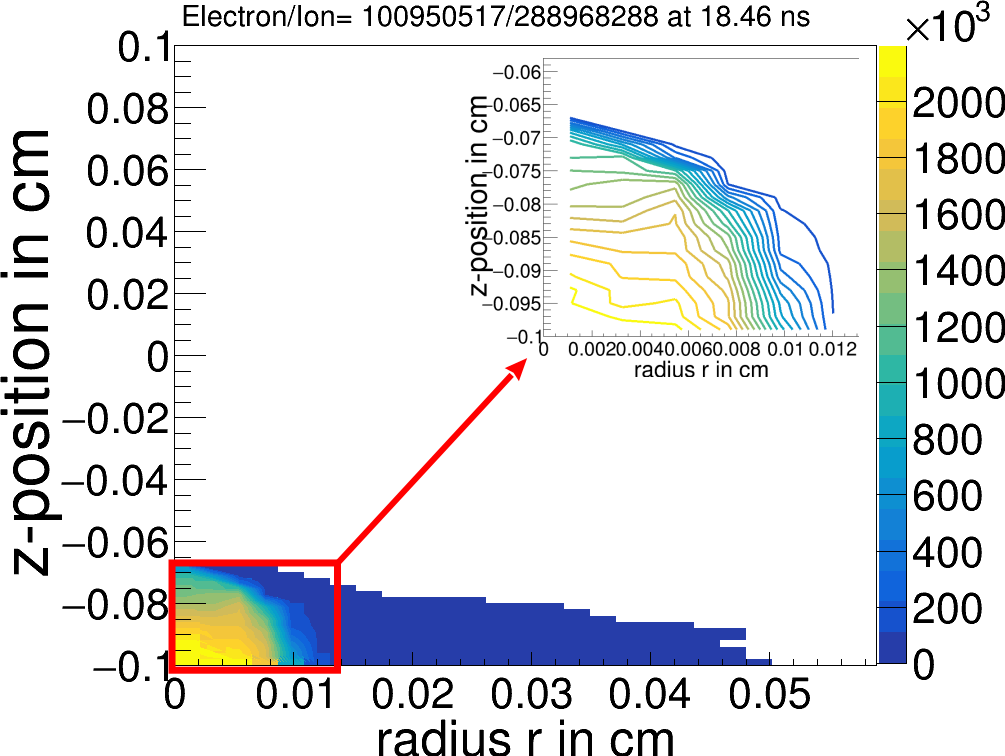}
		
	}~~\subfloat[\label{fig:18.96nsE}]{\includegraphics[scale=0.23]{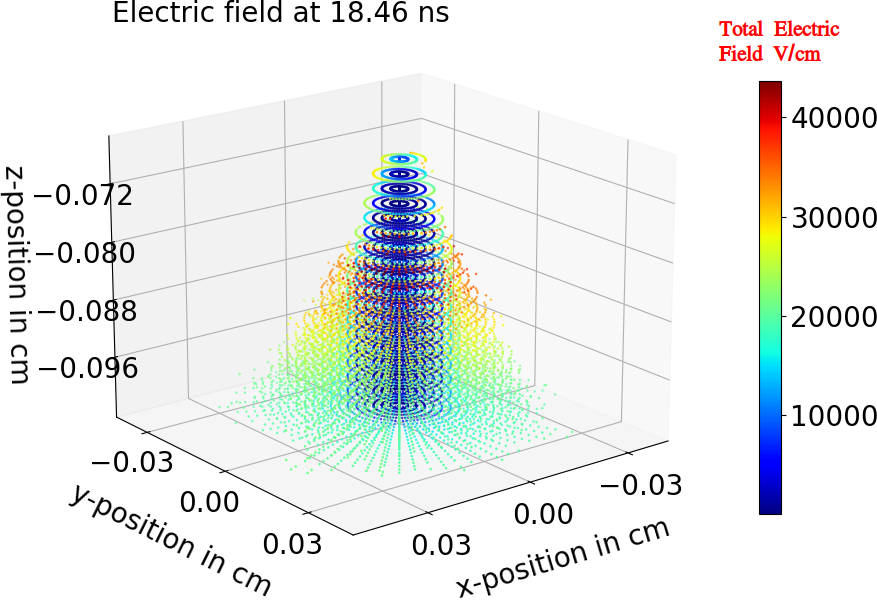}
		
	}
	
	\center\subfloat[\label{fig:23.46}]{\includegraphics[scale=0.17]{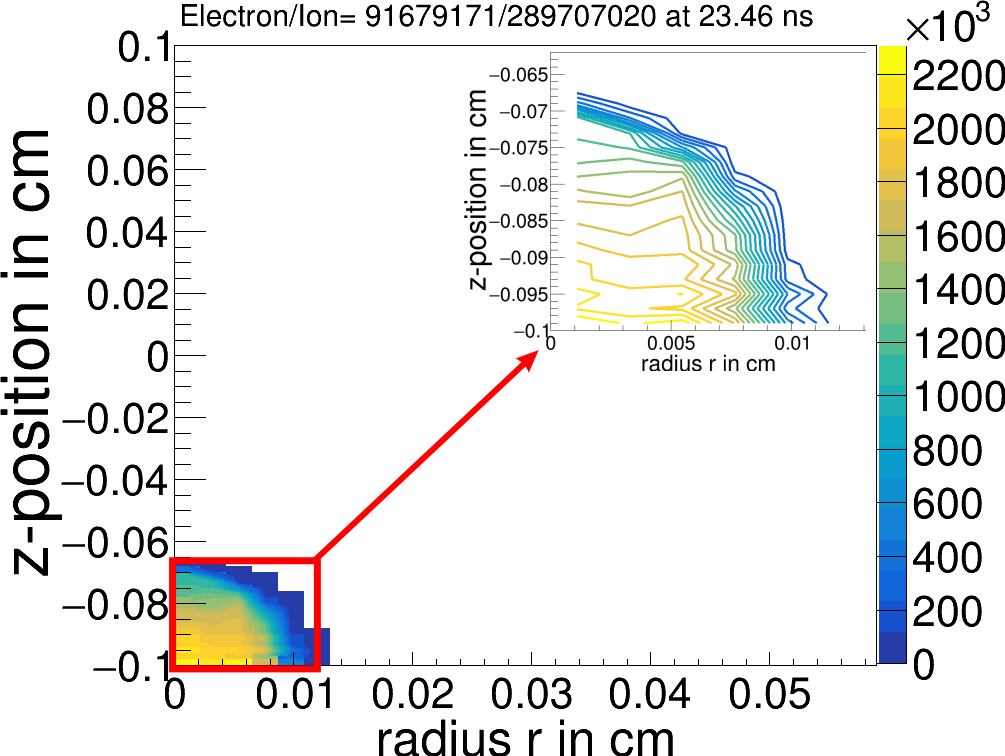}
	}~~\subfloat[\label{fig:23.46nsE}]{\includegraphics[scale=0.23]{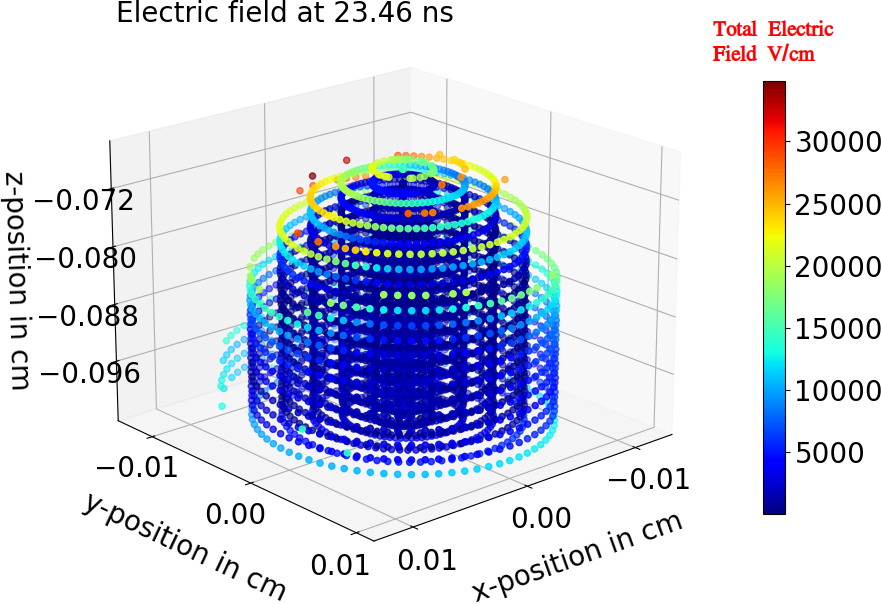}
	}
	
	\caption{(a) Location of electrons in z-r plane at time 18.46 ns, (b) Shape of the electron cloud and electric field magnitude (tE) (color bar) at different grid elements at time 18.46 ns, (c) Location of electrons in z-r plane at time 23.46 ns, (d) Shape of the electron cloud and electric field magnitude (tE) (color bar) at different grid elements  at time 23.46 ns.}\label{fig:trk_aftersp2}
	
\end{figure}

\begin{figure}
	\center\subfloat[\label{fig:23.46nsdrft}]{\includegraphics[scale=0.23]{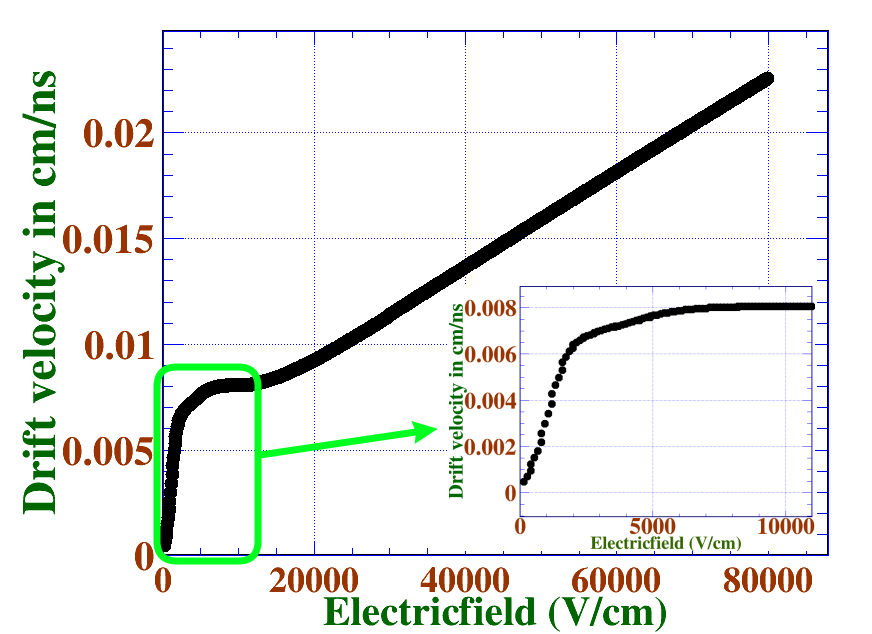}
		
	}\subfloat[\label{fig:23.46nsalpEta}]{\includegraphics[scale=0.23]{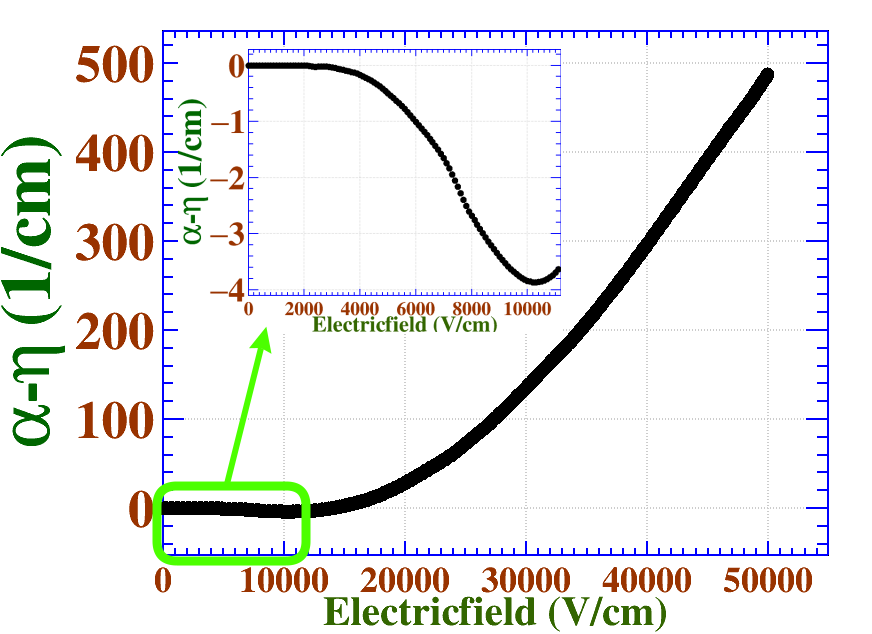}
		
	}
	
	\caption{(a) Drift velocity as a function of electric field, (b) Effective Townsend coefficient as a function of electric field. \label{fig:driftvel}}
	
\end{figure}

\begin{enumerate}[i.]
	\item At 15.46 ns electron gain reaches its maximum value (see Figure  \ref{fig:gain}). All the electron clusters are merged together (see Figure \ref{fig:15.46}). Since already many electrons have left the gas gap, the total number of ions (253,862,910) is far greater than the total number of electrons 
	(160,516,288). From Figure \ref{fig:15.46ns_z}, it is confirmed that due to negative sE$_z$, the net z-directional field (tE$_z$) has reduced by a large amount at the tip and the center (r=0) of the avalanche. Also, due to the significant difference in the electron and ion numbers, the radial field of ions dominates; hence, only a positive radial field exists everywhere. As a result, the remaining electrons feel a strong radial force towards the center, as shown in Figure \ref{fig:15.46ns_rad}. Therefore, the transverse spread is stopped, and all the electrons tend to converge toward the center. The component sE$_\phi$ attains a maximum value of 2 kV/cm approximately (see Figure \ref{fig:15.46ns_phi}) which indicates significant axial asymmetry of the space charge distribution. 
	In Figure \ref{fig:15.46nsE}, the shape of the electron cloud and electric field at different electron grid points has been shown. 
	The maximum percentage of increment and decrement of a total electric field from the applied field at 15.46 ns is 84.4\% and 50.5\%, respectively. 
	
	\item At time 16.96 ns, there are 126,594,692 electrons and 282,514,583 ions left inside the gas gap (see Figure \ref{fig:16.96}). This is the time when the space charge field attains its maximum value. The maximum increment and decrement of the magnitude of the total field (space charge + applied field) from the initially applied field are 132.8\% and 86.2\% respectively. The shape of the electron cloud becomes conical (see Figure \ref{fig:16.96nsE}), where at the tip, the field is reduced drastically, and at the tail, it becomes more than double the initial value of the field. 
	
	\item  At 18.46 ns, the gain curve with space charge effect of Figure \ref{fig:gain} contains a second knee point after which the saturation region starts. At this stage, there are 100,950,517 electrons and 288,968,288 ions. The density contours are not smooth near the center (see Figure \ref{fig:18.46}). There are three possibilities for the reduction of the number of outermost electrons  (i) attachment, (ii) leaving the gas gap, and (iii) at some places, the inward radial field is much strong so that it attracts the electrons towards the center. Therefore, the conical shape gradually modifies into cylindrical (see Figure \ref{fig:18.96nsE}). The maximum increment and decrement of the magnitude of the total field (tE) from the initially applied field are 85.8\% and 99.9\%.
	
	\item At 23.46 ns, the gain of the avalanche is saturated. In most of the places of the electron grid, the total field is less than 1 kV/cm (see Figure \ref{fig:23.46nsE}). Therefore at those places drift velocity of the electron is less than 0.008cm/ns (see Figure \ref{fig:23.46nsdrft}). Also, in that low-field regions, the attachment coefficient ($\eta$) dominates over the first Townsend coefficient (alpha) ($\alpha$). Hence the value of $\alpha - \eta$ becomes negative (see Figure \ref{fig:23.46nsalpEta}). At higher field regions, there are some electrons capable of producing ionisations. However, the number of electrons created in ionisations is compensated by the number of electrons attached; that's why the electron gain saturates, as shown in Figure \ref{fig:gain}. The maximum increment and decrement of the magnitude of the total field (tE) from the initially applied field are 41.5\% and 99.8\%.
\end{enumerate}	

\section{Timing performance of pAvalancheMC using OpenMp}
\label{sec:6_speedUp}
We have used the OpenMP multithreading technique to parallelize the class pAvalancheMC of Garfield++. In this study, a timing RPC of area 30 cm $\times$ 30 cm, 0.3 mm gas gap, and 2 mm thick bakelite electrode has been studied in order to assess the timing performance of the parallelized code. A voltage of $\pm$1720 V  has been applied on the graphite surface. As a result, the average electric field inside the gas gap is 43 kV/cm. A gas mixture of $\ce{C_2H_2F_4}$ (85\%), $\ce{i-C_4H_10}$ (5\%), $\ce{SF_6}$ (10\%) has been used. 
\par
\begin{figure}
	\center\subfloat[\label{fig:timePer_nosp}]{\includegraphics[width=.5\linewidth]{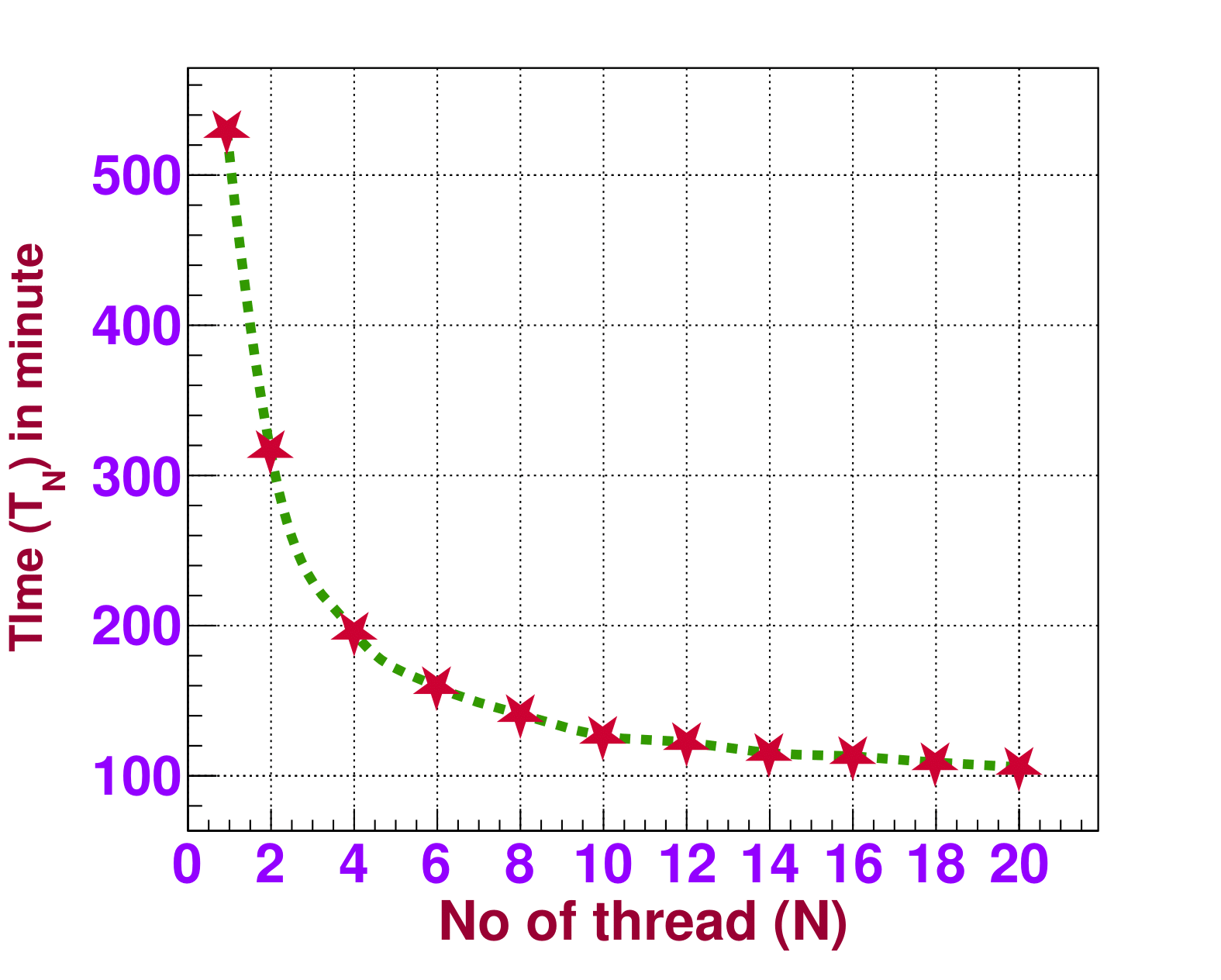}}\subfloat[\label{fig:speedUp_nosp}]{\includegraphics[width=.5\linewidth]{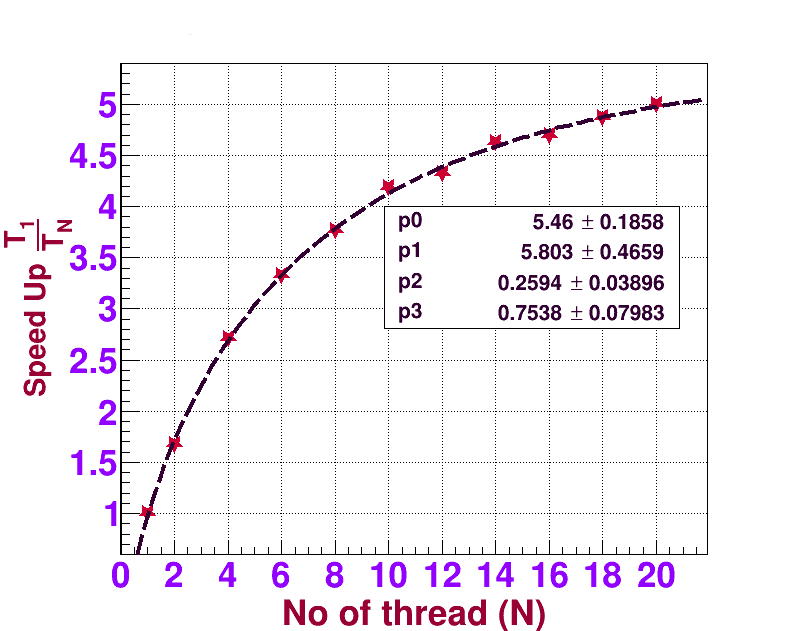}}
	
	\center\subfloat[\label{fig:timePer_wsp}]{\includegraphics[width=.5\linewidth]{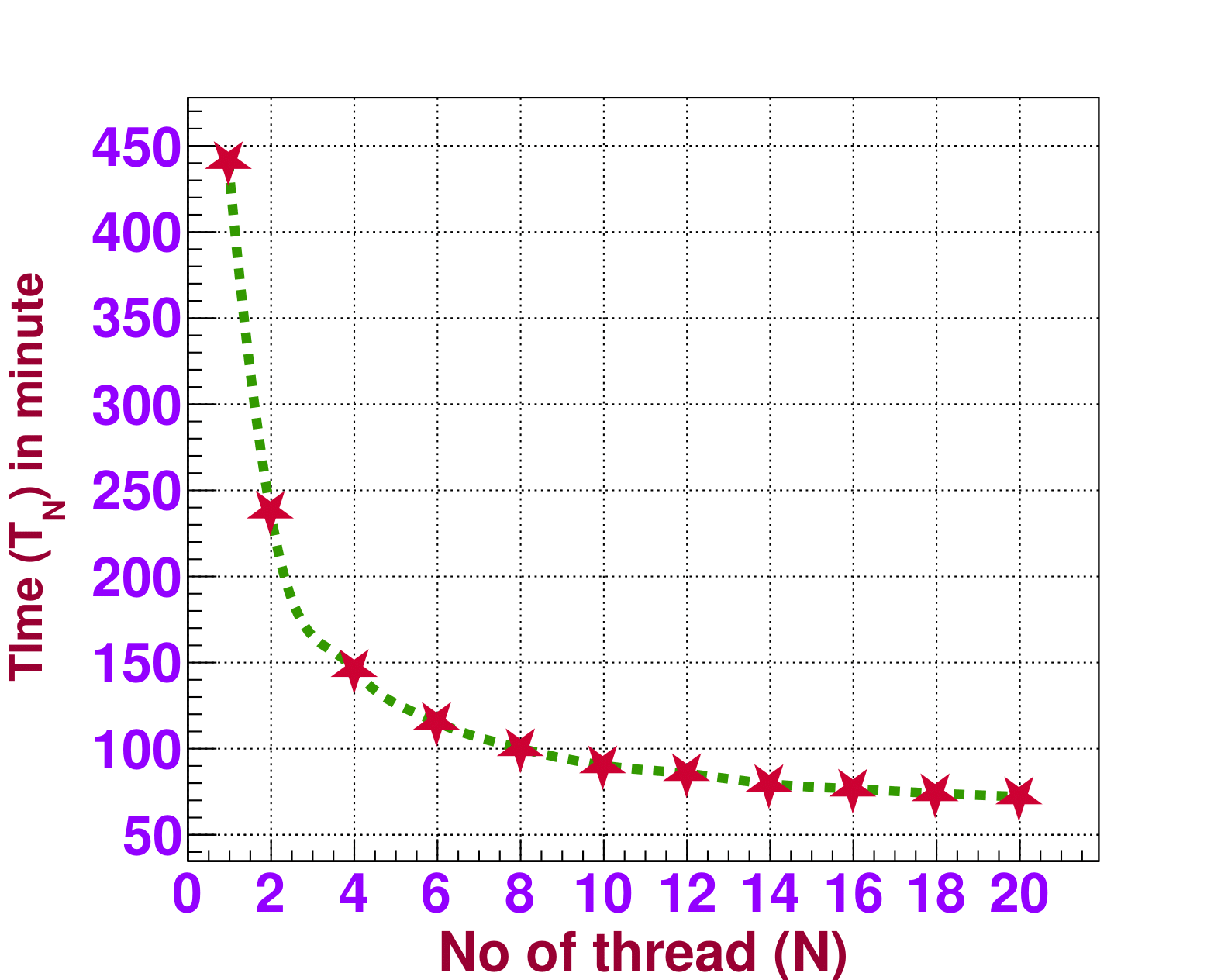}}\subfloat[\label{fig:speedUp_wsp}]{\includegraphics[width=.5\linewidth]{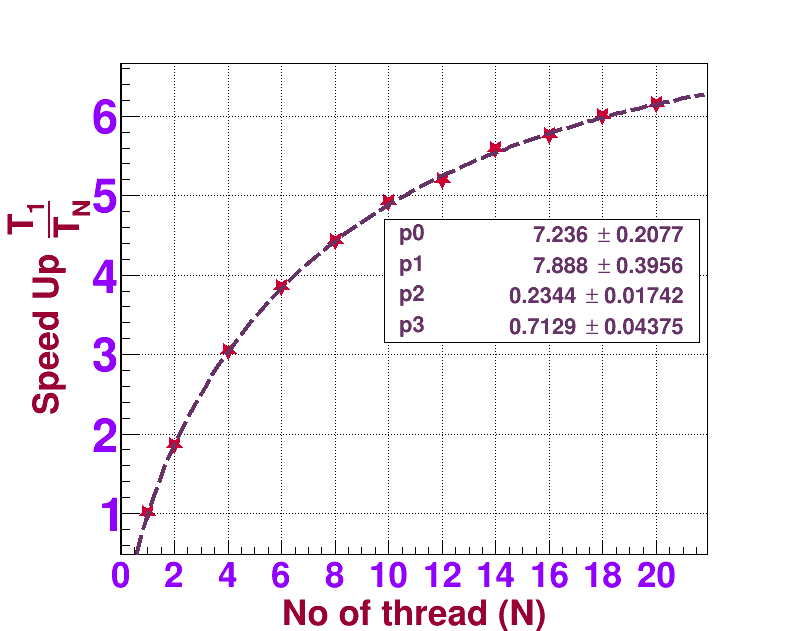}}
	
	\caption{ Variation of execution time to complete $10^4$ avalanches with number of threads (a) without space charge effect and (c) with space charge effect, Speed up performance with number of threads (b) without space charge effect, (d) with space charge effect.}
	
\end{figure}

 A single primary electron is placed near the negative electrode to generate avalanches. A set of $10^4$ avalanches has been generated repeatedly by varying the number of thread values (N) in OpenMP. Let $T_N$ represent the time taken to generate $10^4$ events with a total N number of threads. The variation in total time ($T_N$) with and without space charge effect as a function of N threads has been shown in Figures \ref{fig:timePer_nosp} and \ref{fig:timePer_wsp} respectively. It is clear from the preliminary figures that in both cases, with and without the space charge effect, the $T_N$ decreases with the increase in the total number of N threads. The speed-up performance of the code can be calculated by taking the ratios between sequential time ($T_1$) with $T_N$. Therefore, speed up factor ($v_p$) can be written as:
\begin{equation}
v_p=\frac{T_1}{T_N}.
\end{equation}
The speed up performance $v_p$ as a function of number of threads (N) has been shown in Figures \ref{fig:speedUp_nosp} and \ref{fig:speedUp_wsp} with and without space charge effect respectively. 
The $v_p$ increases with the number of threads (N). The data points of both Figures \ref{fig:speedUp_nosp} and \ref{fig:speedUp_wsp} are fitted with a non linear function: 
\begin{equation}\label{eqn:speed_up}
f(N)=p_0-p_1\,e^{-(p_2\,N^{p_3})},\,(N\ge 1)
\end{equation}
where the fit parameters are shown in the same figures. From equation \ref{eqn:speed_up}, it can be said that for large N, the contribution of the second term is significantly less; hence the parameter $p_0$ signifies the maximum speed-up which can be achieved.
Therefore, based on parameter $p_0$, the maximum speed up of 5.46 and 7.2 have been observed without and with space charge effect, respectively.
\par The saturation in speed-up came from the overhead issues. 
The overhead arises during the communication and synchronization between threads. More details about the overhead in OpenMP can be found in~\cite{openmp_overhead1,openmp_overhead2}. It is also noted that the time plotted in Figures \ref{fig:timePer_nosp} and \ref{fig:timePer_wsp} is the total time to complete all avalanches, which is the combination of time of parallel and non-parallel computations. Therefore, $T_N$ is not the time purely taken by N threads; instead, it is the sum of time by N threads for parallel computations and a single thread for non-parallel or sequential computations. 

\par
\begin{figure}
	\center\subfloat[\label{fig:Elc_gain_nosp}]{\includegraphics[width=.5\linewidth]{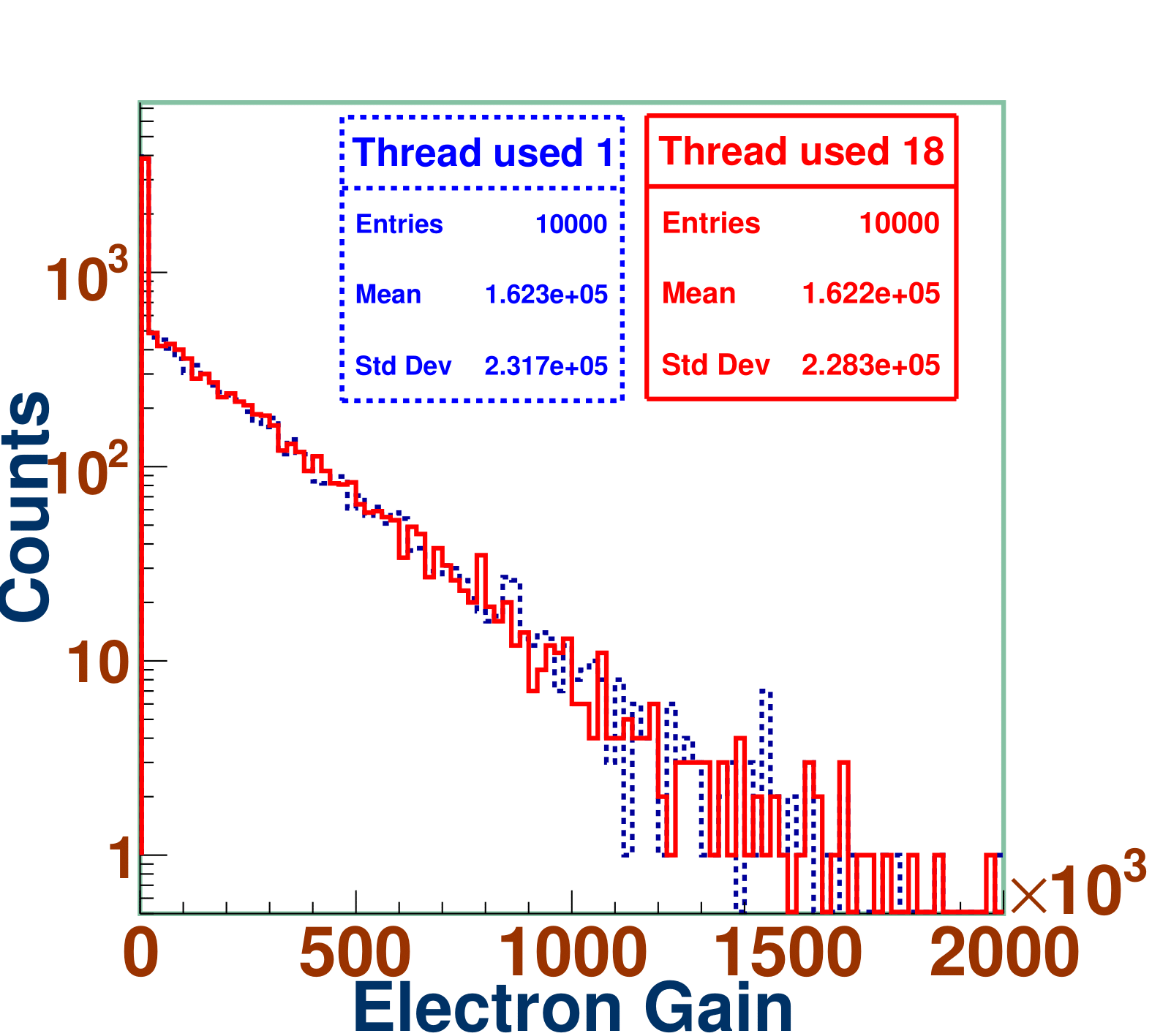}}\subfloat[\label{fig:Elc_gain_wsp}]{\includegraphics[width=.5\linewidth]{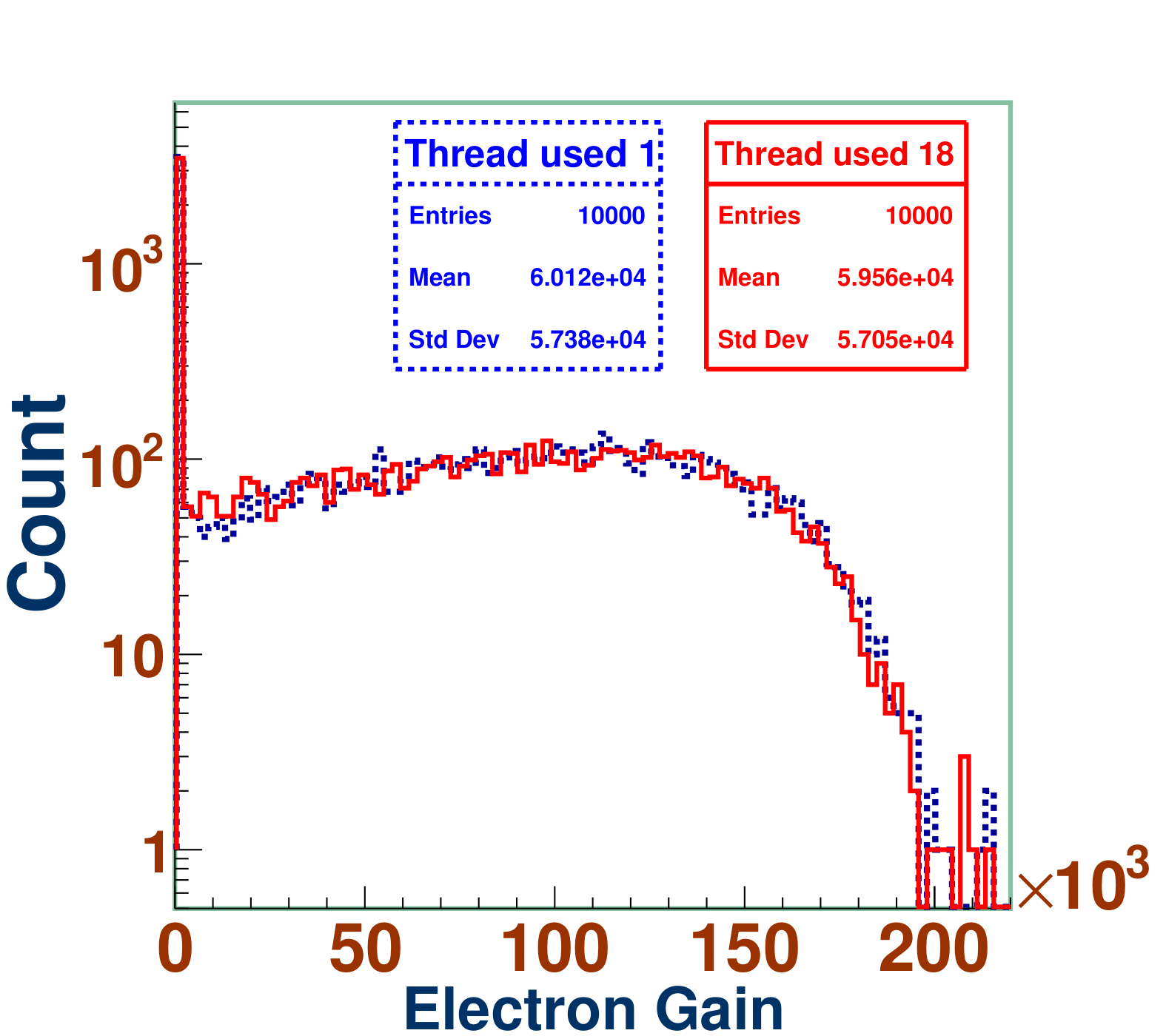}}
	\caption{Distribution of electron gain of $10^4$ avalanches (a) without space charge effect (b) with space charge effect.\label{fig:gain_comp}}
	
\end{figure}

 In Figures \ref{fig:Elc_gain_nosp} and \ref{fig:Elc_gain_wsp}   the distributions of electron gain of $10^4$ avalanches have been compared with and without space charge effect respectively. As a result, it is found that the mean gain had reduced by order of 10 when the space charge effect was considered, and the shape of the distribution also modified significantly.
Again, It is expected that the average electron gain should not vary much with the number of threads (N). In the same Figures, \ref{fig:Elc_gain_nosp} and \ref{fig:Elc_gain_wsp}, the gain calculations have been carried out with single thread and a total of 18 threads, where it can be seen that they are matched within some statistical uncertainties.

\section{Induced charge distribution of avalanches}
\label{sec:7_neBEM}
A Charge can be induced on the readout electrodes due to the movement of the electrons and ions. The induced charge $Q^{ind}$ due to $N_{av}$ number of q point charge moving inside the RPC can be calculated by using Ramo's equation as follows \cite{ShockleyRamo}:
\begin{equation}\label{eqn:induced_charge}
Q^{ind}=\int_{0}^{t} dt \sum_{n=0}^{N_{av}}\, \,q\,({\phi}_w^n({r_f}(t))-{\phi}_w^n({r_i}(t))),
\end{equation}
where ${\phi}_w^n({r_f}(t))$ and $\,{\phi}_w^n({r_i}(t)$ are the weighting potential at initial ($r_i$) and final potition ($r_f$) of the step calculated by using neBEM.

\par
\begin{figure}[H]
	\center{\includegraphics[width=.7\linewidth]{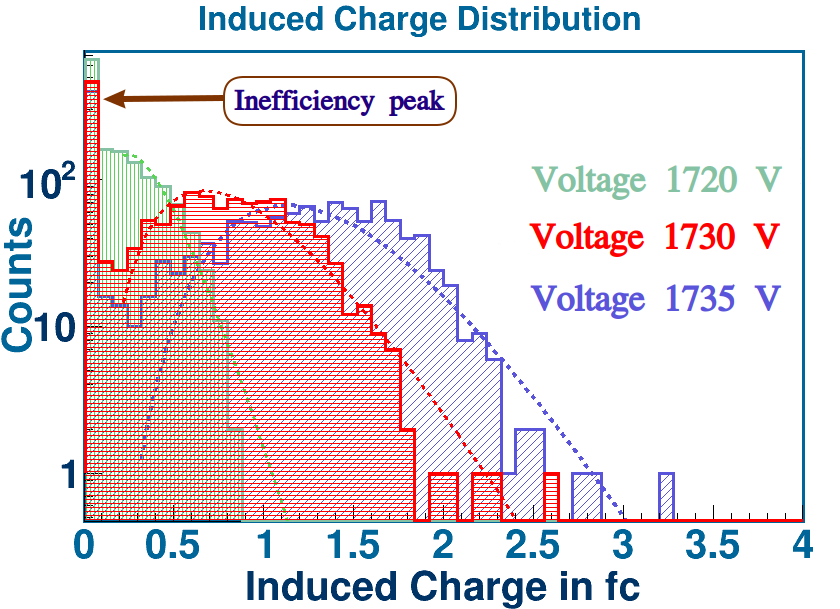}}
	\caption{Comparison between induced charge distributions corresponding to three applied voltages 1720 V, 1730 V and 1735 V.\label{fig:induced_charge}}
	
\end{figure}

 The induced charge distributions of a timing RPC for three different applied voltages, 1720 V, 1730 V, and 1735 V, are shown in Figure \ref{fig:induced_charge}. The RPC's geometry and other configurations remain the same as in section \ref{sec:6_speedUp}. The three distributions of Figure \ref{fig:induced_charge} have been fitted with the Polya function given as follows \cite{polya_charge,KOBAYASHI2006136} :
\begin{equation}\label{eqn:polyadist}
f(Q^{ind})=a(\frac{Q^{ind}\,b}{c})^{b-1}\,e^{-\frac{b}{c}Q^{ind}},
\end{equation}
where parameter $a$ is the scaling factor, $b$ is a free parameter determining the shape of the distribution, and $c$ is the mean charge. In all fitting processes with polya function, the left inefficiency peak of the charge distributions of the Figure \ref{fig:induced_charge} has not been considered. The fitted values of $a,b$, and $c$ for different voltages given in the Table \ref{tab:ind_charge}. From the values of fit parameters of Figure \ref{fig:induced_charge}, it can be said that the value of the parameter $c$ or mean charge increases with the increase in applied voltage. 
Moreover, the value of $b$ also shifted towards a higher value with the voltage increment, determining the broadness of the charge spectrum. From experimental results such as ~\cite{Fonte:491918, LippmanThesis}, the shifting of the mean and broadening of the shape of induced charge distribution are also observed.
\begin{table}
	\center%
	\begin{tabular}{|c|c|c|c|c|}
		\hline 
		Voltage (V) & a & b & c (fc)
		\tabularnewline
		\hline 
		\hline 
		1720 & $\begin{array}{c}
		139.7\\
		\pm\\
		68.7
		\end{array}$ & $\begin{array}{c}
		3.8\\
		\pm\\
		0.5
		\end{array}$ & $\begin{array}{c}
		0.34\\
		\pm\\
		0.01
		\end{array}$\tabularnewline
		\hline 
		1730 & $\begin{array}{c}
		19.8\\
		\pm\\
		7.9
		\end{array}$ & $\begin{array}{c}
		4.9\\
		\pm\\
		0.3
		\end{array}$ & $\begin{array}{c}
		0.8\\
		\pm\\
		0.01
		\end{array}$\tabularnewline
		\hline 
		1735 & $\begin{array}{c}
		0.06\\
		\pm\\
		0.07
		\end{array}$ & $\begin{array}{c}
		8.2\\
		\pm\\
		0.7
		\end{array}$ & $\begin{array}{c}
		1.3\\
		\pm\\
		0.02
		\end{array}$ \tabularnewline
		\hline 
	\end{tabular}
	
	\caption{Fit parameters of induced charge distribution of Figure \ref{fig:induced_charge}\label{tab:ind_charge}.}
	
\end{table}

\section{Variation of induced charge and time resolution with the gas gap of the RPC}
In this section, the main motivation is to discuss the variation of induced charge and time resolution with the gas gap of the RPC. To consider the dynamic space charge field, we have implemented a 3D line charge model inside Garfield++ as discussed in \cite{Dey_2020,Dey_2022}.
To test the working of this model, we have taken two numerical timing RPCs of gas gap 0.02 cm and 0.03 cm, and we discuss a comparison of the results. Another discussion on the simulation of the response of a single gap timing RPC with different space charge models can be found in \cite{LIPPMANN200319}. The experimental results of single gap and multigap timing RPCs for the same gas gaps can be found in ~\cite{BLANCO2002328,BLANCO200370,Fonte:491918}.
\subsection{Estimation of electric field inside RPC using neBEM}\label{sec:field}
\begin{figure}[H]
	\center\includegraphics[scale=0.4]{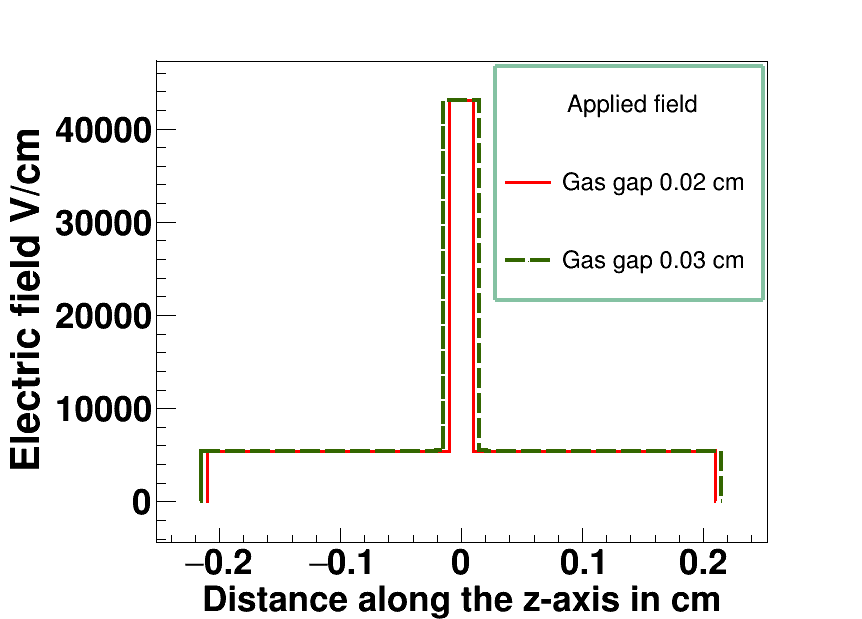}
	\caption{Applied field for RPCs of gas-gaps  0.02 cm and 0.03 cm \label{fig:Applied_field}}
\end{figure}

Two RPCs, RPC1 and RPC2, with gas gaps of 0.02 cm and 0.03 cm respectively, were designed using the geometrical tool of Garfield++. The electrode thickness and surface area were considered as 0.2 cm and 30 cm$^2$, respectively. The applied voltages for RPC1 and RPC2 were chosen so that the electric field inside each RPC remained the same (43 kV/cm) as shown in Figure \ref{fig:Applied_field}. The values of the applied voltages on RPC1 and RPC2 were 1505 V and 1720 V, respectively. A gas mixture of 85\% $\ce{C_2H_2F_4}$, 5\% $\ce{i\text{-}C_4H_{10}}$, and 10\% $\ce{SF_6}$ was used in each RPC.
\subsection{Distribution of primary cluster in RPCs of different gas-gap} \label{sec:track}
The size of the avalanche in an RPC detector depends on the number of primary electrons, making it necessary to understand the distribution of primary electrons in the different gas gaps of the RPCs. In the following discussion, we will explore the distribution of primary electrons inside RPC1 and RPC2.

\begin{figure}[H]
	\center\includegraphics[scale=0.4]{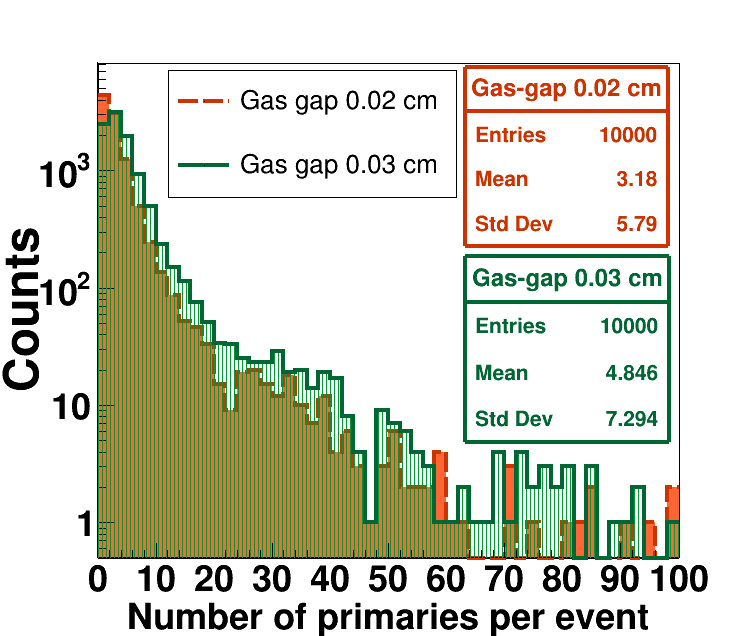}
	\caption{Disribution of number of primary electrons for two RPCs \label{fig:dist_prim}}
\end{figure}

Muon tracks with an energy of 2 GeV were generated using HEED, a built-in Garfield++ feature. The direction of the tracks was chosen perpendicular to the surface of the RPC or along the positive z-direction. The distribution of primary electrons generated from a set of 10$^4$ muon tracks inside the gas gaps of the two RPCs are shown in Figure \ref{fig:dist_prim}, where the integral mean of the primary electron distribution for RPC1 and RPC2 is approximately 3 and 5, respectively. As shown in the figure, as the gap increased, the spectrum was also broadened, which is expected.

\subsection{Induced charge distribution for different gas gaps of timing RPCs}\label{sec:indCh}
A set of 10$^4$ avalanches was generated using the Montecarlo particle tracing model of Garfield++ inside RPC1 and RPC2. The primary electrons were taken from the muon tracks mentioned in Section \ref{sec:track}.
\begin{figure}[H]
	\center\includegraphics[scale=0.4]{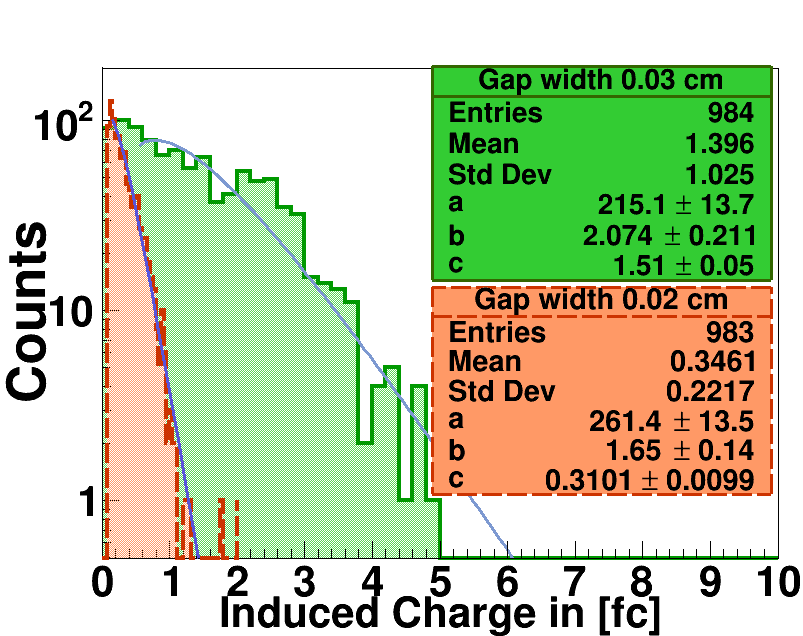}
	\caption{Disribution of number of primary electrons \label{fig:ind_chrg}}

\end{figure}

The induced charge $Q^{ind}$ due to the movement of electrons and ions inside the RPC can be calculated by using  equation \ref{eqn:induced_charge}.
\par The induced charge distributions for RPC1 and RPC2 are shown in Figure \ref{fig:ind_chrg}. In this figure, events are selected from the muon tracks that contain a maximum of eight primaries, and a threshold of 0.1 fC on induced charge has been chosen. The charge distributions shown in Figure \ref{fig:ind_chrg} have been fitted with a Polya function as described in equation \ref{eqn:polyadist}. From the fit parameters in Figure \ref{fig:ind_chrg}, it can be concluded that the mean charge parameter $c$ increases with an increase in gas gap, which is expected. For RPC1 and RPC2, the mean charge values are 0.3 fC and 1.5 fC, respectively. The shape of the distribution is broader with an increase in gas gap, which is reflected in the parameter $b$ value shown in Figure \ref{fig:ind_chrg}.
\subsection{Calculation of signal rise time} \label{sec:risetime}
\begin{figure}[H]
	\center\includegraphics[scale=0.4]{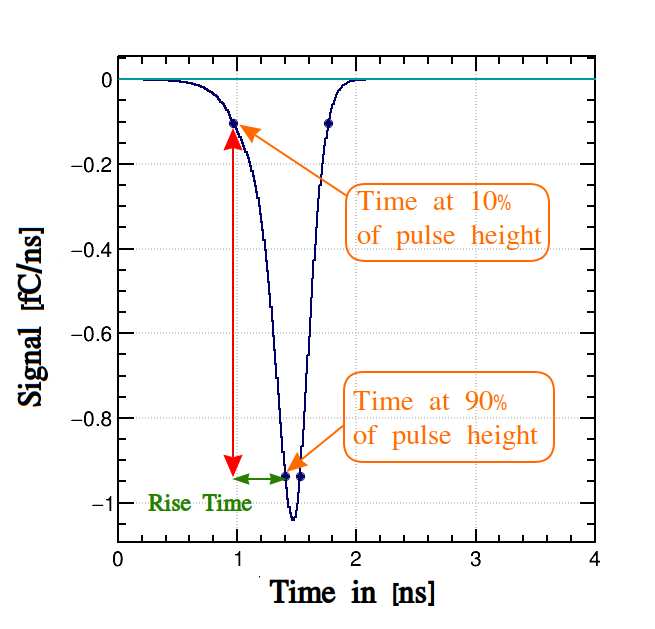}
	\caption{Calculation of rise time of a signal. \label{fig:rise_time}}
	
\end{figure}
\begin{figure}[h]
	\center\includegraphics[scale=0.4]{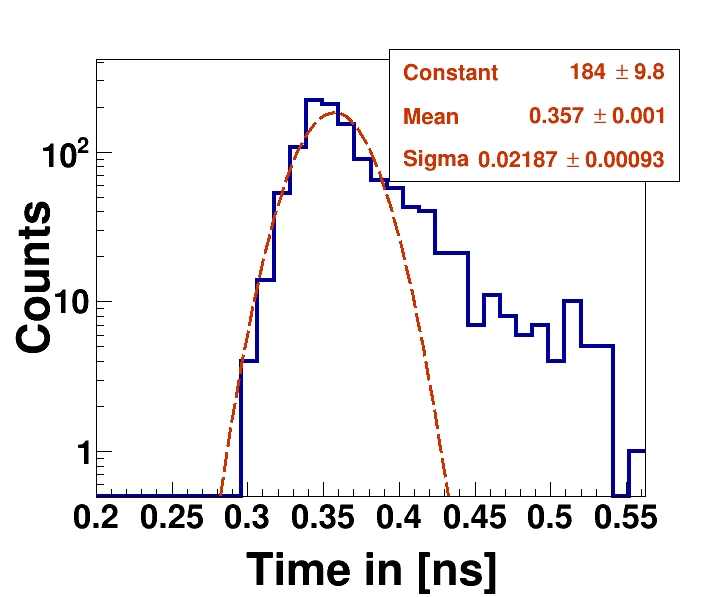}
	\caption{Rise time distribution of RPC1 \label{fig:time_reso_1}}
	
\end{figure}
\begin{figure}[H]
	\center\includegraphics[scale=0.4]{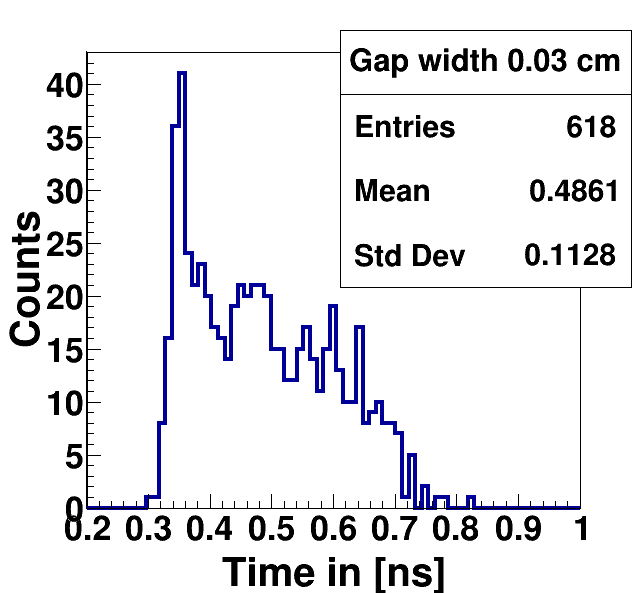}
	\caption{Rise time distribution of RPC2 \label{fig:time_reso_2}}
	
\end{figure}
\begin{figure}[H]
	\center\includegraphics[scale=0.4]{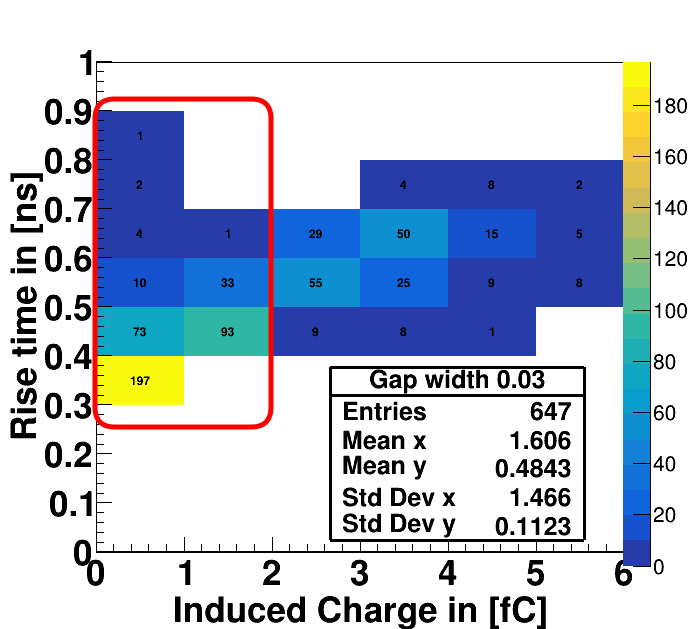}
	\caption{Correlation between Rise time and Induced charge of RPC2  \label{fig:time_reso_3}}
	
\end{figure}
\begin{figure}[H]
	\center\includegraphics[scale=0.4]{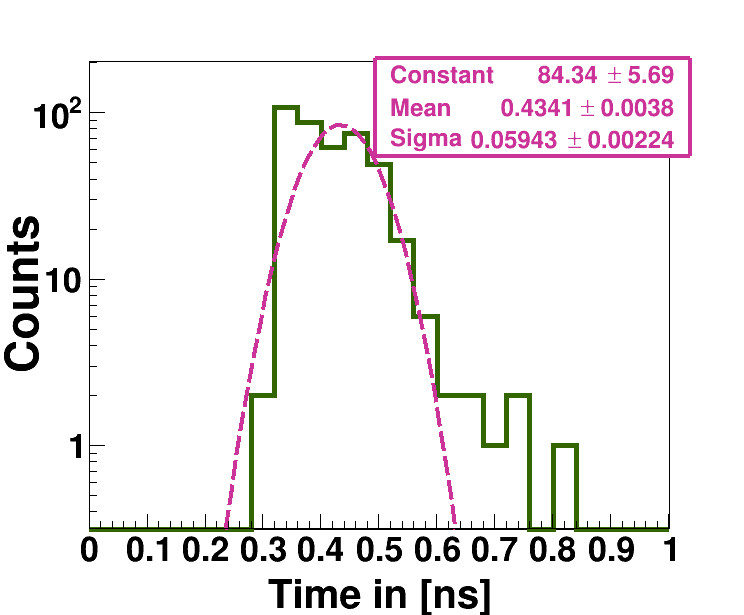}
	\caption{Rise time distribution of RPC2 for selected ($<$ 2 fC) range of induced charge. \label{fig:time_reso_4}}
	
\end{figure}

The signal rise time can be calculated by taking the difference between the time at 90\% and 10\% of pulse height, as shown in Figure \ref{fig:rise_time}. Figures \ref{fig:time_reso_1} and \ref{fig:time_reso_2} show the distribution of rise time for RPC1 and RPC2, respectively. The rise time distribution of RPC1 (shown in Figure \ref{fig:time_reso_1}) is fitted with a Gaussian function, and the resulting time resolution, or the sigma of the fit, is approximately 21.8 ps.
However, due to multiple peaks in the rise time distribution of RPC2 (Figure \ref{fig:time_reso_2}), the calculation of time resolution is not straightforward. By looking at the correlation plot (Figure \ref{fig:time_reso_3}) of rising time vs induced charge and considering the mean induced charge value (1.5 fC) of RPC2, it can be inferred that below 2 fC, the probability of getting a signal is higher than in other higher charge regions. Therefore, the rise time distribution was limited to only those signals, as shown in Figure \ref{fig:time_reso_4}. The distribution was fitted with a Gaussian function, resulting in a sigma of approximately 59.4 ps. This value is consistent with the experimental result shown in \cite{BLANCO200370}.
\section{Summary}
\label{sec:8_summary}
In this chapter, we have discussed the method
of generation of uncorrelated and uniform parallel random numbers. The problem of correlation between the random numbers generated by multiple threads has been successfully solved using different seed values and objects of TRandom3.
\par The steps of implementing the dynamic space charge field inside Garfield++ have been discussed. In this context, the detailed modeling of a saturated avalanche has been examined to understand the space charge effect. The changing of the dynamic space charge field and the evolution of the 3D shape of the avalanche with time have also been shown. As a result, we have found that at the time of saturation, the rate of attachment is dominant in the low-field regions, and at high-field regions rate of ionization is dominant; hence overall, the number of charges remains the same, which brings this saturation.  
\par Parallelization has been acheived by implementing multithreading techniques in Garfield++. With OpenMP, we have achieved approximately 5.4 times and 7.2 times speed up in the avalanche generation process without and with space charge effect, respectively.
\par The induced charge distribution for a timing RPC has been calculated for three different voltages. Spectrum of the induced charge distribution predicted by the present simulations are similar to those observed in other experimental results. 

\par We also have simulated the induced charge distribution and signal rise time distribution, considering the space charge effect, of two different timing RPCs with gas gaps of 0.02 cm (RPC1) and 0.03 cm (RPC2).

It was found that the induced charge distribution of RPC1 and RPC2 follows a non-linear Polya function. The mean induced charge was calculated to be 0.3 fC and 1.5 fC  for RPC1 and RPC2, respectively. The shape of the induced charge distribution broadens as the gas gap increases due to the space charge effect.

It was observed that as the gas gap increases, the timing performance of the RPCs deteriorates due to the increment of time resolution. The mean rise time was also found to shift towards higher values for larger gas gaps. For RPC2, with multiple peaks in the rise time distribution, a cut on induced charge was introduced to select events and calculate time resolution. The simulation results for the time resolution of RPC2 (0.03 cm gas gap) were verified with experimental results.

\par In future, we plan to implement a photon transport model in our code to study the avalanche-to-streamer transition. Moreover, the study of other gaseous detectors will also be possible with the model described in the present paper.

\chapter{Reconstruction Of Cosmic Muon ($\mu^{\pm}$) Track To Determine Scattering angle Using A Stack of RPCs }\label{ch:tomography}
\section{Introduction}
\begin{figure}[H]
	\center\includegraphics[scale=0.25]{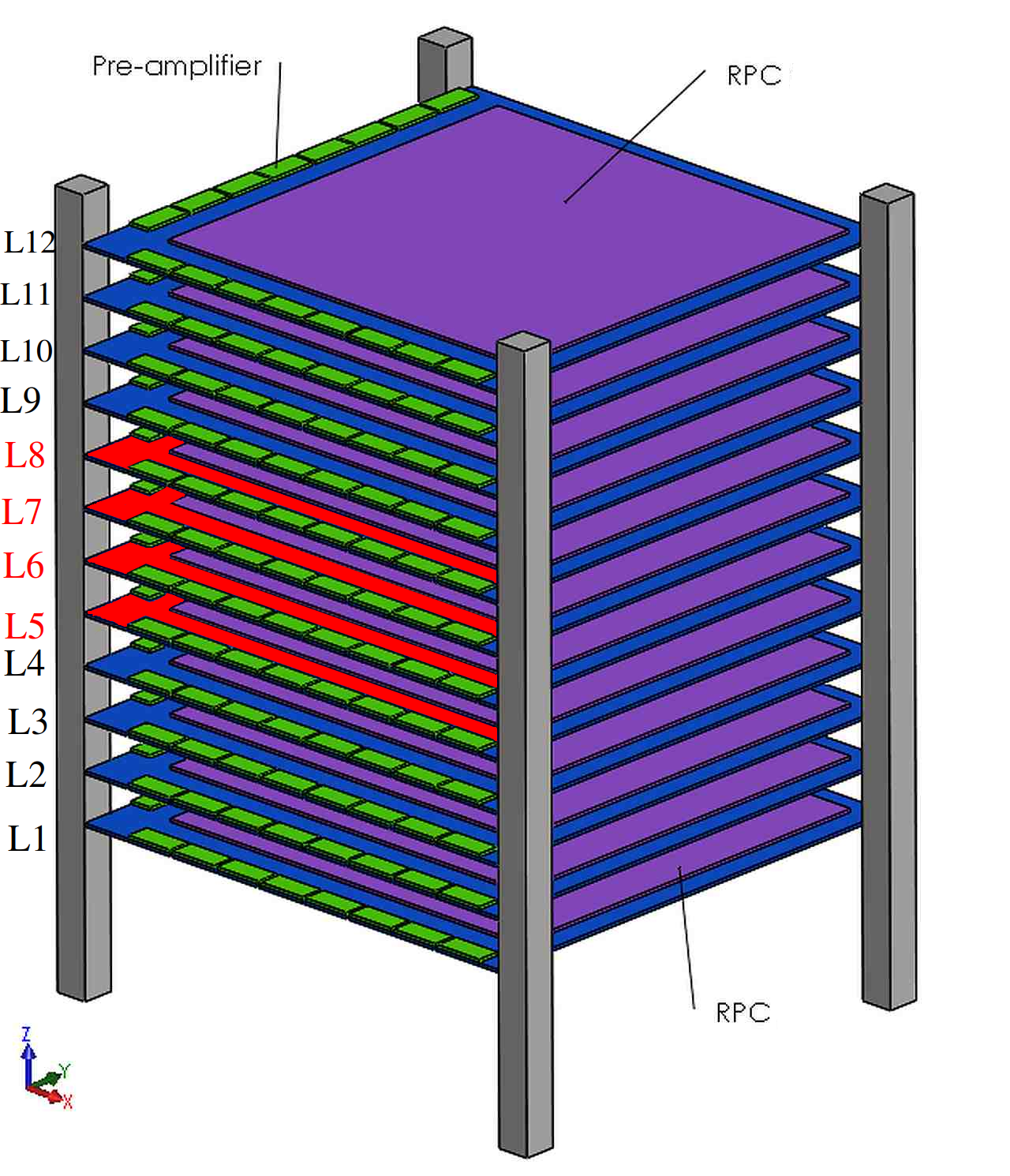}
	
	\caption{schematic of RPC stack built at IICHEP,madurai. The Data from the red shaded RPCs are neglected in the analysis. \label{fig:inostack}}	
\end{figure}

\begin{figure}[H]
	\center\includegraphics[scale=0.1]{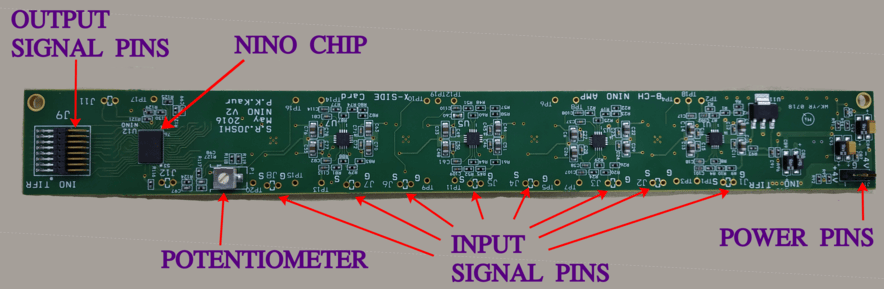}
	
	\caption{NINO board to amplifly and discriminate RPC signals \label{fig:nino}}	
\end{figure}

Cosmic muons leave their signatures on Resistive Plate Chambers (RPCs), which we detect in the form of current signals. These signals can be read and registered as data in computers to analyze the position and timing information of the muons that pass through the RPCs. At the INO prototype, the study of cosmic muons is conducted using a 12-layer stack of RPCs made of glass electrodes, each with an area of $2 \,\mathrm{m} \times 2 \,\mathrm{m}$, as shown in Figure \ref{fig:inostack}. The stack is located at the Institutional Centre for High Energy Physics (IICHEP) in Madurai, which is in southern India. To read the signals, 64 X-side and 64 Y-side copper strips, each of $2.8 \,\mathrm{cm}$ width of, are used of each RPC. All the RPCs are mounted on an aluminum frame, with a gap of approximately $17 \,\mathrm{cm}$ between two RPCs. NINO \cite{NINO} boards (see Figure \ref{fig:nino}) are used to amplify and discriminate the signals from the RPC strips, and an FPGA board-based data acquisition system is used to collect the data. Details of the electronics developed for the stack can be found in \cite{bhatt2019measurement,bhuyan2012vme,pal2014development}.
 
\par At VECC Kolkata, we have installed a prototype stack consisting of eight bakelite RPCs, each with an area of 30 $\times$ 30 $\mathrm{cm}^2$, as shown in Figure \ref{fig:rpcstack}. The stack has the provision of adjusting the gaps between the RPCs. Our goal is to reconstruct the track of cosmic muons and perform offline alignment of the detector using position residual correction of the cosmic tracks. Additionally, we aim to determine the scattering angle of cosmic tracks in the presence of high-density materials such as lead and tungsten.

\section{Experimental Design And Methodology}
\begin{figure}[H]
	\center\includegraphics[scale=0.2]{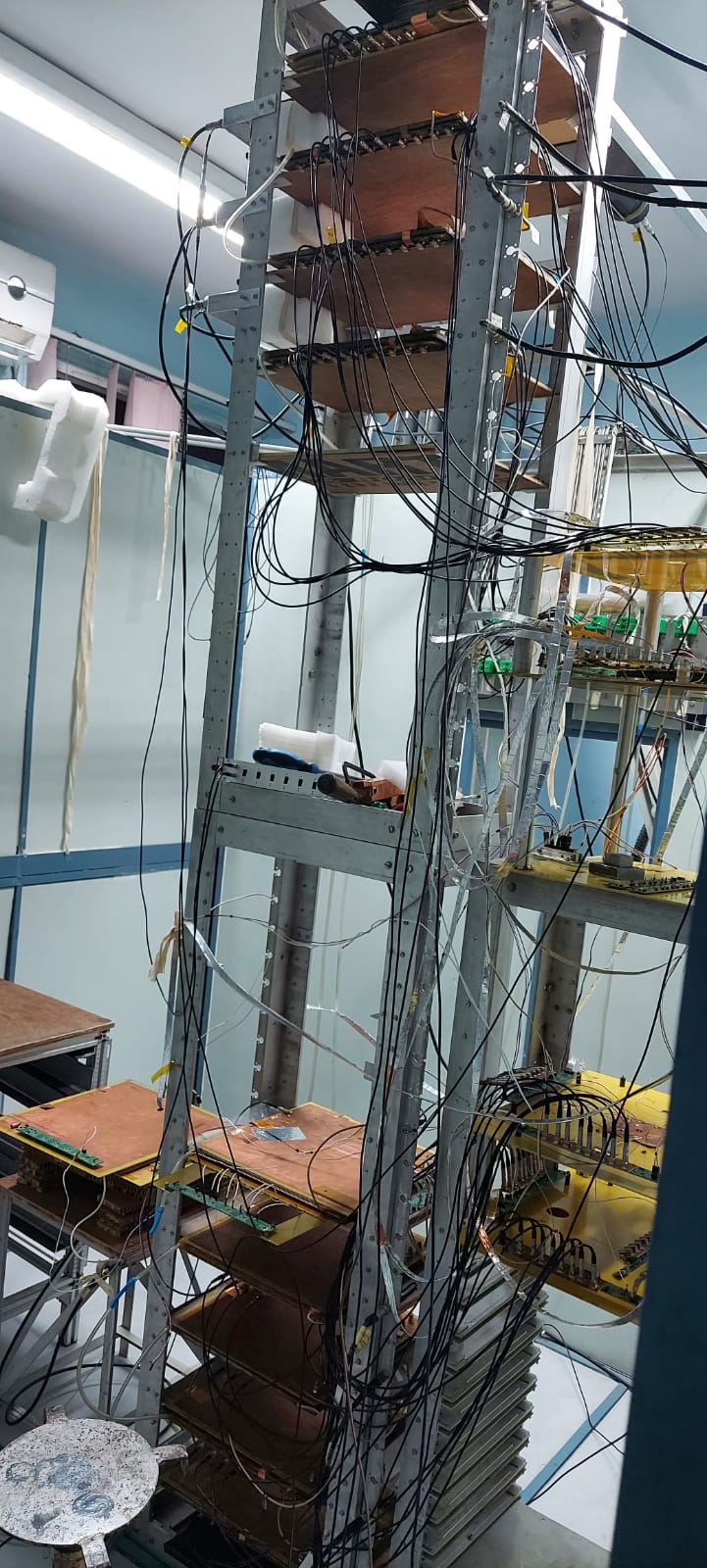}\includegraphics[scale=0.2]{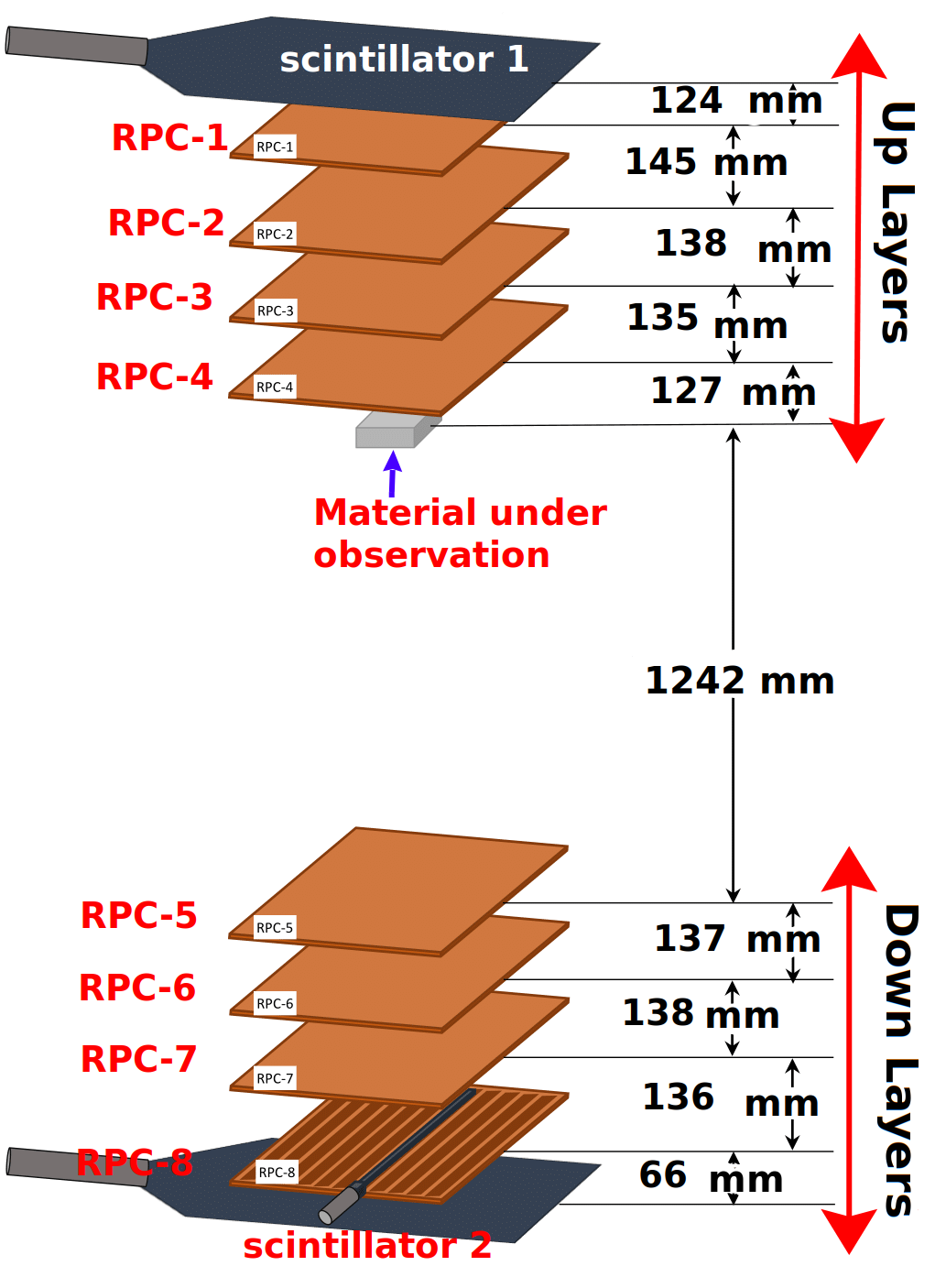}
	
	\caption{Photograph of the setup (left) and the schematic diagram of the setup (right) of VECC stack \label{fig:rpcstack}}	
\end{figure}

The Resistive Plate Chamber (RPC) used in this experiment has dimensions of 30 cm width, 30 cm length, and a thickness of 2 mm for the bakelite material. The distance between the electrodes is 2 mm, and each pickup strip is 2.3 cm wide with a 0.2 cm gap between two strips. This results in strips being present on both the X and Y sides.
 Two scintillators are placed at the top and bottom of the stack to trigger the experiment. The RPCs are divided into two sections: up-layers and down-layers (as shown in Figure \ref{fig:rpcstack}), with the option to insert a block of high-Z material between them. The total height of the stack is 220 cm, and the distance between the up and down layers is 136.9 cm, which can be adjusted to different values.

When charged particles pass through any material, they interact with the nuclei of the material, causing multiple Coulomb scattering. As discussed earlier, the aim of this experiment is to detect the scattering of cosmic muons and calculate their average scattering angle. The scattering angle distribution can be modeled as a Gaussian distribution with the standard deviation determined as follows \cite{Baesso_2014_tomography,Baesso_2013}:
\vspace{-.5cm}
\begin{align}
\sigma&=\frac{13.6}{\beta c p}\sqrt{\frac{L}{X_0}}\Bigg(1+0.038\,\text{ln}\frac{L}{X_0}\Bigg)\\
X_0&=\frac{716.4}{\rho}\frac{A}{Z(Z+1)\,\text{ln}(\frac{287}{\sqrt{Z}})}
\end{align}
where,
$\sigma$= RMS width, $\beta$=Ratio between the velocity\,of muon v to the velocity of light c, $p$= Momentum of the muon (MeV/c),
$X_0$= Radiation length of the material (cm) ,
$L$= Length of the material traversed (cm),
$\rho$= Density of the material (gm-$cm^{-3}$),
$A$= Atomic mass, and
$Z$= Atomic number. By measuring the sigma value, it may be possible to identify the presence of high Z unknown material being used in the experiment. Such an exersise would have a potential application in tomography.

The gas mixture used in this experiment is a combination of R134a (95\%), Isobuten (4.5\%), and SF6 (0.5\%). The operating voltage of each RPC is set to a value between 5200-5600 V to ensure maximum efficiency and minimum crosstalk.
\section{Electronics} \label{sec:electronics}
In our current setup, we are using the NINO chip \cite{NINO} to amplify and discriminate the raw RPC signals, similar to the setup used for the IICHEP Madurai Stack. The raw signals are processed by the NINO, which produces an LVDS signal which is then transmitted to a specialized FPGA board. The FPGA board converts the LVDS signal into digitized binary data that incorporates information about the channel number and timing of the pulses. Finally, this data is saved in a DAQ PC for further processing and analysis. In the following section, we will describe the process of characterizing the RPCs that were built at VECC before they were used in the stack.
\section{Testing of RPCs before installation}

\begin{figure}[H]
	\center\includegraphics[scale=0.3]{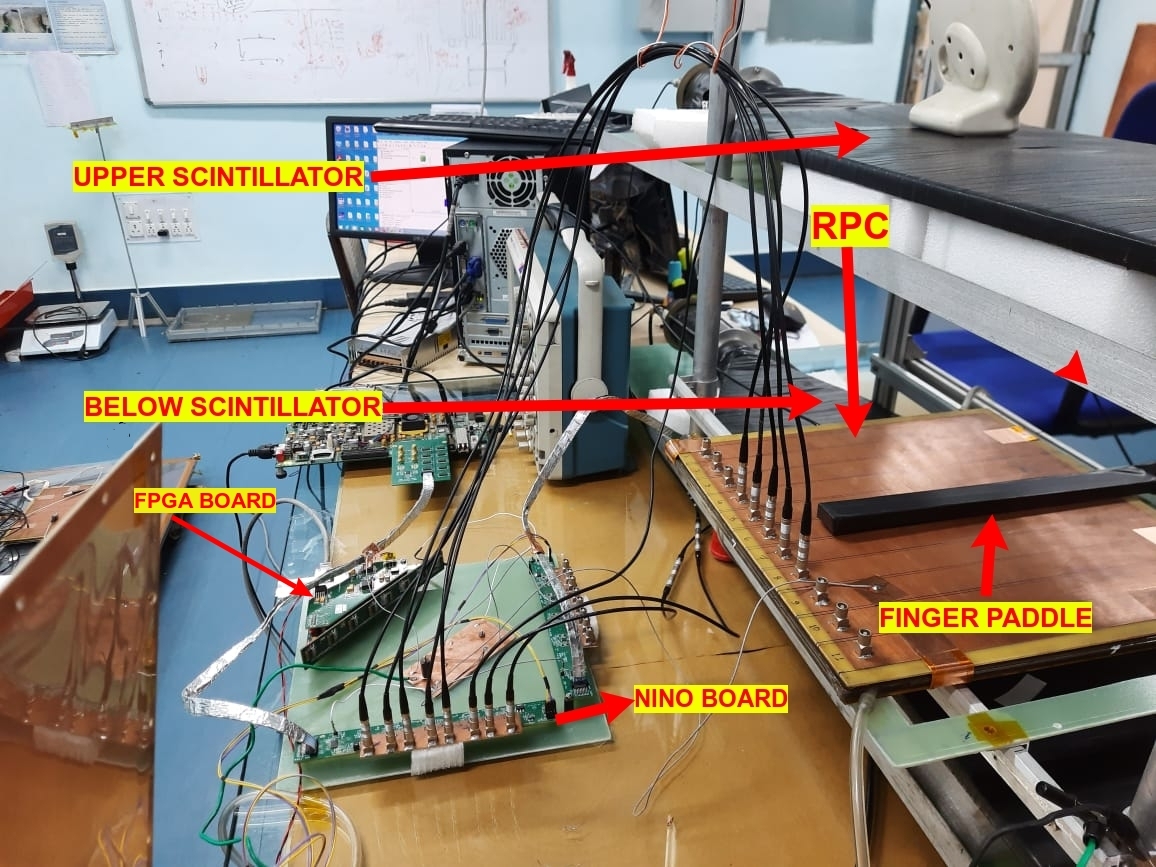}
	
	\caption{Testbench setup to examine RPCs at VECC \label{fig:testbench}}	
\end{figure}
All RPCs are tested on a test bench, as shown in Figure \ref{fig:testbench}, before being installed on the long stack (see Figure \ref{fig:rpcstack}). The test bench consists of two large scintillators, labeled as the upper and below scintillators in Figure \ref{fig:testbench}. The RPC under test is placed between the upper and below scintillators, and an examination of individual strips is performed using a finger paddle. The same electronics described in Section \ref{sec:electronics} are used to record and analyze data. To test an RPC on the test bench, a series of measurements are performed, which are described as follows.
\subsection{IV characteristic}
\begin{figure}[H]
	\center\includegraphics[scale=0.4]{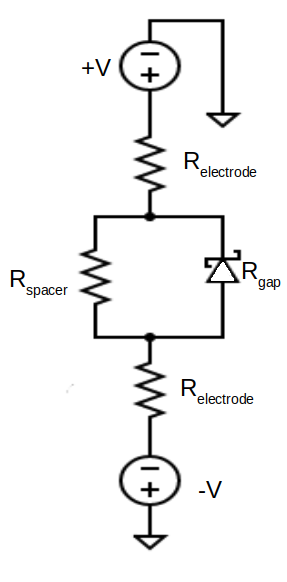}
	
	\caption{Equivalent circuit diagram of an RPC \label{fig:eqcircuit}}	
\end{figure}
The I-V test is a crucial first step in assessing the health of an RPC detector. It examines the current-voltage curve, which typically exhibits two distinct slopes. This phenomenon can be explained by the equivalent circuit of the RPC, as illustrated in Figure \ref{fig:eqcircuit}.
\begin{figure}[H]
	\center\includegraphics[scale=0.45]{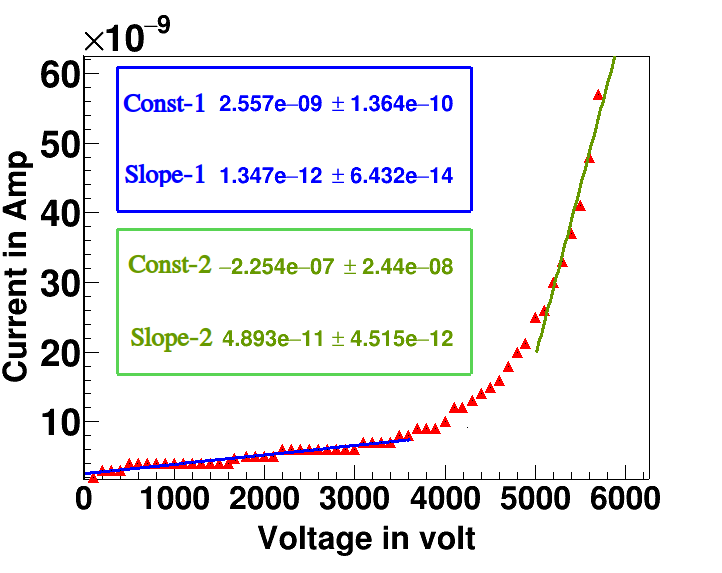}
	
	\caption{Variation of RPC current as a function of applied voltage during ramping up \label{fig:ivchar}}	
\end{figure}
At low voltage regions, where the voltage is below the discharge threshold, the gas gap behaves as an insulator. This means that the gap resistance ($R_{gap}$) is approximately infinite, and no avalanches or streamers are formed. Additionally, it is well-known that $R_{electrode} \ll R_{spacer}$. Therefore, the slope of the I-V characteristic curve for low voltage can be expressed as follows:
\begin{align}
\frac{dI}{dV}\big|_{\text{low voltage}}\propto\frac{1}{R_{spacer}}
\end{align}

Taking the linear fit (blue line) of the data points at low voltage in Figure \ref{fig:ivchar} as an example, the conductance of the spacers ($\frac{1}{R_{spacer}}$) can be expressed as follows:
\begin{align}
\frac{1}{R_{spacer}}&\approx 1.347\times 10^{-12}\frac{1}{\Omega}\\ \nonumber
R_{spacer}&\approx 7.42\times 10^{11}\,\Omega
\end{align}
 On the other hand, at the very high voltage regions where the voltage is above the discharge threshold, the gas gap behaves as a conductor ($R_{gap}\approx 0$) due to the formation of large avalanches or streamers. Hence, the slope of the I-V characteristic curve for high voltage can be expressed as follows:
 \begin{align}
 \frac{dI}{dV}\big|_{\text{high voltage}}\propto \frac{1}{R_{electrode}}.
 \end{align}
Taking the linear fit (green line) of the data points at high voltage in Figure \ref{fig:ivchar} as an example, the conductance of the electrode ($\frac{1}{R_{electrode}}$) can be expressed as follows:
\begin{align}
\frac{1}{R_{electrode}}&\approx 4.893\times  10^{-11} \frac{1}{\Omega}\\ \nonumber
R_{electrode}&\approx 2.04\times 10^{10}\,\Omega
\end{align} 
The breakdown voltage of an RPC, which is the voltage at which an RPC begins to generate signals, is another crucial parameter that can be estimated from the I-V characteristic curve. In Figure \ref{fig:ivchar}, the breakdown voltage can be seen to be approximately 4500 volts. The I-V characterization has been done for all the RPCs. 
\subsection{Noise Monitoring}
\begin{figure}[H]
	\center\includegraphics[scale=0.42]{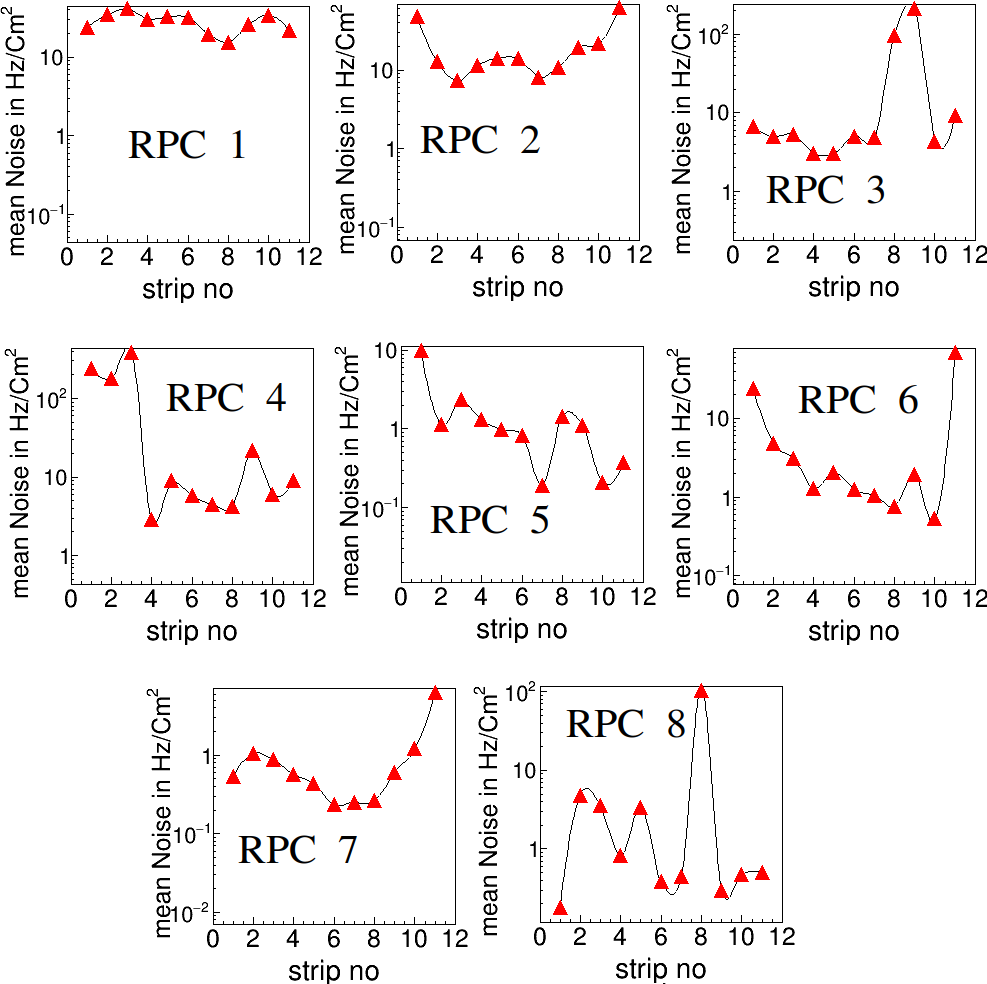}
	
	\caption{X-side strips noise of all RPCs (VECC) \label{fig:xnoise}}	
\end{figure}

\begin{figure}[H]
	\center\includegraphics[scale=0.42]{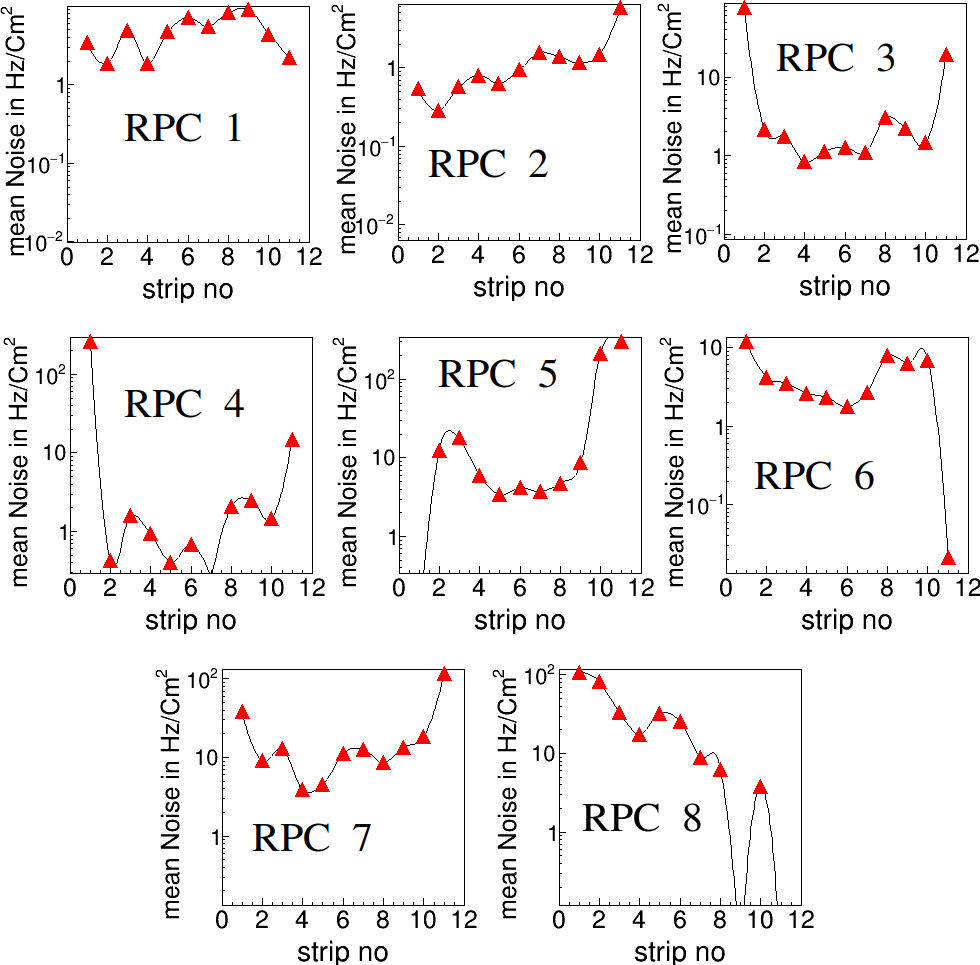}
	
	\caption{Y-side strips noise of all RPCs (VECC) \label{fig:ynoise}}	
\end{figure}

The FPGA generates two data files from the signals of the NINO: the monitoring data and the triggered data. In the monitoring data, signals are captured at one-second intervals and consist of responses from cosmic particles, background radiation, and electronic noise. In the triggered data, the coincidence method (shown in Figure \ref{fig:effilogic}) is used to capture mostly cosmic particles coming from the atmosphere. Using the monitoring data, the noise of both the X and Y side strips of each RPC is measured in Hz/cm$^2$, as shown in Figures \ref{fig:xnoise} and \ref{fig:ynoise}, respectively. It is found that the noise on the strips positioned near the junction between the voltage terminal and the graphite surface is larger as compared to the noise on the strips positioned farther away. However, in a few cases, strips that are far from the terminal point might be noisy. In the present set of RPCs, the noise level in the maximum X and Y side strips varies between 1-10 Hz/cm². However, in some strips, the noise rises beyond the previously specified range, which is mostly near the voltage terminal. It should be noted that for obtaining decent quality data, the noise level should fall between 1-15 Hz/$cm^2$. The interval may vary for different setups. If the noise level falls below 1 Hz/cm$^2$, the efficiency of that particular strip will be low. If it exceeds 15 Hz/cm$^2$, the chances of obtaining fake hits with trigger will be increased, which can ultimately affect the track reconstruction efficiency. Therefore, it is essential to mask the strips whose noise falls outside the selected interval.
\subsection{Characterization of the RPCs by efficiency measurement}\label{sec:effical}
\begin{figure}[H]
	\center\includegraphics[scale=0.4]{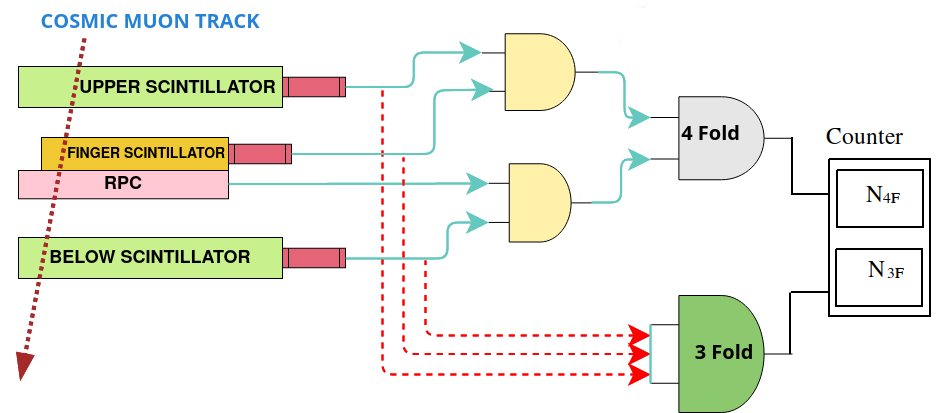}
	
	\caption{Schematic Logic diagram to calculate efficiency of an RPC \label{fig:effilogic}}	
\end{figure}

The efficiency of an RPC for cosmic muons with respect to the scintillator signals can be calculated by using the coincidence method. In Figure \ref{fig:effilogic} the logic diagram to calculate number of 3 fold trigger count ($N_{3F}$) and number of 4 fold ($N_{4F}$) trigger count is shown. Therefore the efficiency $\varepsilon_{rpc}^{eff}$ of an RPC can be determined as follows:
\begin{equation}
\varepsilon_{rpc}^{eff}=\frac{N_{4F}}{N_{3F}}
\end{equation}
The diagram in Figure \ref{fig:effilogic} is to show the concept behind the calculation of efficiency. As discussed before, the pulses from an RPC are directly fed to the NINO board and then from the NINO to the FPGA board. The number 0 and 1 pins of the FPGA are programmed as trigger pins, which means that data will be registered only when those pins detect signals. The trigger pins are connected to the upper and lower scintillators. The rest of the calculations are done in offline data analysis.

\begin{figure}[H]
	\center\includegraphics[scale=0.3]{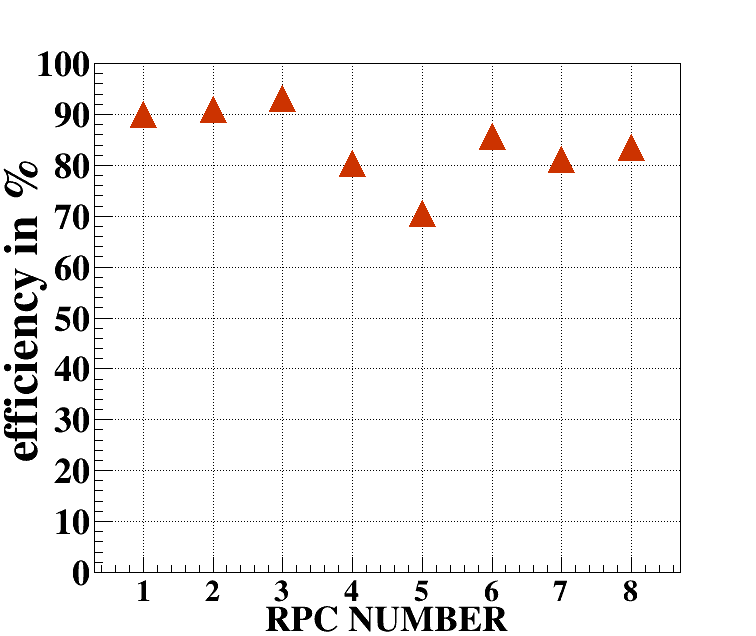}
	
	\caption{Average efficiencies of RPCs used in the VECC stack\label{fig:sateffi}}	
\end{figure}

We tested 12 RPCs, and among them, we chose 8 good RPCs to be placed in the stack. The average maximum or saturated efficiencies of those 8 RPCs over a 376 mV threshold voltage set at the pot in NINO are shown in Figure \ref{fig:sateffi}, where most of RPCs are working above 80\% efficiency, with a few working between 70\%-80\% efficiency.

\subsection{Testing of crosstalk efficiency between strips}\label{sec:crosseffi}
\begin{figure}[H]
	\center\includegraphics[scale=0.3]{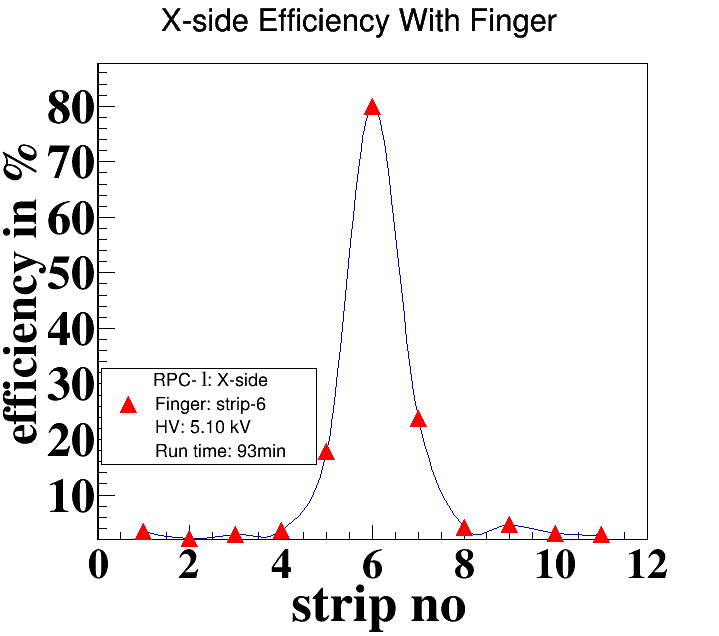}
	
	\caption{Calulation of crosstalk efficiencies for different strips (VECC)\label{fig:crosstalk}}	
\end{figure}
The phenomenon of crosstalk must be studied in order to accurately locate cosmic muons. A maximum of three strips on both sides of the Resistive Plate Chamber (RPC) can be considered valid hits for a muon. It is expected that a muon is only hit one strip at a time. Therefore, if we trigger a strip using a finger scintillator, the efficiency of getting a hit on the strip just below the scintillator is expected to be higher than the efficiency of getting a hit on the strips adjacent to it. However, due to charge sharing, adjacent strips may also register a hit, albeit with lower efficiency.

For instance, we placed the finger paddle on the 6th X-side Strip of the RPC-1 and calculated the efficiencies of all the strips using the method described in section \ref{sec:effical}. The results of this study are presented in Figure \ref{fig:crosstalk}, where the efficiency of the middle strip is around 80\%, and the efficiencies of the adjacent two strips range from 20\% to 30\%. Additionally, the efficiencies of the remaining strips are below 4\%.Therefore, spread of crosstalk lies mostly between three strips, which is our selection criteria during the track fitting, which will be discussed in section \ref{sec:trackalgo}.
\subsection{Testing of position uniformity in RPCs}
Measurement with hits from all over the detector region are necessary to optimize the tracking performance. The distribution of applied voltage over the surface depends on the resistivity of the graphite surface. Additionally, the electric field inside the gas gap depends on the smoothness of the surface of the bakelite electrode. The standard surface resistivity of the graphite coating is approximately 1 M$\Omega/\Box$. However, it has been observed that the uniformity of the graphite coating may not be the same all over the surface, resulting in different surface resistivity at different locations. Therefore, the electric field inside the gas gap may be different for different places, and the efficiency of getting hits for different places is different.

\begin{figure}[H]
\subfloat[RPC 1]{\includegraphics[scale=0.33]{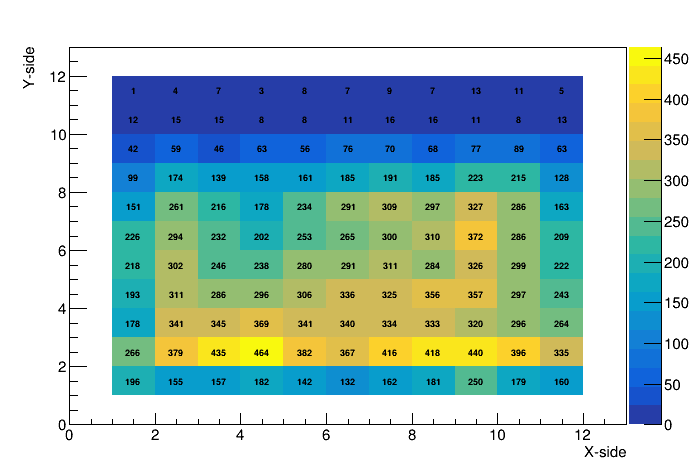}}\subfloat[RPC \label{fig:rpc2} 2]{\includegraphics[scale=0.33]{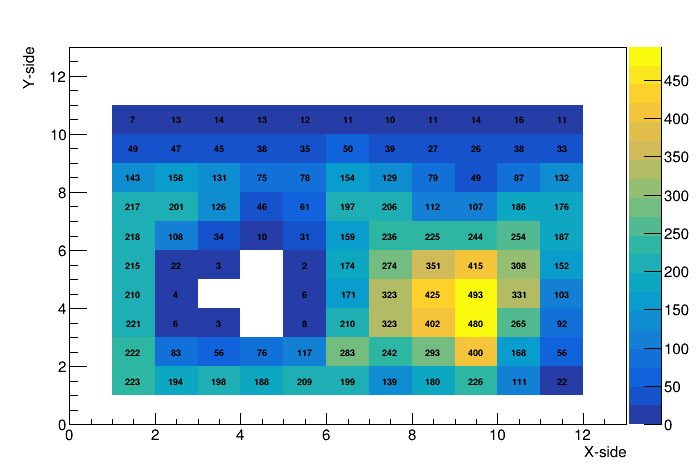}}

\subfloat[RPC 3]{\includegraphics[scale=0.33]{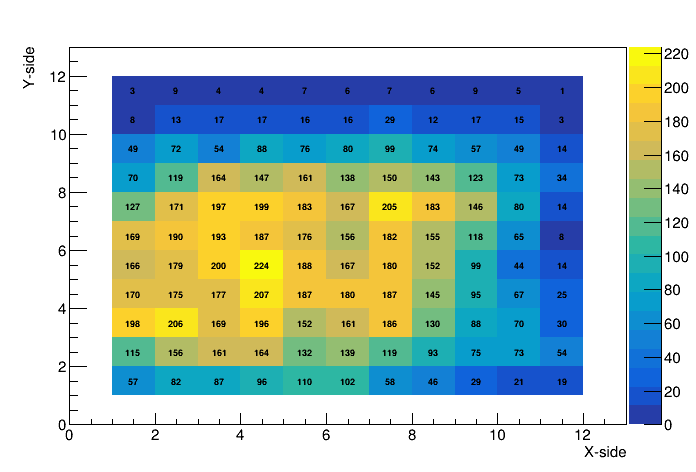}}\subfloat[RPC 4]{\includegraphics[scale=0.33]{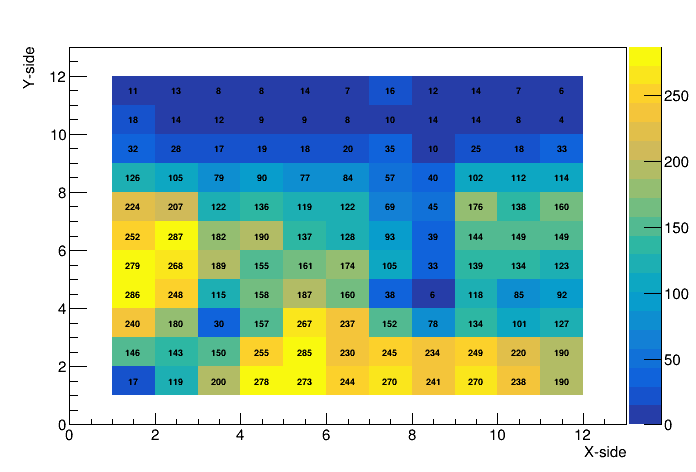}}
\caption{Correlation between x-y positions of the up layer RPCs (VECC) \label{fig:hituniformity1}}
\end{figure}

\begin{figure}[H]
	\subfloat[RPC 5]{\includegraphics[scale=0.33]{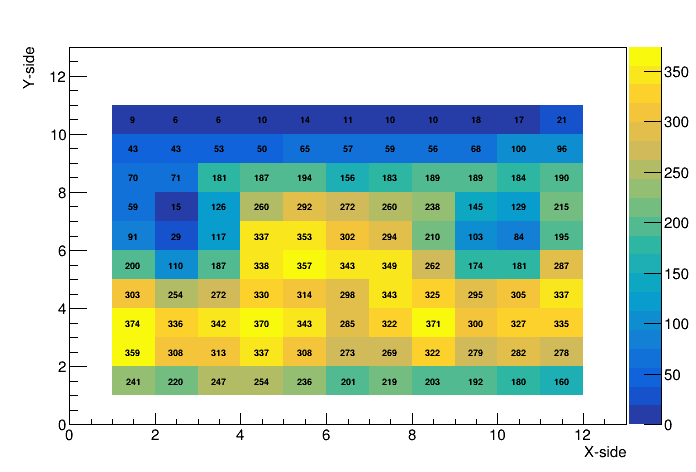}}\subfloat[RPC 6]{\includegraphics[scale=0.33]{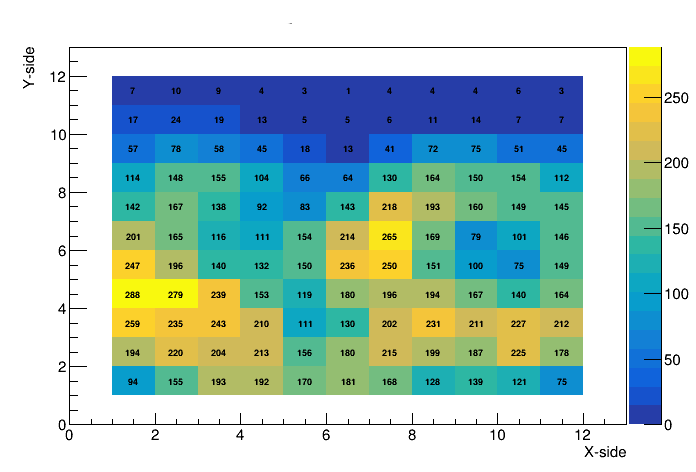}}
	
	\subfloat[RPC 7]{\includegraphics[scale=0.33]{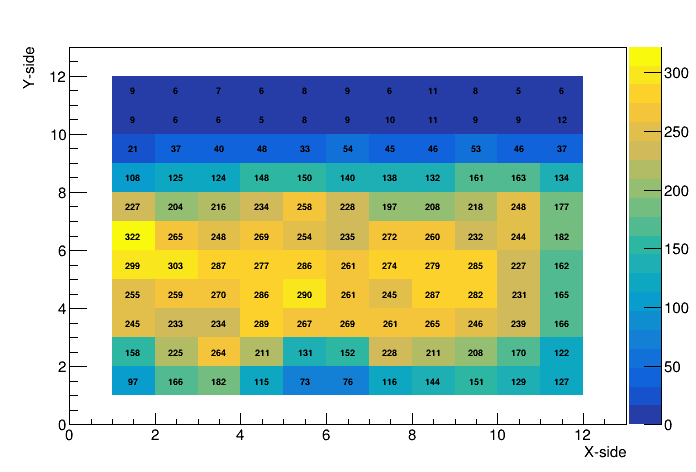}}\subfloat[RPC 8]{\includegraphics[scale=0.33]{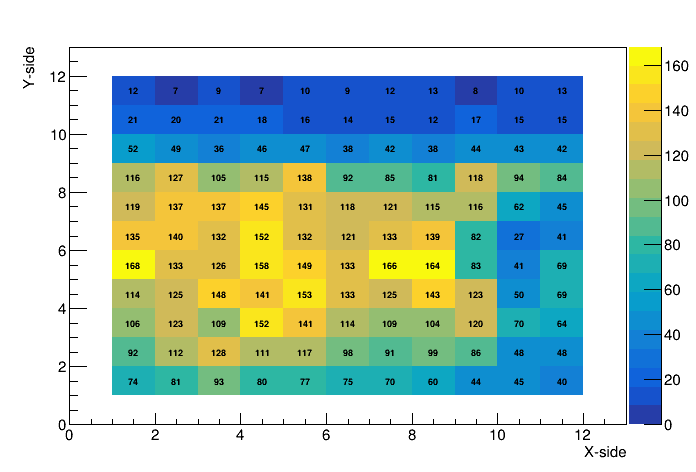}}
	\caption{Correlation between x-y positions of the down layer RPCs (VECC) \label{fig:hituniformity2}}
\end{figure}

The uniformity of hit distribution of all RPCs has been studied with a 2D distribution of the correlation between the x and y positions in the strip unit, as shown in Figures \ref{fig:hituniformity1} and \ref{fig:hituniformity2}. The color signifies the number of correlated hits at different x, y locations. From the former figures, it can be concluded that the probability of getting hits at the middle strips of the detector is higher than the edge strips. We detected a small inefficient region in RPC 2, as shown in Figure \ref{fig:rpc2}. This could be due to insufficient electric field strength in that area. Another possible cause of this is a disturbance in the graphite coating. Therefore, we repaired the graphite coating to achieve a more uniform hit distribution.

\section{Reconstruction method of tracks of cosmic muons}\label{sec:trackalgo}
\begin{figure}[H]
	\center\includegraphics[scale=0.4]{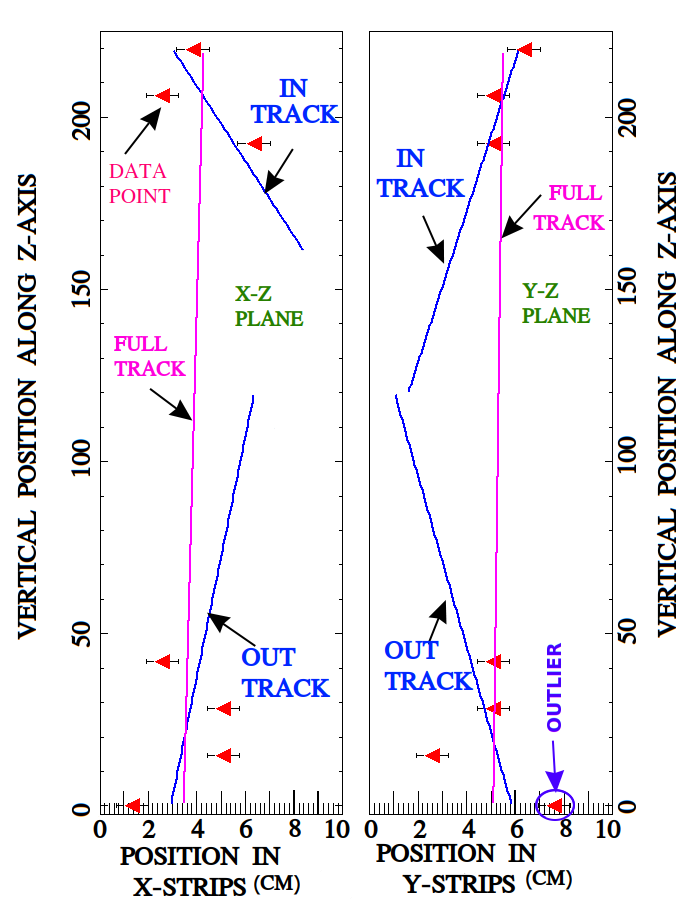}
	
	\caption{Example of cosmic muon track reconstructed in the VECC stack. \label{fig:muondatatrack}}	
\end{figure}

When a muon passes through an RPC, it is expected to hit only one strip on each side (X and Y) of the RPC. However, the induced charge can be shared between several strips, which is known as the cross-talk effect. Sometimes electronic noise can also mimic the cross-talk effect, generating hits on several strips. It has been seen in Section \ref{sec:crosseffi} that the probability of signal reaching the neighbouring cells is non zero. Therefore, we only consider a maximum of three strip hits, or events with multiplicity three, as a possible muon hit. Figure \ref{fig:muondatatrack} shows an example of the X-Z and Y-Z projection of a muon track through the RPC stack at VECC. In Figure \ref{fig:muondatatrack}, "IN-TRACK" and "OUT-TRACK" represent the linear fit of the data points from the up-layers and down-layers of the RPC stack shown in Figure \ref{fig:rpcstack}, respectively. "Full TRACK" represents the combined fit of data points from the up-layers and the down-layers. The mathematical representation of the X-Z and Y-Z projection of a linear muon track can be defined as follows:
\begin{align}
x(/y)=m\,z+c,
\end{align}
where $m$ is the slope, $c$ is the constant, $z$ is the vertical position  and $x/y$ is the horizontal positions on the RPCs. The error matrix for linear regression is as follows \cite{lyons_1986}:
\begin{align}\label{eqn:errlinear}
\frac{1}{D}\begin{pmatrix}
[z^2] & -[z]\\
-[z] & [1] \\
\end{pmatrix},
\end{align}
now with ``n" number of data points and uncertainty $\sigma_i$ in the horizontal co-ordinates $x$ and $y$, a general expression of terms inside the square brackets and D of matrix \ref{eqn:errlinear} is,
\begin{align}
[\Lambda]&=\sum_{i=1}^{n}\frac{\Lambda_i}{\sigma_i^2}  \\
      D&=[z^2][1]-[z][z]
\end{align}
If $x^\prime/y^\prime$ is the fitted point then the errors on the fit points ($\epsilon_{x/y}$) can be estimated as follows:
\begin{align}
\epsilon_{x/y}=\zeta_m^2+2\,z\,\text{Cov(m,c)}+z^2 \zeta_c^2,
\end{align}
where,
\begin{align}
	\zeta_m^2&=\frac{[z^2]}{D},\\\nonumber
	\zeta_c^2&=\frac{[1]}{D},\\\nonumber
	\text{Cov(m,c)}&=-\frac{[z]}{D}
\end{align}
A data point is considerd as outlier (see Figure \ref{fig:muondatatrack}) if the  deviation between  actual data point $x(/y)$ and fitted point $x^{\prime}(/y^\prime$) is larger than the one strip pitch. The position residual $\Delta R$ can be defined as:
\begin{align}
\Delta R=x(/y)-x^\prime(/y^\prime).
\end{align}
 Therefore, if $\Delta R>$one strip pitch that data point is rejected from itterative fitting procedure and a new fitting is done with the remaining data points. After 3-4 itterations all outliers are possible to be removed.

\subsection{Method of detector alignment using position residual correction}\label{sec:methodResidual}
Before proceeding with any physics analysis, it is crucial to ensure the proper alignment of the detectors. Even a slight shift in any of the detectors can significantly affect the fit parameters and lead to incorrect physics results. The initial alignment of the detector is done mechanically, but there may still be small misalignments that cannot be detected by our mechanical instruments. Therefore, further alignment is performed offline by analyzing the position residuals $\Delta R$ from the muon data. In this case, the offset in the mean of the residual distribution is calculated and corrected in
the next iteration. The steps of the iterative algorithm for position residual correction for all layers are given below:
\begin{enumerate}
	\item[] \textbf{step 1:} In each iteration, the layer ($L^{th}$) under observation is excluded from the track fitting analysis to avoid bias in calculation. In the first iteration, the uncertainty in the position is taken as $\frac{1}{\sqrt{12}}$ in strip units.
	\item[] \textbf{step 2:} The position residue for the $L^{th}$ layer is estimated using all data, and the distribution is fitted with a Gaussian. The mean and sigma of the distribution for the $n^{th}$ iteration and $L^{th}$ layer are recorded. The same calculation is done for all layers.
	\item[] \textbf{step 3:} In the next iteration, the offsets or means of the residual distributions are subtracted from all x or y positions of the corresponding layers. Starting from the second iteration, the sigma ($\sigma_{pos,x/y}$) from the Gaussian fit of the residual distribution is used as the position resolution in the fitting algorithm. However, the true resolution $\sigma_{true,x/y}$ is:
	\begin{align}
	\sigma_{true,x/y}=\sqrt{\sigma^2_{pos,x/y}-\epsilon^2_{x/y}},\,\,\text{where}\,\sigma^2_{pos,x/y}>\epsilon^2_{x/y}
	\end{align}
	
	 The $\sigma_{true,x/y}$ should be used as the new updated position uncertainty in the linear track fitting. It should be noted that the $\sigma_{true,x/y}$ for events with multiplicities of one, two, and three may be different, and therefore, we should give different weightage to each case. Therefore, three separate residual distributions are created corresponding to each case to calculate three different sigmas. However, the mean of the combined residual distribution is used for the offset corrections.
	 
	 Steps 1 to 3 are repeated approximately 5-6 times, and in the last few iterations, no changes will be found. The values of the mean and $\sigma_{true,x/y}$ from the last iteration are stored and will be used in further analysis.
	 
	 To calculate the final position residuals, all layers are considered in the fitting algorithm.
\end{enumerate}
\subsection{Event selection criteria}\label{sec:Eventselection}
The selection criteria of a data point during linear fitting is discussed in section \ref{sec:trackalgo}. Similarly, the selection criteria to select an event during the detector alignment proccedure is follows:
\begin{enumerate}
	\item An event will be selected if there is a valid hit and hit multiplicity is less than three in that layer which is under calibration and the fired strips should be consecutive.
	\item The interpolation/extrapolation error $\epsilon_{x/y}<$0.2.
	\item Consider that $x^\prime$ is the fitted position in the $n^{th}$ strip of the $L^{th}$ layer. The deviation $\delta$ of the fitted position from the middle of the strip is calculated. Now, the condition to select that fit point is $|\delta|+\epsilon_{x/y}<1.25$ times the strip pitch. An additional quarter of a strip is taken into consideration to account for higher multiplicity events.
	\item To select the best track, two additional criteria may be used: $\frac{\chi^2}{\text{Ndf}-2}<2$ and $\text{Ndf}-2>2$, where Ndf is the number of selected layers for the fit.
\end{enumerate}

\subsection{Results from IICHEP, Madurai stack data}
\begin{figure}[H]
	\centering\includegraphics[scale=0.25]{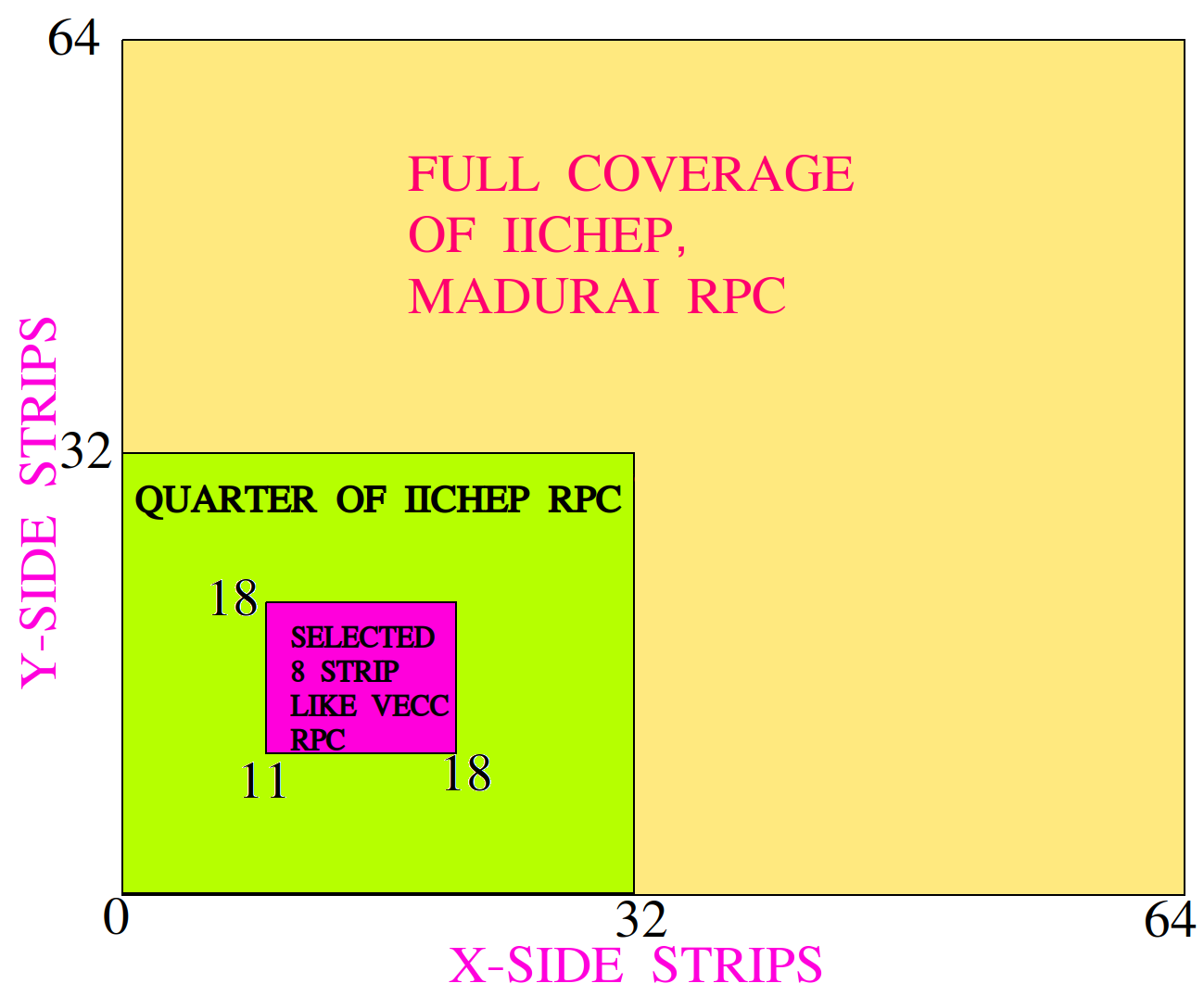}
	
	\caption{Selection of data range for each RPCs of IICHEP stack, where the first region consists of 32 strips and the second region consists of 8 strips from both the X and Y sides. \label{fig:rpcsize}}	
\end{figure}
To analyze the data from both the VECC and IICHEP stacks, we developed a C++ code. However, the data from the VECC stack remains further noise reduction before track fitting. Therefore, to test the code, we used data from the IICHEP stack. For this analysis, we focused on selected regions of the IICHEP stack and neglected data from the four middle layers (i.e., layers 5, 6, 7, and 8) as shown in red in Figure \ref{fig:inostack}. As a result, we now have two sections in this stack, similar to our VECC stack. The first section contains layers numbered 1, 2, 3, and 4, while the second section contains layers numbered 9, 10, 11, and 12.
\par In order to compare the data, we had to consider position information from two different sets of strips in the RPCs of the IICHEP stack. We chose a maximum of 32 strips from the available 64 strips in the first set, and a maximum of 8 strips from the same 64 strips in the second set, which closely resembled the configuration of the VECC stack, as depicted in Figure \ref{fig:rpcsize}. It's important to note that the strip width of the RPCs in the VECC stack was 2.3 cm, while that of the IICHEP stack was 2.8 cm. Additionally, there was a gap of approximately 14 cm and 16 cm between the layers of the VECC and IICHEP stacks, respectively. While there were minor differences in strip width and layer gap between the RPCs of the two stacks, we still expected qualitatively similar results from both. Nevertheless, at the moment, we only had results available for the IICHEP stack, which we would discuss in the following sections.

In view of this, we present the results from the IICHEP stack, which when optimized can be applied on the cleaned up VECC stack data.

\subsubsection{Results for selected 8 layers and 32 strips of each RPCs from IICHEP stack}\label{sec:32by8rpc}
 The data were collected from eight layers of RPCs, each containing 32 strips. It is worth noting that the first four layers (L1-L4) are considered trigger layers. In this analysis, a maximum of three multiplicity events are considered, as four or greater multiplicity events can be generated from the streamer pulse or from the electronics noise.
 
\begin{figure}[H]
 	\center\subfloat[\label{fig:itter_mean}]{\includegraphics[scale=0.35]{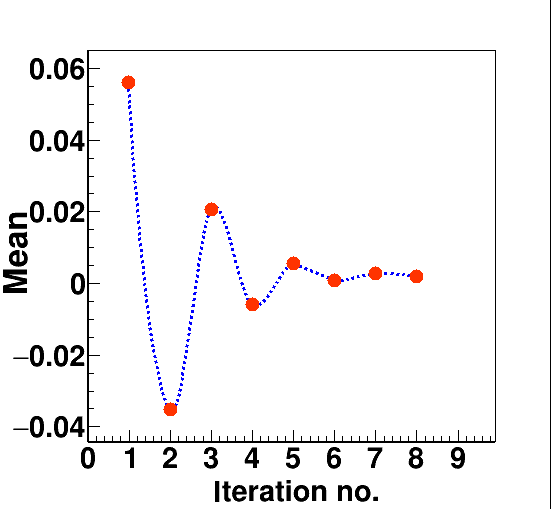}}\subfloat[\label{fig:itter_rms}]{\includegraphics[scale=0.35]{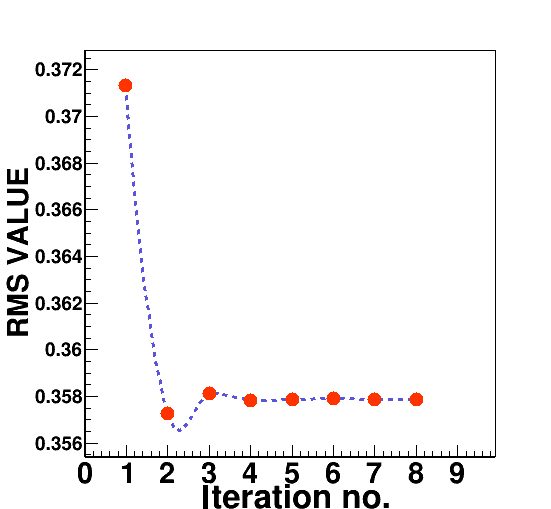}}
 	
 	\caption{(a) Oscillation of position residual mean about zero during iteration (b) Changing of RMS value of position residual during iteration}	
 \end{figure}

\begin{figure}[H]
	\centering\includegraphics[scale=0.32]{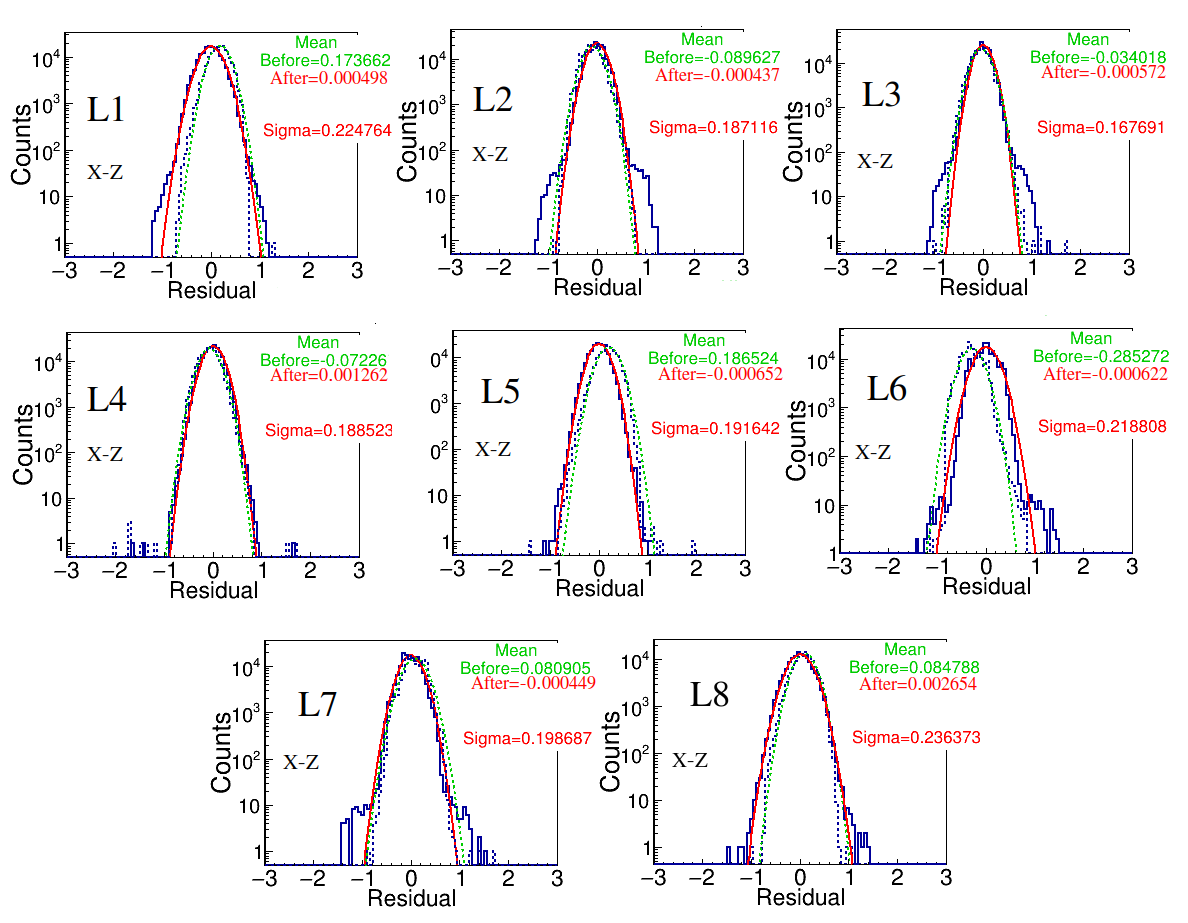}
	
	\caption{The position residual distribution of 8 layers of RPCs with 32 strips each in the X-Z plane is shown, where the green Gaussian distribution represents the data before any offset correction and the red Gaussian distribution represents the data after offset correction.\label{fig:resiINO}}	
\end{figure}

\begin{figure}[H]
	\centering\includegraphics[scale=0.4]{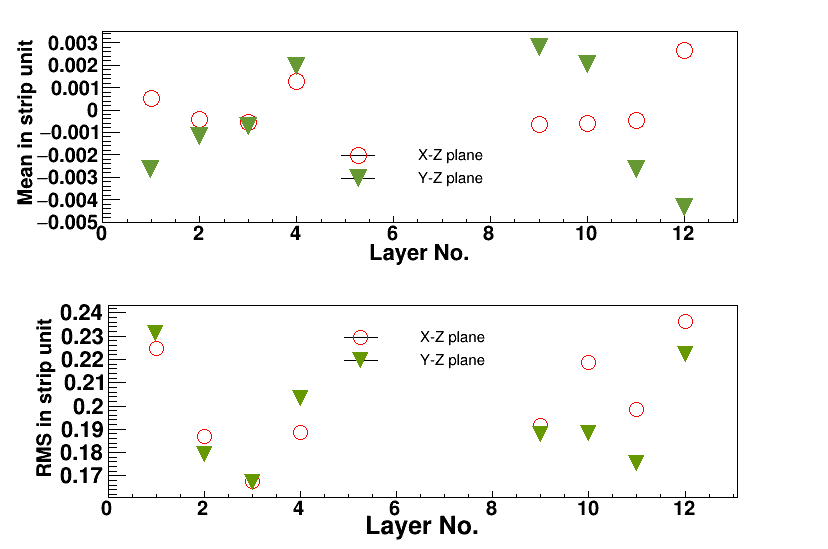}
	
	\caption{Mean and RMS values of 8 layers of RPCs with 32 strips each after offset corrections in X-Z and Y-Z planes. \label{fig:allmeanrms}}	
\end{figure}

\begin{figure}[H]
	\centering\includegraphics[scale=0.37]{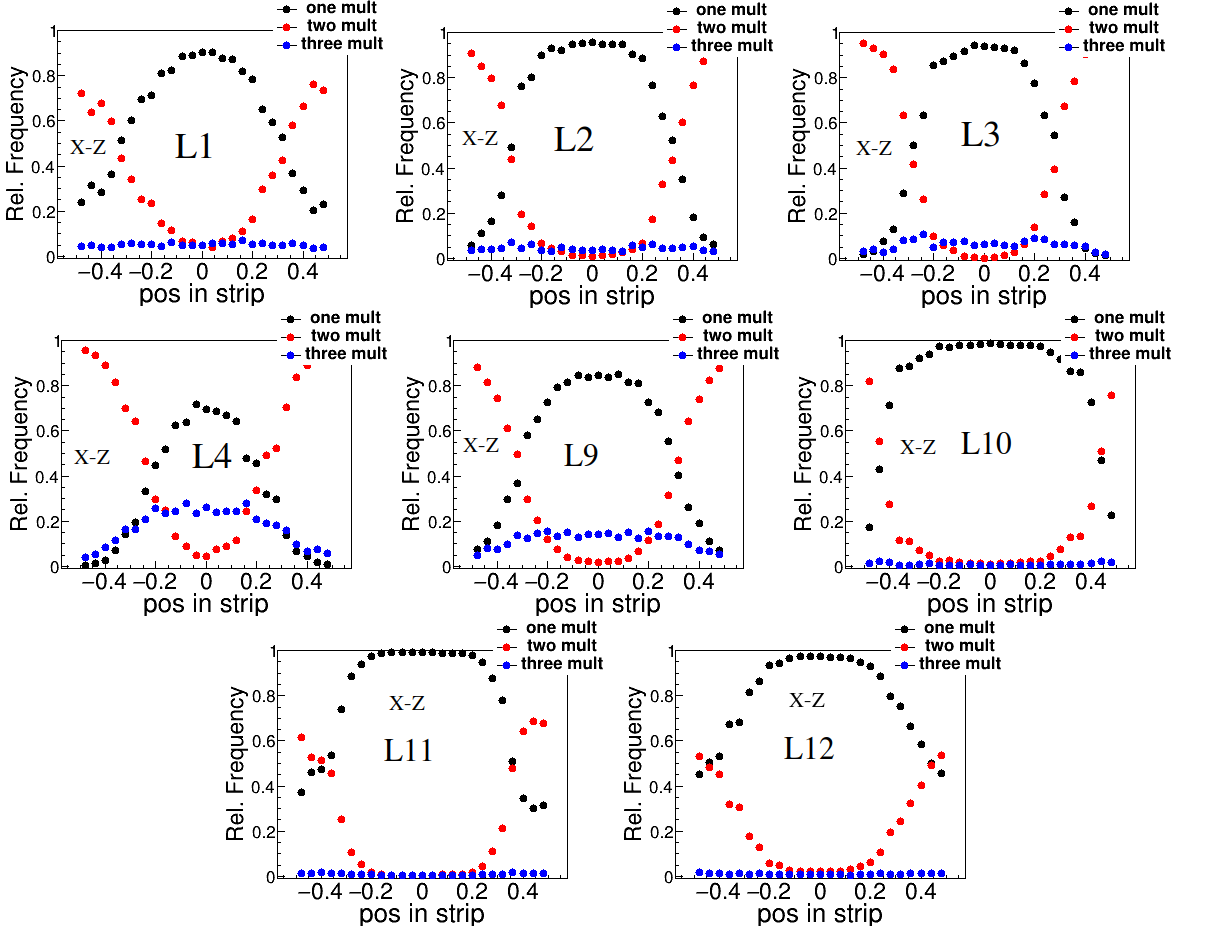}
	
	\caption{Relative Strip multiplicity as a function of muon position in a strip with respect to center of the strip for the case of 8 layers of RPCs with 32 strips each and in X-Z plane.\label{fig:multINO}}
\end{figure}

\begin{figure}[H]
		\centering\includegraphics[scale=0.37]{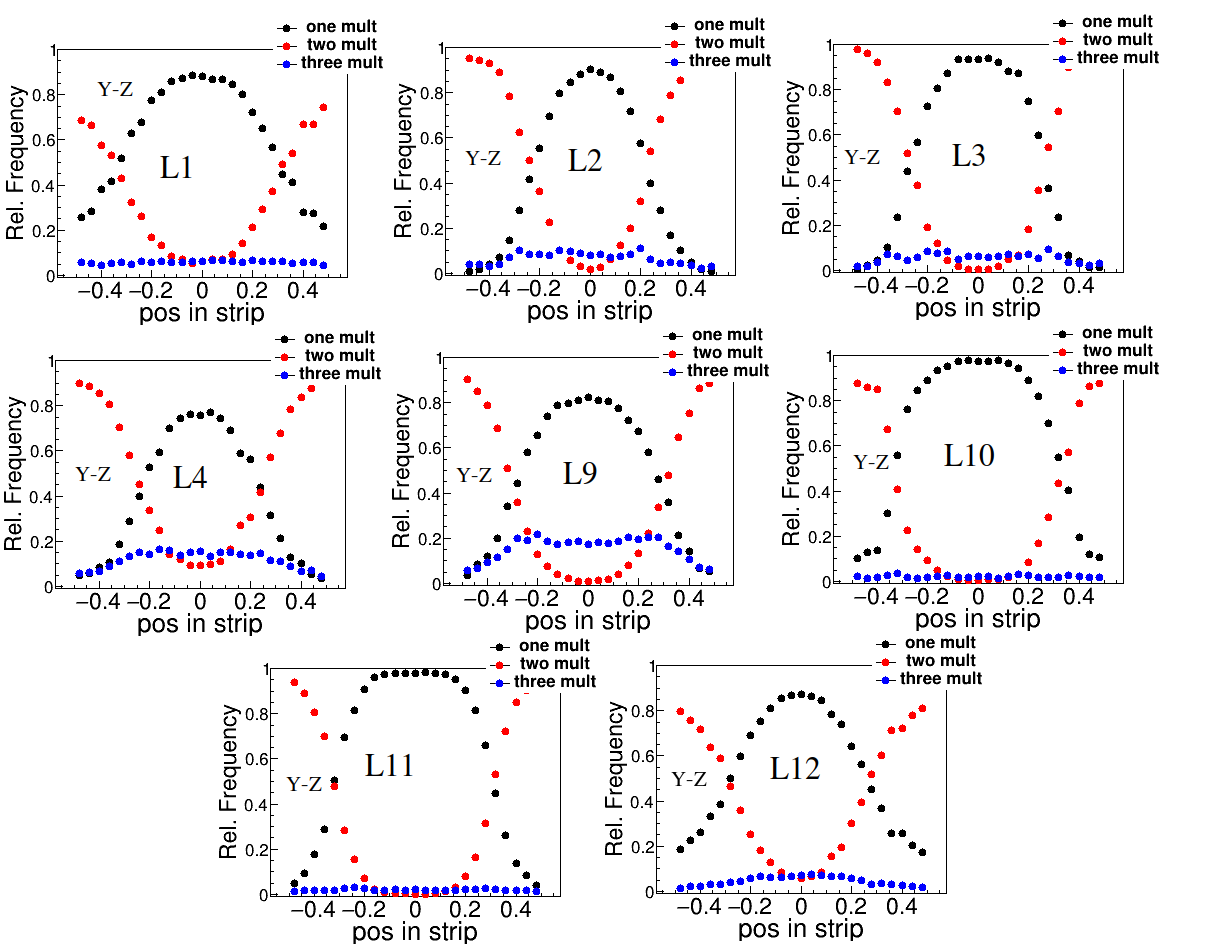}
		
		\caption{Relative Strip multiplicity as a function of muon position in a strip with respect to center of the strip for the case of 8 layers of RPCs with 32 strips each and in Y-Z plane.\label{fig:multINOy}}	
\end{figure}

 Figure \ref{fig:itter_mean} and \ref{fig:itter_rms} illustrate the variation of the mean and RMS with iteration for layer 12, respectively. During the iteration process, the mean of the residual distribution oscillates around zero, with a decreasing amplitude, and finally saturates after the 6th iteration. The RMS of the outer layer is larger than that of the inner layer because the respective extrapolation error is large for the outer layer. Figure \ref{fig:resiINO} shows the distribution of position residuals in the X-Z plane for all multiplicity events. Two distributions fitted with a Gaussian function are shown in each figure. The green Gaussian fit represents the distribution before the offset corrections, and the red Gaussian fit represents the distribution after the offset corrections. In contrast to the iteration process, all layers are considered in the residual distribution. The x and y positions are corrected with corresponding offsets calculated from the iteration process and shown in the corrected residual distributions (red Gaussian in Figure \ref{fig:resiINO}). As seen from Figure \ref{fig:resiINO}, for each layer, the mean of the residual distribution shifts closer to zero after correction. The final shifted mean and RMS values for all layers and both X-Z and Y-Z planes are summarized in Figure \ref{fig:allmeanrms}. Figures \ref{fig:multINO} and \ref{fig:multINOy} show the relative frequency of one, two, and three multiplicity events as a function of the muon's deflection from the center of the strip (i.e., $\delta$) for all layers in the X-Z and Y-Z planes, respectively. The transverse extent of the avalanche for small pulses is typically limited to an area of only a few square millimeters, as confirmed by the simulation results presented in Figure \ref{fig:aval} in Chapter 2. Therefore, the width of a strip is much greater than the spatial distribution of the avalanche. As a result,
  \noindent when a muon passes through the center of a strip, the chances of the strip firing more than once are low, resulting in more single multiplicities, as shown in Figures \ref{fig:multINO} and \ref{fig:multINOy}. The figures also indicate that as the muon passes through a point deflected from the center, the probability of double multiplicities increases. With relatively large pulses, triple or greater multiplicities are usually observed. In such cases, the probability of a muon passing through a strip is nearly uniform.
	
\subsubsection{Results for selected 8 layers and 8 strips of each RPCs from IICHEP stack}
As a next step towards analysing VECC stack like data for the available IICHEP data. We selected eight strips and eight layers of RPCs to simulate the VECC stack.
\begin{figure}[H]
	\centering\includegraphics[scale=0.35]{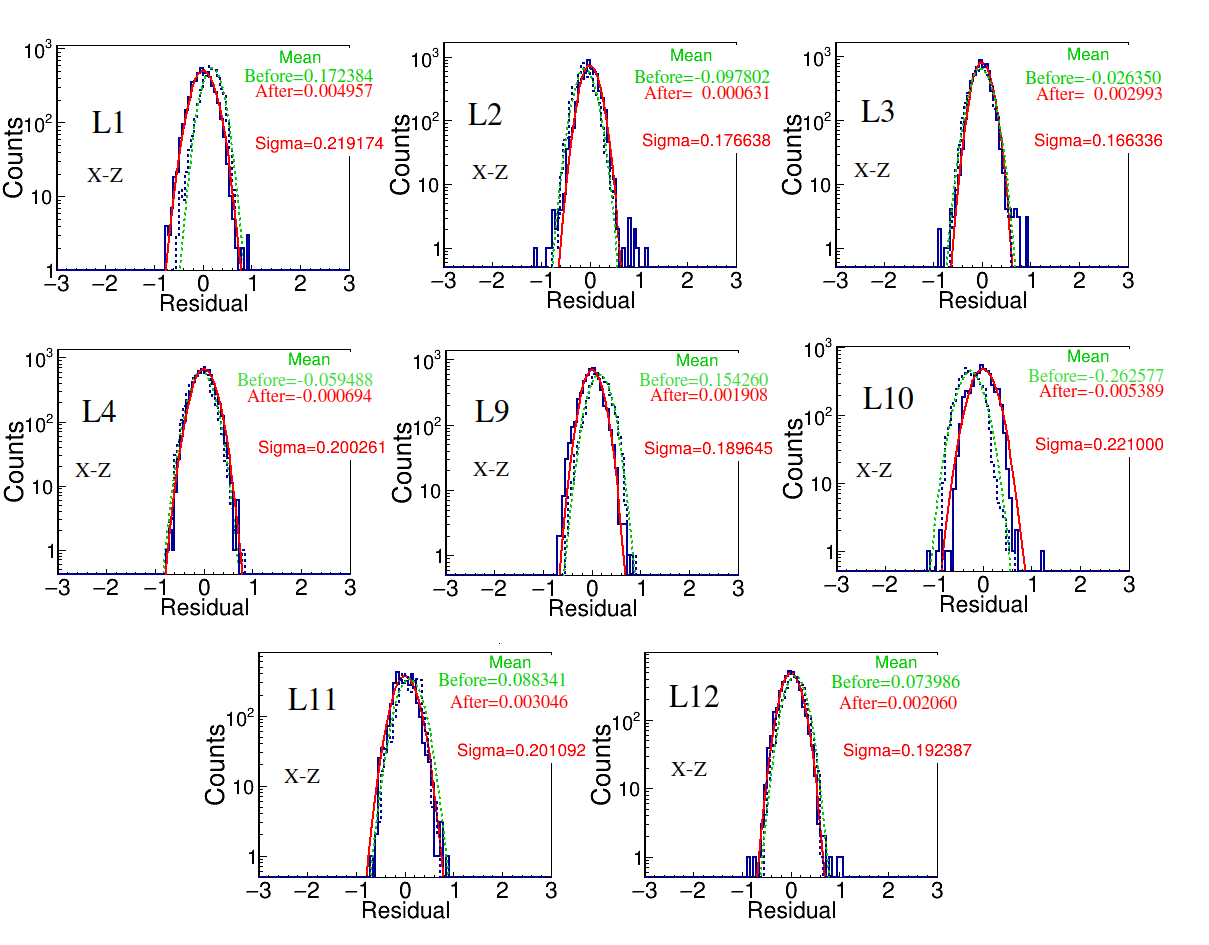}
	
	\caption{The position residual distribution of 8 layers of RPCs with 8 strips each in the X-Z plane is shown, where the green Gaussian distribution represents the data before any offset correction and the red Gaussian distribution represents the data after offset correction.\label{fig:resi-ino-vecc}}	
\end{figure}
\begin{figure}[H]
	\centering\includegraphics[scale=0.4]{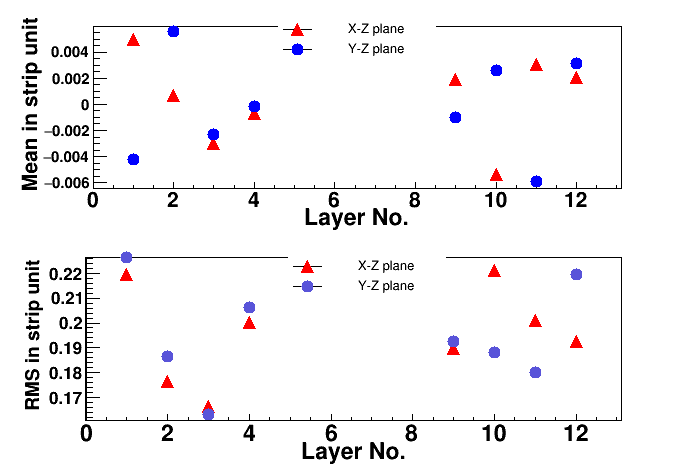}
	
	\caption{Mean and RMS values of 8 layers of RPCs with 8 strips each after offset corrections in X-Z and Y-Z planes.\label{fig:meanerr-11-18}}	
\end{figure}
\noindent  IICHEP stack. In Figure \ref{fig:resi-ino-vecc}, the position residual distribution before and after correction in the X-Z plane is shown. In the same figure, the green Gaussian fit represents the data before residual correction, while the red Gaussian fit represents the data after residual correction. The mean and RMS values after residual correction are close to the values discussed in section \ref{sec:32by8rpc}. Figure \ref{fig:meanerr-11-18} displays the mean and RMS values for all eight layers in the X-Z and Y-Z planes. It is evident that all mean values are very close to zero, and the RMS of outer layers is larger than that of inner layers, as expected due to the large extrapolation error.

\begin{figure}[H]
	\centering\includegraphics[scale=0.37]{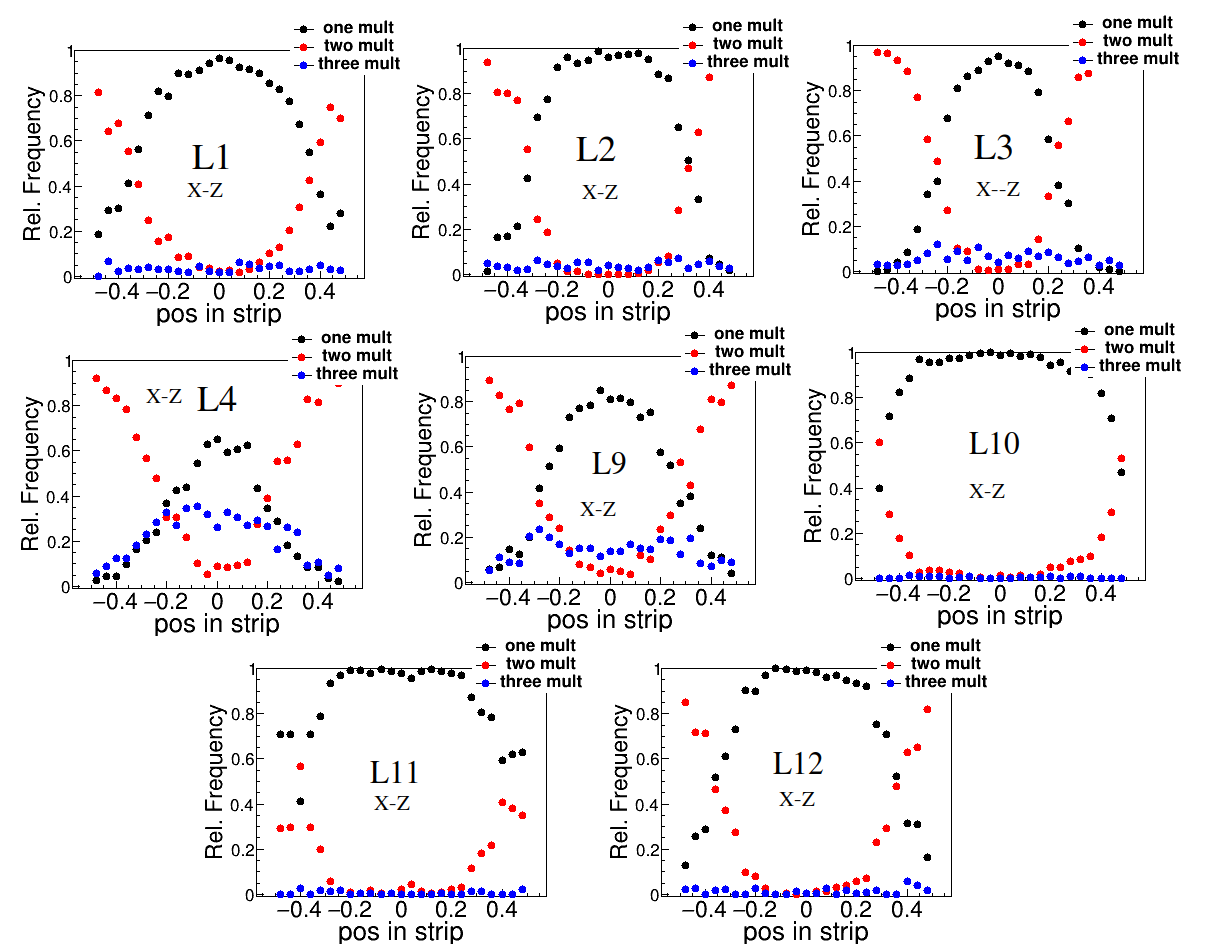}
	
	\caption{Relative Strip multiplicity as a function of muon position in a strip with respect to center of the strip for the case of 8 layers of RPCs with 8 strips each and in X-Z plane.\label{fig:multvecc}}	
\end{figure}
\par
\begin{figure}[H]
	\centering\includegraphics[scale=0.37]{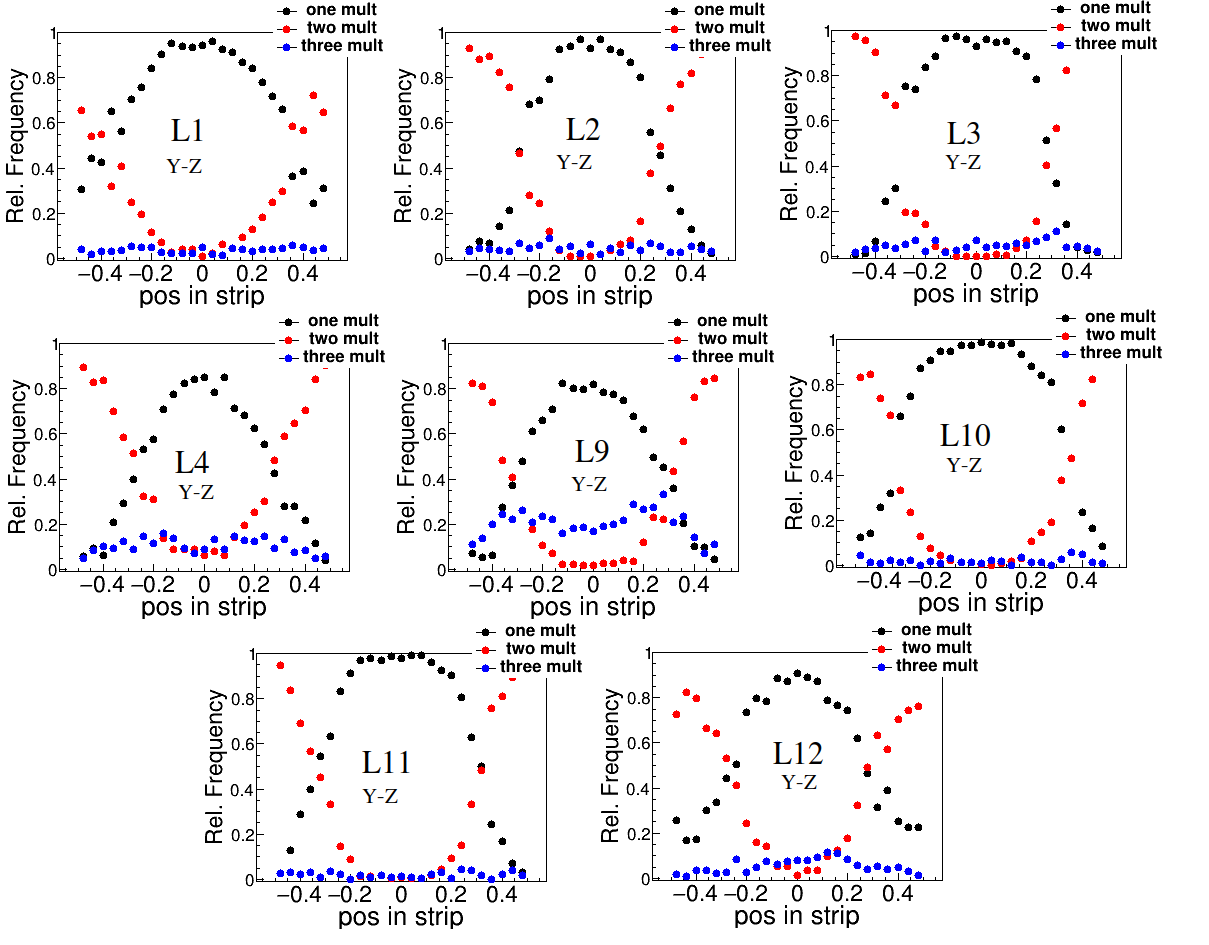}
	
	\caption{Relative Strip multiplicity as a function of muon position in a strip with respect to center of the strip for the case of 8 layers of RPCs with 8 strips each and in Y-Z plane.\label{fig:multveccY}}	
\end{figure}

 In this study, the method described in section \ref{sec:32by8rpc} is employed to determine the relative frequencies of events with one, two, and three multiplicities as a function of position in a strip relative to its center. The obtained results are presented in Figures \ref{fig:multvecc} and \ref{fig:multveccY} for all layers and in both the X-Z and Y-Z planes, respectively. The results are consistent with the observations made in section \ref{sec:32by8rpc} (Figures \ref{fig:multINO} and \ref{fig:multINOy}), where the probability of single multiplicity events is higher at the center and decreases towards the edges. Conversely, the relative frequency of double multiplicity
\noindent  events is higher at the edges and lower at the center, and that of triple multiplicity events is almost uniform. However, in the case of 8 strips and 8 layers, the number of events is fewer than in the 32 strips and 8 layers case. Hence, the points in Figures \ref{fig:multvecc} and \ref{fig:multveccY} are not placed smoothly, unlike in Figures \ref{fig:multINO} and \ref{fig:multINOy}. 

\section{Scattering Angle Calculations}
\begin{figure}[H]
	\centering\includegraphics[scale=0.35]{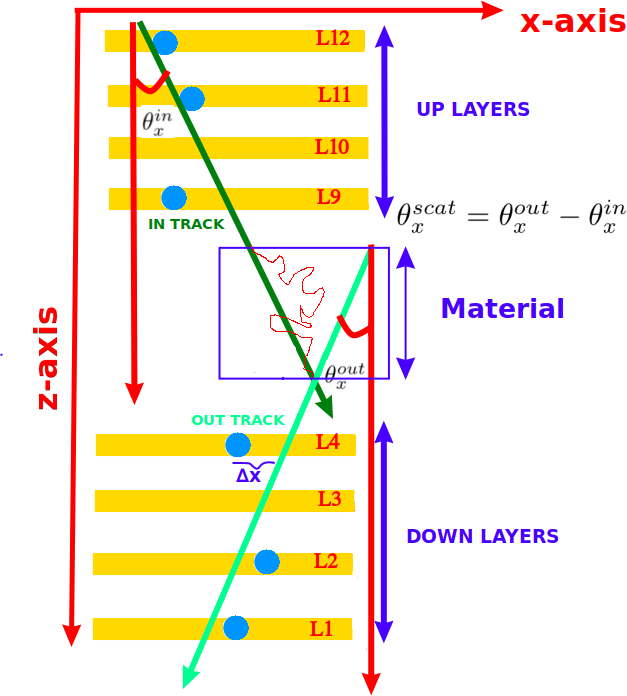}
	
	\caption{Illustration of the calculation of the scattering angle of muons in the presence of different materials between RPC layers. \label{fig:anglecal}}	
\end{figure}
After performing the offline alignment, we obtain the layer offsets and position resolutions or RMS for all layers. Then, two tracklets are formed using data points from the up-layers and down-layers, as shown in Figure \ref{fig:anglecal}. The layer offsets are added to the data points, and the RMS values are used in straight line fitting to calculate the angle between the tracklets and the vertical z-axis. Therefore, if $\vec{P}_{in}$ and $\vec{P}_{out}$ are the incoming and outgoing tracklets corresponding to the up-layers and down-layers respectively, then the vertical scattering angle in 3D space can be expressed as follows:
\begin{align}
\theta^{scat}=\cos^{-1}\left(\frac{\vec{P}_{out}.\hat{z}}{\mid \vec{P}_{out}\mid}\)-\cos^{-1}\(\frac{\vec{P}_{in}.\hat{z}}{\mid \vec{P}_{in}\mid}\right)
\end{align}
\begin{figure}[H]
	\center\subfloat[\label{fig:scat-0-32}]{\includegraphics[scale=0.32]{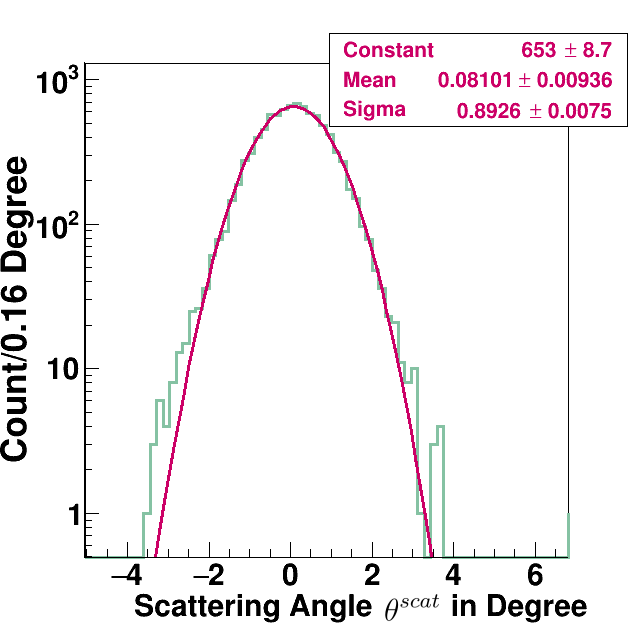}}\subfloat[\label{fig:scat-11-18}]{\includegraphics[scale=0.32]{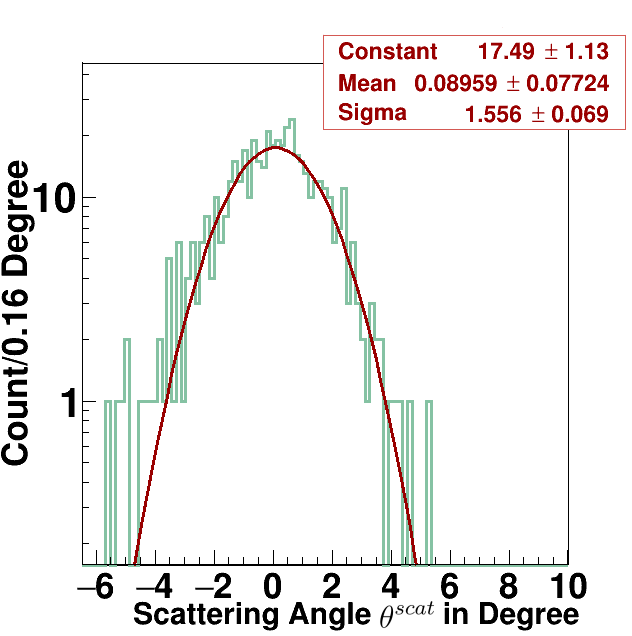}}
	
	\caption{Scattering angle distribution in the presence of air is shown for (a) the case of 8 layers of RPCs with 32 strips each, and (b) the case of 8 layers of RPCs with 8 strips each.}	
\end{figure}
\par The aim of this study is to calculate the scattering angle in the presence of a high-density material, as shown in Figure \ref{fig:anglecal}. However, data with material for IICHEP stack is currently not available. Therefore, we begin by calculating the scattering angle without material, or in air. Ideally, in the air, the scattering angle should be zero. However, due to the finite detector resolution, the scattering angle may not be exactly zero, resulting in a Gaussian distribution of angles. In such cases, the mean of the distribution can be considered as offsets. Figures \ref{fig:scat-0-32} and \ref{fig:scat-11-18} depict the scattering angle distribution without material for the IICHEP stack for a total of 32 and 8 strips, respectively. The results indicate that, in both cases, the mean of the distribution is close to zero. However, the sigma of the distribution is slightly higher for the 8 strips and 8 layers case (Figure \ref{fig:scat-11-18}).
\section{Chapter Summary}
In this chapter, we discuss the installation and data-taking process of an 8-layer RPC stack at VECC, as well as the development of a track reconstruction and scattering angle algorithm for data-taking by the 12-layer IICHEP stack. The chapter covers the necessary tests that have been performed on the eight RPC detectors before they are stacked together. To conduct these tests, we set up a test bench, and the first test performed was to check the resistivity of the graphite coating. Next, a range of voltages was applied to the RPCs and measurements were taken to determine the noise, efficiency, and crosstalk efficiencies. After that, the uniformity of hits in both the X and Y directions were checked for all RPCs. Finally, the RPCs were manually aligned into the stack.

\par Furthermore, a discussion took place regarding the utilization of the cosmic muon track reconstruction technique for RPC stacks at both VECC and IICHEP. However, as the VECC data requires further noise processing, we developed necessary algorithms to analyse IICHEP stack data. Thus, the focus of the discussion in this thesis is on the offline alignment procedure for the IICHEP stack. In addition, the discussion also touched upon the methodology employed in calculating the scattering angle of cosmic muons when high-density materials are not present.  
\chapter{Summary and Outlook}\label{ch:summary}
\section{Summary of the thesis}
Neutrinos are tiny particles that hold great interest for scientists. Despite being incredibly abundant throughout the universe, they are notoriously difficult to detect because of their weak interactions with other matter. However, a number of experiments, such as Super-Kamiokande, Daya Bay, and Ice Cube, are working to uncover the mysteries of neutrinos by studying their properties, such as mass and oscillations.
\par The study of neutrino oscillations in the atmospheric sector is a primary focus of the neutrino physics community in India. To facilitate this research, a new facility called INO is being developed in the southern region of the country. INO will house several experiments, including ICAL, NDBD, and DINO, which will examine different aspects of neutrinos. Among these three experiments, ICAL will search for neutrino properties, NDBD will search for neutrinoless double beta decay, and DINO will search for dark matter.
\par One instrument of particular interest is ICAL, which will be a large detector composed of active detectors constructed from RPCs sandwiched between magnetized iron plates used as passive converters. ICAL's primary objective is to determine the genuine mass hierarchy of neutrinos by analyzing how they interact with the Earth's matter. Additionally, ICAL will contribute to the precise measurement of $\theta_{23}$ and $|\Delta m^2_{32}|$, while also potentially leading to the discovery of new physics regarding neutrinos. More details on ICAL and neutrino physics can be found in Chapter \ref{ch1.Introduction}.
\par The Resistive Plate Chamber (RPC) is the active detector element of the ICAL detector. Hence, for it's development of it a detailed simulation techniques is needed. The working of an RPC depends on the development of signals from an Avalanche or streamer process. Therefore, a detailed simulation model of avalanche or streamer development is needed. However, streamer simulation or avalanche to streamer transition is beyond the scope of the thesis. A short review on Gaseaous detctors and the working principle of RPC can be found in Chapter \ref{ch2}.
\par The electron-ion generation during an avalanche inside an RPC is strongly dependent on the  electric-field inside the RPC. The total electric-field inside the RPC is continuously changing during an avalanche due to dynamic electric-field of electrons and ions which is called the space charge effect. A straight line model has been developed to calculate the space charge field. The benefits of this model is it does not include any numerical integrations which results a fast estimation of the electric field. Furthermore, with a proper choice of length of the line element the precision in electric field calculation can be improved. The details of this model is discussed in Chapter \ref{space_charge_field_calculation}.
\par For more precise calculation the electric  field due to the polarisation in the dielectric electrodes have been included. This polarisation field is calculated by using the method of image  which can be found in Chapter \ref{ch4}, where we have developed a technique to find image charge field for five layer dielectric case. Additionally, we have discussed how the electric field inside a Resistive Plate Chamber (RPC) depends on various electrode parameters such as permittivity, thickness of electrodes, and gas gap. We have observed that, when the applied voltage, electrode thickness, and gas gap are fixed, the electric field increases rapidly with increasing relative permittivity. However, the electric field starts to saturate after a certain value of relative permittivity (around 20). We have also observed that for smaller electrode thickness, the variation of electric field with permittivity decreases. On the other hand, when the permittivity and the gas gap are fixed, increasing the electrode thickness leads to a reduction in the electric field. The electric field inside the middle of the gas gap decreases with an increase in the gas gap.

\par A new class pAvalancheMC is inroduced in the Garfield++ software to add the above straight line and image charge model to consider the space charge effect during avalanche generation.  To reduce the time to generate electron-ion pairs along with space charge field calculation during the development avalanche parallel computing is used using OpenMP in the class pAvalancheMC. As a result a speed up of 5.46 and 7.2 have been observed
without and with space charge effect, respectively. More details can be found in Chapter \ref{chp:chap_fast_aval}    

\par As an application of RPC detectors, we have developed a stack of eight RPCs at VECC that can be used to determine the scattering angle of cosmic muons in the presence of high-density materials. We also developed a testbench to perform necessary tests, such as efficiency,  crosstalk, and hit uniformity, on the RPCs before putting them in the stack. We developed a reconstruction algorithm and tested it using data obtained from the IICHEP stack in Madurai, which had twelve RPCs. We plan to use this algorithm for the VECC stack. More details can be found in Chapter \ref{ch:tomography}.
\section{Future Outlook}
\begin{enumerate}
	\item \textbf{Simulation activities}:
	\begin{enumerate}
	\item It is discused that the Monte Carlo method of Garfield++ to simulate an avalanche in an RPC is time consuming, which is a major issue. We have solved to a certain extent with the help of parallel computations using OpenMP. However, using more efficient resources of computing like GPU, MPI etc. it can be optimised further. 
	\item In this thesis Yule-Furry Monte Carlo model is used to generate avalanche. In future another algorithm like Riegler Model can also be utilized to compare the performance.
	\item Currently with space-charge-effect we only can simulate avalanche to saturated avalanche transition. In future with the implementation of photon transport model the avalanche to streamer transition can also be possible to simulate in Garfield++.
	\item With the above improvements in the future, it will also be possible to simulate the precise rate handling capacity of an RPC.
	\item Here, the line model is used to calculate the space-charge effect inside an RPC. However, it is a general method; hence can be used in another gaseous detectors. 
	\end{enumerate}
\item \textbf{Experimental activities:}
\begin{enumerate}
	\item It is discussed that the VECC data require noise clean up; hence we have not use it for track reconstruction. Therefore, next our task is to find the source of noise and store a clean data as much as possible with VECC stack.
	\item A more better track reconstruction algorithm can be found for better tracking efficiency.
	\item Precise calculation of scattering angle.
	\item Distinguise materials with the help of spread in scattering angle. 
	\item Implementation of Point of closest approach (Poca) algorithm.
	\item Try to determine shape of the observed materials.
\end{enumerate}	

\end{enumerate}

\begin{appendices}
\end{appendices}

\backmatter

\backmatter
\bibliographystyle{unsrt}
\bibliography{Image_issues_inRPC.bib}



\end{spacing}

\end{document}